\documentclass{article}

\usepackage{amsmath}
\usepackage{amssymb}
\usepackage{graphicx}
\usepackage{authblk}
\usepackage{hyperref}
\usepackage{setspace}
\usepackage{tikz}
\usepackage{tikz-cd}
\usepackage{fullpage}
\usepackage{proof}
\usepackage{physics}
\usepackage{xcolor}

\newtheorem{definition}{Definition}

\title{ZX-Calculus and Extended Wolfram Model Systems II: Fast Diagrammatic Reasoning with an Application to Quantum Circuit Simplification}
\author[1]{Jonathan Gorard\footnote{\url{jg865@cam.ac.uk}}}
\author[2]{Manojna Namuduri\footnote{\url{manon@wolfram.com}}}
\author[3]{Xerxes D. Arsiwalla\footnote{\url{x.d.arsiwalla@gmail.com}}}
\affil[1]{\small University of Cambridge, Cambridge, UK}
\affil[2]{Wolfram Research, USA}
\affil[3]{Pompeu Fabra University, Barcelona, Spain}

\begin{document}
\maketitle

\begin{abstract}
This article presents a novel algorithmic methodology for performing automated diagrammatic deductions over combinatorial structures, using a combination of modified equational theorem-proving techniques and the extended Wolfram model hypergraph rewriting formalism developed by the authors in previous work. We focus especially upon the application of this new algorithm to the problem of automated circuit simplification in quantum information theory, using Wolfram model multiway operator systems combined with the ZX-calculus formalism of Coecke and Duncan for enacting fast diagrammatic reasoning over linear transformations between qubits. We show how the techniques of Bachmair and Ganzinger can be used to construct a generalization of the deductive inference rules for Knuth-Bendix completion in which equation matches are selected on the basis of causal edge density in the associated multiway system, before proceeding to demonstrate how the methods of Mycielski and Kerber can be used to embed the higher-order logic of the ZX-calculus rules within this first-order equational framework. After showing explicitly how the (hyper)graph rewritings of both Wolfram model systems and the ZX-calculus can be effectively realized within this formalism, we proceed to exhibit comparisons of time complexity vs. proof complexity for this new algorithmic approach when simplifying randomly-generated Clifford circuits down to pseudo-normal form, as well as when reducing the number of T-gates in randomly-generated non-Clifford circuits, with circuit sizes ranging up to 3000 gates, illustrating that the method performs favorably in comparison with existing diagrammatic theorem-proving and circuit simplification frameworks such as \textit{Quantomatic} and \textit{PyZX}, and also exhibiting the approximately quadratic speedup obtained by employing the causal edge density optimization. Finally, we present a worked example of an automated proof of correctness for a simple quantum teleportation protocol, in order to demonstrate more clearly the internal operations of the theorem-proving procedure.
\end{abstract}

\tableofcontents

\clearpage

\section{Introduction}

In previous work\cite{gorard}, the authors detailed the formal relationship between the \textit{ZX-calculus} - a diagrammatic reasoning language for linear maps between qubits developed by Coecke and Duncan in 2008\cite{coecke}\cite{coecke2} in the context of Abramsky and Coecke's \textit{categorical quantum mechanics} program\cite{abramsky}\cite{abramsky2} - and \textit{the Wolfram model} - a combinatorial model for the fundamental structure of spacetime\cite{wolfram}\cite{wolfram2} whose dynamics is based upon diagrammatic rewriting systems over hypergraphs\cite{gorard2}\cite{gorard3}. In particular, we employed the formalism of \textit{multiway operator systems} to show that the (extended) Wolfram model can be formalized as a double-pushout rewriting system over a (selective) adhesive category, such that both the category of \textit{multiway evolution graphs} \textbf{MuGraph} and the category of \textit{branchial graphs} \textbf{BrGraph} inherit the structure of a dagger symmetric monoidal category (with the dagger structure being given by inversion of multiway evolution edges, and the monoidal structure being given by the disjoint union of multiway evolution rules). Due to the known completeness and soundness of the ZX-calculus for stabilizer quantum mechanics\cite{backens}\cite{backens2}, Clifford+T quantum mechanics\cite{jeandel}\cite{hadzihasanovic}, etc.\cite{jeandel2}, this result therefore allows us to translate, in a reasonably efficient way, problems in quantum information theory regarding equivalences between quantum circuits into problems in (extended) Wolfram model evolution regarding equivalences between (generalized) hypergraph states.

The objective of the present article is to make use of this effective translation to develop a novel algorithm for performing fast diagrammatic reasoning over general multiway operator systems (which include both the ZX-calculus and the Wolfram model as special cases), with a particular focus on the application of this new technique to the problem of automated simplification of quantum circuits. To this end, in Section \ref{sec:Section1}, we begin by extending the refutation-complete proof calculus for first-order logic with equality predicates developed by Bachmair and Ganzinger\cite{bachmair}, introducing a new selection function for the resolution and superposition inference rules based upon causal edge density in multiway operator systems. In Section \ref{sec:Section2}, we proceed to show how arbitrary theorems in full first-order (multisorted) logic can be proved using this same general approach, building upon the work of Mycielski\cite{mycielski} and Burris\cite{burris} on the reduction of predicate formulas to their Skolem normal forms; moreover, we show how arbitrary theorems in higher-order predicate logics can be converted to equivalent theorems in first-order multisorted logic, using the (provably complete and sound) collection of higher-order logic morphisms introduced by Kerber\cite{kerber}. This is particularly significant for our purposes, since many of the ``rules'' of the ZX-calculus (such as the Z- and X-spider fusion rules) are in fact infinite rule schemas - one rule for each possible choice of spider arities - which may in turn be treated as formulas in a second-order logic (in contrast with the \textit{bang-box} notation\cite{quick} approach adopted within software frameworks such as \textit{Quantomatic}\cite{kissinger}).

We continue in Section \ref{sec:Section3} with an illustration of how the extended Wolfram model can be formulated as a double-pushout (DPO) rewriting system over (partial) adhesive categories, and we build upon previous work of the authors in showing that both the categories ${\textbf{MuGraph}}$ (the category of multiway evolution graphs) and ${\textbf{BrGraph}}$ (the category of branchial graphs) are dagger symmetric monoidal categories\cite{gorard}. In particular, we combine these results with the \textit{causal category} formalism of Coecke and Lal\cite{coecke3}, which consequently allows us to formulate extended Wolfram model causal graphs as symmetric strict partial monoidal categories, thus yielding a purely category-theoretic description of the entire extended Wolfram model rewriting system (including its causal structure). In Section \ref{sec:Section4}, we show how this formulation therefore enables us to perform efficient diagrammatic rewriting and theorem-proving over extended Wolfram model systems, including the ZX-calculus as a special case. Lastly, in Section \ref{sec:Section5}, we show how this new fast diagrammatic theorem-proving algorithm (when combined with the causal optimization procedure previously outlined) can be applied to the general problem of circuit simplification in quantum information theory, applying the algorithm first to the problem of reducing randomly-generated Clifford circuits (stabilizer circuits) to a pseudo-normal form comprising graph states with local Clifford gates, before proceeding to apply the algorithm to the more computationally demanding task of reducing the total number of T-gates present within randomly-generated non-Clifford circuits, with random circuit sizes up to 3000 gates in each case. In both cases, we find an approximately quadratic speedup in the time complexity of the theorem-prover, coupled with an approximately quadratic reduction in the generated proof complexity, whenever the causal optimization procedure is applied, yielding efficiencies that compare favorably with existing software frameworks such as \textit{Quantomatic}\cite{kissinger} and \textit{PyZX}\cite{kissinger2}. We also present a fully worked example of an automatically-generated proof of correctness for a simple quantum teleportation protocol in the ZX-calculus, illustrating explicitly how the individual ZX-calculus rules (and higher-order rule schemas) are used by the theorem-prover in the construction of the resulting proof.

It should be noted that all of the code necessary to reproduce all of the computations presented within this paper is available for free on the \textit{Wolfram Function Repository}. In particular, the function \textit{MakeZXDiagram} (\url{https://resources.wolframcloud.com/FunctionRepository/resources/MakeZXDiagram}) can be used to construct the ZX-diagrams, the function \textit{MultiwayOperatorSystem} (\url{https://resources.wolframcloud.com/FunctionRepository/resources/MultiwayOperatorSystem}) can be used to evolve the resulting multiway operator system, the function \textit{FindWolframModelProof} (\url{https://resources.wolframcloud.com/FunctionRepository/resources/FindWolframModelProof}) can be used to reproduce the automatically-generated proofs of equivalence between Wolfram model hypergraphs, etc. Moreover, the ZX-calculus and automated theorem-proving frameworks are fully compatible with the prototype version of the Wolfram Language's own open source quantum computing/discrete-state quantum mechanics framework, currently comprising functions such as \textit{QuantumBasis} (\url{https://resources.wolframcloud.com/FunctionRepository/resources/QuantumBasis/}), \textit{QuantumDiscreteState} (\url{https://resources.wolframcloud.com/FunctionRepository/resources/QuantumDiscreteState/}), \textit{QuantumDiscreteOperator} (\url{https://resources.wolframcloud.com/FunctionRepository/resources/QuantumDiscreteOperator/}), etc.

\section{A Generalized Knuth-Bendix Completion Procedure}
\label{sec:Section1}

Following Bachmair and Ganzinger\cite{bachmair}\cite{bachmair2}, we begin by presenting a refutation-complete proof calculus for first-order logic equipped with an equality predicate; this proof calculus effectively generalizes the standard completion procedure of Knuth and Bendix\cite{knuth} used previously in the context of Wolfram model multiway operator systems, as well as its subsequent extension to an unfailing and refutation-complete completion procedure by Lankford\cite{lankford}, Hsiang and Rusinowitch\cite{hsiang}\cite{hsiang2}, and Bachmair, Dershowitz and Plaisted\cite{bachmair3}. This generalization hinges upon the crucial observation that all forms of completion are based around the use of \textit{superposition} as their primary deductive inference rule, which may in turn be thought of as a restricted form of the \textit{paramodulation} rule for first-order clauses. \textit{Simplification by rewriting} then represents a supplementary deductive inference rule that allows any redundant equations that are generated by the \textit{superposition} rule to be deleted as necessary.

\begin{definition}
If $T$ denotes a set of terms and ${\Sigma}$ denotes a set of substitutions (such that if ${A \left[ s \right]}$ indicates that the expression $A$ contains the expression $s$ as a subexpression, and ${A \left[ t \right]}$ indicates that an occurrence of $s$ in $A$ has been replaced by $t$, then we use ${A \sigma}$ to denote the substitution instance of $t$ obtained by applying the substitution ${\sigma}$ to the expression $A$, where ${\sigma \in \Sigma}$), then a ``rewrite relation'', denoted ${\to}$, is a binary relation satisfying the property that:

\begin{equation}
\forall s, t, u \in T, \sigma \in \Sigma, \qquad s \to t \implies u \left[ s \sigma \right] \to u \left[ t \sigma \right].
\end{equation}
\end{definition}

\begin{definition}
A ``reduction ordering'', denoted ${\succ}$, is a rewrite relation ${\to}$ that is both transitive:

\begin{equation}
\forall x, y, z \in T, \qquad x \to y \text{ and } y \to z \implies x \to z,
\end{equation}
and well-founded:

\begin{equation}
\forall X \subseteq T, \qquad X \neq \emptyset \implies \exists x \in X \text{ such that } \forall y \in X, \qquad y \not\to x.
\end{equation}
\end{definition}

\begin{definition}
A ``clause'', denoted ${\Gamma \implies \Delta}$, is an ordered pair of multisets of equations, i.e. an ordered pair ${\left( \Gamma, \Delta \right)}$, for two multisets ${\Gamma}$ and ${\Delta}$.
\end{definition}
In the above, each \textit{equation} is simply an expression of the form ${s \approx t}$, with ${s, t \in T}$ being first-order terms (constructed from variables and function symbols), which can in turn be represented as a single multiset ${\left\lbrace s, t \right\rbrace}$. Within the clause ${\Gamma \implies \Delta}$, we use the phrase \textit{antecedent} to refer to the multiset ${\Gamma}$, and the phrase \textit{succedent} to refer to the multiset ${\Delta}$.

We may start by defining a simple inference rule which asserts that the equality predicate ${\approx}$ is reflexive, namely the rule of \textit{equality resolution}:

\begin{definition}
``Equality resolution'' is a deductive inference rule which asserts that, for any terms ${u, v \in T}$ whose most general unifier is ${\sigma \in \Sigma}$, and arbitrary multisets of equations ${\Lambda}$ and ${\Pi}$:

\begin{equation}
\infer{ \Lambda \sigma \implies \Pi \sigma}{\Lambda \cup \left\lbrace u \approx v \right\rbrace \implies \Pi},
\end{equation}
whenever ${u \sigma \approx v \sigma}$ is an occurrence of an equation within the clause:

\begin{equation}
\Lambda \sigma \cup \left\lbrace u \sigma \approx v \sigma \right\rbrace \implies \Pi \sigma,
\end{equation}
that is maximal with respect to the reduction ordering ${\succ}$.
\end{definition}
In conventional resolution-based theorem-proving systems, it is also standard to incorporate a \textit{factoring} rule which effectively unifies a pair of terms within the same clause, thus allowing resolution with factoring to be a refutation-complete inference system, whereas resolution alone would not be. Here, we introduce a restricted form of the factoring inference rule, known as \textit{ordered factoring}, which applies only to succedents of clauses:

\begin{definition}
``Ordered factoring'' is a deductive inference rule which asserts that, for any pair of equations $A$ and $B$ whose most general unifier is ${\sigma \in \Sigma}$, and arbitrary multisets of equations ${\Gamma}$ and ${\Delta}$:

\begin{equation}
\infer{\Gamma \sigma \implies \Delta \sigma \cup \left\lbrace A \sigma \right\rbrace}{\Gamma \implies \Delta \cup \left\lbrace A, B \right\rbrace},
\end{equation}
whenever ${A \sigma}$ is an occurrence of an equation within the clause:

\begin{equation}
\Gamma \sigma \to \Delta \sigma \cup \left\lbrace A \sigma, B \sigma \right\rbrace,
\end{equation}
that is maximal with respect to the reduction ordering ${\succ}$.
\end{definition}

The simplest case of the \textit{paramodulation} inference rule (specifically for the case of purely variable-free terms and expressions) is now traditionally given as, for any terms ${u, v, s, t \in T}$, and arbitrary multisets of equations ${\Gamma}$, ${\Delta}$, ${\Lambda}$ and ${\Pi}$:

\begin{equation}
\infer{\Gamma \cup \Lambda \implies \Delta \cup \Pi \cup \left\lbrace u \left[ t \right] \approx v \right\rbrace}{\Gamma \implies \Delta \cup \left\lbrace s \approx t \right\rbrace \qquad \Lambda \implies \Pi \cup \left\lbrace u \left[ s \right] \approx v \right\rbrace}.
\end{equation}
Assuming that the reduction order ${\succ}$ is total, such a paramodulation inference would then be described as \textit{ordered} whenever ${s \succ t}$, ${s \approx t}$ is an occurrence of an equation within the clause ${\Gamma \cup \Delta}$ that is strictly maximal with respect to the reduction ordering ${\succ}$, and ${u \left[ s \right] \approx v}$ is an occurrence of an equation within the clause ${\Lambda \cup \Pi}$ that is strictly maximal with respect to ${\succ}$; such inference systems based on ordered paramodulation were proved by Hsiang and Rusinowitch\cite{hsiang}\cite{hsiang2} to be refutation-complete. On the other hand, if we relax the requirement that ${s \approx t}$ is an occurrence of an equation within the clause ${\Gamma \cup \Delta}$ that is strictly maximal with respect to ${\succ}$, and replace it with the condition that ${u \left[ s \right] \succ v}$, then we obtain a \textit{weak superposition} inference; such inference systems based on weak superposition were subsequently proved by Rusinowitch\cite{rusinowitch} to be refutation-complete also. However, for our present purposes, we opt instead to introduce a pair of inference rules, namely \textit{left superposition} and \textit{right superposition}, that may be considered to constitute restricted cases of the general paramodulation rule, which in turn requires introducing a notion of \textit{reductivity} on clauses:

\begin{definition}
A clause $C$ of the general form:

\begin{equation}
C = \Gamma \to \Delta \cup \left\lbrace s \approx t \right\rbrace,
\end{equation}
for any terms ${s, t \in T}$ and arbitrary multisets of equations ${\Gamma}$ and ${\Delta}$, is \textit{reductive} for the equation ${s \approx t}$ (with respect to the reduction ordering ${\succ}$) if and only if ${t \not\succeq s}$, and ${s \approx t}$ is an occurrence of an equation within clause $C$ that is maximal with respect to ${\succ}$.
\end{definition}

\begin{definition}
``Left superposition'' is a deductive inference rule which asserts that, for any terms ${u, v, s, s^{\prime}, t \in T}$ such that the most general unifier of $s$ and ${s^{\prime}}$ is ${\sigma \in \Sigma}$, ${s^{\prime}}$ is not a variable, and ${v \sigma \not\succ u \sigma}$, and arbitrary multisets of equations ${\Gamma}$, ${\Delta}$, ${\Lambda}$ and ${\Pi}$:

\begin{equation}
\infer{\left\lbrace u \left[ t \right] \sigma \approx v \sigma \right\rbrace \cup \Gamma \sigma \cup \Lambda \sigma \implies \Lambda \sigma \cup \Pi \sigma}{\Gamma \implies \Delta \cup \left\lbrace s \approx t \right\rbrace \qquad \left\lbrace u \left[ s^{\prime} \right] \approx v \right\rbrace \cup \Lambda \implies \Pi},
\end{equation}
whenever the clause:

\begin{equation}
\Gamma \sigma \implies \Delta \sigma \cup \left\lbrace s \sigma \approx t \sigma \right\rbrace,
\end{equation}
is reductive for the equation ${s \sigma \approx t \sigma}$, and ${u \sigma \approx v \sigma}$ is an occurrence of an equation within the clause:

\begin{equation}
\left\lbrace u \sigma \approx v \sigma \right\rbrace \cup \Lambda \sigma \implies \Pi \sigma,
\end{equation}
that is maximal with respect to the reduction ordering ${\succ}$.
\end{definition}

\begin{definition}
``Right superposition'' is a deductive inference rule which asserts that, for any terms ${u, v, s, s^{\prime}, t \in T}$ such that the most general unifier of $s$ and ${s^{\prime}}$ is ${\sigma \in \Sigma}$, and ${s^{\prime}}$ is not a variable, and arbitrary multisets of equations ${\Gamma}$, ${\Delta}$, ${\Lambda}$ and ${\Pi}$:

\begin{equation}
\infer{\Gamma \sigma \cup \Lambda \sigma \implies \left\lbrace u \left[ t \right] \sigma \approx v \sigma \right\rbrace \cup \Delta \sigma \cup \Pi \sigma}{\Gamma \implies \Delta \cup \left\lbrace s \approx t \right\rbrace \qquad \Lambda \implies \left\lbrace u \left[ s^{\prime} \right] \approx v \right\rbrace \cup \Pi},
\end{equation}
whenever the clause:

\begin{equation}
\Gamma \sigma \implies \Delta \sigma \cup \left\lbrace s \sigma \approx t \sigma \right\rbrace,
\end{equation}
is reductive for the equation ${s \sigma \approx t \sigma}$, and the clause:

\begin{equation}
\Lambda \sigma \implies \left\lbrace u \sigma \approx v \sigma \right\rbrace \cup \Pi \sigma,
\end{equation}
is reductive for the equation ${u \sigma \approx v \sigma}$.
\end{definition}

For the case of \textit{strict superposition} inferences, i.e. ordered paramodulation inferences of the form outlined above, but where ${u \left[ s \right] \succ v}$, and where the term $s$ does not appear anywhere in the multiset of equations ${\Gamma}$, such inference systems were shown by Bachmair and Ganzinger\cite{bachmair2} not to be refutation-complete, indicating that we must combine the above rules with some additional inferences. One possibility is to generalize the ordered factoring rule, which usually applies only to succedents of clauses, and apply it instead to arbitrary equations, thus yielding the \textit{equality factoring} inference rule:

\begin{definition}
``Equality factoring'' is a deductive inference rule which asserts that, for any terms ${s, s^{\prime}, t, t^{\prime} \in T}$ such that the most general unifier of $s$ and ${s^{\prime}}$ is ${\sigma \in \Sigma}$, ${t \sigma \not\succ s \sigma}$, and ${t^{\prime} \sigma \not\succ s^{\prime} \sigma}$, and arbitrary multisets of equations ${\Gamma}$ and ${\Delta}$:

\begin{equation}
\infer{\Gamma \sigma \cup \left\lbrace t \sigma \approx t^{\prime} \sigma \right\rbrace \implies \Delta \sigma \cup \left\lbrace s^{\prime} \sigma \approx t^{\prime} \sigma \right\rbrace}{\Gamma \implies \Delta \cup \left\lbrace s \approx t, s^{\prime} \approx t^{\prime} \right\rbrace},
\end{equation}
whenever ${s \sigma \approx t \sigma}$ is an occurrence of an equation within the clause:

\begin{equation}
\Gamma \sigma \implies \Delta \sigma \cup \left\lbrace s \sigma \approx t \sigma, s^{\prime} \sigma \approx t^{\prime} \sigma \right\rbrace,
\end{equation}
that is maximal with respect to the reduction ordering ${\succ}$.
\end{definition}
Alternatively, we could choose to render the system refutation-complete by introducing a stronger form of the paramodulation rule which, when applied repeatedly to a ground clause (i.e. a clause containing no variables) and combined with the ordered factoring rule above, merges any atomic expressions that exist within whichever succedent contains a maximal term with respect to the reduction order ${\succ}$. We call this the \textit{merging paramodulation} inference rule:

\begin{definition}
``Merging paramodulation'' is a deductive inference rule which asserts that, for any terms ${u, u^{\prime}, v, v^{\prime}, s, s^{\prime}, t \in T}$ such that the most general unifier of $s$ and ${s^{\prime}}$ is ${\tau \in \Sigma}$, the most general unifier of ${u \tau}$ and ${u^{\prime} \tau}$ is ${\rho \in \Sigma}$, the composition of the most general unifiers ${\tau}$ and ${\rho}$ is ${\sigma = \tau \rho \in \Sigma}$, ${s^{\prime}}$ is not a variable, ${u \tau \succ v \tau}$, and ${v^{\prime} \sigma \not\succeq v \sigma}$, and arbitrary multisets of equations ${\Gamma}$, ${\Delta}$, ${\Lambda}$ and ${\Pi}$:

\begin{equation}
\infer{\Gamma \sigma \cup \Lambda \sigma \implies \left\lbrace u \sigma \approx v \left[ t \right] \sigma, u \sigma \approx v^{\prime} \sigma \right\rbrace \cup \Delta \sigma \cup \Pi \sigma}{\Gamma \implies \Delta \cup \left\lbrace s \approx t \right\rbrace \qquad \Lambda \implies \left\lbrace u \approx v \left[ s^{\prime} \right], u^{\prime} \approx v^{\prime} \right\rbrace \cup \Pi},
\end{equation}
whenever the clause:

\begin{equation}
\Gamma \sigma \implies \Delta \sigma \cup \left\lbrace s \sigma \approx t \sigma \right\rbrace,
\end{equation}
is reductive for the equation ${s \sigma \approx t \sigma}$, and the clause:

\begin{equation}
\Lambda \sigma \implies \Pi \sigma \cup \left\lbrace u \sigma \approx v \sigma, u^{\prime} \sigma \approx v^{\prime} \sigma \right\rbrace,
\end{equation}
is reductive for the equation ${u \sigma \approx v \sigma}$.
\end{definition}

Since it becomes relevant to the various optimizations that we present subsequently in this article, it is worth also considering variations of the above rules in which inferences do not always select occurrences of equations that are necessarily maximal with respect to the reduction ordering ${\succ}$. In particular, we may introduce a \textit{selection function} $S$ that maps from clauses to (potentially empty) multisets of occurrences of equations in the antecedents of those clauses, such that any element of ${S \left( C \right)}$ for some clause $C$ is thought to be \textit{selected}, and ${S \left( C \right) = \emptyset}$ corresponds to no equation being selected. A canonical ordering is imposed by $S$ whenever multiple equations are selected. This allows us to define selective generalizations of the equality resolution and left superposition rules, namely the \textit{selective resolution} and \textit{selective superposition} inference rules, respectively:

\begin{definition}
``Selective resolution'' is a deductive inference rule which asserts that, for any terms ${u, v \in T}$ whose most general unifier is ${\sigma \in \Sigma}$, and arbitrary multisets of equations ${\Lambda}$ and ${\Pi}$:

\begin{equation}
\infer{\Lambda \sigma \implies \Pi \sigma}{\Lambda \cup \left\lbrace u \approx c \right\rbrace \implies \Pi},
\end{equation}
whenever ${u \approx v}$ is an occurrence of an equation within the clause:

\begin{equation}
\Lambda \cup \left\lbrace u \approx v \right\rbrace \implies \Pi,
\end{equation}
that has been selected by the selection function $S$.
\end{definition}

\begin{definition}
``Selective superposition'' is a deductive inference rule which asserts that, for any terms ${u, v, s, s^{\prime}, t \in T}$ such that the most general unifier of $s$ and ${s^{\prime}}$ is ${\sigma \in \Sigma}$, ${s^{\prime}}$ is not a variable, and ${v \sigma \not\succeq u \sigma}$, and arbitrary multisets of equations ${\Gamma}$, ${\Delta}$, ${\Lambda}$ and ${\Pi}$:

\begin{equation}
\infer{\left\lbrace u \left[ t \right] \sigma \approx v \sigma \right\rbrace \cup \Gamma \sigma \cup \Lambda \sigma \implies \Delta \sigma \cup \Pi \sigma}{\Gamma \implies \Delta \cup \left\lbrace s \approx t \right\rbrace \qquad \left\lbrace u \left[ s^{\prime} \right] \approx v \right\rbrace \cup \Lambda \implies \Pi},
\end{equation}
whenever the clause:

\begin{equation}
C = \left( \Gamma \implies \Delta \cup \left\lbrace s \approx t \right\rbrace \right),
\end{equation}
does not contain any equations selected by the selection function $S$, the clause ${C \sigma}$ is reductive for the equation ${s \sigma \approx t \sigma}$, and ${u \approx v}$ is an occurrence of an equation within the clause:

\begin{equation}
\left\lbrace u \approx v \right\rbrace \cup \Lambda \implies \Pi,
\end{equation}
that has been selected by the selection function $S$.
\end{definition}

By allowing for the existence of arbitrary predicate symbols in addition to ordinary function symbols, we can therefore construct expressions of the form ${P \left( t_1, \dots, t_n \right)}$, for some predicate symbol $P$ and non-predicate terms ${t_1, \dots, t_n \in T}$ (i.e. terms constructed from variables and function symbols, but not arbitrary predicate symbols). One can therefore draw a distinction between \textit{function equations} of the form ${s \approx t}$, for non-predicate terms ${s, t \in T}$, and \textit{predicate equations} of the form ${P \left( t_1, \dots, t_n \right) = tt}$, for an arbitrary predicate symbol $P$ and a unary predicate symbol $tt$ that is distinguished by being minimal with respect to the reduction ordering ${\succ}$. Now, the inference rules of equality resolution and selective resolution presented above clearly cannot be applied to any clause of the form:

\begin{equation}
\Gamma \cup \left\lbrace P \left( t_1, \dots, t_n \right) \approx tt \right\rbrace \implies \Delta, \qquad \text{ or } \qquad \Gamma \implies \Delta \cup \left\lbrace P \left( t_1, \dots, t_n \right) \approx tt \right\rbrace,
\end{equation}
within which ${P \left( t_1, \dots, t_n \right) \approx tt}$ is an occurrence of an equation that is either maximal with respect to the reduction ordering ${\succ}$ (in the equality resolution case), or that has been selected by the selection function $S$ (in the selective resolution case). Moreover, any attempt to superpose a clause of the form:

\begin{equation}
\Gamma \implies \Delta \cup \left\lbrace P \left( s_1, \dots s_n \right) \approx t \right\rbrace,
\end{equation}
onto a second clause of the form:

\begin{equation}
\Lambda \implies \Pi \cup \left\lbrace P \left( t_1, \dots, t_n \right) \approx tt \right\rbrace,
\end{equation}
will be a redundant inference, since it simply yields the following tautology:

\begin{equation}
\Gamma \sigma \cup \Lambda \sigma \implies \Delta \sigma \cup \Pi \sigma \cup \left\lbrace tt \approx tt \right\rbrace.
\end{equation}
The only remaining inference rules are those of left superposition and selective superposition, which both take on the general form:

\begin{equation}
\infer{\Gamma \sigma \cup \Lambda \sigma \cup \left\lbrace tt \approx tt \right\rbrace \implies \Delta \sigma \cup \Pi \sigma}{\Gamma \implies \Delta \cup \left\lbrace P \left( s_1, \dots, s_n \right) \approx tt \right\rbrace \qquad \left\lbrace P \left( t_1, \dots, t_n \right) \approx tt \right\rbrace \cup \Lambda \implies \Pi},
\end{equation}
in which the equation ${tt \approx tt}$ is trivial, and hence can be eliminated by resolution, thus yielding the derived \textit{ordered resolution} rule:

\begin{definition}
``Ordered resolution'' is a deductive inference rule which asserts that, for any terms ${u, v, s, s^{\prime}, t \in T}$ such that the most general unifier of $s$ and ${s^{\prime}}$ is ${\sigma \in \Sigma}$, ${s^{\prime}}$ is not a variable, and ${v \sigma \not\succ u \sigma}$, and arbitrary multisets of equations ${\Gamma}$, ${\Delta}$, ${\Lambda}$ and ${\Pi}$:

\begin{equation}
\infer{\Gamma \sigma \cup \Lambda \sigma \implies \Delta \sigma \cup \Pi \sigma}{\Gamma \implies \Delta \cup \left\lbrace P \left( s_1, \dots, s_n \right) \approx tt \right\rbrace \qquad \left\lbrace P \left( t_1, \dots, t_n \right) \approx tt \right\rbrace \cup \Lambda \implies \Pi},
\end{equation}
either whenever the clause:

\begin{equation}
\Gamma \sigma \implies \Delta \sigma \cup \left\lbrace s \sigma \approx t \sigma \right\rbrace,
\end{equation}
is reductive for the equation ${s \sigma \approx t \sigma}$, and also ${u \sigma \approx v \sigma}$ is an occurrence of an equation within the clause:

\begin{equation}
\left\lbrace u \sigma \approx v \sigma \right\rbrace \cup \Lambda \sigma \implies \Pi \sigma,
\end{equation}
that is maximal with respect to the reduction ordering ${\succ}$ (in the non-selective superposition case), or otherwise whenever the clause:

\begin{equation}
C = \left( \Gamma \implies \Delta \cup \left\lbrace s \approx t \right\rbrace \right),
\end{equation}
does not contain any equations selected by the selection function $S$, the clause ${C \sigma}$ is reductive for the equation ${s \sigma \approx t \sigma}$, and also ${u \approx v}$ is an occurrence of an equation within the clause:

\begin{equation}
\left\lbrace u \approx v \right\rbrace \cup \Lambda \implies \Pi,
\end{equation}
that has been selected by the selection function $S$.
\end{definition}
We shall subsequently define the selection function $S$ in such a way that terms are ordered on the basis of their causal edge density in the corresponding multiway operator system.

\begin{figure}[ht]
\centering
\includegraphics[width=0.795\textwidth]{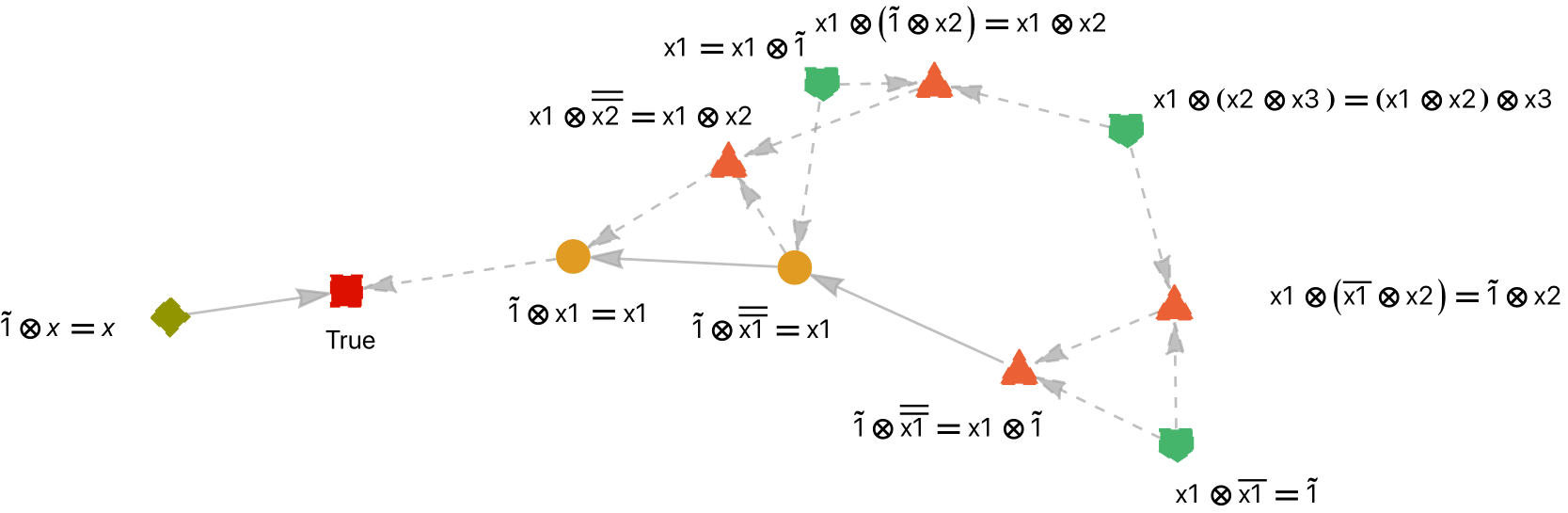}
\caption{The proof graph corresponding to the proof of the equational proposition ${\forall x : \tilde{1} \otimes x = x}$ (existence of left identity) using the axioms of group theory, namely ${\forall x_1, x_2, x_3 : x_1 \otimes \left( x_2 \otimes x_3 \right) = \left( x_1 \otimes x_2 \right) \otimes x_3}$ (associativity), ${\forall x_1 : x_1 = x_1 \otimes \tilde{1}}$ (existence of right identity), and ${\forall x_1 : x_1 \otimes \overline{x_1} = \tilde{1}}$ (existence of right inverse). Here, pointed light green boxes represent axioms, dark orange triangles represent critical pair lemmas (i.e. instances of completions/superpositions/paramodulations), light orange circles represent substitution lemmas (i.e. instances of resolutions/factorings), and dark green diamonds represent hypotheses. Solid lines represent substitutions, and dashed lines represent derived inference rules.}
\label{fig:Figure1}
\end{figure}

\begin{figure}[ht]
\centering
\includegraphics[width=0.895\textwidth]{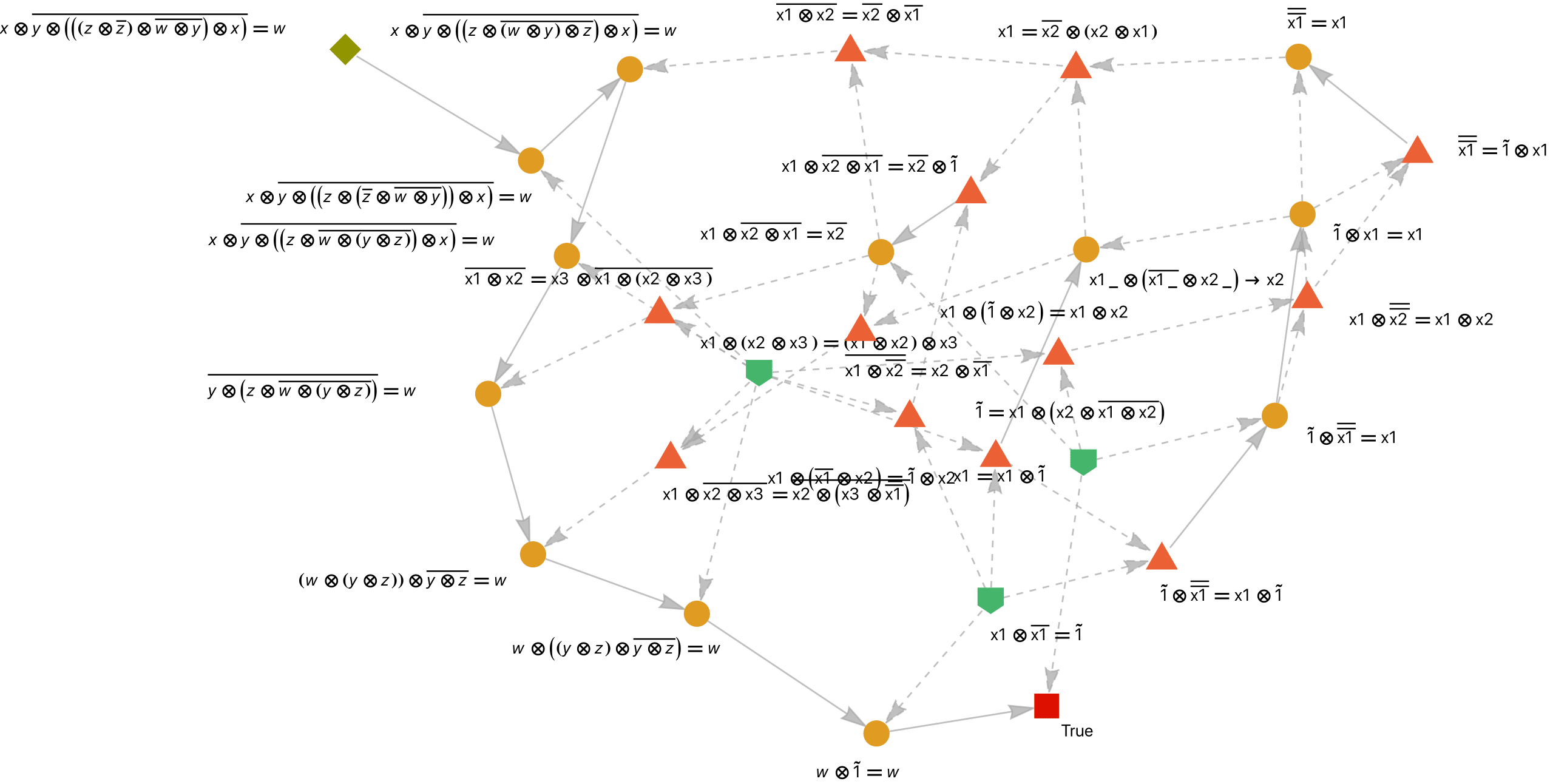}
\caption{The proof graph corresponding to the proof of the equational proposition ${\forall x, y, z, w : x \otimes \overline{y \otimes \left( \left( \left( z \otimes \overline{z} \right) \otimes \overline{w \otimes y} \right) \otimes x \right)} = w}$ (McCune's single group theory axiom) using the axioms of group theory, namely ${\forall x_1, x_2, x_3 : x_1 \otimes \left( x_2 \otimes x_3 \right) = \left( x_1 \otimes x_2 \right) \otimes x_3}$ (associativity), ${\forall x_1 : x_1 = x_1 \otimes \tilde{1}}$ (existence of right identity), and ${\forall x_1 : x_1 \otimes \overline{x_1} = \tilde{1}}$ (existence of right inverse). Here, pointed light green boxes represent axioms, dark orange triangles represent critical pair lemmas (i.e. instances of completions/superpositions/paramodulations), light orange circles represent substitution lemmas (i.e. instances of resolutions/factorings), and dark green diamonds represent hypotheses. Solid lines represent substitutions, and dashed lines represent derived inference rules.}
\label{fig:Figure2}
\end{figure}

We show an example of the automatically-generated proof graph for a simple equational theorem, namely the existence of a left identity that is equal to the right identity (${\forall x : \tilde{1} \otimes x = x}$), using the standard axioms of group theory, namely associativity (${\forall x_1, x_2, x_3 : x_1 \otimes \left( x_2 \otimes x_3 \right) = \left( x_1 \otimes x_2 \right) \otimes x_3}$), existence of a right identity (${\forall x_1 : x_1 = x_1 \otimes \tilde{1}}$), and existence of a right inverse (${\forall x_1 : x_1 = x_1 \otimes \overline{x_1} = \tilde{1}}$), in Figure \ref{fig:Figure1}. This proof was generated using the deductive inference rules of equality resolution, ordered factoring, left superposition, right superposition and merging paramodulation (without any selection function specified); the pointed light green boxes represent axioms of the theory, dark orange triangles represent critical pair lemmas (i.e. lemmas generated via completion/superposition/paramodulation inferences), light orange circles represent substitution lemmas (i.e. lemmas generated via resolution/factoring instances), and dark green diamonds represent the hypotheses being proven. Solid lines indicate that a substitution has taken place between two equations, and dashed lines indicate that one equation is being used as a derived inference rule in the proof of a different equation. An automatically-generated proof graph of a slightly more sophisticated equational theorem, namely the validity of McCune's single axiom\cite{mccune} for group theory (${\forall x, y, z, w : x \otimes \overline{y \otimes \left( \left( \left( z \otimes \overline{z} \right) \otimes \overline{ w \otimes y} \right) \otimes x \right)} = w}$), using the same standard axioms of group theory as above, is shown in Figure \ref{fig:Figure2}.

\clearpage

\section{Equational Treatment of Higher-Order Logics}
\label{sec:Section2}

The ZX-calculus is, by its nature, a higher-order diagrammatic reasoning system, due to the existence of infinite rule schemas (such as the Z- and X-spider fusion rules). As an initial step towards reducing arbitrary higher-order predicate logics to an equational form, we begin by presenting an equational treatment of full first-order predicate logic based on the methods of Mycielski\cite{mycielski}\cite{mycielski2}, building upon previous work by Burris\cite{burris}. The core of this treatment rests upon the reduction of arbitrary predicate formulas ${\varphi}$ in the first-order language $L$ to their respective \textit{Skolem normal forms}:

\begin{definition}
The ``Skolem normal form'' of an arbitrary predicate formula ${\varphi}$ in the first-order language $L$ is a quantifier-free formula ${\varphi^{\prime}}$ that is obtained by repeated replacement of subformulas of ${\varphi}$ in accordance with the following replacement rules:

\begin{equation}
\exists x_0 : \psi \left( x_0, x_1, \dots, x_n \right) \qquad \to \qquad \psi \left( f \left( x_1, \dots, x_n \right), x_1, \dots, x_n \right),
\end{equation}
whenever the subformula ${\exists x_0 : \psi \left( x_0, x_1, \dots, x_n \right)}$ does not lie within the scope of any other quantifier and is not negated (i.e. there exist an even number of negation operators preceding it);

\begin{equation}
\exists x_0 : \psi \left( x_0, x_1, \dots, x_n \right) \qquad \to \qquad \psi \left( x_k, x_1, \dots, x_n \right),
\end{equation}
whenever the subformula ${\exists x_0 : \psi \left( x_0, x_1, \dots, x_n \right)}$ does not lie within the scope of any other quantifier and is negated (i.e. there exist an odd number of negation operators preceding it);

\begin{equation}
\forall x_0 : \psi \left( x_0, \dots, x_n \right) \qquad \to \qquad \psi \left( x_k, x_1, \dots, x_n \right),
\end{equation}
whenever the subformula ${\forall x_0 : \psi \left( x_0, x_1, \dots, x_n \right)}$ does not lie within the scope of any other quantifier and is not negated (i.e. there exist an even number of negation operators preceding it); and:

\begin{equation}
\forall x_0 : \psi \left( x_0, \dots, x_n \right) \qquad \to \qquad \psi \left( f \left( x_1, \dots, x_n \right), x_1, \dots, x_n \right),
\end{equation}
whenever the subformula ${\forall x_0 : \psi \left( x_0, x_1, \dots x_n \right)}$ does not lie within the scope of any other quantifier and is negated (i.e. there exist an odd number of negation operators preceding it). In the above, $f$ is a newly-introduced function symbol (i.e. a ``Skolem function''), and $k$ is chosen to be the lowest natural number such that the variable ${x_k}$ does not appear within the formula ${\varphi}$.
\end{definition}
Note that this particular definition of the \textit{Skolemization} procedure is slightly more general than the conventional one, since it does not assume that the formula ${\varphi}$ has already been reduced to prenex normal form; for instance, if the formula ${\varphi \left( z \right)}$ is of the form:

\begin{equation}
\varphi \left( z \right) = \left( \forall x : \exists y : P \left( x, y, z \right) \implies \forall x : \exists y : Q \left( x, y, z \right) \right),
\end{equation}
then its Skolem normal form, denoted ${S \left( \varphi \right)}$, will be of the form:

\begin{equation}
S \left( \varphi \right) = \left( P \left( f \left( z \right), y, z \right) \implies Q \left( x, g \left( x, z \right), z \right) \right),
\end{equation}
with newly-introduced Skolem functions $f$ and $g$.

If ${\varphi^{\prime}}$ denotes the Skolem normal form of the formula ${\varphi}$, and ${\varphi^{\prime \prime}}$ denotes a copy of ${\varphi^{\prime}}$ with some of the Skolem function symbols replaced with different symbols (up to equality), then clearly by G\"odel's completeness theorem it is necessarily the case that ${\varphi^{\prime \prime} \vdash \varphi}$, even though it is not necessarily the case that ${\varphi^{\prime \prime} \vdash \varphi^{\prime}}$, or that ${\varphi^{\prime} \vdash \varphi^{\prime \prime}}$. This new formula ${\varphi^{\prime \prime}}$ is sometimes referred to as an \textit{s-Skolemization} of the original first-order formula ${\varphi}$. By induction on the depth of the original formula ${\varphi}$, it is also easy to see that, whenever ${\varphi^{\prime}}$ and ${\neg \varphi^{\prime \prime}}$ are s-Skolemizations of the first-order formula ${\varphi}$ and its negation ${\neg \varphi}$, respectively, there must exist a pair of substitutions ${s_1}$ and ${s_2}$ that replace terms with variables, with the property that the free variables in the original formula ${\varphi}$ are not modified by either ${s_1}$ or ${s_2}$, but rather by:

\begin{equation}
s_1 \left( \varphi^{\prime} \right) = s_2 \left( \varphi^{\prime \prime} \right).
\end{equation}
For instance, if the formula ${\varphi}$ corresponds to a slight modification of the example formula given above, namely:

\begin{equation}
\varphi = \left( \forall x : \exists y : P \left( x, y, z \right) \implies \forall x: \exists y : Q \left( x, y, u \right) \right),
\end{equation}
then the two s-Skolemizations ${\varphi^{\prime}}$ and ${\neg \varphi^{\prime \prime}}$ are given by:

\begin{equation}
\varphi^{\prime} = \left( P \left( f \left( z \right), y, z \right) \implies Q \left( x, g \left( x, u \right), u \right) \right),
\end{equation}
and:

\begin{equation}
\neg \varphi^{\prime \prime} = \neg \left( P \left( x, h \left( x, z \right), z \right) \implies Q \left( k \left( u \right), y, u \right) \right),
\end{equation}
respectively, such that:

\begin{equation}
s_1 \left( \varphi^{\prime} \right) = s_2 \left( \varphi^{\prime \prime} \right) = \left( P \left( f \left( z \right), h \left( f \left( z \right), z \right), z \right) \implies Q \left( k \left( u \right), g \left( k \left( u \right), u \right), u \right) \right),
\end{equation}
as required.

Supposing that ${\mathcal{A}}$ denotes a set of axioms in our first-order language $L$, we will use the notation ${S \left( \mathcal{A} \right)}$ to designate the Skolemization of ${\mathcal{A}}$ (i.e. the set of Skolem normal forms for axioms in ${\mathcal{A}}$); our first objective is therefore to supplement the set ${S \left( \mathcal{A} \right)}$ by adding appropriate rules of proof. We will henceforth also adopt the notation:

\begin{equation}
S \left( \neg \varphi \right) = S \left( \neg \varphi \right) \left( \bar{x}, \bar{y} \right), \qquad \text{ and } \qquad S \left( \varphi \right) = S \left( \varphi \right) \left( \bar{x}, \bar{z} \right),
\end{equation}
where ${\bar{x}}$ designates the sequence of free variables in the original formula ${\varphi}$, and ${\bar{y}}$ and ${\bar{z}}$ designate the sequences of free variables that are introduced by the Skolemization procedure. A somewhat tedious argument\cite{mycielski}\cite{mycielski2} then yields the fact that there must exist a pair of unique sequences of terms with variables taken from the sequence ${\bar{x}}$, namely ${\bar{s} \left( \bar{x} \right)}$ and ${\bar{t} \left( \bar{x} \right)}$, for which the following equivalence between formulas holds:

\begin{equation}
S \left( \neg \varphi \right) \left( \bar{x}, \bar{s} \left( \bar{x} \right) \right) = \neg S \left( \varphi \right) \left( \bar{x}, \bar{t} \left( \bar{x} \right) \right).
\end{equation}
This lemma then enables us to prove the following pair of more significant theorems regarding \textit{equalizers} of Skolemizations:

\begin{definition}
The ``equalizer'' of the Skolemization ${S \left( \varphi \implies \varphi \right)}$, for some arbitrary predicate formula ${\varphi}$ in the first-order language $L$, with Skolemizations ${S \left( \varphi \right)}$ and ${S \left( \neg \varphi \right)}$ chosen such that:

\begin{equation}
\vdash S \left( \varphi \implies \varphi \right) \Leftrightarrow \left( S \left( \neg \varphi \right) \vee S \left( \varphi \right) \right),
\end{equation}
is a formula ${\varphi^{*}}$ of the form:

\begin{equation}
\varphi^{*} = \varphi^{*} \left( \bar{x} \right) = S \left( \varphi \right) \left( \bar{x}, \bar{t} \left( \bar{x} \right) \right).
\end{equation}
\end{definition}
Clearly, if we denote the set of all formulas in language $L$ of the form ${\varphi \implies \varphi}$ by ${\mathcal{B}}$, with ${S \left( \mathcal{B} \right)}$ being the set of Skolem normal forms of formulas in ${\mathcal{B}}$, then the equalizer ${\varphi^{*}}$ will be a quantifier-free element of ${L^{*}}$ (where ${L^{*}}$ denotes the language of ${S \left( \mathcal{B} \right)}$) containing the same free variables as the original formula ${\varphi}$. For instance, if ${\varphi_{0} \left( z \right)}$ designates the example formula given previously:

\begin{equation}
\varphi_{0} \left( z \right) = \left( \forall x : \exists y : P \left( x, y, z \right) \implies \forall x : \exists y : Q \left( x, y, z \right) \right),
\end{equation}
then its equalizer ${\varphi_{0}^{*} \left( z \right)}$ will be of the form:

\begin{equation}
\varphi_{0}^{*} \left( z \right) = \left( P \left( f \left( z \right), h \left( f \left( z \right), h \left( f \left( z \right), z \right) \right), z \right) \implies Q \left( k \left( z \right), g \left( k \left( z \right), z \right), z \right) \right).
\end{equation}
From here, it is easy to see that, whenever ${\varphi^{*}}$ is the equalizer of a formula of the form ${S \left( \varphi \implies \varphi \right)}$, then:

\begin{equation}
S \left( \varphi \implies \varphi \right) \vdash \varphi \Leftrightarrow \varphi^{*},
\end{equation}
and, moreover, whenever ${\varphi^{*}}$ and ${\varphi^{* \prime}}$ are the equalizers of the pair of Skolemizations ${S \left( \varphi \implies \varphi \right)}$ and ${S^{\prime} \left( \varphi \implies \varphi \right)}$ of the proposition ${\varphi \implies \varphi}$, respectively, then:

\begin{equation}
\left\lbrace S \left( \varphi \implies \varphi \right), S^{\prime} \left( \varphi \implies \varphi \right) \right\rbrace \vdash \varphi^{*} \Leftrightarrow \varphi^{* \prime}.
\end{equation}

These two theorems collectively suggest a new set of rules of inference with respect to the quantifier-free language ${L^{*}}$ associated with the set of Skolem normal forms ${S \left( \mathcal{B} \right)}$. Namely, one has the three standard \L{}ukasiewicz axioms\cite{cignoli} for the implication symbol in propositional logic, such that any substitution instance of the following tautologies:

\begin{equation}
\left( p \implies q \right) = \left[ \left( q \implies r \right) \implies \left( p \implies r \right) \right], \qquad p \implies \left( \neg p \implies q \right), \qquad \left( \neg p \implies p \right) \implies p,
\end{equation}
is necessarily a theorem; one has the standard modus ponens rule, such that if ${\alpha \implies \beta}$ and ${\alpha}$ are both theorems, then ${\beta}$ is necessarily a theorem; any theorem in which terms are substituted for free variables is necessarily a theorem; any formula in the set of Skolem normal forms ${S \left( \mathcal{B} \right)}$ is necessarily a theorem; and finally, all axioms of equality hold within the quantifier-free language ${L^{*}}$. If ${x \vdash^{*} y}$ indicates that $y$ is provable from $x$ using only these inference rules, and if ${\mathcal{A}^{*}}$ designates the set of equalizers of the axioms ${\mathcal{A}}$:

\begin{equation}
\mathcal{A}^{*} = \left\lbrace \alpha^{*} : \alpha \in \mathcal{A} \right\rbrace,
\end{equation}
then one immediately has the following theorem:

\begin{equation}
\mathcal{A} \vdash \varphi \qquad \Leftrightarrow \qquad \mathcal{A}^{*} \vdash^{*} \varphi^{*},
\end{equation}
with the property that any Hilbert-style natural deduction argument for the formula ${\varphi}$ using the axioms ${\mathcal{A}}$ can be translated into a natural deduction argument of the same length for the quantifier-free formula ${\varphi^{*}}$ using the axioms ${\mathcal{A}^{*}}$ and the inference rules above. Moreover, note that one necessarily has:

\begin{equation}
S \left( \mathcal{A} \right) \vdash^{*} \mathcal{A}^{*}, \qquad \text{ and } \qquad \mathcal{A}^{*} \vdash^{*} S \left( \mathcal{A} \right).
\end{equation}
This theorem is essentially a variant of G\"odel's completeness theorem for the quantifier-free language ${L^{*}}$.

With the completeness of the inference rules for the quantifier-free language ${L^{*}}$ thus established, all that remains is to reduce the formulas in this language into a purely equational form, which we now proceed to do by means of the formalism of \textit{multisorted algebras}:

\begin{definition}
A ``multisorted algebra'', denoted ${\mathcal{U}}$, is an algebraic structure of the general form:

\begin{equation}
\left\langle A_i, f_j \right\rangle_{i \in I, j \in J},
\end{equation}
for ``sorts'' ${A_i}$ (i.e. disjoint, non-empty sets), and symbols ${f_j}$, with the property that:

\begin{equation}
\forall j \in J, \qquad \exists \left( i_0, \dots, i_{n \left( j \right)} \right) \in I^{n \left( j \right) + 1}, \qquad \text{ with } \qquad n \left( j \right) < \omega,
\end{equation}
such that:

\begin{equation}
n \left( j \right) > 0 \qquad \implies \qquad f_j : A_{i_1} \times \cdots \times A_{i_{n \left( j \right)}} \to A_{i_0},
\end{equation}
and:

\begin{equation}
n \left( j \right) = 0, \qquad \implies \qquad f_j \in A_{i_0}.
\end{equation}
\end{definition}

\begin{definition}
The ``type'' of a multisorted algebra ${\mathcal{U}}$ is a map from the index $j$ to the sequence of indices from ${i_0}$ to ${i_{n \left( j \right)}}$:

\begin{equation}
j \mapsto \left( i_0, \dots, i_{n \left( j \right)} \right).
\end{equation}
\end{definition}
If ${\tau}$ and ${\theta}$ now denote arbitrary terms (i.e. functions obtained by means of legal composition of the ${f_j}$ functions) that both take values in the same sort ${A_i}$, then an \textit{equation} is any expression of the form:

\begin{equation}
\tau = \theta.
\end{equation}
One then has the standard rules of inference for the equality predicate, namely reflexivity, symmetry and transitivity:

\begin{equation}
\forall i \in I : \qquad x_{0}^{i} = x_{0}^{i}, \qquad \left( \tau = \theta \right) \implies \left( \theta = \tau \right), \qquad \left( \left( \tau = \theta \right) \wedge \left( \theta = \rho \right) \right) \implies \left( \tau = \rho \right);
\end{equation}
one also has that:

\begin{equation}
\tau \left( x_{n}^{i} \right) = \theta \left( x_{n}^{i} \right) \qquad \implies \qquad \tau \left( \sigma \right) = \theta \left( \sigma \right),
\end{equation}
whenever the equation ${\tau \left( \sigma \right) = \theta \left( \sigma \right)}$ is well-formed; finally, one has that:

\begin{equation}
\left( \tau_1 = \theta_1 \right) \wedge \cdots \wedge \left( \tau_{n \left( j \right)} = \theta_{n \left( j \right)} \right) \qquad \implies \qquad f_j \left( \tau_1, \dots, \tau_{n \left( j \right)} \right) = f_j \left( \theta_1, \dots, \theta_{n \left( j \right)} \right),
\end{equation}
whenever the equation ${f_j \left( \tau_1, \dots, \tau_{n \left( j \right)} \right) = f_j \left( \theta_1, \dots, \theta_{n \left( j \right)} \right)}$ is well-formed. If we now fix the language $L$, then we have that, for any set of equations ${\epsilon}$, there exists a multisorted algebra ${\mathcal{U}}$ (whose type is determined by the choice of language $L$) with the property that, for any equation of the form ${\tau = \theta}$, the algebra ${\mathcal{U}}$ satisfies this equation if and only if it is derivable from the original set of equations ${\epsilon}$ by means of the inference rules above. This theorem is essentially a variant of Birkhoff's completeness theorem\cite{birkhoff} for multisorted algebras.

\begin{figure}[ht]
\centering
\includegraphics[width=0.995\textwidth]{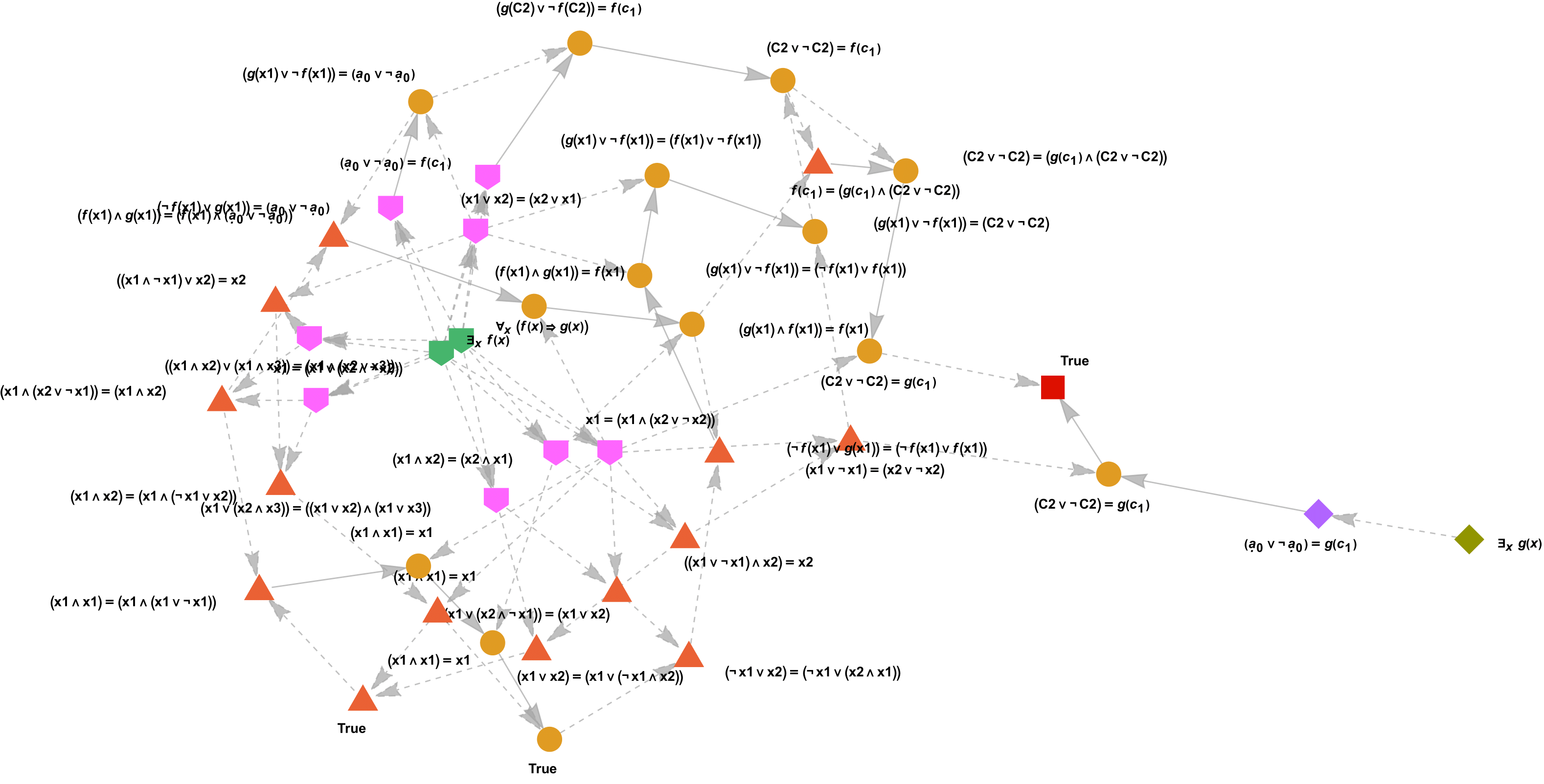}
\caption{The proof graph corresponding to the proof of the first-order proposition ${\exists x : g \left( x \right)}$ (derived by modus ponens) using the axioms ${\forall x : f \left( x \right) \implies g \left( x \right)}$ and ${\exists x : f \left( x \right)}$. Here, pointed light green boxes represent axioms, pointed pink boxes represent \textit{Skolemized} axioms, dark orange triangles represent critical pair lemmas (i.e. instances of completions/superpositions/paramodulations), light orange circles represent substitution lemmas (i.e. instances of resolutions/factorings), and dark green diamonds represent hypotheses. Solid lines represent substitutions, and dashed lines represent derived inference rules.}
\label{fig:Figure3}
\end{figure}

We show a simple example of an automatically-generated proof graph for an elementary theorem in full first-order logic, namely an application of the modus ponens rule, deriving the formula ${\exists x : g \left( x \right)}$ using the axioms ${\forall x : f \left( x \right) \implies g \left( x \right)}$ and ${\exists x : f \left( x \right)}$, in Figure \ref{fig:Figure3}. The core of the proof was generated using the deductive inference rules of equality resolution, ordered factoring, left superposition, right superposition and merging paramodulation (without any selection function specified), as above. As previously, the pointed light green boxes designate the underlying axioms, dark orange triangles designate critical pair lemmas (generated via the application of completion/superposition/paramodulation inferences), light orange circles designate substitution lemmas (generated via the application of resolution/factoring instances), and dark green diamonds represent the hypotheses being proven. However, we now also introduce pointed pink boxes to designate the \textit{Skolemized} versions of the underlying axioms. As usual, solid lines indicate the application of a substitution between a pair of equations, and dashed lines indicate the use of one equation as a derived inference rule in the proof of a second equation. It is worth noting that Skolem symbols are here denoted by ${c_i}$ and ${C_i}$ for ${i \in \mathbb{N}}$, depending on context.

From here, we follow the approach of Kerber\cite{kerber} by defining an appropriate set of \textit{morphisms} from arbitrary higher-order predicate logics to first-order (multisorted) logic, such that these morphisms are provably both complete and sound. Below, we shall denote the $n$th-order predicate logic by ${\mathcal{L}^n}$, each of which is treated as a subset of the ordinal logic ${\mathcal{L}^{\omega}}$, as defined by Turing\cite{turing}. We also assume the restricted type structure of Church\cite{church}, in which function and predicate symbols are typed, but sorts are not.

\begin{definition}
The ``signature'', denoted $S$, of a logic, denoted ${\mathcal{L}^{\omega} \left( S \right)}$, contained as a subset of the ordinal logic ${\mathcal{L}^{\omega}}$ (since all such logics may be defined by their signatures), is a set of the form:

\begin{equation}
S = \left(\bigcup_{\tau} S_{\tau}^{const} \right) \cup \left( \bigcup_{\tau} S_{\tau}^{var} \right),
\end{equation}
where ${S_{\tau}^{const}}$ denotes a (possibly empty) set of constant symbols, each of type ${\tau}$, and ${S_{\tau}^{var}}$ denotes a countably infinite set of variable symbols, each of type ${\tau}$, with the assumption that all sets ${S_{\tau}^{const}}$ and ${S_{\tau}^{var}}$ are disjoint.
\end{definition}

\begin{definition}
For a logic ${\mathcal{L}^{\omega} \left( S \right)}$, contained as a subset of the ordinal logic ${\mathcal{L}^{\omega}}$, the set of ``terms'' is generated by the set of constant and variable symbols of type ${\tau}$, with the rule that if the terms ${f_{\left( \tau_1 \times \cdots \times \tau_m \to \sigma \right)}}$, ${t_{\tau_1}}$, \dots, ${t_{\tau_m}}$ are of type ${\tau_1 \times \cdots \tau_m \to \sigma}$, ${\tau_1}$, \dots, ${\tau_m}$, respectively, then:

\begin{equation}
f_{\left( \tau_1 \times \cdots \times \tau_m \to \sigma \right)} \left( t_{\tau_1}, \dots, t_{\tau_m} \right),
\end{equation}
is a term of type ${\sigma}$.
\end{definition}

\begin{definition}
For a logic ${\mathcal{L}^{\omega} \left( S \right)}$, contained as a subset of the ordinal logic ${\mathcal{L}^{\omega}}$, the set of ``formulas'' is generated by the set of terms of type ${\mathbf{0}}$ (where ${\mathbf{0}}$ is the type of order 1 that designates the type of truth values), with the rule that if ${\varphi}$ and ${\psi}$ are both formulas, and $x$ is a variable symbol (of any type), then:

\begin{equation}
\left( \neg \varphi \right), \qquad \left( \varphi \wedge \psi \right), \qquad \left( \forall x : \varphi \right),
\end{equation}
are all also formulas.
\end{definition}

\begin{definition}
A ``higher-order predicate logic'', denoted ${\mathcal{L}^{n}}$ for ${n \geq 1}$ (note that, for this purpose, we treat first-order predicate logic as a ``higher-order'' predicate logic), is a subset of the ordinal logic ${\mathcal{L}^{\omega}}$ satisfying the rules that ${\mathcal{L}^{2n}}$ must always be a subset of ${\mathcal{L}^{\omega}}$ in which the order of every constant and variable symbol is less than or equal to $n$, and ${\mathcal{L}^{2n - 1}}$ must always be a subset of ${\mathcal{L}^{2n}}$ in which the order of every quantified variable symbol is less than or equal to $n$.
\end{definition}
Henceforth, we adopt the notation ${\mathcal{C} \left( \mathcal{L}^{n} \right)}$ to designate the set of constant symbols, ${\mathcal{V} \left( \mathcal{L}^{n} \right)}$ to designate the set of variable symbols, ${\mathcal{T} \left( \mathcal{L}^{n} \right)}$ to designate the set of terms, and ${\mathcal{F} \left( \mathcal{L}^{n} \right)}$ to designate the set of formulas, in the logic ${\mathcal{L}^{n}}$. We also use the notation ${\mathcal{L}_{sort}^{1}}$ to designate the first-order multisorted logic whose signature ${S_{\Sigma}}$ is given by the following union:

\begin{equation}
S_{\Sigma} = \left( \bigcup_{s_1, \dots, s_m, s \in \Sigma} S^{\left( s_1, \dots, s_m \right) : s} \right) \cup \left( \bigcup_{s_1, \dots, s_n \in \Sigma} S^{\left( s_1, \dots, s_n \right)} \right) \cup \left( \bigcup_{s \in \Sigma} S_{const}^{s} \right) \cup \left( \bigcup_{s \in \Sigma} S_{var}^{s} \right),
\end{equation}
where ${\Sigma}$ denotes a finite set of sorts, ${S^{\left( s_1, \dots, s_m \right) : s}}$ denotes a (possibly empty) set of $m$-ary function constants, ${S^{\left( s_1, \dots, s_m \right)}}$ denotes a (possibly empty) set of $m$-ary predicate constants, ${S_{const}^{s}}$ denotes a (possibly empty) set of object constants, and ${S_{var}^{s}}$ denotes a countably infinite set of object variables. The binary equality predicate symbol ${=^{\left( s, s \right)}}$ is associated to each set ${S^{\left( s, s \right)}}$ with sort ${s \in \Sigma}$, and we index elements of ${S_{\Sigma}}$ by their respective sorts.

We are now able to introduce the required notions of \textit{morphisms} and (quasi-)\textit{homomorphisms} between logical systems, as follows:

\begin{definition}
A ``morphism'', denoted ${\Theta}$, between logical systems ${\mathcal{F}^{1}}$ and ${\mathcal{F}^{2}}$ (which may represent either the ordinal logic ${\mathcal{L}^{\omega}}$, the higher-order logic ${\mathcal{L}^{n}}$ or the first-order multisorted logic ${\mathcal{L}_{sort}^{1}}$), is a map from the signature of a logic in ${\mathcal{F}^{1}}$ to the signature of a logic in ${\mathcal{F}^{2}}$:

\begin{equation}
\Theta : \mathcal{F}^{1} \left( S \right) \to \mathcal{F}^{2} \left( \Theta \left( S \right) \right),
\end{equation}
that also maps sets of formulas in the logic ${\mathcal{F}^{1} \left( S \right)}$ to sets of formulas in the logic ${\mathcal{F}^{2} \left( \Theta \left( S \right) \right)}$:

\begin{equation}
\Theta : \mathcal{F} \left( \mathcal{F}^{1} \left( S \right) \right) \to \mathcal{F} \left( \mathcal{F}^{2} \left( \Theta \left( S \right) \right) \right),
\end{equation}
where a single formula ${\varphi}$ is treated, for these purposes, as being a formula set containing a single element, i.e:

\begin{equation}
\Theta \left( \varphi \right) = \Theta \left( \left\lbrace \varphi \right\rbrace \right).
\end{equation}
\end{definition}

\begin{definition}
A ``quasi-homomorphism'', denoted ${\Theta}$, is a morphism mapping between two logics, denoted ${\mathcal{F}^{1} \left( S_1 \right)}$ and ${\mathcal{F}^{2} \left( S_2 \right)}$, with the property that every formula (respectively, every term) in logic ${\mathcal{F}^{1} \left( S_1 \right)}$ is mapped to a corresponding formula (respectively, a corresponding term) in logic ${\mathcal{F}^{2} \left( S_2 \right)}$:

\begin{equation}
\Theta : \mathcal{F} \left( \mathcal{F}^{1} \left( S_1 \right) \right) \to \mathcal{F} \left( \mathcal{F}^{2} \left( S_2 \right) \right), \qquad \Theta : \mathcal{T} \left( \mathcal{F}^{1} \left( S_1 \right) \right) \to \mathcal{T} \left( \mathcal{F}^{2} \left( S_2 \right) \right),
\end{equation}
such that the following conditions hold for all formulas and variables in the logic ${\mathcal{F}^{1} \left( S_1 \right)}$:

\begin{equation}
\forall \varphi_1, \varphi_2 \in \mathcal{F} \left( \mathcal{F}^{1} \left( S_1 \right) \right), \qquad \Theta \left( \varphi_1 \wedge \varphi_2 \right) = \Theta \left( \varphi_1 \right) \wedge \Theta \left( \varphi_2 \right),
\end{equation}
\begin{equation}
\forall \varphi \in \mathcal{F} \left( \mathcal{F}^{1} \left( S_1 \right) \right), \qquad \Theta \left( \neg \varphi \right) = \neg \Theta \left( \varphi \right),
\end{equation}
and:

\begin{equation}
\forall \varphi \in \mathcal{F} \left( \mathcal{F}^{1} \left( S_1 \right) \right), x \in \mathcal{V} \left( \mathcal{F}^{1} \left( S_1 \right) \right), \qquad \Theta \left( \forall x : \varphi \right) = \forall \Theta \left( x \right) \Theta \left( \varphi \right);
\end{equation}
the following conditions hold for all terms in the logic ${\mathcal{F}^{1} \left( S_1 \right)}$:

\begin{equation}
x \in \mathcal{V} \left( \mathcal{F}^{1} \left( S_1 \right) \right) \qquad \implies \qquad \Theta \left( x \right) \in \mathcal{V} \left( \mathcal{F}^{2} \left( S_2 \right) \right),
\end{equation}
\begin{equation}
c \in \mathcal{C} \left( \mathcal{F}^{1} \left( S_1 \right) \right) \qquad \implies \qquad \Theta \left( c \right) \in \mathcal{C} \left( \mathcal{F}^{2} \left( S_2 \right) \right),
\end{equation}
and:

\begin{equation}
f \left( t_1, \dots, t_m \right) \in \mathcal{T} \left( \mathcal{F}^{1} \left( S_1 \right) \right) \qquad \implies \qquad \Theta \left( f \left( t_1, \dots, t_m \right) \right) = \theta \left( \Theta \left( f \right), \Theta \left( t_1 \right), \dots, \Theta \left( t_m \right) \right),
\end{equation}
where we have introduced the function:

\begin{equation}
\theta \left( a, a_1, \dots, a_m \right) = \begin{cases}
a \left( a_1, \dots, a_m \right)\\
\boldsymbol\alpha_a \left( a, a_1, \dots, a_m \right)
\end{cases},
\end{equation}
with cases chosen depending on the symbol $a$, and where the application symbol ${\boldsymbol\alpha}$ has the property that all generated constants in the signature ${S_2}$ are unique, such that there cannot exist some ${\alpha^{\prime} \in S_1}$ with the property that:

\begin{equation}
\boldsymbol\alpha_a = \Theta \left( \alpha^{\prime} \right);
\end{equation}
and finally the condition that all terms in the logic ${\mathcal{F}^{1} \left( S_1 \right)}$ that are not formulas must be mapped to corresponding terms in the logic ${\mathcal{F}^{2} \left( S_2 \right)}$ that are also not formulas:

\begin{equation}
\Theta : \mathcal{T} \left( \mathcal{F}^{1} \left( S_1 \right) \right) \setminus \mathcal{F} \left( \mathcal{F}^{1} \left( S_1 \right) \right) \to \mathcal{T} \left( \mathcal{F}^{2} \left( S_2 \right) \right) \setminus \mathcal{F} \left( \mathcal{F}^{2} \left( S_2 \right) \right).
\end{equation}
\end{definition}
Henceforth, we also adopt the notation ${\mathcal{PC} \left( \mathcal{L}^{n} \right) \subseteq \mathcal{C} \left( \mathcal{L}^{n} \right)}$ to designate the set of predicate constant symbols, ${\mathcal{FC} \left( \mathcal{L}^{n} \right) \subseteq \mathcal{C} \left( \mathcal{L}^{n} \right)}$ to designate the set of function constant symbols, ${\mathcal{PT} \left( \mathcal{L}^{n} \right) \subseteq \mathcal{T} \left( \mathcal{L}^{n} \right)}$ to designate the set of predicate terms, ${\mathcal{FT} \left( \mathcal{L}^{n} \right) \subseteq \mathcal{T} \left( \mathcal{L}^{n} \right)}$ to designate the set of function terms, ${\mathbf{O} \left( e \right)}$ to designate the order of expression $e$, ${\mathbf{A} \left( e \right)}$ to designate the arity of expression $e$, ${\mathbf{T} \left( e \right)}$ to designate the type of expression $e$, ${\mathbf{S} \left( e \right)}$ to designate the sort of expression $e$, ${\mathbf{AF} \left( f \right)}$ where ${f \in \mathcal{F} \left( \mathcal{L}^{n} \right)}$ to designate that formula $f$ is atomic, and ${\mathbf{TE} \left( e \right)}$ to designate the top-level expression in expression $e$.

Our objective is therefore to define a complete and sound family of morphisms ${\hat{\Theta}_n}$:

\begin{equation}
\hat{\Theta}_n : \mathcal{L}^{n} \to \mathcal{L}_{sort}^{1},
\end{equation}
for which it suffices to define ${\hat{\Theta}_n}$ for all odd $n$, since the morphism for any even $n$ can be obtained by an appropriate restriction of the next higher morphism, i.e:

\begin{equation}
\hat{\Theta}_{2n} = \left. \hat{\Theta}_{2n + 1} \right|_{\mathcal{L}^{2n}}.
\end{equation}
Following the approach of Kerber\cite{kerber}, we note that such morphisms can be expressed in the general form:

\begin{equation}
\hat{\Theta}_{n} \left( \varphi \right) = \hat{\Theta}_{n}^{\prime} \left( \varphi \right) \cup EXT_{n},
\end{equation}
for some appropriate quasi-homomorphism ${\hat{\Theta}_{n}^{\prime}}$, and a set of signature-dependent extensionality axioms ${EXT_{n}}$. In order to define the morphism ${\hat{\Theta}_{2n - 1}}$ for a higher-order predicate logic in ${\mathcal{L}^{2n - 1}}$, whose signature ${S^{2n - 1}}$ is of the form:

\begin{equation}
S^{2n - 1} = \cup_{\tau} S_{\tau},
\end{equation}
we begin by constructing the signature ${S_{\Sigma}}$ for a corresponding first-order (multisorted) logic in ${\mathcal{L}_{sort}^{1}}$, by initially associating every predicate constant symbol $c$ of order $n$ in the logic ${\mathcal{L}^{2n - 1}}$:

\begin{equation}
c \in \mathcal{PC} \left( \mathcal{L}^{2n - 1} \right), \qquad \mathbf{O} \left( c \right) = n, \qquad \mathbf{A} \left( c \right) = m, \qquad \mathbf{T} \left( c \right) = \tau = \left( \tau_1 \times \cdots \times \tau_m \to \mathbf{0} \right),
\end{equation}
with a corresponding predicate constant symbol ${c^{\prime}}$ of order 1 in the logic ${\mathcal{L}_{sort}^{1}}$:

\begin{equation}
c^{\prime} \in \mathcal{PC} \left( \mathcal{L}_{sort}^{1} \right), \qquad \mathbf{O} \left( c^{\prime} \right) = 1, \qquad \mathbf{A} \left( c^{\prime} \right) = m, \qquad \mathbf{T} \left( c^{\prime} \right) = \tau^{\prime} = \left( \mathbf{1} \times \cdots \times \mathbf{1} \to \mathbf{0} \right),
\end{equation}
such that:

\begin{equation}
\mathbf{S} \left( c^{\prime} \right) = \left( \text{``} \tau_1 \text{''}, \dots, \text{``} \tau_m \text{''} \right),
\end{equation}
where, as above, ${\mathbf{0}}$ is the type of order 1 designating the type of \textit{truth values}, and ${\mathbf{1}}$ is the type of order 0 designating the type of \textit{individuals}. Moreover, we associate all other constant and variable symbols $c$ whose orders are less than $n$ in the logic ${\mathcal{L}^{2n - 1}}$:

\begin{equation}
c \in \mathcal{C} \left( \mathcal{L}^{2n - 1} \right) \cup \mathcal{V} \left( \mathcal{L}^{2n - 1} \right), \qquad \mathbf{O} \left( c \right) < n, \qquad \mathbf{T} \left( c \right) = \sigma,
\end{equation}
with corresponding constant and variable symbols ${c^{\prime}}$ of order 1 in the logic ${\mathcal{L}_{sort}^{1}}$:

\begin{equation}
c^{\prime} \in \mathcal{C} \left( \mathcal{L}_{sort}^{1} \right) \cup \mathcal{V} \left( \mathcal{L}_{sort}^{1} \right), \qquad \mathbf{O} \left( c^{\prime} \right) = 1, \qquad \mathbf{T} \left( c^{\prime} \right) = \mathbf{1}, \qquad \mathbf{S} \left( c^{\prime} \right) = \text{``} \sigma \text{''}.
\end{equation}
In fact, since (by definition) all sets in the signature ${S^{2n - 1}}$ are disjoint, there is actually no need to use different symbols $c$ and ${c^{\prime}}$ for the inputs and outputs of ${\hat{\Theta}_{2n - 1}}$, since there is no risk of ambiguity. We also associate all types ${\tau}$ whose orders are less than $n$ in the logic ${\mathcal{L}^{2n - 1}}$ of the form:

\begin{equation}
\mathbf{O} \left( \tau \right) = \mathbf{O} \left( \left( \tau_1 \times \cdots \times \tau_m \to \mathbf{0} \right) \right) < n,
\end{equation}
with corresponding predicate constants ${\boldsymbol\alpha^{\text{``} \tau \text{''}}}$ (with ${\boldsymbol\alpha}$ designating the application symbol, as usual) of order 1 in the logic ${\mathcal{L}_{sort}^{1}}$:

\begin{equation}
\boldsymbol\alpha^{\text{``} \tau \text{''}} \in \mathcal{PC} \left( \mathcal{L}_{sort}^{1} \right), \qquad \mathbf{O} \left( \boldsymbol\alpha^{\text{``} \tau \text{''}} \right) = 1, \qquad \mathbf{A} \left( \boldsymbol\alpha^{\text{``} \tau \text{''}} \right) = m + 1, \qquad \mathbf{S} \left( \boldsymbol\alpha^{\text{``} \tau \text{''}} \right) = \left( \text{``} \tau \text{''}, \text{``} \tau_1 \text{''}, \dots, \text{``} \tau_m \text{''} \right),
\end{equation}
as well as all types ${\tau}$ whose orders are less than $n$ in the logic ${\mathcal{L}^{2n - 1}}$ of the alternative form:

\begin{equation}
\mathbf{O} \left( \tau \right) = \mathbf{O} \left( \left( \tau_1 \times \cdots \tau_m \to \sigma \right) \right) < n, \qquad \text{ where } \sigma \neq \mathbf{0},
\end{equation}
with corresponding function constants ${\boldsymbol\alpha^{\text{``} \tau \text{''}}}$ of order 1 in the logic ${\mathcal{L}_{sort}^{1}}$:

\begin{equation}
\boldsymbol\alpha^{\text{``} \tau \text{''}} \in \mathcal{FC} \left( \mathcal{L}_{sort}^{1} \right), \qquad \mathbf{O} \left( \boldsymbol\alpha^{\text{``} \tau \text{''}} \right), \qquad \mathbf{A} \left( \boldsymbol\alpha^{\text{``} \tau \text{''}} \right) = m + 1, \qquad \mathbf{S} \left( \boldsymbol\alpha^{\text{``} \tau \text{''}} \right) = \left( \text{``} \tau \text{''}, \text{``} \tau_1 \text{''}, \dots, \text{``} \tau_m \text{''} \right) : \text{``} \sigma \text{''}.
\end{equation}

We are now able to construct the quasi-homomorphism ${\hat{\Theta}_{2n - 1}^{\prime}}$ by means of the following inductively-defined rules, firstly for all terms in the logic ${\mathcal{L}^{2n - 1}}$:

\begin{equation}
\forall x_{\tau} \in \mathcal{V} \left( \mathcal{L}^{2n - 1} \right), \qquad \hat{\Theta}_{2n - 1}^{\prime} \left( x_{\tau} \right) = x^{\text{``} \tau \text{''}},
\end{equation}
\begin{multline}
\forall c_{\tau} \in \mathcal{C} \left( \mathcal{L}^{2n - 1} \right), \text{ such that } \mathbf{O} \left( c_{\tau} \right) = n \text{ and } \tau = \left( \tau_1 \times \cdots \times \tau_m \to \mathbf{0} \right),\\
\hat{\Theta}_{2n - 1}^{\prime} \left( c_{\tau} \right) = c^{\left( \text{``} \tau_1 \text{''}, \dots, \text{``} \tau_m \text{''} \right)},
\end{multline}
\begin{equation}
\forall c_{\tau} \in \mathcal{C} \left( \mathcal{L}^{2n - 1} \right), \text{ such that } \mathbf{O} \left( c_{\tau} \right) < n, \qquad \hat{\Theta}_{2n - 1}^{\prime} \left( c_{\tau} \right) = c^{\text{``} \tau \text{''}},
\end{equation}
and:

\begin{multline}
\forall t \in \mathcal{T} \left( \mathcal{L}^{2n - 1} \right), \text{ such that } \mathbf{TE} \left( t \right) = f, f \in \mathcal{FT} \left( \mathcal{L}^{2n - 1} \right) \text{ and } \mathbf{A} \left( f \right) = m,\\
\hat{\Theta}_{2n - 1}^{\prime} \left( f \left( t_1, \dots, t_n \right) \right) = \boldsymbol\alpha^{\text{``} \tau \text{''}} \left( \hat{\Theta}_{2n - 1}^{\prime} \left( f \right), \hat{\Theta}_{2n - 1}^{\prime} \left( t_1 \right), \dots, \hat{\Theta}_{2n - 1}^{\prime} \left( t_m \right) \right);
\end{multline}
and secondly for all formulas in the logic ${\mathcal{L}^{2n - 1}}$:

\begin{multline}
\forall f \in \mathcal{F} \left( \mathcal{L}^{2n - 1} \right), \text{ such that } \mathbf{AF} \left( f \right), \mathbf{TE} \left( f \right) = p, p \in \mathcal{PC} \left( \mathcal{L}^{2n - 1} \right) \text{ and } \mathbf{O} \left( p \right) = n,\\
\hat{\Theta}_{2n - 1}^{\prime} \left( p \left( t_1, \dots, t_m \right) \right) = \hat{\Theta}_{2n - 1}^{\prime} \left( p \right) \left( \hat{\Theta}_{2n - 1}^{\prime} \left( t_1 \right), \dots, \hat{\Theta}_{2n - 1}^{\prime} \left( t_m \right) \right),
\end{multline}
and:

\begin{multline}
\forall t \in \mathcal{T} \left( \mathcal{L}^{2n - 1} \right), \text{ such that } \mathbf{TE} \left( t \right) = p, p \in \mathcal{PT} \left( \mathcal{L}^{2n - 1} \right), \mathbf{O} \left( p \right) < n, \mathbf{A} \left( p \right) = m, \mathbf{T} \left( p \right) = \tau,\\
\hat{\Theta}_{2n - 1}^{\prime} \left( p \left( t_1, \dots, t_m \right) \right) = \boldsymbol\alpha^{\text{``} \tau \text{''}} \left( \hat{\Theta}_{2n - 1}^{\prime} \left( p \right), \hat{\Theta}_{2n - 1}^{\prime} \left( t_1 \right), \dots, \hat{\Theta}_{2n - 1}^{\prime} \left( t_m \right) \right),
\end{multline}
with the value of ${\hat{\Theta}_{2n - 1}^{\prime}}$ for all other formulas in the logic ${\mathcal{L}^{2n - 1}}$ defined by means of homomorphism using the above rules. The signature-dependent extensionality axioms ${EXT_{2n - 1}}$  are then given by the following pair of formulas in the first-order (multisorted) logic ${\mathcal{L}_{sort}^{1}}$:

\begin{multline}
\forall \boldsymbol\alpha^{\text{``} \tau \text{''}} \in \mathcal{FC} \left( \mathcal{L}_{sort}^{1} \right), \text{ such that } \tau = \left( \tau_1 \times \cdots \times \tau_m \to \sigma \right),\\
\forall f^{\text{``} \tau \text{''}} : \forall g^{\text{``} \tau \text{''}} : \left( \forall x_{1}^{\text{``} \tau_1 \text{''}} : \cdots : \forall x_{m}^{\text{``} \tau_m \text{''}} : \boldsymbol\alpha^{\text{``} \tau \text{''}} \left( f, x_1, \dots, x_m \right) =^{\left( \text{``} \sigma \text{''}, \text{``} \sigma \text{''} \right)} \boldsymbol\alpha^{\text{``} \tau \text{''}} \left( g, x_1, \dots, x_m \right) \right)\\
\implies f =^{\left( \text{``} \tau \text{''}, \text{``} \tau \text{''} \right)} g,
\end{multline}
and:

\begin{multline}
\forall \boldsymbol\alpha^{\text{``} \tau \text{''}} \in \mathcal{PC} \left( \mathcal{L}_{sort}^{1} \right), \text{ such that } \tau = \left( \tau_1 \times \cdots \times \tau_m \to \sigma \right),\\
\forall p^{\text{``} \tau \text{''}} : \forall q^{\text{``} \tau \text{''}} : \left( \forall x_{1}^{\text{``} \tau_1 \text{''}} : \cdots : \forall x_{m}^{\text{``} \tau_m \text{''}} : \boldsymbol\alpha^{\text{``} \tau \text{''}} \left( p, x_1, \dots, x_m \right) \Leftrightarrow \boldsymbol\alpha^{\text{``} \tau \text{''}} \left( q, x_1, \dots, x_m \right) \right)\\
\implies p =^{\left( \text{``} \tau \text{''}, \text{``} \tau \text{''} \right)} q,
\end{multline}
such that our desired morphism ${\hat{\Theta}_{2n - 1}}$ is indeed given by:

\begin{equation}
\hat{\Theta}_{2n - 1} \left( \varphi \right) = \hat{\Theta}_{2n - 1}^{\prime} \left( \varphi \right) \cup EXT_{2n - 1},
\end{equation}
as required.

Finally, we need to define precisely what it means for this family of morphisms ${\hat{\Theta}_{n}}$ to be complete and sound, which requires also providing a formal definition of \textit{models}, otherwise known as \textit{interpretations}, for higher-order (ordinal) logics, as follows:

\begin{definition}
A ``frame'', denoted ${\left\lbrace \mathcal{D}_{\tau} \right\rbrace_{\tau}}$, is a collection of non-empty sets ${\mathcal{D}_{\tau}}$ of type ${\tau}$, such that:

\begin{equation}
\mathcal{D}_{\left( \tau_1 \times \cdots \times \tau_m \to \sigma \right)} \subseteq \mathbf{F} \left( \mathcal{D}_{\tau_1}, \dots, \mathcal{D}_{\tau_m} ; \mathcal{D}_{\sigma} \right),
\end{equation}
in which ${\mathbf{F} \left( A_1, \dots, A_m ; B \right)}$ denotes the set of all functions from ${A_1 \times \cdots \times A_m}$ to $B$ (for sets ${A_1, \dots, A_m, B}$), ${\mathcal{D}_{\mathbf{1}}}$ denotes the set of individuals, and:

\begin{equation}
\mathcal{D}_{\mathbf{0}} = \left\lbrace T, F \right\rbrace,
\end{equation}
denotes the set of truth values.
\end{definition}

\begin{definition}
A ``model'', denoted ${\mathcal{M} = \left\langle \left\lbrace \mathcal{D}_{\tau} \right\rbrace_{\tau}, \mathcal{J} \right\rangle}$, of the ordinal logic ${\mathcal{L}^{\omega}}$, is an ordered pair consisting of a frame ${\left\langle \mathcal{D}_{\tau} \right\rbrace_{\tau}}$ and a function ${\mathcal{J}}$ of the form:

\begin{equation}
\mathcal{J} : \mathcal{C} \left( \mathcal{L}^{\omega} \right) \to \mathcal{D}_{\tau},
\end{equation}
mapping constants of type ${\tau}$ in the logic ${\mathcal{L}^{\omega}}$ to elements of the set ${\mathcal{D}_{\tau}}$.
\end{definition}

\begin{definition}
An ``assignment'', denoted ${\xi}$, for a frame ${\left\lbrace \mathcal{D}_{\tau} \right\rbrace_{\tau}}$ is a function of the form:

\begin{equation}
\xi : \mathcal{V} \left( \mathcal{L}^{\omega} \right) \to \mathcal{D}_{\tau},
\end{equation}
mapping variables of type ${\tau}$ to elements of the set ${\mathcal{D}_{\tau}}$.
\end{definition}
For the sake of expressive convenience, we henceforth adopt the notation ${\xi \left[ x_{\tau} \leftarrow d \right]}$, where ${\xi}$ denotes an arbitrary assignment, ${x_{\tau}}$ denotes an arbitrary variable of type ${\tau}$, and $d$ denotes an arbitrary element of the set ${\mathcal{D}_{\tau}}$, to designate a function which is equal to ${\xi}$ everywhere except for ${x_{\tau}}$, at which point it is equal to $d$.

\begin{definition}
A model ${\mathcal{M} = \left\langle \left\lbrace \mathcal{D}_{\tau} \right\rbrace_{\tau}, \mathcal{J} \right\rangle}$ is a ``weak model'' of the ordinal logic ${\mathcal{L}^{\omega}}$, if and only if there exists a binary function ${\mathcal{V}^{\mathcal{M}}}$ mapping assignments ${\xi}$ and terms $t$ of type ${\tau}$ to elements of the set ${D_{\tau}}$, i.e:

\begin{equation}
\mathcal{V}_{\xi}^{\mathcal{M}} \left( t \right) \in \mathcal{D}_{\tau},
\end{equation}
such that the following constraints are satisfied for all assignments ${\xi}$ and all terms $t$ of type ${\tau}$:

\begin{equation}
\forall x_{\tau} \in \mathcal{V} \left( \mathcal{L}^{\omega} \right), \qquad \mathcal{V}_{\xi}^{\mathcal{M}} \left( x_{\tau} \right) = \xi \left( x_{\tau} \right),
\end{equation}
\begin{equation}
\forall c_{\tau} \in \mathcal{C} \left( \mathcal{L}^{\omega} \right), \qquad \mathcal{V}_{\xi}^{\mathcal{M}} \left( c_{\tau} \right) = \mathcal{J} \left( c_{\tau} \right),
\end{equation}
\begin{equation}
\forall \varphi_1, \varphi_2 \in \mathcal{F} \left( \mathcal{L}^{\omega} \right), \qquad \mathcal{V}_{\xi}^{\mathcal{M}} \left( \varphi_1 \wedge \varphi_2 \right) = \mathcal{V}_{\xi}^{\mathcal{M}} \left( \varphi_1 \right) \wedge \mathcal{V}_{\xi}^{\mathcal{M}} \left( \varphi_2 \right),
\end{equation}
\begin{equation}
\forall \varphi \in \mathcal{F} \left( \mathcal{L}^{\omega} \right), \qquad \mathcal{V}_{\xi}^{\mathcal{M}} \left( \neg \varphi \right) = \neg \mathcal{V}_{\xi}^{\mathcal{M}} \left( \varphi \right),
\end{equation}
and:

\begin{equation}
\forall \varphi \in \mathcal{F} \left( \mathcal{L}^{\omega} \right), \qquad \mathcal{V}_{\xi}^{\mathcal{M}} \left( \forall x_{\tau} : \varphi \right) = \forall d \in \mathcal{D}_{\tau} : \mathcal{V}_{\xi \left[ x_{\tau} \leftarrow d \right]}^{\mathcal{M}} \left( \varphi \right),
\end{equation}
and, for any composite term of the form ${f_{\left( \tau_1 \times \cdots \times \tau_n \to \sigma \right)} \left( t_{\tau_1}, \dots, t_{\tau_m} \right)}$, one has:

\begin{equation}
\mathcal{V}_{\xi}^{\mathcal{M}} \left( f_{\left( \tau_1 \times \dots \times \tau_n \to \sigma \right)} \left( t_{\tau_1}, \dots, t_{\tau_m} \right) \right) = \mathcal{V}_{\xi}^{\mathcal{M}} \left( f_{\left( \tau_1 \times \cdots \times \tau_n \to \sigma \right)} \right) \left( \mathcal{V}_{\xi}^{\mathcal{M}} \left( t_{\tau_1} \right), \dots, \mathcal{V}_{\xi}^{\mathcal{M}} \left( t_{\tau_m} \right) \right).
\end{equation}
\end{definition}
Weak models may otherwise be known as \textit{weak interpretations} or \textit{general models}.

\begin{definition}
A model ${\mathcal{M} = \left\langle \left\lbrace \mathcal{D}_{\tau} \right\rbrace_{\tau}, \mathcal{J} \right\rangle}$ is a ``strong model'' if it satisfies all of the requisite conditions of a weak model, and moreover, for any type ${\tau}$ of the form ${\tau = \left( \tau_1 \times \cdots \times \tau_m \to \sigma \right)}$, one has:

\begin{equation}
\mathcal{D}_{\tau} = \mathbf{F} \left( \mathcal{D}_{\tau_1}, \dots \mathcal{D}_{\tau_m} ; \mathcal{D}_{\sigma} \right).
\end{equation}
\end{definition}
Strong models may otherwise be known as \textit{strong interpretations} or \textit{standard models}.

\begin{definition}
A morphism ${\Theta : \mathcal{F}^1 \to \mathcal{F}^2}$ between two logical systems ${\mathcal{F}^1}$ and ${\mathcal{F}^2}$ is ``weakly sound'' if and only if, for every set of formulas ${\Gamma}$ in system ${\mathcal{F}^1}$, the existence of a weak model for ${\Gamma}$ in system ${\mathcal{F}^1}$ guarantees the existence of a weak model for ${\Theta \left( \Gamma \right)}$ in system ${\mathcal{F}^2}$.
\end{definition}

\begin{definition}
A morphism ${\Theta : \mathcal{F}^1 \to \mathcal{F}^2}$ between two logical systems ${\mathcal{F}^1}$ and ${\mathcal{F}^2}$ is ``strongly sound'' if and only if, for every set of formulas ${\Gamma}$ in system ${\mathcal{F}^1}$, the existence of a strong model for ${\Gamma}$ in system ${\mathcal{F}^1}$ guarantees the existence of a strong model for ${\Theta \left( \Gamma \right)}$ in system ${\mathcal{F}^2}$.
\end{definition}

\begin{definition}
A morphism ${\Theta : \mathcal{F}^1 \to \mathcal{F}^2}$ between two logical systems ${\mathcal{F}^1}$ and ${\mathcal{F}^2}$ is ``weakly complete'' if and only if, for every set of formulas ${\Gamma}$ in system ${\mathcal{F}^1}$, the existence of a weak model for ${\Theta \left( \Gamma \right)}$ in system ${\mathcal{F}^2}$ guarantees the existence of a weak model for ${\Gamma}$ in system ${\mathcal{F}^1}$.
\end{definition}

\begin{definition}
A morphism ${\Theta : \mathcal{F}^1 \to \mathcal{F}^2}$ between two logical systems ${\mathcal{F}^1}$ and ${\mathcal{F}^2}$ is ``strongly complete'' if and only if, for every set of formulas ${\Gamma}$ in system ${\mathcal{F}^1}$, the existence of a strong model for ${\Theta \left( \Gamma \right)}$ in system ${\mathcal{F}^2}$ guarantees the existence of a strong model for ${\Gamma}$ in system ${\mathcal{F}^1}$.
\end{definition}
Following the approach of Kerber\cite{kerber}, we now use the fact that any injective quasi-homomorphism ${\Theta}$:

\begin{equation}
\Theta : \mathcal{L}^{n} \left( S \right) \to \mathcal{L}_{sort}^{1} \left( S_{\Sigma} \right),
\end{equation}
from the higher-order predicate logic ${\mathcal{L}^{n} \left( S \right)}$ to the first-order (multisorted) logic ${\mathcal{L}_{sort}^{1} \left( S_{\Sigma} \right)}$, is necessarily weakly sound, in order to conclude that our family of morphisms ${\hat{\Theta}_{n}}$ must indeed be both weakly and strongly sound, as well as weakly complete, as required.

\begin{figure}[ht]
\centering
\includegraphics[width=0.995\textwidth]{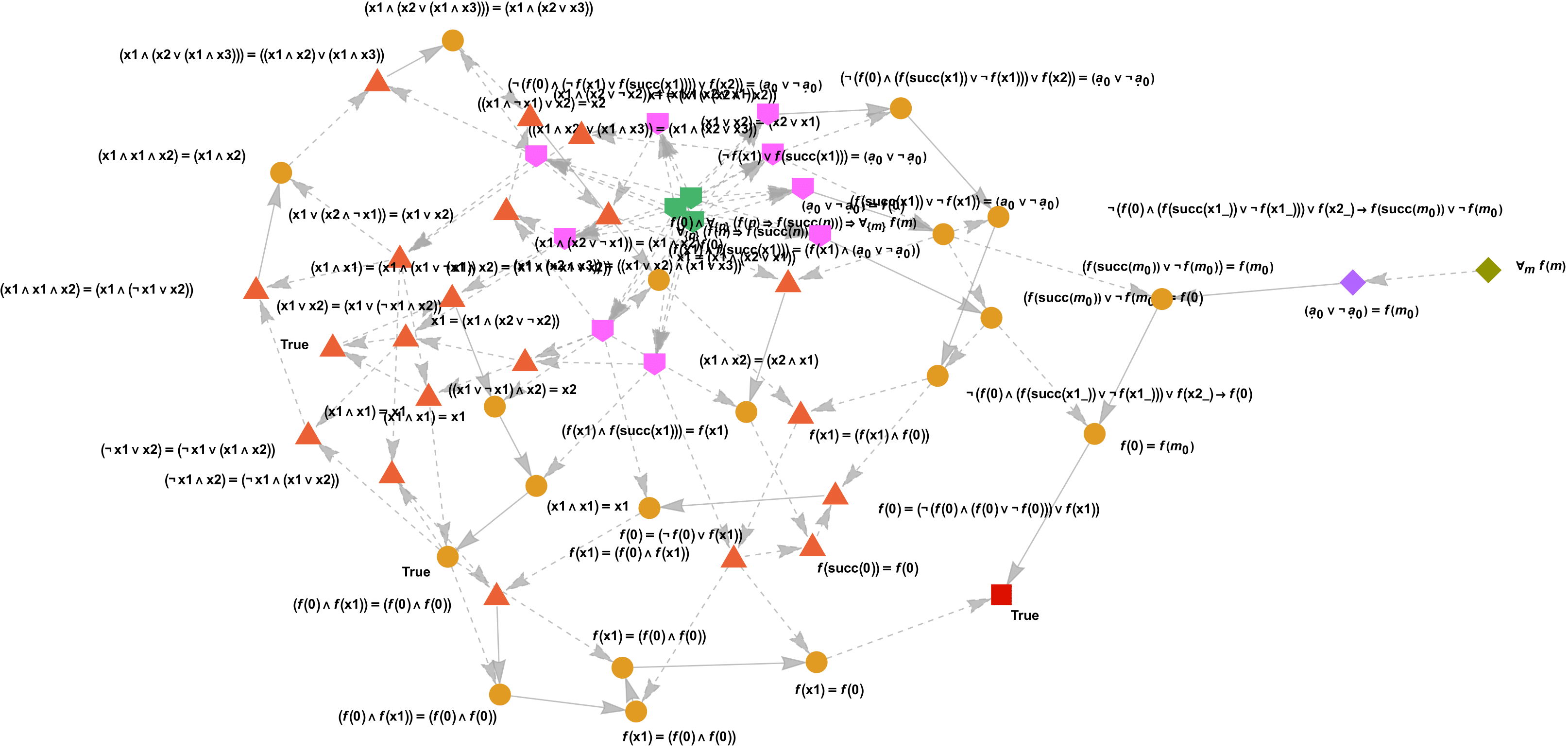}
\caption{The proof graph corresponding to the proof of the higher-order (second-order) proposition ${\forall m : f \left( m \right)}$ (derived by induction) using the higher-order (second-order) induction axiom ${\forall \phi : \left( \phi \left( 0 \right) \wedge \left( \forall n : \left( \phi \left( n \right) \implies \phi \left( succ \left( n \right) \right) \right) \right) \right) \implies \forall m : \phi \left( m \right)}$, along with the ordinary first-order axioms ${f \left( 0 \right)}$ and ${\forall n : \left( f \left( n \right) \implies f \left( succ \left( n \right) \right) \right)}$. Here, pointed light green boxes represent axioms, pointed pink boxes represent \textit{Skolemized} axioms, dark green orangle triangles represent critical pair lemmas (i.e. instances of completions/superpositions/paramodulations), light orange circles represent substitution lemmas (i.e. instances of resolutions/factorings), and dark green diamonds represent hypotheses. Solid lines represent substitutions, and dashed lines represent derived inference rules.}
\label{fig:Figure4}
\end{figure}

An example of an automatically-generated proof graph for a simple theorem in a full higher-order (specifically, second-order) logic, namely an application of the induction axiom from Peano's axioms, deriving the formula ${\forall m : f \left( x \right)}$ using the higher-order (second-order) induction axiom:

\begin{equation}
\forall \phi : \left( \phi \left( 0 \right) \wedge \left( \forall n : \left( \phi \left( n \right) \implies \phi \left( succ \left( n \right) \right) \right) \right) \right) \implies \forall m : \phi \left( m \right),
\end{equation}
combined with the ordinary first-order axioms ${f \left( 0 \right)}$ and ${\forall n : \left( f \left( n \right) \implies f \left( succ \left( n \right) \right) \right)}$, is shown in Figure \ref{fig:Figure4}. As previously, the main proof has been generated using the rules of equality resolution, ordered factoring, left superposition, right superposition and merging paramodulation (without selection), with pointed light green boxes designating underlying axioms, dark orange triangles designating critical pair lemmas (applications of completions/superpositions/paramodulations), light orange circles designating substitution lemmas (applications of resolutions/factoring instances), dark green diamonds representing hypotheses being proven, and pointed pink boxes to designate \textit{Skolemized} versions of the axioms after they have been pre-converted from higher-order axioms in ${\mathcal{L}^{2}}$ to axioms in the first-order (multisorted) logic ${\mathcal{L}_{sort}^{1}}$. As above, solid lines are used to indicate the application of a substitution lemma, with dashed lines indicating the use of a derived inference rule, and Skolem symbols denoted by ${c_i}$ or ${C_i}$ with ${i \in \mathbb{N}}$.

\clearpage

\section{The Wolfram Model and Categorical Hypergraph Rewriting}
\label{sec:Section3}

\begin{figure}[ht]
\centering
\includegraphics[width=0.295\textwidth]{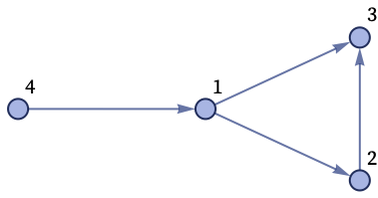}\hspace{0.1\textwidth}
\includegraphics[width=0.295\textwidth]{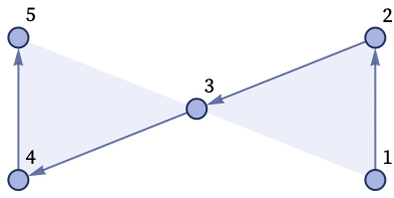}
\caption{Two spatial hypergraphs, represented as the finite collections of (un)ordered relations ${\left\lbrace \left\lbrace 1, 2 \right\rbrace, \left\lbrace 1, 3 \right\rbrace, \left\lbrace 2, 3 \right\rbrace, \left\lbrace 4, 1 \right\rbrace \right\rbrace}$ and ${\left\lbrace \left\lbrace 1, 2, 3 \right\rbrace, \left\lbrace 3, 4, 5 \right\rbrace \right\rbrace}$, respectively. Example taken from \cite{wolfram2}.}
\label{fig:Figure5}
\end{figure}

\begin{figure}[ht]
\centering
\includegraphics[width=0.395\textwidth]{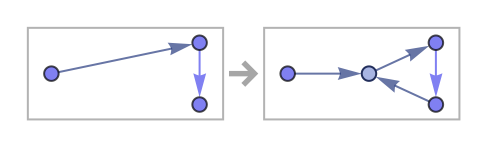}
\caption{A simple hypergraph rewriting rule, represented as the set substitution rule ${\left\lbrace \left\lbrace x, y \right\rbrace, \left\lbrace y, z \right\rbrace \right\rbrace \to \left\lbrace \left\lbrace w, y \right\rbrace, \left\lbrace y, z \right\rbrace, \left\lbrace z, w \right\rbrace, \left\lbrace x, w \right\rbrace \right\rbrace}$. Example taken from \cite{wolfram2}.}
\label{fig:Figure6}
\end{figure}

\begin{figure}[ht]
\centering
\includegraphics[width=0.695\textwidth]{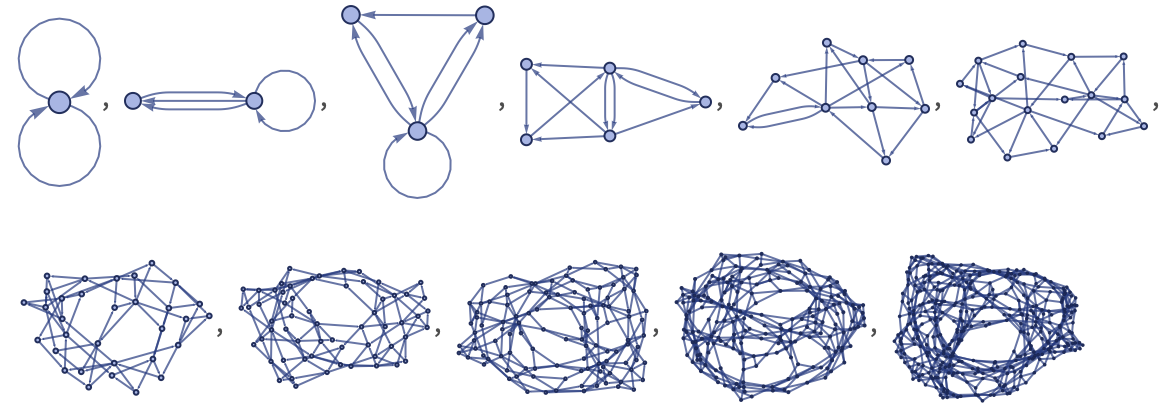}
\caption{The first 10 steps in the evolution history for the set substitution rule ${\left\lbrace \left\lbrace x, y \right\rbrace, \left\lbrace y, z \right\rbrace \right\rbrace \to \left\lbrace \left\lbrace w, y \right\rbrace, \left\lbrace y, z \right\rbrace, \left\lbrace z, w \right\rbrace, \left\lbrace x, w \right\rbrace \right\rbrace}$, assuming an initial condition consisting of a double self-loop. Example taken from \cite{wolfram2}.}
\label{fig:Figure7}
\end{figure}

\begin{figure}[ht]
\centering
\includegraphics[width=0.495\textwidth]{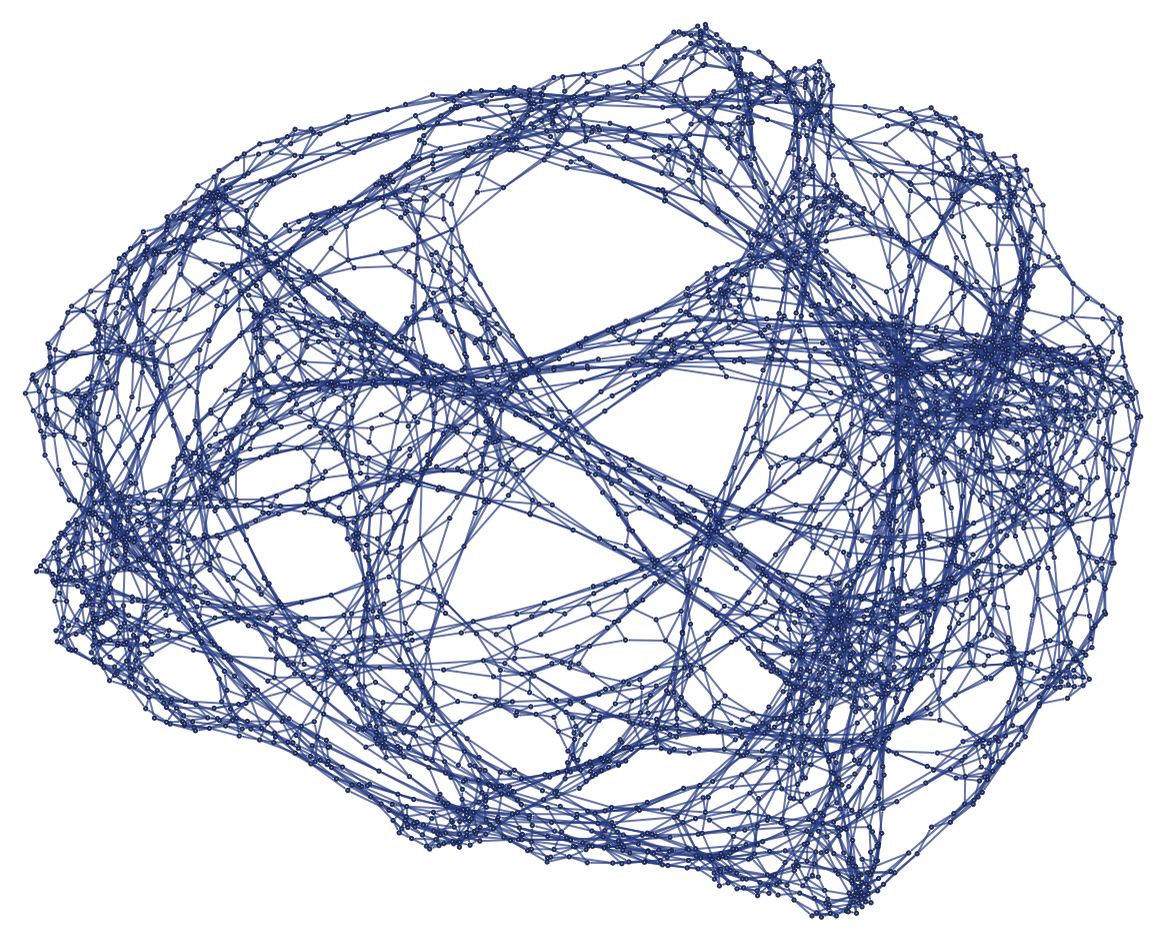}
\caption{The 14th step in the evolution history for the set substitution rule ${\left\lbrace \left\lbrace x, y \right\rbrace, \left\lbrace y, z \right\rbrace \right\rbrace \to \left\lbrace \left\lbrace w, y \right\rbrace, \left\lbrace y, z \right\rbrace, \left\lbrace z, w \right\rbrace, \left\lbrace x, w \right\rbrace \right\rbrace}$, assuming an initial condition consisting of a double self-loop. Example taken from \cite{wolfram2}.}
\label{fig:Figure8}
\end{figure}

The term rewriting techniques presented above (which effectively take place on rooted expression trees) can then be extended to the more general case of arbitrary graph and hypergraph rewriting using the formalism of the \textit{Wolfram model}, in which the basic objects of investigation are finite, undirected \textit{spatial hypergraphs} of the form ${H = \left( V, E \right)}$\cite{gorard}\cite{gorard2}, with:

\begin{equation}
E \subset \mathcal{P} \left( V \right) \setminus \left\lbrace \emptyset \right\rbrace,
\end{equation}
for ${\mathcal{P} \left( V \right)}$ the power set of the vertex set $V$. An illustrative example of a representation of two spatial hypergraphs as finite collections of (potentially ordered) relations between abstract elements is shown in Figure \ref{fig:Figure5}. The dynamics of a Wolfram model system can then be defined in terms of hypergraph rewriting rules (\textit{update rules}) of the form ${H_1 = \left( V_1, E_1 \right) \to H_2 = \left(  V_2, E_2 \right)}$, in which a subhypergraph matching the pattern ${H_1 = \left( V_1, E_1 \right)}$ is replaced by a distinct subhypergraph matching the pattern ${H_2 = \left( V_2, E_2 \right)}$\cite{gorard3}\cite{gorard4}; a concrete representation of a simple hypergraph rewriting rule as a set substitution rule, in which subsets of ordered relations matching particular patterns are replaced by distinct subsets of ordered relations matching different patterns, is shown in Figure \ref{fig:Figure6}. An example evolution of such a Wolfram model system, in which the substitution rule is applied to every possible matching and non-overlapping subhypergraph simultaneously, is shown in Figures \ref{fig:Figure7} and \ref{fig:Figure8}.

In order to represent such rewriting rules within a purely category-theoretic framework, we begin by introducing the category ${\mathbf{G}}$, to allow for the description of directed (multi-)graphs:

\begin{equation}
\begin{tikzcd}
E \arrow[r, bend left, "s"] \arrow[r, swap, bend right, "t"] & V
\end{tikzcd},
\end{equation}
where $E$ and $V$ are objects in ${\mathrm{ob} \left( \mathbf{G} \right)}$ representing edges and vertices respectively, and $s$ and $t$ are morphisms in ${\mathrm{hom} \left( \mathbf{G} \right)}$ mapping from a given edge to its source and target vertices, respectively. For simplicity, we shall present the construction for the case of (multi-)graphs, although it is easy to see how these techniques can be straightforwardly extended to the more general (multi-)hypergraph case by allowing for an arbitrary number of morphisms between objects $E$ and $C$ in ${\mathrm{ob} \left( \mathbf{G} \right)}$, corresponding to the description of hyperedges of arbitrary arity. Any given (multi-)graph $G$ is then simply a functor of the form:

\begin{equation}
G : \mathbf{G} \to \mathbf{Set},
\end{equation}
for ${\mathbf{Set}}$ the category of sets, such that the category of directed (multi-)graphs, denoted ${\mathbf{Graph}}$, is given by the \textit{functor category} ${\left[ \mathbf{G}, \mathbf{Set} \right]}$:

\begin{definition}
A ``functor category'', denoted ${\left[ \mathbf{C}, \mathbf{D} \right]}$, given categories ${\mathbf{C}}$ and ${\mathbf{D}}$, is the category whose class of objects ${\mathrm{ob} \left( \left[ \mathbf{C}, \mathbf{D} \right] \right)}$ is given by the class of functors $F$ of the form:

\begin{equation}
F : \mathbf{C} \to \mathbf{D},
\end{equation}
and whose class of morphisms ${\mathrm{hom} \left( \left[ \mathbf{C}, \mathbf{D} \right] \right)}$ is given by the class of natural transformations ${\eta}$ between such functors, of the form:

\begin{equation}
\eta : F \to G,
\end{equation}
where the functor $G$:

\begin{equation}
G : \mathbf{C} \to \mathbf{D},
\end{equation}
is taken to be an alternative object in ${\mathrm{ob} \left( \left[ \mathbf{C}, \mathbf{D} \right] \right)}$.
\end{definition}
We can see that the object ${\left[ \mathbf{C}, \mathbf{D} \right]}$ in the above definition does indeed satisfy the requisite axioms of a category, from the fact that a pair of natural transformations ${\mu}$, ${\epsilon}$:

\begin{equation}
\mu \left( A \right) : F \left( A \right) \to G \left( A \right), \qquad \text{ and } \qquad \eta \left( A \right) : G \left( A \right) \to H \left( A \right),
\end{equation}
between the pairs of functors ${F, G}$ and ${G, H}$:

\begin{equation}
F, G : \mathbf{C} \to \mathbf{D}, \qquad \text{ and } \qquad G, H : \mathbf{C} \to \mathbf{D},
\end{equation}
can be composed associatively to yield a natural transformation ${\eta \mu}$:

\begin{equation}
\eta \left( A \right) \mu \left( A \right) : F \left( A \right) \to H \left( A \right),
\end{equation}
between functors $F$ and $H$, as required. Note, however, that we must restrict ${\mathbf{C}}$ to being a \textit{small category}, in which the collections of objects ${\mathrm{ob} \left( \mathbf{C} \right)}$ and morphisms ${\mathrm{hom} \left( \mathbf{C} \right)}$ are restricted to being sets, rather than arbitrary proper classes.

As indicated by Kissinger\cite{kissinger3}\cite{kissinger4}\cite{kissinger5}, the two key conceptual difficulties that one encounters when attempting to extend term rewriting approaches to the case of arbitrary (hyper)graphs are, firstly, that it is not clear how to identify where the left-hand-side of the rule ${H_1 = \left( V_1, E_1 \right) \to H_2 = \left( V_2, E_2 \right)}$ has been applied (in the absence of any distinguished root vertex), and, secondly, that it is not clear where the right-hand-side of the rule should be ``glued'' into the background (hyper)graph. The Wolfram model approach of treating hypergraph transformations as substitution operations on sets resolves both of these difficulties in a rather elegant fashion. The solution to the first problem lies in the fact that a \textit{rule match} is always a strictly injective homomorphism of the form ${m : L \to G}$ on graphs $L$ and $G$. Any graph homomorphism of the form ${f : G \to H}$ is ultimately just a natural transformation, so it can be represented as a pair of morphisms, denoted ${f_V : V \left( G \right) \to V \left( H \right)}$ and ${f_E : E \left( G \right) \to E \left( H \right)}$ (where ${V \left( G \right)}$ and ${V \left( H \right)}$ are the vertex objects of $G$ and $H$, respectively, and ${E \left( G \right)}$ and ${E \left( H \right)}$ are the edge objects of $G$ and $H$, respectively), such that the following pair (or more, in the case of higher-arity hypergraphs) of diagrams commute:

\begin{equation}
\begin{tikzcd}
E \left( G \right) \arrow[r, "s"] \arrow[d, "f_E"] & V \left( G \right)\arrow[d, "f_V"]\\
E \left( H \right) \arrow[r, "s"] & V \left( H \right)
\end{tikzcd}, \qquad \text{ and} \qquad
\begin{tikzcd}
E \left( G \right) \arrow[r, "t"] \arrow[d, "f_E"] & V \left( G \right) \arrow[d, "f_V"]\\
E \left( H \right) \arrow[r, "t"] & V \left( H \right)
\end{tikzcd}.
\end{equation}
We are able to formalize the notion of such a \textit{rule match} in terms of \textit{spans} of \textit{monomorphisms}, as follows:

\begin{definition}
A ``span'' is any diagram consisting of two maps $f$ and $g$ which share a common domain $A$:

\begin{equation}
\begin{tikzcd}
B & A \arrow[l, "f"] \arrow[r, "g"] & C
\end{tikzcd}.
\end{equation}
\end{definition}
Therefore, a span naturally generalizes the concept of a binary relation between a pair of objects in ${\mathrm{ob} \left( \mathbf{C} \right)}$ for some category ${\mathbf{C}}$, by considering instead a triple of objects, $A$, $B$ and $C$ in ${\mathrm{ob} \left( \mathbf{C} \right)}$, along with the following pair of morphisms:

\begin{equation}
f : A \to B, \qquad \text{ and } \qquad g : A \to C,
\end{equation}
in ${\mathrm{hom} \left( \mathbf{C} \right)}$.

\begin{definition}
A ``monomorphism'' is any morphism that is left-cancellative under composition.
\end{definition}
In other words, a monomorphism is a morphism $f$ in ${\mathrm{hom} \left( \mathbf{C} \right)}$ for some category ${\mathbf{C}}$, of the form:

\begin{equation}
f : A \to B,
\end{equation}
that generalizes the concept of $f$ being an injective function, such that, for every object $C$ in ${\mathrm{ob} \left( \mathbf{C} \right)}$, and for every pair of morphisms:

\begin{equation}
g_1, g_2 : C \to A,
\end{equation}
in ${\mathrm{hom} \left( \mathbf{C} \right)}$, one has:

\begin{equation}
f \circ g_1 = f \circ g_2 \qquad \implies \qquad g_1 = g_2.
\end{equation}

\begin{definition}
A ``rewrite rule'' is a span of monorphisms ${\rho}$ of the general form:

\begin{equation}
\rho = \left( l : K \to L, r : K \to R \right).
\end{equation}
\end{definition}
Here, the left- and right-hand-sides of the rewrite rule are given by the objects $L$ and $R$, respectively.

\begin{definition}
A ``rule match'' is a morphism $m$ of the general form:

\begin{equation}
m : L \to G.
\end{equation}
\end{definition}
In other words, for a rewrite rule ${\rho}$, applied to an object (such as a graph or a hypergraph) $G$, a rule match is a morphism from the left-hand-side of the rule ${\rho}$ to the object $G$.

The second problem can be overcome by noticing that the object $K$ in the definition of the rewrite rule ${\rho}$ can be interpreted as playing the role of an \textit{invariant subgraph}, which can then be used to ``glue'' the right-hand-side $R$ to the object $G$, once the non-invariant part of object $L$ has been subtracted. This procedure can be illustrated neatly using the formalism of \textit{double-pushout} rewriting\cite{ehrig}\cite{habel}:

\begin{definition}
The ``pushout'' of a pair of morphisms $f$ and $g$ sharing a common domain:

\begin{equation}
f : C \to A, \qquad \text{ and } \qquad g : C \to B,
\end{equation}
in ${\mathrm{hom} \left( \mathbf{C} \right)}$ for some category ${\mathbf{C}}$, which we denote ${P = A +_{C} B}$, is defined by an object $P$ in ${\mathrm{ob} \left( \mathbf{C} \right)}$ and a pair of morphisms ${p_1}$ and ${p_2}$ sharing a common codomain:

\begin{equation}
p_1 : A \to P, \qquad \text{ and } \qquad p_2 : B \to P,
\end{equation}
in ${\mathrm{hom} \left( \mathbf{C} \right)}$, such that the following diagram commutes:

\begin{equation}
\begin{tikzcd}
P & B \arrow[l, "p_2"]\\
A \arrow[u, "p_1"] & C \arrow[l, "f"] \arrow[u, "g"]
\end{tikzcd},
\end{equation}
and such that the pushout triple ${\left( P, p_1, p_2 \right)}$ is universal with respect to this diagram.
\end{definition}
More concretely, the universal property states that, for any other triple ${\left( Q, q_1, q_2 \right)}$ with morphisms ${q_1}$ and ${q_2}$ sharing the same codomain of the form:

\begin{equation}
q_1 : A \to Q, \qquad \text{ and } \qquad q_2 : B \to Q,
\end{equation}
in ${\mathrm{hom} \left( \mathbf{C} \right)}$, satisfying the following compositional property:

\begin{equation}
q_1 \circ f = q_2 \circ g,
\end{equation}
there must exist a unique morphism $u$ of the form:

\begin{equation}
u : P \to Q,
\end{equation}
in ${\mathrm{hom} \left( \mathbf{C} \right)}$, such that the following pair of compositional equations are satisfied:

\begin{equation}
u \circ p_2 = q_2, \qquad \text{ and } \qquad u \circ p_1 = q_1.
\end{equation}
These compositional equations for the universal property may be summarized as the statement that, for any triple ${\left( Q, q_1, q_2 \right)}$ for which the following diagram commutes:

\begin{equation}
\begin{tikzcd}
Q\\
& P \arrow[ul, dashed, "u"] & B \arrow[l, "p_2"] \arrow[ull, bend right, "q_2"]\\
& A \arrow[u, "p_1"] \arrow[uul, bend left, "q_1"] & C \arrow[u, "g"] \arrow[l, "f"]
\end{tikzcd},
\end{equation}
there necessarily exists a unique morphism ${u : P \to Q}$ for which the same diagram also commutes.

\begin{definition}
A rewrite rule ${\rho}$ is ``applicable'' at a match $m$ if there exist a pair of pushout diagrams of the form:

\begin{equation}
\begin{tikzcd}
L \arrow[d, "m"] & K \arrow[l, "l"] \arrow[d, "n"] \arrow[r, "r"] & R \arrow[d, "p"]\\
G & D \arrow[l, "g"] \arrow[r, "h"] & H
\end{tikzcd}.
\end{equation}
\end{definition}
In other words, the rule ${\rho}$ can be applied at the match $m$ if and only if the pairs of morphisms $m$ and $g$, and $p$ and $h$, of the form:

\begin{equation}
m : L \to G, g : D \to G, \qquad \text{ and } \qquad p : R \to H, h : D \to H,
\end{equation}
in ${\mathrm{hom} \left( \mathbf{C} \right)}$, constitute pushouts of the pairs of morphisms $l$ and $n$, and $r$ and $n$, of the form:

\begin{equation}
l : K \to L, n : K \to D, \qquad \text{ and } \qquad r : K \to R, n : K \to D,
\end{equation}
in ${\mathrm{hom} \left( \mathbf{C} \right)}$, respectively. To see more concretely how this approach works, we follow Kissinger\cite{kissinger3} by observing that, if one removes the \textit{interior} of graph $K$ (where the \textit{interior} designates the graph-theoretic subtraction ${L - K}$ consisting of graph $G$ with graph $K$ excised, but with all edges between graph ${L - K}$ and graph $K$ still remaining), then one obtains the unique graph $D$ such that the following diagram is a pushout:

\begin{equation}
\begin{tikzcd}
K \arrow[r, "l"] \arrow[d, "n"] & L \arrow[d, "m"]\\
D \arrow[r, "g"] & G
\end{tikzcd},
\end{equation}
i.e. the graph $D$ is the unique graph such that $G$ can be obtained by ``gluing'' together graphs $L$ and $D$ along graph $K$. Graph $D$ is therefore the \textit{pushout complement} of the pair of morphisms ${l : K \to L}$ and ${m : L \to G}$:

\begin{definition}
The ``pushout complement'' of a pair of morphisms $m$ and $g$ with overlapping codomain and domain:

\begin{equation}
m : C \to A, \qquad \text{ and } \qquad g : A \to D,
\end{equation}
in ${\mathrm{hom} \left( \mathbf{C} \right)}$ for some category ${\mathbf{C}}$, is a pair of morphisms $f$ and $n$ with overlapping codomain and domain:

\begin{equation}
f : C \to B, \qquad \text{ and } \qquad n : B \to D,
\end{equation}
in ${\mathrm{hom} \left( \mathbf{C} \right)}$, such that the following diagram commutes, and is a pushout:

\begin{equation}
\begin{tikzcd}
C \arrow[r, "m"] \arrow[d, "f"] & A \arrow[d, "g"]\\
B \arrow[r, "n"] & D
\end{tikzcd}.
\end{equation}
\end{definition}
Note the strong formal analogy between pushout complements of pairs of morphisms and Knuth-Bendix completions of critical pairs. With the pushout complement graph $D$ and the morphism ${n : K \to D}$ thus defined, we can ``glue'' the right-hand-side of the rule $R$ to graph $D$ by means of a second pushout of the form:

\begin{equation}
\begin{tikzcd}
K \arrow[r, "r"] \arrow[d, "n"] & R \arrow[d, "p"]\\
D \arrow[r, "h"] & H
\end{tikzcd},
\end{equation}
thus yielding the full double-pushout diagram (consisting of both the initial pushout complement, and the subsequent ordinary pushout):

\begin{equation}
\begin{tikzcd}
L \arrow[d, "m"] & K \arrow[l ,"l"] \arrow[d, "n"] \arrow[r, "r"] & R \arrow[d, "p"]\\
G & D \arrow[l, "g"] \arrow[r, "h"] & H
\end{tikzcd}.
\end{equation}
Since these pushouts and their complements are only defined up to the graph isomorphism relation ${\cong}$, we see that if ${\to_{R}}$ denotes the rewrite relation on graphs, then one has:

\begin{equation}
G \cong G^{\prime} \text{ and } H \cong H^{\prime} \qquad \implies \qquad \left( \left( G \to_{R} H \right) \Leftrightarrow \left( G^{\prime} \to_{R} H^{\prime} \right) \right).
\end{equation}

Although the double-pushout rewriting approach is developed above for the category of directed (multi-)graphs ${\mathbf{Graph}}$, we also wish to perform rewritings across more general combinatorial objects, such as hypergraphs (in the case of ordinary Wolfram model evolution) and labeled open graphs (in the case of ZX-diagram rewriting). For this purpose, it is helpful to abstract certain properties of the category ${\mathbf{Graph}}$ in order to obtain the more general concept of an \textit{adhesive category}\cite{lack}\cite{ehrig2}, as defined in terms of \text{van-Kampen squares} and compatibility conditions for \textit{pullbacks}:

\begin{definition}
The ``pullback'' of a pair of morphisms $f$ and $g$ sharing a common codomain:

\begin{equation}
f : A \to C, \qquad \text{ and } \qquad g : B \to C,
\end{equation}
in ${\mathrm{hom} \left( \mathbf{C} \right)}$ for some category ${\mathbf{C}}$, which we denote ${P = A \times_{C} B}$, is defined by an object $P$ in ${\mathrm{ob} \left( \mathbf{C} \right)}$ and a pair of morphisms ${p_1}$ and ${p_2}$ sharing a common domain:

\begin{equation}
p_1 : P \to A, \qquad \text{ and } \qquad p_2 : P \to B,
\end{equation}
in ${\mathrm{hom} \left( \mathbf{C} \right)}$, such that the following diagram commutes:

\begin{equation}
\begin{tikzcd}
P \arrow[r, "p_2"] \arrow[d, "p_1"] & B \arrow[d, "g"]\\
A \arrow[r, "f"] & C
\end{tikzcd},
\end{equation}
and such that the pullback triple ${\left( P, p_1, p_2 \right)}$ is universal with respect to this diagram.
\end{definition}
In this regard, a pullback is just the categorical dual of a pushout. More concretely, the universal property states that, for any other triple ${\left( Q, q_1, q_2 \right)}$ with morphisms ${q_1}$ and ${q_2}$ sharing the same domain of the form:

\begin{equation}
q_1 : Q \to A, \qquad \text{ and } \qquad q_2 : Q \to B,
\end{equation}
in ${\mathrm{hom} \left( \mathbf{C} \right)}$, satisfying the compositional property:

\begin{equation}
f \circ q_1 = g \circ q_2,
\end{equation}
there must exist a unique morphism $u$ of the form:

\begin{equation}
u : Q \to P,
\end{equation}
in ${\mathrm{hom} \left( \mathbf{C} \right)}$, such that the following pair of compositional equations are satisfied:

\begin{equation}
p_2 \circ u = q_2, \qquad \text{ and } \qquad p_1 \circ u = q_1.
\end{equation}
These compositional equations for the universal property may be summarized as the statement that, for any triple ${\left( Q, q_1, q_2 \right)}$ for which the following diagram commutes:

\begin{equation}
\begin{tikzcd}
Q \arrow[rd, dashed, "u"] \arrow[rrd, bend left, "q_2"] \arrow[rdd, bend right, "q_1"]\\
& P \arrow[r, "p_2"] \arrow[d, "p_1"] & B \arrow[d, "g"]\\
& A \arrow[r, "f"] & C
\end{tikzcd},
\end{equation}
there necessarily exists a unique morphism ${u : Q \to P}$ for which the same diagram also commutes.

\begin{definition}
A ``van-Kampen square'' is a pushout of a span (i.e. a pair of morphisms $g$ and $f$ sharing a common domain) of the form:

\begin{equation}
g : A \to B, \qquad \text{ and } \qquad f : A \to C,
\end{equation}
in ${\mathrm{hom} \left( \mathbf{C} \right)}$, i.e. it is a pair of morphisms ${f^{\prime}}$ and ${g^{\prime}}$ sharing a common codomain:

\begin{equation}
f^{\prime} : B \to D, \qquad \text{ and } \qquad g^{\prime} : C \to D,
\end{equation}
in ${\mathrm{hom} \left( \mathbf{C} \right)}$, such that, for every commutative diagram of the form:

\begin{equation}
\begin{tikzcd}
B^{\prime} \arrow[dr, "h_B"] \arrow[ddd, "f_{h}^{\prime}"] & & & A^{\prime} \arrow[lll, "g_h"] \arrow[dl, "h_A"] \arrow[ddd, "f_h"]\\
& B \arrow[d, "f^{\prime}"] & A \arrow[l, "g"] \arrow[d, "f"] &\\
& D & C \arrow[l, "g^{\prime}"] &\\
D^{\prime} \arrow[ur, "h_D"] & & & C^{\prime} \arrow[lll, "g_{h}^{\prime}"] \arrow[ul, "h_C"]
\end{tikzcd},
\end{equation}
in which the pair of subdiagrams:

\begin{equation}
\begin{tikzcd}
B^{\prime} \arrow[d, "h_B"] & A^{\prime} \arrow[l, "g_H"] \arrow[d, "h_A"]\\
B & A \arrow[l, "g"]
\end{tikzcd}, \qquad \text{ and } \qquad
\begin{tikzcd}
A \arrow[d, "f"] & A^{\prime} \arrow[l, "h_A"] \arrow[d, "f_h"]\\
C & C^{\prime} \arrow[l, "h_C"]
\end{tikzcd},
\end{equation}
are both pullbacks, the pushouts and pullbacks are compatible.
\end{definition}
More concretely, the compatibility condition states that, for any pair of morphisms ${f_{h}^{\prime}}$ and ${g_{h}^{\prime}}$ sharing the same codomain of the form:

\begin{equation}
f_{h}^{\prime} : B^{\prime} \to D^{\prime}, \qquad \text{ and } \qquad g_{h}^{\prime} : C^{\prime} \to D^{\prime},
\end{equation}
in ${\mathrm{hom} \left( \mathbf{C} \right)}$, these morphisms constitute the pushout of a span (i.e. a pair of morphisms ${g_h}$ and ${f_h}$ sharing a common domain) of the form:

\begin{equation}
g_h : A^{\prime} \to B^{\prime}, \qquad \text{ and } \qquad f_h : A^{\prime} \to C^{\prime},
\end{equation}
in ${\mathrm{hom} \left( \mathbf{C} \right)}$, if and only if the pair of subdiagrams:

\begin{equation}
\begin{tikzcd}
B^{\prime} \arrow[d, "f_{h}^{\prime}"] \arrow[r, "h_B"] & B \arrow[d, "f^{\prime}"]\\
D^{\prime} \arrow[r, "h_D"] & D
\end{tikzcd}, \qquad \text{ and } \qquad
\begin{tikzcd}
D & C \arrow[l, "g^{\prime}"]\\
D^{\prime} \arrow[u, "h_D"] & C^{\prime} \arrow[l, "g_{h}^{\prime}"] \arrow[u, "h_C"]
\end{tikzcd},
\end{equation}
are both pullbacks. If one has a span of the form:

\begin{equation}
\begin{tikzcd}
B & A \arrow[l, "g"] \arrow[r, "f"] & C
\end{tikzcd},
\end{equation}
in which either $f$ and $g$ is a monomorphism, then we shall refer to its pushout:

\begin{equation}
\begin{tikzcd}
A \arrow[r, "g"] \arrow[d, "f"] & B \arrow[d, "f^{\prime}"]\\
C \arrow[r, "g^{\prime}"] & D
\end{tikzcd},
\end{equation}
as a \textit{pushout along a monomorphism}.

\begin{definition}
An ``adhesive category'' is a category ${\mathbf{C}}$ that has pushouts along monomorphisms, that has pullbacks, and in which all pushouts along monomorphisms are van-Kampen squares.
\end{definition}

Following Kissinger\cite{kissinger3}, we can see how the van-Kampen square condition (and hence the associated adhesivity condition) is derived, by considering the following simple but highly instructive example. If we have an injective map $f$ from a set ${X^{\prime}}$ to a set $X$, where set $X$ can be partitioned into (not necessarily disjoint) subsets $A$ and $B$, i.e:

\begin{equation}
f : X^{\prime} \to X, \qquad \text{ where } \qquad X = A \cup B,
\end{equation}
then we can construct a pair of maps ${f_A}$ and ${f_B}$ of the form:

\begin{equation}
f_A : A^{\prime} \to A, \qquad \text{ and } \qquad f_B : B^{\prime} \to B,
\end{equation}
for sets ${A^{\prime}}$ and ${B^{\prime}}$, in such a way that ${f_A}$ and ${f_B}$ share the same value when restricted to the intersection ${A \cap B}$:

\begin{equation}
\left. f_A \right| \left( A \cap B \right) : K \to A \cap B, \qquad \text{ and } \qquad \left. f_B \right| \left( A \cap B \right) : K^{\prime} \to A \cap B,
\end{equation}
such that:

\begin{equation}
K = K^{\prime}, \qquad \text{ and } \qquad \left. f_A \right| \left( A \cap B \right) = \left. f_B \right| \left( A \cap B \right).
\end{equation}
To achieve this, we simply restrict the map ${f : X^{\prime} \to X}$ to the subsets $A$ and $B$, thus yielding maps ${f_A}$ and ${f_B}$, respectively; since these are restrictions of functions to subsets of their codomains, this operation can be described by a pair of pullbacks, such that the following diagram commutes:

\begin{equation}
\begin{tikzcd}
A^{\prime} \arrow[r, "g^{\prime}"] \arrow[d, "f_A"] & X^{\prime} \arrow[d, "f"] & B^{\prime} \arrow[l, "h^{\prime}"] \arrow[d, "f_B"]\\
A \arrow[r, "g"] & X & B \arrow[l, "h"]
\end{tikzcd}.
\end{equation}
On the other hand, to construct the map ${f : X^{\prime} \to X}$ from the restricted maps ${f_A}$ and ${f_B}$, where:

\begin{equation}
\left. f_A \right| \left( A \cap B \right) = \left. f_B \right| \left( A \cap B \right),
\end{equation}
we can again describe these restrictions in terms of a pair of pullbacks, as represented by the following commutative diagram:

\begin{equation}
\begin{tikzcd}
A^{\prime} \arrow[d, "f_A"] & A^{\prime} \arrow[l, "i^{\prime}"] \arrow[d, "f^{\prime}"] \arrow[r, "j^{\prime}"] \cap B^{\prime} & B^{\prime} \arrow[d, "f_B"]\\
A & A \cap B \arrow[l, "i"] \arrow[r, "j"] & B
\end{tikzcd},
\end{equation}
where ${K = A^{\prime} \cap B^{\prime}}$. Since we have defined the set ${X^{\prime} = A^{\prime} \cup B^{\prime}}$, we can therefore construct the map $f$ using:

\begin{equation}
f \left( x \right) = \begin{cases}
f_A \left( x \right), \qquad & \text{ if } x \in A,\\
f_B \left( x \right), \qquad & \text{ if } x \in B,
\end{cases}
\end{equation}
thus ensuring that $f$ is perfectly well-defined (since in cases where ${x \in A}$ and ${x \in B}$, ${f_A \left( x \right) = f_B \left( x \right)}$ by definition). As the map $f$ is effectively a ``continuation'' of maps ${f_A}$ and ${f_B}$ to supersets of their codomains, this operation can be described elegantly as a pair of pushouts, such that the following diagram commutes:

\begin{equation}
\begin{tikzcd}
A^{\prime} \cap B^{\prime} \arrow[r, "j^{\prime}"] \arrow[d, "i^{\prime}"] & B^{\prime} \arrow[d, "h^{\prime}"] \arrow[dr, "f_B"]\\
A^{\prime} \arrow[r, "g^{\prime}"] \arrow[dr, "f_A"] & X^{\prime} \arrow[dr, dashed, "f"] & B \arrow[d, "h"]\\
& A \arrow[r, "g"] & X
\end{tikzcd},
\end{equation}
with $f$ therefore being the induced map of the diagram. Since these two operations are mutual inverses, we can combine the three above diagrams together to obtain the following commutative cube:

\begin{equation}
\begin{tikzcd}
A^{\prime} \arrow[rrr, "g^{\prime}"] \arrow[ddd, "f_A"] & & & X^{\prime} \arrow[ddd, "f"]\\
& A^{\prime} \cap B^{\prime} \arrow[ul, "i^{\prime}"] \arrow[d, "f^{\prime}"] \arrow[r, "j^{\prime}"] & B^{\prime} \arrow[ur, "h^{\prime}"] \arrow[d, "f_B"]\\
& A \cap B \arrow[dl, "i"] \arrow[r, "j"] & B \arrow[dr, "h"]\\
A \arrow[rrr, "g"] & & & X
\end{tikzcd},
\end{equation}
in which the bottom face is a pushout:

\begin{equation}
\begin{tikzcd}
A \cap B \arrow[r, "j"] \arrow[d, "i"] & B \arrow[d, "h"]\\
A \arrow[r, "g"] & X
\end{tikzcd},
\end{equation}
since the set ${X = A \cup B}$, and the left and back faces are both pullbacks:

\begin{equation}
\begin{tikzcd}
A^{\prime} \cap B^{\prime} \arrow[r, "f^{\prime}"] \arrow[d, "i^{\prime}"] & A \cap B \arrow[d, "i"]\\
A^{\prime} \arrow[r, "f_A"] & A
\end{tikzcd}, \qquad \text{ and } \qquad
\begin{tikzcd}
A^{\prime} \arrow[r, "g^{\prime}"] \arrow[d, "f_A"] & X^{\prime} \arrow[d, "f"]\\
A \arrow[r, "g"] & X
\end{tikzcd},
\end{equation}
since the restricted maps ${f_A}$ and ${f_B}$ are such that ${\left. f_A \right| \left( A \cap B \right) = \left. f_B \right| \left( A \cap B \right)}$. Since we know that the map $f$ will restrict to maps ${f_A}$ and ${f_B}$ on subsets $A$ and $B$ respectively, if and only if we define the set ${X^{\prime} = A^{\prime} \cup B^{\prime}}$, it follows that the front and right faces will both be pullbacks:

\begin{equation}
\begin{tikzcd}
A^{\prime} \cap B^{\prime} \arrow[r, "j^{\prime}"] \arrow[d, "f^{\prime}"] & B^{\prime} \arrow[d, "f_B"]\\
A \cap B \arrow[r, "j"] & B
\end{tikzcd}, \qquad \text{ and } \qquad
\begin{tikzcd}
B^{\prime} \arrow[d, "f_B"] \arrow[r, "h^{\prime}"] & X^{\prime} \arrow[d, "f"]\\
B \arrow[r, "h"] & X
\end{tikzcd},
\end{equation}
if and only if the top face is a pushout:

\begin{equation}
\begin{tikzcd}
A^{\prime} \cap B^{\prime} \arrow[r, "i^{\prime}"] \arrow[d, "j^{\prime}"] & A^{\prime} \arrow[d, "g^{\prime}"]\\
B^{\prime} \arrow[r, "h^{\prime}"] & X^{\prime}
\end{tikzcd},
\end{equation}
which are precisely the van-Kampen square conditions, as required.

Slice, coslice and functor categories of adhesive categories all inherit the property of adhesivity, although it is not the case in general that a full subcategory ${\mathbf{C}^{\prime}}$ of an adhesive category ${\mathbf{C}}$, i.e. a subcategory in which, for every pair of objects $A$ and $B$ in ${\mathrm{ob} \left( \mathbf{C}^{\prime} \right)}$, one has:

\begin{equation}
\mathrm{hom}_{\mathbf{C}^{\prime}} \left( A, B \right) = \mathrm{hom}_{\mathbf{C}} \left( A, B \right),
\end{equation}
will necessarily also be adhesive. We can nevertheless characterize those full subcategories that inherit sufficient adhesivity for double-pushout rewriting systems to remain definable on certain classes of spans (namely \textit{partial adhesive categories}) as follows:

\begin{definition}
A ``partial adhesive category'' is a full subcategory ${\mathbf{C}^{\prime}}$ of an adhesive category ${\mathbf{C}}$ for which the embedding functor $S$:

\begin{equation}
S : \mathbf{C}^{\prime} \to \mathbf{C},
\end{equation}
preserves monomorphisms.
\end{definition}
The category of Wolfram model hypergraphs is easily shown to be partial adhesive (though it is clearly not adhesive, since the arbitrary connectivity of vertices in a hypergraph implies that not all pushouts along monomorphisms are guaranteed to exist). We can then define the associated concepts of \textit{S-spans} and \textit{S-pushouts}, as follows:

\begin{definition}
An ``S-span'', for a partial adhesive category ${\mathbf{C}^{\prime}}$ with embedding functor ${S : \mathbf{C}^{\prime} \to \mathbf{C}}$, is any diagram consisting of two maps $f$ and $g$ which share a common domain:

\begin{equation}
\begin{tikzcd}
B & A \arrow[l, "f"] \arrow[r, "g"] & C
\end{tikzcd},
\end{equation}
for which a pushout diagram of the form:

\begin{equation}
\begin{tikzcd}
A \arrow[r, "f"] \arrow[d, "g"] & B \arrow[d, "p_1"]\\
C \arrow[r, "p_2"] & P
\end{tikzcd},
\end{equation}
exists, and is preserved by the embedding functor $S$.
\end{definition}
Such a pushout diagram for an \textit{S-span} is therefore known as an \textit{S-pushout}. The significance of \textit{S-pushouts} in partial adhesive categories is that they guarantee the uniqueness of \textit{S-pushout complements}:

\begin{definition}
An ``S-pushout complement'', for a partial adhesive category ${\mathbf{C}^{\prime}}$ with embedding functor ${S : \mathbf{C}^{\prime} \to \mathbf{C}}$, of a pair of morphisms $m$ and $g$ with overlapping domain and codomain:

\begin{equation}
m : C \to A, \qquad \text{ and } \qquad g : A \to D,
\end{equation}
in ${\mathrm{hom} \left( \mathbf{C}^{\prime} \right)}$, is a pair of morphisms $f$ and $n$ with overlapping codomain and domain:

\begin{equation}
f : C \to B, \qquad \text{ and } \qquad n : B \to D,
\end{equation}
in ${\mathrm{hom} \left( \mathbf{C}^{\prime} \right)}$, such that the following diagram commutes, and is an S-pushout:

\begin{equation}
\begin{tikzcd}
C \arrow[r, "m"] \arrow[d, "f"] & A \arrow[d, "g"]\\
B \arrow[r, "n"] & D
\end{tikzcd}.
\end{equation}
\end{definition}
From here, it is very straightforward to extend the standard double-pushout rewriting formalism to the partial adhesive case:

\begin{definition}
An ``S-rule match'', for a partial adhesive category ${\mathbf{C}^{\prime}}$ with embedding functor ${S : \mathbf{C}^{\prime} \to \mathbf{C}}$, of a rewrite rule ${\rho = \left( l : K \to L, r : K \to R \right)}$, is a monomorphism $m$ of the general form:

\begin{equation}
m : L \to G,
\end{equation}
in ${\mathrm{hom} \left( \mathbf{C}^{\prime} \right)}$, for which the following pair of morphisms $l$ and $m$ with overlapping codomain and domain:

\begin{equation}
l : K \to L, \qquad \text{ and } \qquad m : L \to G,
\end{equation}
in ${\mathrm{hom} \left( \mathbf{C}^{\prime} \right)}$, have an S-pushout complement $D$ of the form:

\begin{equation}
\begin{tikzcd}
K \arrow[r, "l"] \arrow[d, "n"] & L \arrow[d, "m"]\\
D \arrow[r, "g"] & G
\end{tikzcd},
\end{equation}
for morphisms $n$ and $g$ with overlapping codomain and domain:

\begin{equation}
n : K \to D, \qquad \text{ and } \qquad g : D \to G,
\end{equation}
in ${\mathrm{hom} \left( \mathbf{C}^{\prime} \right)}$.
\end{definition}

\begin{definition}
A rewrite rule ${\rho}$ is ``S-applicable'' at an S-rule match $m$, for a partial adhesive category ${\mathbf{C}^{\prime}}$ with embedding functor ${S : \mathbf{C}^{\prime} \to \mathbf{C}}$, if there exist a pair of S-pushout diagrams of the form:

\begin{equation}
\begin{tikzcd}
L \arrow[d, "m"] & K \arrow[l, "l"] \arrow[d, "n"] \arrow[r, "r"] & R \arrow[d, "p"]\\
G & D \arrow[l, "g"] \arrow[r, "h"] & H
\end{tikzcd}.
\end{equation}
\end{definition}

Next, we note that the evolution of an arbitrary spatial hypergraph will, in general, be non-deterministic, since Wolfram model systems lack a canonical updating order (i.e. there will generically exist many possible sets of maximally non-overlapping subhypergraphs to which the rewrite rules can be applied, and different choices of such sets will generically yield non-isomorphic sequences of spatial hypergraphs). We can parametrize this non-deterministic evolution using the formalism of \text{abstract rewriting systems}\cite{baader}\cite{bezem} in mathematical logic and \textit{multiway systems}\cite{gorard}\cite{gorard4} in Wolfram model evolution:

\begin{definition}
An ``abstract rewriting system'' (``ARS'') is a set, denoted $A$, equipped with a binary relation, denoted ${\to_{R}}$.
\end{definition}
The elements of $A$ are known as \textit{objects} of the abstract rewriting system, with the relation ${\to_{R}}$ known as the \textit{rewrite relation}.

\begin{figure}[ht]
\centering
\includegraphics[width=0.695\textwidth]{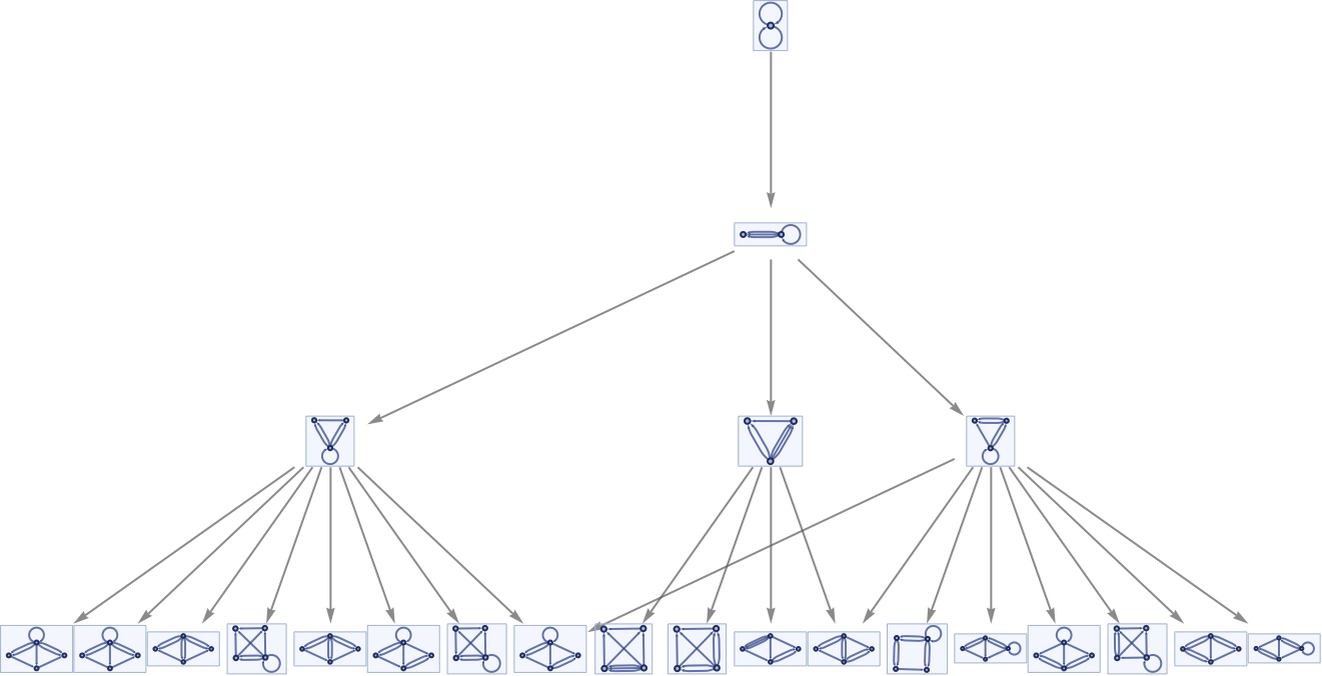}
\caption{The first 3 steps in the non-deterministic evolution history for the set substitution rule ${\left\lbrace \left\lbrace x, y \right\rbrace, \left\lbrace y, z \right\rbrace \right\rbrace \to \left\lbrace \left\lbrace w, y \right\rbrace, \left\lbrace y, z \right\rbrace, \left\lbrace z, w \right\rbrace, \left\lbrace x, w \right\rbrace \right\rbrace}$, as represented by a multiway evolution graph. Example taken from \cite{wolfram2}.}
\label{fig:Figure9}
\end{figure}

\begin{figure}[ht]
\centering
\includegraphics[width=0.495\textwidth]{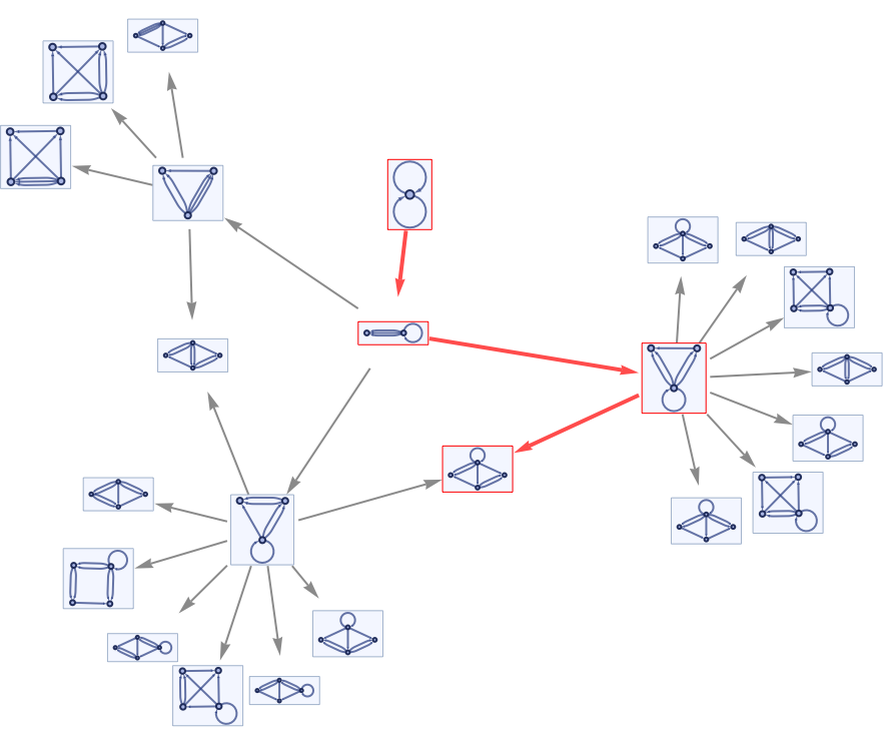}
\caption{The first 3 steps in the canonical evolution history (i.e. the evolution history with canonical updating order) for the set substitution rule ${\left\lbrace \left\lbrace x, y \right\rbrace, \left\lbrace y, z \right\rbrace \right\rbrace \to \left\lbrace \left\lbrace w, y \right\rbrace, \left\lbrace y, z \right\rbrace, \left\lbrace z, w \right\rbrace, \left\lbrace x, w \right\rbrace \right\rbrace}$, as represented by a single path in the associated multiway evolution graph. Example taken from \cite{wolfram2}.}
\label{fig:Figure10}
\end{figure}

\begin{definition}
The relation ${\to_{R}^{*}}$ denotes the reflexive transitive closure of the relation ${\to_{R}}$.
\end{definition}
More specifically, ${\to_{R}^{*}}$ is the transitive closure of the binary relation ${\to \cup =}$, where $=$ denotes the identity relation, and therefore ${\to_{R}^{*}}$ is also the smallest preorder that contains ${\to_{R}}$, i.e. ${t_{R}^{*}}$ is the smallest binary relation that contains ${\to_{R}}$ and also satisfies the axioms of reflexivity and transitivity:

\begin{equation}
\forall a, b, c \in A, \qquad a \to_{R}^{*} a, \qquad \text{ and } \qquad \left( a \to_{R}^{*} b, b \to_{R}^{*} c \right) \implies \left( a \to_{R}^{*} c \right).
\end{equation}
Concretely, we can represent the abstract rewriting structure of a Wolfram model system by means of a \textit{multiway system}\cite{gorard2}\cite{gorard3}:

\begin{definition}
A ``multiway evolution graph'', denoted ${G_{multiway} = \left( V_{multiway}, E_{multiway} \right)}$, is a directed, acyclic graph associated to an abstract rewriting system $A$, in which every vertex in ${V_{multiway}}$ corresponds to an object ${a \in A}$, and in which the directed edge ${a \to b}$ only exists in ${E_{multiway}}$ (for ${a, b \in A}$) if ${a \to_{R} b}$.
\end{definition}
In other words, directed multiway edges ${a \to b}$ indicate the existence of a single rewrite rule application that transforms object $a$ to object $b$, and directed multiway paths ${a \to^{*} b}$ indicate the existence of a finite rewrite sequence that transforms object $a$ to object $b$. Generic Wolfram model evolutions are therefore described in terms of multiway evolution graphs, with the canonical updating order shown previously hence corresponding to a single path through this graph, as demonstrated in Figures \ref{fig:Figure9} and \ref{fig:Figure10}, with state vertices merged on the basis of hypergraph isomorphism, using a generalization of the algorithm presented in \cite{gorard5}.

Such abstract rewriting systems (and, by extension, multiway evolution graphs) can be described in a purely category-theoretic fashion as \textit{F-coalgebras}, as follows. Suppose that the rewrite relation ${\to_{R}}$ for the abstract rewriting system ${\left( A, \to_{R} \right)}$ can be represented more generally as an indexed union of subrelations, i.e:

\begin{equation}
\to_{R} = \bigcup_{i} \left( \to_{i \in \Lambda} \right), \qquad \text{ such as } \qquad \to_{R} = \to_{1} \cup \to_{2},
\end{equation}
for some index set ${\Lambda}$, since there could in principle exist multiple rewrite rules within a single rewrite system. We thus obtain a labeled state transition system ${\left( A, \Lambda, \to_{R} \right)}$, which is simply a bijective map from the set $A$ to a subset of the power set of $A$, indexed by set ${\Lambda}$; namely, it is the map ${\mathcal{P} \left( \Lambda \times A \right)}$ defined by:

\begin{equation}
p \mapsto \left\lbrace \left( \alpha, q \right) \in \Lambda \times A : p \to_{R}^{\alpha} q \right\rbrace.
\end{equation}
Then, for an \textit{endofunctor} ${F : \mathbf{C} \to \mathbf{C}}$ mapping from a category ${\mathbf{C}}$ to itself, one has:

\begin{definition}
The ``F-coalgebra'', denoted ${\left( A, \alpha \right)}$, for the endofunctor $F$ of the form:

\begin{equation}
F: \mathbf{C} \to \mathbf{C},
\end{equation}
for some category ${\mathbf{C}}$, is an object $A$ in ${\mathrm{ob} \left( \mathbf{C} \right)}$, equipped with a morphism ${\alpha}$ of the form:

\begin{equation}
\alpha : A \to F A,
\end{equation}
in ${\mathrm{hom} \left( \mathbf{C} \right)}$.
\end{definition}
Any such labeled state transition system is therefore described by an F-coalgebra for the power set functor ${\mathcal{P} \left( \Lambda \times \left( - \right) \right)}$, since we can represent the power set construction on the category ${\mathbf{Set}}$ of sets as a \textit{covariant} endofunctor ${\mathcal{P}}$ (i.e. an endofunctor preserving both the identity morphisms and the composition of morphisms in ${\mathrm{hom} \left( \mathbf{C} \right)}$) of the form:

\begin{equation}
\mathcal{P} : \mathbf{Set} \to \mathbf{Set}.
\end{equation}
Thus, the abstract rewriting system ${\left( A, \to_{R} \right)}$ is simply an object $A$ equipped with an additional morphism ${\to_{R}}$ of the category ${\mathbf{Set}}$ (i.e. the rewrite relation) of the form:

\begin{equation}
\to_{R} : A \to \mathcal{P} A,
\end{equation}
as required.

This construction now allows us to generalize concepts such as \textit{confluence} and \textit{(strong) normalization}\cite{dershowitz}\cite{huet} from the theory of ordinary term rewriting systems to the much more general case of arbitrary (hyper)graph rewriting systems, albeit with the notion of equality between terms replaced with the notion of isomorphism between (hyper)graphs\cite{gorard5}:

\begin{definition}
An object ${a \in A}$ is ``confluent'' if and only if:

\begin{equation}
\forall b, c \in A, \text{ such that } a \to^{*} b \text{ and } a \to^{*} c, \qquad \exists d \in A \text{ such that } b \to^{*} d \text{ and } c \to^{*} d.
\end{equation}
\end{definition}

\begin{definition}
An abstract rewriting system is (globally) ``confluent'' if and only if every object ${a \in A}$ is confluent.
\end{definition}
For historical reasons relating to the confluence properties of certain variants of the ${\lambda}$-calculus (i.e. the \textit{Church-Rosser theorem}), confluent rewriting systems are occasionally said to exhibit the \textit{Church-Rosser property}.

\begin{definition}
An object ${a \in A}$ is a ``normal form'' if and only if:

\begin{equation}
\nexists b \in A, \qquad \text{ such that } a \to b.
\end{equation}
\end{definition}
Thus, \textit{normal forms} are defined as objects which cannot be rewritten further.

\begin{definition}
An object ${a \in A}$ is ``weakly normalizing'' if and only if:

\begin{equation}
\exists b \in A, \qquad \text{ such that } a \to^{*} b,
\end{equation}
where $b$ is a normal form.
\end{definition}
In other words, \textit{weak normalization} of an object indicates that there exists a finite rewrite sequence which reduces that object to some normal form.

\begin{definition}
An object ${a \in A}$ is ``strongly normalizing'' if and only if every finite rewrite sequence ${a \to^{*} \dots}$ eventually terminates at some normal form.
\end{definition}

\begin{definition}
An abstract rewriting system $A$ is ``weakly normalizing'' if and only if every object ${a \in A}$ is weakly normalizing.
\end{definition}

\begin{figure}[ht]
\centering
\includegraphics[width=0.395\textwidth]{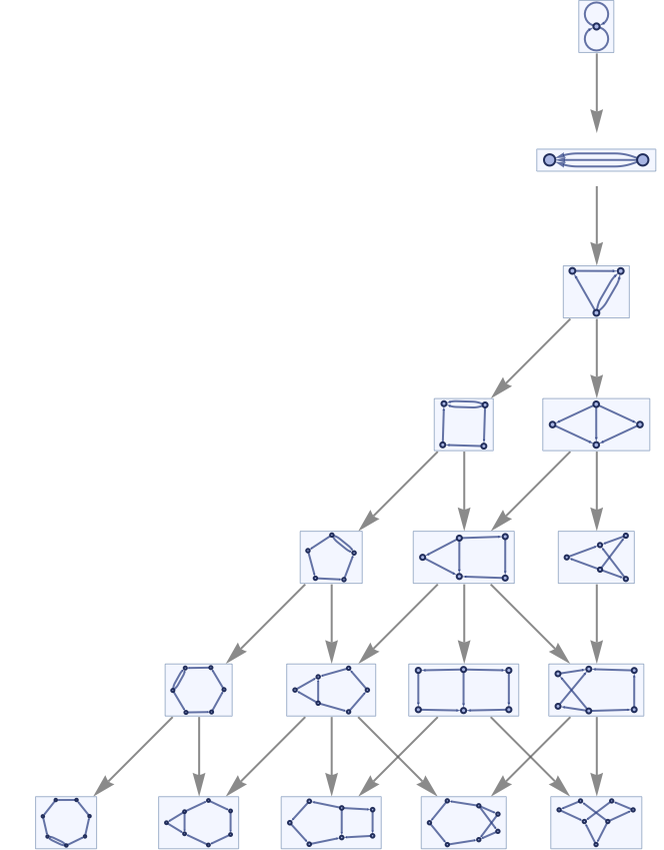}
\caption{The first 6 steps in the non-deterministic evolution history for the (globally) confluent set substitution rule ${\left\lbrace \left\lbrace x, y \right\rbrace, \left\lbrace z, y \right\rbrace \right\rbrace \to \left\lbrace \left\lbrace x, w \right\rbrace, \left\lbrace y, w \right\rbrace, \left\lbrace z, w \right\rbrace \right\rbrace}$, as represented by a multiway evolution graph in which every bifurcation merges after a single step. Example taken from \cite{wolfram2}.}
\label{fig:Figure11}
\end{figure}

\begin{definition}
An abstract rewriting system $A$ is ``strongly normalizing'' if and only if every object ${a \in A}$ is strongly normalizing.
\end{definition}
When represented as a multiway evolution graph, (global) confluence therefore implies that every bifurcation in the evolution history must eventually converge, as illustrated in Figure \ref{fig:Figure11}.

As previously discussed by the authors\cite{gorard}, the multiway evolution graph for a Wolfram model system is naturally equipped with the structure of a \textit{dagger compact closed category}, much like the category ${\mathbf{FdHilb}}$ of finite-dimensional Hilbert spaces (with morphisms described by linear maps)\cite{selinger}\cite{hasegawa}, in which the \textit{monoidal structure} is given by the disjoint union of multiway system rules, and the \textit{dagger structure} is given by the inversion of multiway evolution edges:

\begin{definition}
A ``monoidal category''\cite{kelly}\cite{kelly2}, denoted ${\left( \mathbf{C}, \otimes, I \right)}$, is a category ${\mathbf{C}}$ equipped with a bifunctor ${\otimes}$ of the form:

\begin{equation}
\otimes : \mathbf{C} \times \mathbf{C} \to \mathbf{C},
\end{equation}
satisfying the axioms of associativity (up to natural isomorphism), and also equipped with a distinguished object $I$ that acts as both a left and right identity for the bifunctor ${\otimes}$ (again, up to natural isomorphism).
\end{definition}
The bifunctor ${\otimes}$ is generally referred to as the \textit{monoidal product} or the \textit{monoidal structure}, and it generalizes the notion of a tensor product of finite-dimensional vector spaces\cite{baez}\cite{joyal}; the distinguished object $I$ is generally referred to as the \textit{identity object} of the monoidal structure, and generalizes the notion of an identity matrix. More concretely, we require that, for any triple of objects $A$, $B$ and $C$ in ${\mathrm{ob} \left( \mathbf{C} \right)}$, there exists a natural isomorphism ${\alpha}$ in ${\mathrm{hom} \left( \mathbf{C} \right)}$, with components of the form:

\begin{equation}
\alpha_{A, B, C} : A \otimes \left( B \otimes C \right) \cong \left( A \otimes B \right) \otimes C,
\end{equation}
which we can summarize neatly by means of the following commutative diagram for all tuples of objects $A$, $B$, $C$ and $D$ in ${\mathrm{ob} \left( \mathbf{C} \right)}$, corresponding to the coherence condition for the ${\alpha}$ isomorphism:

\begin{equation}
\begin{tikzcd}
A \otimes \left( B \otimes \left( C \otimes D \right) \right) \arrow[r, "\alpha_{A, B, C \otimes D}"] \arrow[d, "id_{A} \otimes \alpha_{B, C, D}"] & \left( A \otimes B \right) \otimes \left( C \otimes D \right) \arrow[r, "\alpha_{A \otimes B, C, D}"] & \left( \left( A \otimes B \right) \otimes C \right) \otimes D\\
A \otimes \left( \left( B \otimes C \right) \otimes D \right) \arrow[rr, "\alpha_{A, B \otimes C, D}"] & & \left( A \otimes \left( B \otimes C \right) \right) \otimes D \arrow[u, "\alpha_{A, B, C} \otimes id_D"]
\end{tikzcd},
\end{equation}
where ${id_X}$ denotes the identity morphism of object $X$ in ${\mathrm{ob} \left( \mathbf{C} \right)}$; this natural isomorphism ${\alpha}$ is designated the \textit{associator} isomorphism. Moreover, we require that, for every object $A$ in ${\mathrm{ob} \left( \mathbf{C} \right)}$, there exist a pair of natural isomorphisms ${\lambda}$ and ${\rho}$ in ${\mathrm{hom} \left( \mathbf{C} \right)}$, with components of the form:

\begin{equation}
\lambda_{A} : I \otimes A \cong A, \qquad \text{ and } \qquad \rho_A : A \otimes I \cong A,
\end{equation}
which we can also summarize by means of the following commutative diagram for every pair of objects $A$ and $B$ in ${\mathrm{ob} \left( \mathbf{C} \right)}$, corresponding to the coherence conditions for the ${\lambda}$ and ${\rho}$ isomorphisms:

\begin{equation}
\begin{tikzcd}
A \otimes \left( I \otimes B \right) \arrow[rr, "\alpha_{A, I, B}"] \arrow[dr, "id_{A} \otimes \lambda_{B}"] & & \left( A\otimes I \right) \otimes B \arrow[dl, "\rho_{A} \otimes id_{B}"]\\
& A \otimes B
\end{tikzcd};
\end{equation}
these natural isomorphisms ${\lambda}$ and ${\rho}$ are designated the \textit{left unitor} and \textit{right unitor} isomorphisms, respectively.

\begin{definition}
A ``symmetric monoidal category'' is a monoidal category ${\left( \mathbf{C}, \otimes, I \right)}$ in which, for every pair of objects $A$ and $B$ in ${\mathrm{ob} \left( \mathbf{C} \right)}$, there exists a natural isomorphism ${\sigma}$ in ${\mathrm{hom} \left( \mathbf{C} \right)}$ whose components are of the following form:

\begin{equation}
\sigma_{A, B} : A \otimes B \cong B \otimes A,
\end{equation}
such that ${\sigma}$ obeys an inverse law, and is suitably coherent with the associator and left and right unitor isomorphisms.
\end{definition}
The natural isomorphism ${\sigma}$ is generally referred to as the \textit{symmetry} isomorphism of the symmetric monoidal category ${\mathbf{C}}$, and it generalizes the intuitive notion that the monoidal structure (i.e. the tensor product) should be ``as commutative as possible''; its coherence condition with the associator isomorphism ${\alpha}$ may be summarized by means of the following commutative diagram for all triples of objects $A$, $B$ and $C$ in ${\mathrm{ob} \left( \mathbf{C} \right)}$:

\begin{equation}
\begin{tikzcd}
\left( A \otimes B \right) \otimes C \arrow[r, "\sigma_{A, B} \otimes id_{C}"] \arrow[d, "\alpha_{A, B, C}"] & \left( B \otimes A \right) \otimes C \arrow[d, "\alpha_{B, A, C}"]\\
A \otimes \left( B \otimes C \right) \arrow[d, "\sigma_{A, B \otimes C}"] & B \otimes \left( A \otimes C \right) \arrow[d, "id_{B} \otimes \sigma_{A, C}"]\\
\left( B \otimes C \right) \otimes A \arrow[r, "\alpha_{B, C, A}"] & B \otimes \left( C \otimes A \right)
\end{tikzcd},
\end{equation}
whilst the coherence condition with the left and right unitor isomorphisms ${\lambda}$ and ${\rho}$ may be summarized by means of the following pair of commutative diagrams for all pairs of objects $A$ and $B$ in ${\mathrm{ob} \left( \mathbf{C} \right)}$:

\begin{equation}
\begin{tikzcd}
A \otimes I \arrow[rr, "\sigma_{A, I}"] \arrow[dr, "\rho_{A}"] & & I \otimes A \arrow[dl, "\lambda_{A}"]\\
& A
\end{tikzcd} \qquad \text{ and } \qquad
\begin{tikzcd}
& B \otimes A \arrow[dr, "\sigma_{B, A}"]\\
A \otimes B \arrow[ur, "\sigma_{A, B}"] \arrow[rr, equal, "id_{A \otimes B}"] & & A \otimes B
\end{tikzcd}.
\end{equation}

\begin{definition}
A ``dagger category''\cite{burgin}\cite{lambek}, denoted ${\left( \mathbf{C}, \dag \right)}$, is a category ${\mathbf{C}}$ equipped with an involutive functor ${\dag}$ acting on the opposite/dual category ${\mathbf{C}^{op}}$, of the form:

\begin{equation}
\dag : \mathbf{C}^{op} \to \mathbf{C}.
\end{equation}
\end{definition}
The involutive functor ${\dag}$ is generally referred to as the \textit{dagger structure}, and it generalizes the operation of taking the Hermitian adjoint of a linear operator on a finite-dimensional Hilbert space. More concretely, we require that, for every morphism $f$ in ${\mathrm{hom} \left( \mathbf{C} \right)}$, we associate an adjoint morphism ${f^{\dag}}$ in ${\mathrm{hom} \left( \mathbf{C} \right)}$:

\begin{equation}
f^ : A \to B, \qquad \implies \qquad f^{\dag} : B \to A,
\end{equation}
such that the adjoint of the identity morphism ${id_{A}}$ (for any object $A$ in ${\mathrm{ob} \left( \mathbf{C} \right)}$) is always itself:

\begin{equation}
id_{A} = id_{A}^{\dag} : A \to A;
\end{equation}
the dagger structure reverses the order of composition for every pair of morphisms $f$ and $g$ in ${\mathrm{hom} \left( \mathbf{C} \right)}$:

\begin{equation}
f : A \to B \text{ and } g : B \to C, \qquad \implies \qquad \left( g \circ f \right)^{\dag} = f^{\dag} \circ g^{\dag} : C \to A;
\end{equation}
and finally such that the dagger operation is an involution for every morphism $f$ in ${\mathrm{hom} \left( \mathbf{C} \right)}$:

\begin{equation}
f : A \to B, \qquad \implies \qquad f^{\dag \dag} = f : A \to B.
\end{equation}

\begin{definition}
A ``dagger symmetric monoidal category'', denoted ${\left( \mathbf{C}, \otimes, I, \dag \right)}$, is a symmetric monoidal category ${\left( \mathbf{C}, \otimes, I \right)}$ equipped with a dagger structure ${\dag}$ that is compatible with the monoidal structure ${\otimes}$.
\end{definition}
The compatibility condition between the dagger structure ${\dag}$ and the monoidal structure ${\otimes}$ can be summarized as the requirement that, for every pair of morphisms $f$ and $g$ in ${\mathrm{hom} \left( \mathbf{C} \right)}$, one has:

\begin{equation}
f : A \to B \text{ and } g : C \to D, \qquad \implies \qquad \left( f \otimes g \right)^{\dag} = f^{\dag} \otimes g^{\dag} : B \otimes D \to A \otimes C,
\end{equation}
with additional compatibility conditions for the associator isomorphism ${\alpha}$ (for every triple of objects $A$, $B$ and $C$ in ${\mathrm{ob} \left( \mathbf{C} \right)}$):

\begin{equation}
\alpha_{A, B, C}^{\dag} = \alpha_{A, B, C}^{-1} : A \otimes \left( B \otimes  C \right) \cong \left( A \otimes B \right) \otimes C,
\end{equation}
the left and right unitor isomorphisms ${\lambda}$ and ${\rho}$ (for every object $A$ in ${\mathrm{ob} \left( \mathbf{C} \right)}$):

\begin{equation}
\lambda_{A}^{\dag} = \lambda_{A}^{-1} : A \cong I \otimes A, \qquad \text{ and } \qquad \rho_{A}^{\dag} \rho_{A}^{-1} : A \cong A \otimes I,
\end{equation}
and the symmetry isomorphism ${\sigma}$:

\begin{equation}
\sigma_{A, B}^{\dag} = \sigma_{A, B}^{-1} : B \otimes A \cong A \otimes B.
\end{equation}

\begin{definition}
A ``compact closed''\cite{kelly3} symmetric monoidal category is a symmetric monoidal category ${\left( \mathbf{C}, \otimes, I \right)}$ in which, for every object $A$ in ${\mathrm{ob} \left( \mathbf{C} \right)}$, there exists a corresponding dual object ${A^{*}}$ which is unique up to a canonical (unique) isomorphism.
\end{definition}
The dual object ${A^{*}}$ generalizes the concept of a dual vector space in linear algebra, and is assumed to be equipped with an additional pair of morphisms, namely the \textit{unit} ${\eta_{A}}$ and the \textit{counit} ${\epsilon_{A}}$ in ${\mathrm{hom} \left( \mathbf{C} \right)}$, of the general form:

\begin{equation}
\eta_{A} : I \to A^{*} \otimes A, \qquad \text{ and } \qquad \epsilon_{A} : A \otimes A^{*} \to I,
\end{equation}
such that the following pair of compositional equations are satisfied with respect to the associator ${\alpha}$, left unitor ${\lambda}$ and right unitor ${\rho}$ isomorphisms:

\begin{equation}
\lambda_{A} \circ \left( \epsilon_{A} \otimes A \right) \circ \alpha_{A, A^{*}, A}^{-1} \circ \left( A \otimes \eta_{A} \right) \circ \rho_{A}^{-1} = id_{A},
\end{equation}
which we can summarize neatly as the diagrammatic requirement that the following composition of morphisms must equal ${id_{A}}$:

\begin{equation}
\begin{tikzcd}
A^{*} \arrow[r, "\cong"] & I \otimes A^{*} \arrow[r, "A \otimes \eta"] & A \otimes \left( A^{*} \otimes A \right) \arrow[r, "\cong"] & \left( A \otimes A^{*} \right) \otimes A \arrow[r, "\epsilon \otimes A"] & I \otimes A \arrow[r, "\cong"] & A,
\end{tikzcd}
\end{equation}
and also:

\begin{equation}
\rho_{A^{*}} \circ \left( A^{*} \otimes \epsilon_{A} \right) \otimes \alpha_{A^{*}, A, A} \circ \left( \eta_{A} \otimes A^{*} \right) \circ \lambda_{A^{*}}^{-1} = id_{A},
\end{equation}
which we can further summarize as the diagrammatic requirement that the following composition of morphisms must equal ${id_{A^{*}}}$:

\begin{equation}
\begin{tikzcd}
A^{*} \arrow[r, "\cong"] & I \otimes A^{*} \arrow[r, "\eta \otimes A^{*}"] & \left( A^{*} \otimes A \right) \otimes A^{*} \arrow[r, "\cong"] & A^{*} \otimes \left( A \otimes A^{*} \right) \arrow[r, "A^{*} \otimes \epsilon"] & A^{*} \otimes I \arrow[r, "\cong"] & A^{*}.
\end{tikzcd}
\end{equation}

\begin{definition}
A ``dagger compact closed category''\cite{doplicher}\cite{baez2} is a dagger symmetric monoidal category ${\left( \mathbf{C}, \otimes, I, \dag \right)}$ that is also compact closed, in such a way that the dagger structure ${\dag}$ and the compact structure (with unit ${\eta}$ and counit ${\epsilon}$) are compatible.
\end{definition}
The compatibility condition between the dagger structure ${\dag}$ and the compact structure (defined in terms of the unit ${\eta}$ and counit ${\epsilon}$) can be summarized as the requirement that, for every object $A$ in ${\mathrm{ob} \left( \mathbf{C} \right)}$, the following diagram commutes:

\begin{equation}
\begin{tikzcd}
I \arrow[r, "\epsilon_{A}^{\dag}"] \arrow[dr, "\eta_{A}"] & A \otimes A^{*} \arrow[d, "\sigma_{A \otimes A^{*}}"]\\
& A^{*} \otimes A
\end{tikzcd}.
\end{equation}

\begin{figure}[ht]
\centering
\includegraphics[width=0.395\textwidth]{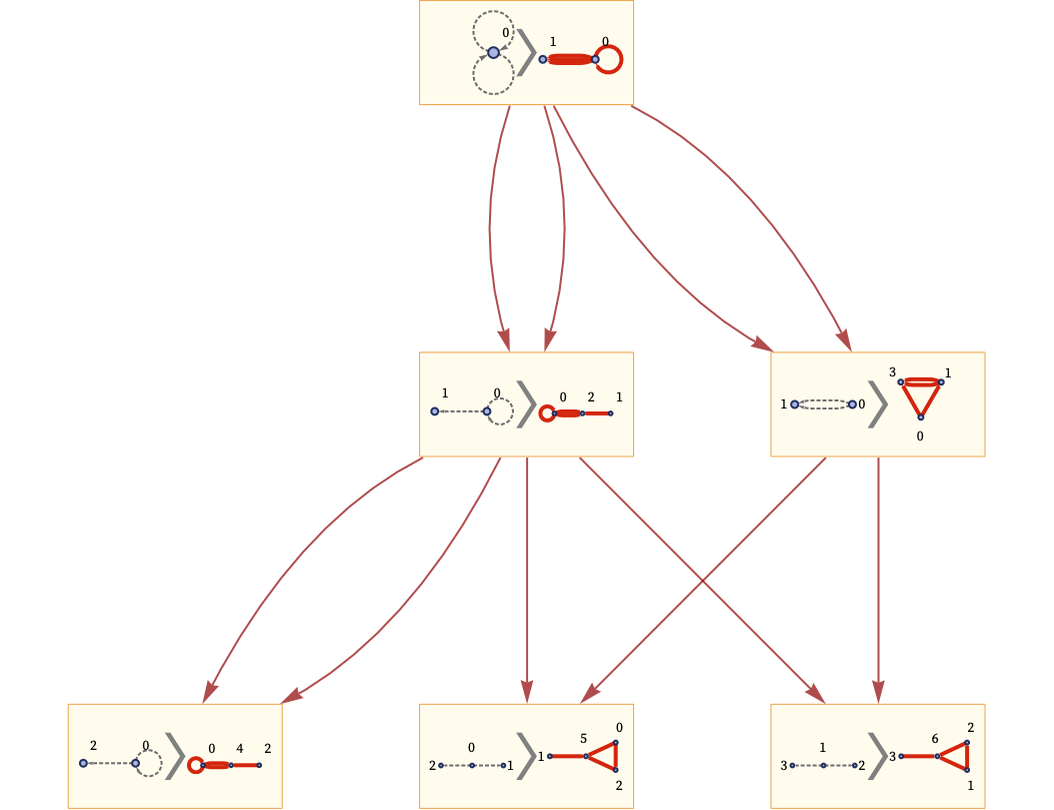}
\includegraphics[width=0.595\textwidth]{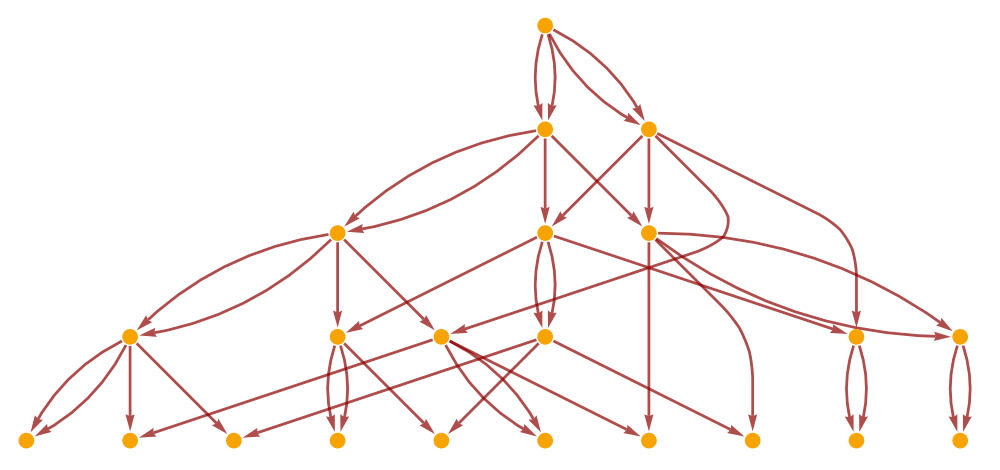}
\caption{The causal graphs corresponding to the first 3 and the first 5 steps in the deterministic evolution history for the set substitution rule ${\left\lbrace \left\lbrace x, y \right\rbrace, \left\lbrace x, z \right\rbrace \right\rbrace \to \left\lbrace \left\lbrace x, y \right\rbrace, \left\lbrace x, w \right\rbrace, \left\lbrace y, w \right\rbrace, \left\lbrace z, w \right\rbrace \right\rbrace}$, respectively. Example taken from \cite{wolfram2}.}
\label{fig:Figure12}
\end{figure}

\begin{figure}[ht]
\centering
\includegraphics[width=0.395\textwidth]{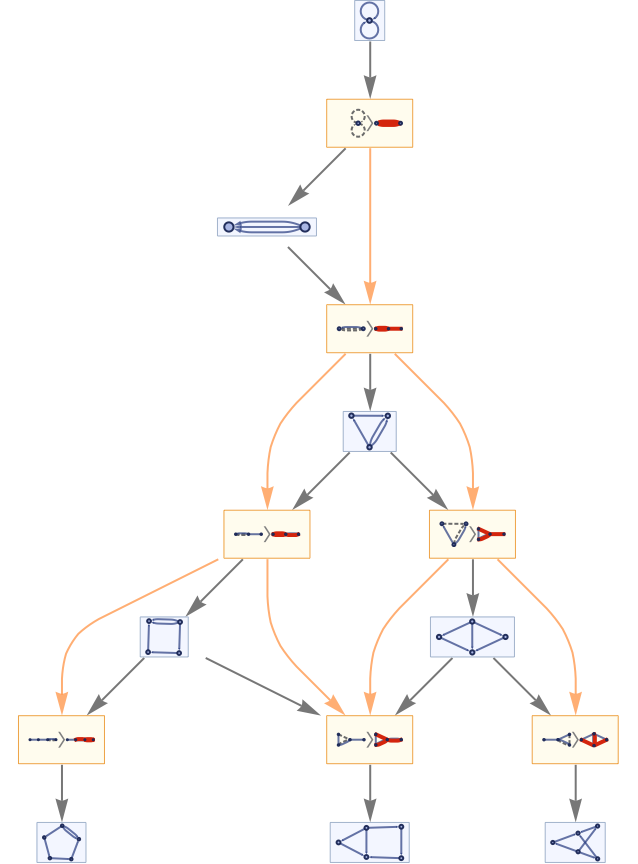}
\caption{The multiway evolution causal graph corresponding to the first 4 steps in the non-deterministic evolution history for the set substitution rule ${\left\lbrace \left\lbrace x, y \right\rbrace, \left\lbrace z, y \right\rbrace \right\rbrace \to \left\lbrace \left\lbrace x, w \right\rbrace, \left\lbrace y, w \right\rbrace, \left\lbrace z, w \right\rbrace \right\rbrace}$, illustrating trivial causal invariance (with state vertices shown in blue, updating event vertices shown in yellow, evolution edges shown in gray and causal edges shown in orange). Example taken from \cite{wolfram2}.}
\label{fig:Figure13}
\end{figure}

In addition to constructing a multiway evolution graph, one can also construct a \textit{causal graph} associated with a given Wolfram model evolution history, as a means of illustrating the pattern of causal dependencies between applications of the update rules:

\begin{definition}
A ``causal graph'', denoted ${G_{causal} = \left( V_{causal}, E_{causal} \right)}$, is a directed, acyclic graph associated to a given Wolfram model evolution history, in which every vertex in ${V_{causal}}$ corresponds to an application of an update rule, and in which the directed edge ${a \to b}$ only exists in ${E_{causal}}$ (for ${a, b \in V_{causal}}$) if:

\begin{equation}
\mathrm{In} \left( b \right) \cap \mathrm{Out} \left( a \right) \neq \emptyset.
\end{equation}
\end{definition}
Each such application of an update rule (i.e. each vertex of the associated causal graph) is known as an \textit{update event}. In other words, a directed edge ${a \to b}$ in the definition of a Wolfram model causal graph formalizes the notion that the input for the update event $b$ makes use of hyperedges that were produced by the output of the update event $a$, and therefore event $b$ could only have been applied if event $a$ has previously been applied. Examples of causal graphs yielded by the first 3 and the first 5 evolution steps of a simple Wolfram model system are shown in Figure \ref{fig:Figure12}. Global confluence of the abstract rewriting system is a necessary but not sufficient condition for \textit{causal invariance}, which states that the causal graphs associated with all possible paths through the multiway system eventually become isomorphic (as directed, acyclic graphs), as illustrated by the \textit{multiway evolution causal graph} for a confluent set substitution system shown in Figure \ref{fig:Figure13} (with state vertices shown in blue, updating event vertices shown in yellow, evolution edges shown in gray and causal edges shown in orange).

Thus, the transitive reduction of a causal graph yields a Hasse diagram for a causal partial order relation ${\prec}$\cite{gorard4}, i.e. a binary relation satisfying the axioms of acyclicity/antisymmetry:

\begin{equation}
\forall x, y \in \mathcal{C}, \qquad x \prec y \text{ and } y \prec x \implies x = y,
\end{equation}
and transitivity:

\begin{equation}
\forall x, y, z \in \mathcal{C}, \qquad x \prec y \text{ and } y \prec z \implies x \prec z,
\end{equation}
where ${\mathcal{C}}$ denotes the set of update events (i.e. the vertex set ${V_{causal}}$ of the causal graph). Since the causal graph does not (conventionally) contain self-loops at each vertex, the binary relation does not strictly satisfy the full axioms of a partial order relation, since it does not obey reflexivity:

\begin{equation}
\forall x \in \mathcal{C}, \qquad x \prec x,
\end{equation}
such that the acyclicity/antisymmetry condition must instead be replaced with:

\begin{equation}
\nexists x, y \in \mathcal{C}, \qquad \text{ such that } x \prec y \text{ and } y \prec x.
\end{equation}
We can reformulate the causal graph (and the induced causal partial order) category-theoretically by means of the \textit{causal category} formalism of Coecke and Lal\cite{coecke3}, based on \textit{partial monoidal categories}:

\begin{definition}
A ``partial functor'', denoted ${F : \mathbf{B} \to \mathbf{C}}$, between categories ${\mathbf{B}}$ and ${\mathbf{C}}$, is a functor ${\hat{F} : \mathbf{A} \to \mathbf{C}}$, where ${\mathbf{A}}$ denotes a subcategory of ${\mathbf{B}}$.
\end{definition}
The category ${\mathbf{B}}$ is conventionally interpreted as the \textit{domain} of the functor $F$, whilst the category ${\mathbf{A}}$ is interpreted as the \textit{domain of definition} of the functor $F$. Whenever the domain ${\mathbf{A}}$ is a product category, $F$ is interpreted as a \textit{partial bifunctor}.

\begin{definition}
A ``symmetric strict monoidal category'', denoted ${\left( \mathbf{C}, \otimes, I \right)}$, is a symmetric monoidal category ${\mathbf{C}}$ in which the associator, left unitor and right unitor isomorphisms (i.e. ${\alpha}$, ${\lambda}$ and ${\rho}$, respectively) are all identity isomorphisms.
\end{definition}

\begin{definition}
A ``symmetric strict partial monoidal category'', denoted ${\left( \mathbf{C}, \otimes, I \right)}$, is a category ${\mathbf{C}}$ equipped with a partial bifunctor ${\otimes}$ of the form:

\begin{equation}
\otimes : \mathbf{C} \times \mathbf{C} \to \mathbf{C},
\end{equation}
whose domain of definition (which we henceforth denote ${\mathrm{dd} \left( \otimes \right)}$) is a full subcategory of its domain, along with a distinguished unit object $I$, such that for every object $A$ in ${\mathrm{ob} \left( \mathbf{C} \right)}$, one has:

\begin{equation}
\left( A, I \right) \in \mathrm{dd} \left( \otimes \right), \qquad \left( I, A \right) \in \mathrm{dd} \left( \otimes \right), \qquad \text{ and } \qquad A \otimes I = A = I \otimes A;
\end{equation}
for every triple of objects $A$, $B$ and $C$ in ${\mathrm{ob} \left( \mathbf{C} \right)}$, one has:

\begin{equation}
\left( A, B \right), \left( A \otimes B, C \right) \in \mathrm{dd} \left( \otimes \right) \qquad \iff \qquad \left( B, C \right), \left( A, B \otimes C \right) \in \mathrm{dd} \left( \otimes \right),
\end{equation}
and moreover, whenever the objects ${A \otimes \left( B \otimes C \right)}$ and ${\left( A \otimes B \right) \otimes C}$ both exist, one has:

\begin{equation}
A \otimes \left( B \otimes C \right) = \left( A \otimes B \right) \otimes C;
\end{equation}
for every triple of morphisms $f$, $g$ and $h$ in ${\mathrm{hom} \left( \mathbf{C} \right)}$, whenever the morphisms ${f \otimes \left( g \otimes h \right)}$ and ${\left( f \otimes g \right) \otimes h}$ exist, one has:

\begin{equation}
f \otimes \left( g \otimes h \right) = \left( f \otimes g \right) \otimes h;
\end{equation}
and finally for every pair of objects $A$ and $B$ in ${\mathrm{ob} \left( \mathbf{C} \right)}$, such that:

\begin{equation}
\left( A, B \right), \left( B, A \right) \in \mathrm{dd} \left( \otimes \right),
\end{equation}
there exists a symmetry morphism ${\sigma_{A, B}}$ of the form:

\begin{equation}
\sigma_{A, B} : A \otimes B \to B \otimes A,
\end{equation}
with the property that:

\begin{equation}
\sigma_{A, B} \circ \sigma_{B, A} = \mathrm{id}_{A \otimes B}.
\end{equation}
\end{definition}

\begin{definition}
A ``terminal object'' in a category ${\mathbf{C}}$, denoted $A$, is a distinguished object in ${\mathrm{ob} \left( \mathbf{C} \right)}$ such that, for every object $B$ in ${\mathrm{ob} \left( \mathbf{C} \right)}$, there exists a unique morphism ${\top_{B}}$ in ${\mathrm{hom} \left( \mathbf{C} \right)}$ of the form:

\begin{equation}
\top_{B} : B \to A.
\end{equation}
\end{definition}

\begin{definition}
A ``causal category'', denoted ${\mathbf{CC}}$, is a symmetric strict partial monoidal category ${\left( \mathbf{CC}, \otimes, I \right)}$ for which every object $A$ in ${\mathrm{ob} \left( \mathbf{CC} \right)}$ has at least one element, i.e:

\begin{equation}
\mathbf{CC} \left( I, A \right) \neq \emptyset,
\end{equation}
for which the unit object $I$ in ${\mathrm{ob} \left( \mathbf{CC} \right)}$ is terminal, i.e. for every object $A$ in ${\mathrm{hom} \left( \mathbf{CC} \right)}$, there exists a unique morphism ${\top_{A}}$ of the form:

\begin{equation}
\top_{A} : A \to I,
\end{equation}
and, moreover, for every pair of objects $A$ and $B$ in ${\mathrm{ob} \left( \mathbf{CC} \right)}$, the monoidal product ${A \otimes B}$ exists if and only if the following pair of conditions are both satisfied:

\begin{equation}
\mathbf{CC} \left( A, B \right) = \left[ \mathbf{CC} \left( I, B \right) \right] \circ \top_{A}, \qquad \text{ and } \qquad \mathbf{CC} \left( B, A \right) = \left[ \mathbf{CC} \left( I, A \right) \right] \circ \top_{B}.
\end{equation}
\end{definition}

The basic intuition behind this causal category construction is that pairs of causally-related events (i.e. pairs of events that are either timelike-separated or lightlike-separated in the context of a continuum spacetime model) are treated as sequential compositions of morphisms, whilst pairs of causally-unrelated events (i.e. pairs of events that are spacelike-separated within in the context of a continuum spacetime model) are treated as partial monoidal products of morphisms. Note that, since the domain of definition of ${\otimes}$ forms a full subcategory of the domain of ${\otimes}$, for a pair of morphisms $f$ and $g$ of the form:

\begin{equation}
f : A \to D, \qquad \text{ and } \qquad g : B \to E,
\end{equation}
in ${\mathrm{hom} \left( \mathbf{CC} \right)}$, the partial monoidal product ${f \otimes g}$ exists if and only if the monoidal products ${A \otimes B}$ and ${D \otimes E}$ both exist. Therefore, the partial monoidal product ${\left( f \otimes g \right) \otimes h}$ exists if and only if the partial monoidal product ${f \otimes \left( g \otimes h \right)}$ exists, for a third morphism $h$ of the form:

\begin{equation}
h : C \to F,
\end{equation}
in ${\mathrm{hom} \left( \mathbf{CC} \right)}$, since the full monoidal product ${\left( A \otimes B \right) \otimes C}$ exists if and if the full monoidal product ${A \otimes \left( B \otimes C \right)}$ exists. Therefore, associativity of the partial monoidal product ${\otimes}$ is inherited from the associativity of the full monoidal product. Similarly, since the domain of definition for partial bifunctor:

\begin{equation}
\otimes : \mathbf{CC} \otimes \mathbf{CC} \to \mathbf{CC},
\end{equation}
forms a full subcategory of the domain of the partial bifunctor ${\otimes}$, it follows that the partial monoidal products of the sequential compositions of morphisms ${h \circ f}$ and ${k \circ g}$, for arbitrary $f$, $g$, $h$, $k$ in ${\mathrm{hom} \left( \mathbf{CC} \right)}$, must always exist, yielding:

\begin{equation}
\left( h \otimes k \right) \circ \left( f \otimes g \right) = \left( h \circ f \right) \otimes \left( k \circ g \right).
\end{equation}
Therefore, bifunctoriality of the partial monoidal product ${\otimes}$ is inherited from the bifunctoriality of the full monoidal product in much the same way as associativity. Altogether, this yields a fully categorical description of Wolfram model multiway evolution, including its causal structure.

\clearpage

\section{Theorem-Proving in the Wolfram Model and the ZX-Calculus}
\label{sec:Section4}

\begin{figure}[ht]
\centering
\includegraphics[width=0.495\textwidth]{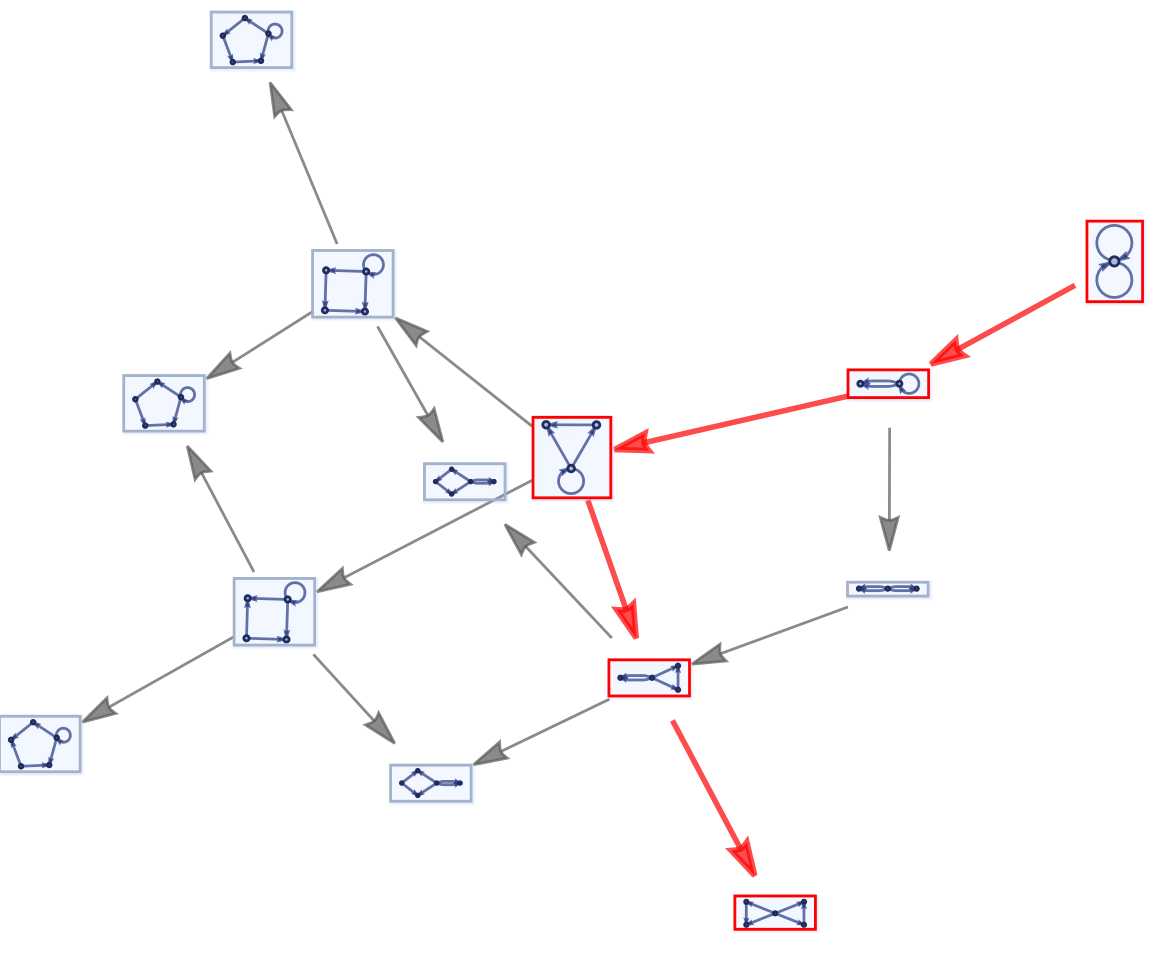}
\caption{The first 4 steps in the non-deterministic evolution history for the set substitution rule ${\left\lbrace \left\lbrace x, y \right\rbrace, \left\lbrace x, z \right\rbrace \right\rbrace \to \left\lbrace \left\lbrace x, z \right\rbrace, \left\lbrace x, w \right\rbrace, \left\lbrace y, w \right\rbrace \right\rbrace}$, as represented by a multiway evolution graph, with the path between state vertices ${\left\lbrace \left\lbrace 0, 0 \right\rbrace, \left\lbrace 0, 0 \right\rbrace \right\rbrace}$ and ${\left\lbrace \left\lbrace 0, 1 \right\rbrace, \left\lbrace 0, 3 \right\rbrace, \left\lbrace 1, 3 \right\rbrace, \left\lbrace 0, 2 \right\rbrace, \left\lbrace 0, 4 \right\rbrace, \left\lbrace 2, 4 \right\rbrace \right\rbrace}$ highlighted.}
\label{fig:Figure14}
\end{figure}

\begin{figure}[ht]
\centering
\includegraphics[width=0.795\textwidth]{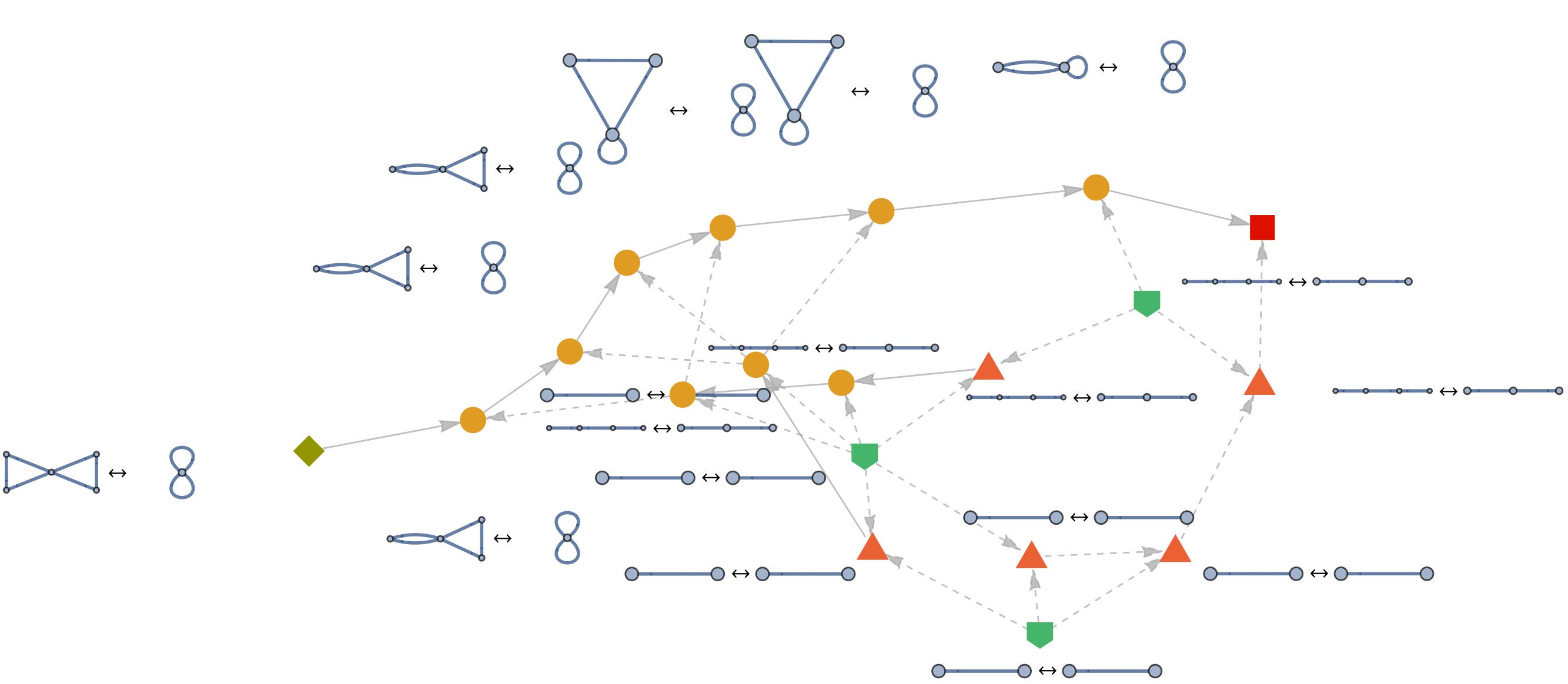}
\caption{The proof graph corresponding to the proof of the proposition that state ${\left\lbrace \left\lbrace 0, 1 \right\rbrace, \left\lbrace 0, 3 \right\rbrace, \left\lbrace 1, 3 \right\rbrace, \left\lbrace 0, 2 \right\rbrace, \left\lbrace 0, 4 \right\rbrace, \left\lbrace 2, 4 \right\rbrace \right\rbrace}$ is reachable from initial state ${\left\lbrace \left\lbrace 0, 0 \right\rbrace, \left\lbrace 0, 0 \right\rbrace \right\rbrace}$, subject to the set substitution rule ${\left\lbrace \left\lbrace x, y \right\rbrace, \left\lbrace x, z \right\rbrace \right\rbrace \to \left\lbrace \left\lbrace x, z \right\rbrace, \left\lbrace x, w \right\rbrace, \left\lbrace y, w \right\rbrace \right\rbrace}$. Here, pointed light green boxes represent axioms, dark orange triangles represent critical pair lemmas (i.e. instances of completions/superpositions/paramodulations), light orange circles represent substitution lemmas (i.e. instances of resolutions/factorings), and dark green diamonds represent hypotheses. Solid lines represent substitutions, and dashed lines represent derived inference rules.}
\label{fig:Figure15}
\end{figure}

\begin{figure}[ht]
\centering
\includegraphics[width=0.495\textwidth]{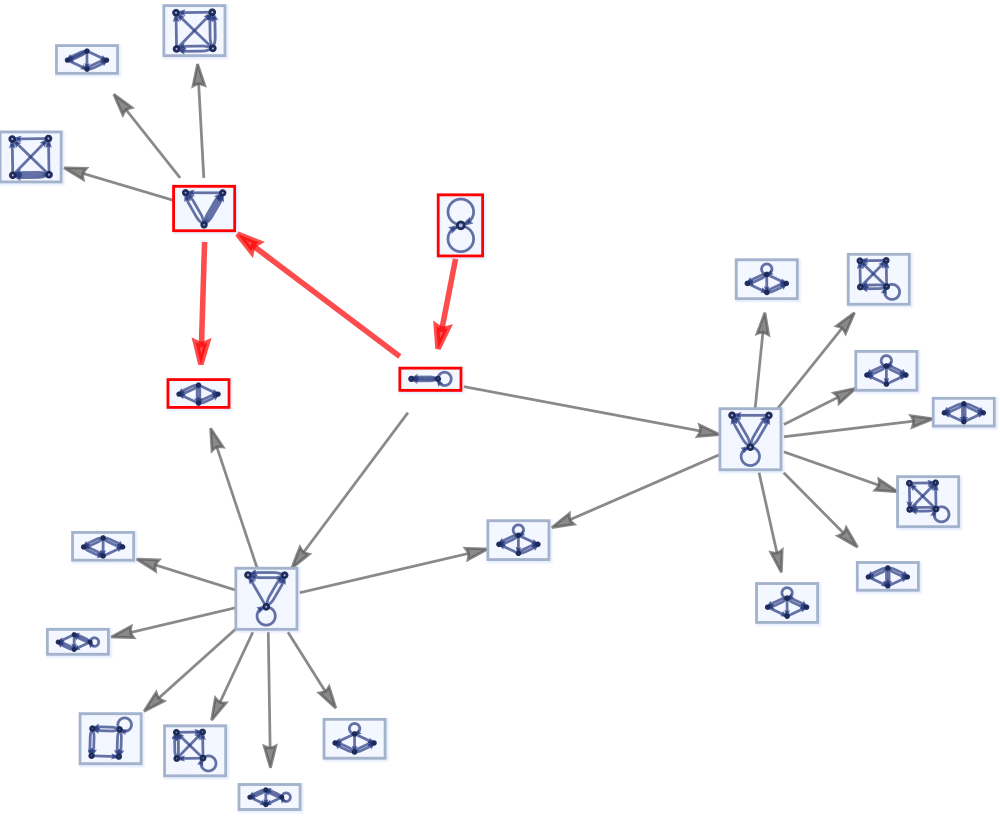}
\caption{The first 3 steps in the non-deterministic evolution history for the set substitution rule ${\left\lbrace \left\lbrace x, y \right\rbrace, \left\lbrace x, z \right\rbrace \right\rbrace \to \left\lbrace \left\lbrace x, z \right\rbrace, \left\lbrace x, w \right\rbrace, \left\lbrace y, w \right\rbrace, \left\lbrace z, w \right\rbrace \right\rbrace}$, as represented by a multiway evolution graph, with the path between state vertices ${\left\lbrace \left\lbrace 0, 0 \right\rbrace \right\rbrace}$ and ${\left\lbrace \left\lbrace 0, 1 \right\rbrace, \left\lbrace 0, 2 \right\rbrace, \left\lbrace 0, 2 \right\rbrace, \left\lbrace 1, 2 \right\rbrace, \left\lbrace 0, 1 \right\rbrace, \left\lbrace 0, 3 \right\rbrace, \left\lbrace 1, 3 \right\rbrace, \left\lbrace 1, 3 \right\rbrace \right\rbrace}$ highlighted.}
\label{fig:Figure16}
\end{figure}

\begin{figure}[ht]
\centering
\includegraphics[width=0.995\textwidth]{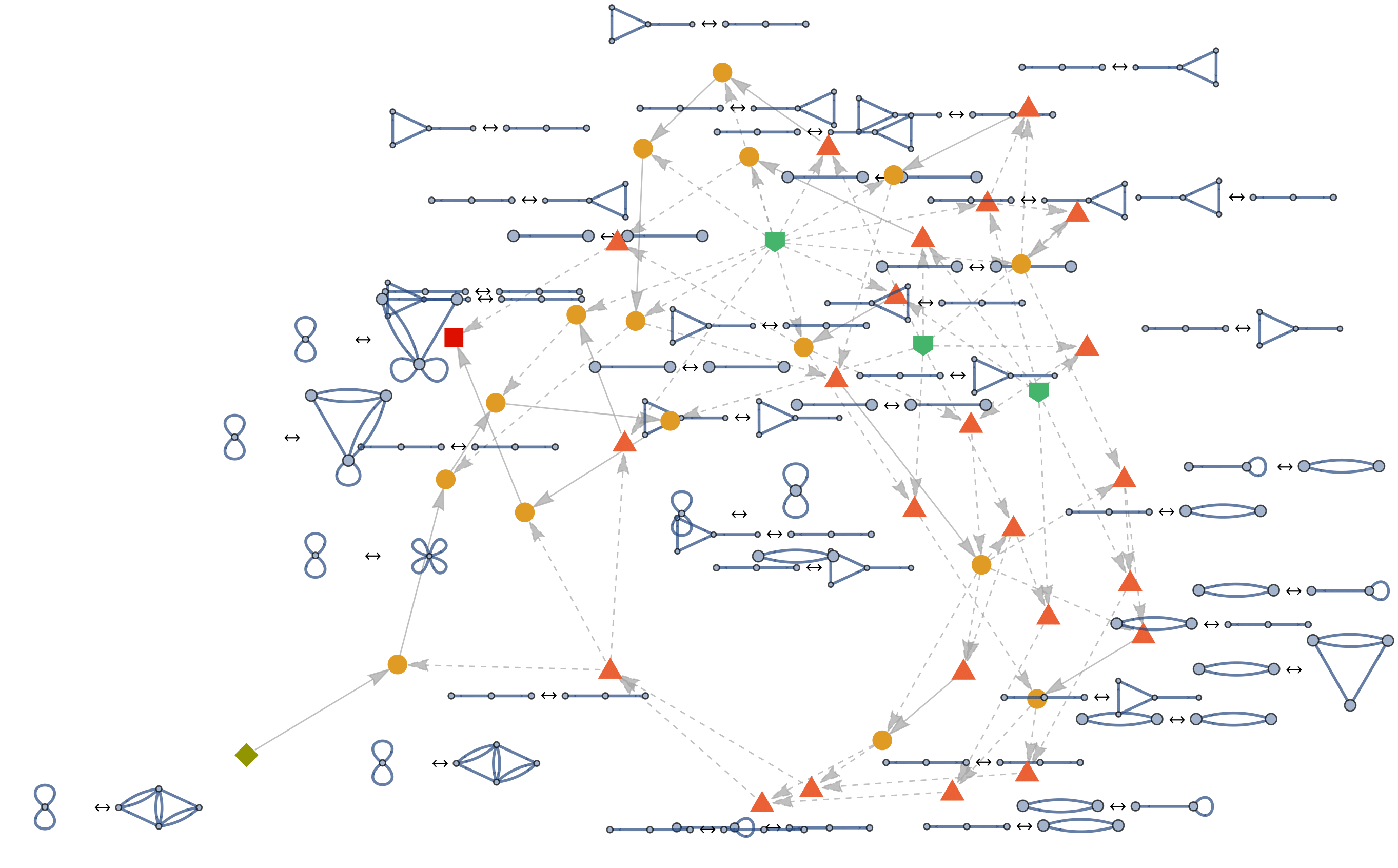}
\caption{The proof graph corresponding to the proof of the proposition that state ${\left\lbrace \left\lbrace 0, 1 \right\rbrace, \left\lbrace 0, 2 \right\rbrace, \left\lbrace 0, 2 \right\rbrace, \left\lbrace 1, 2 \right\rbrace, \left\lbrace 0, 1 \right\rbrace, \left\lbrace 0, 3 \right\rbrace, \left\lbrace 1, 3 \right\rbrace, \left\lbrace 1, 3 \right\rbrace \right\rbrace}$ is reachable from initial state ${\left\lbrace \left\lbrace 0, 0 \right\rbrace, \left\lbrace 0, 0 \right\rbrace \right\rbrace}$, subject to the set substitution rule ${\left\lbrace \left\lbrace x, y \right\rbrace, \left\lbrace x, z \right\rbrace \right\rbrace \to \left\lbrace \left\lbrace x, z \right\rbrace, \left\lbrace x, w \right\rbrace, \left\lbrace y, w \right\rbrace, \left\lbrace z, w \right\rbrace \right\rbrace}$. Here, pointed light green boxes represent axioms, dark orange triangles represent critical pair lemmas (i.e. instances of completions/superpositions/paramodulations), light orange circles represent substitution lemmas (i.e. instances of resolutions/factorings), and dark green diamonds represent hypotheses. Solid lines represent substitutions, and dashed lines represent derived inference rules.}
\label{fig:Figure17}
\end{figure}

We begin by presenting an example of an automatically-generated proof graph for a simple theorem relating two hypergraphs within a multiway Wolfram model evolution, namely that the set substitution rule:

\begin{equation}
\left\lbrace \left\lbrace x, y \right\rbrace, \left\lbrace x, z \right\rbrace \right\rbrace \to \left\lbrace \left\lbrace x, z \right\rbrace, \left\lbrace x, w \right\rbrace, \left\lbrace y, w \right\rbrace \right\rbrace,
\end{equation}
when applied to the double self-loop initial condition ${\left\lbrace \left\lbrace 0, 0 \right\rbrace, \left\lbrace 0, 0 \right\rbrace \right\rbrace}$, will eventually yield a hypergraph isomorphic to ${\left\lbrace \left\lbrace 0, 1 \right\rbrace, \left\lbrace 0, 3 \right\rbrace, \left\lbrace 1, 3 \right\rbrace, \left\lbrace 0, 2 \right\rbrace, \left\lbrace 0, 4 \right\rbrace, \left\lbrace 2, 4 \right\rbrace \right\rbrace}$; we make use here of the standard translation of the Wolfram model into a \textit{multiway operator system}\cite{gorard}, such that the theorem in question can be proved using the same first-order logic theorem-proving techniques presented within the preceding sections. We can determine that this theorem is actually true by first isolating the corresponding path through the multiway operator system, as shown in Figure \ref{fig:Figure14}. The theorem is translated into operator-theoretic form as:

\begin{equation}
\left( 0 \oplus 0 \right) \odot \left( 0 \oplus 0 \right) = \left( 0 \oplus 1 \right) \odot \left( \left( 0 \oplus 3 \right) \odot \left( \left( 1 \oplus 3 \right) \odot \left( \left( 0 \oplus 2 \right) \odot \left( \left( 0 \oplus 4 \right) \odot \left( 2 \oplus 4 \right) \right) \right) \right) \right),
\end{equation}
subject to the axiom representing the operator-theoretic form of the set substitution rule:

\begin{equation}
\forall x, y, z, w : \left( x \oplus y \right) \odot \left( x \oplus z \right) = \left( \left( x \oplus w \right) \odot \left( y \oplus w \right) \right),
\end{equation}
as well as additional axioms representing the associativity and commutativity of the underlying hyperedge concatenation operator ${\odot}$:

\begin{equation}
\forall a, b, c : a \odot \left( b \odot c \right) = \left( a \odot b \right) \odot c, \qquad \text{ and } \qquad \forall a, b : a \odot b = b \odot a.
\end{equation}
The associated proof graph is shown in Figure \ref{fig:Figure15}. The multiway path and corresponding automatically-generated proof graph for a slightly more sophisticated theorem relating two hypergraphs within a Wolfram model evolution, namely hypergraphs isomorphic to ${\left\lbrace \left\lbrace 0, 0 \right\rbrace, \left\lbrace 0, 0 \right\rbrace \right\rbrace}$ and:

\begin{equation}
\left\lbrace \left\lbrace 0, 1 \right\rbrace, \left\lbrace 0, 2 \right\rbrace, \left\lbrace 0, 2 \right\rbrace, \left\lbrace 0, 2 \right\rbrace, \left\lbrace 1, 2 \right\rbrace, \left\lbrace 0, 1 \right\rbrace, \left\lbrace 0, 3 \right\rbrace, \left\lbrace 1, 3 \right\rbrace, \left\lbrace 1, 3 \right\rbrace \right\rbrace,
\end{equation}
subject to the set substitution rule:

\begin{equation}
\left\lbrace \left\lbrace x, y \right\rbrace, \left\lbrace x, z \right\rbrace \right\rbrace \to \left\lbrace \left\lbrace x, z \right\rbrace, \left\lbrace x, w \right\rbrace, \left\lbrace y, w \right\rbrace, \left\lbrace z, w \right\rbrace \right\rbrace,
\end{equation}
are shown in Figures \ref{fig:Figure16} and \ref{fig:Figure17}, respectively. As above, this theorem is translated into operator-theoretic form as:

\begin{multline}
\left( 0 \oplus 0 \right) \odot \left( 0 \oplus 0 \right) = \left( 0 \oplus 1 \right) \odot \left( \left( 0 \oplus 2 \right) \odot \left( \left( 0 \oplus 2 \right) \odot \left( \left( 1 \oplus 2 \right) \odot \left( \left( 0 \oplus 1 \right) \odot \left( \left( 0 \oplus 3 \right) \odot\right. \right. \right. \right. \right.\\
\left. \left. \left. \left. \left. \left( \left( 1 \oplus 3 \right) \odot \left( \left( 1 \oplus 3 \right) \right) \right) \right) \right) \right) \right) \right),
\end{multline}
subject to the set substitution axiom:

\begin{equation}
\forall x, y, z, w : \left( x \oplus y \right) \odot \left( x \oplus z \right) = \left( x \oplus z \right) \odot \left( \left( x \oplus w \right) \odot \left( \left( y \oplus w \right) \odot \left( z \oplus w \right) \right) \right),
\end{equation}
plus the same associativity and commutativity axioms previously shown.

\begin{figure}[ht]
\centering
\includegraphics[width=0.495\textwidth]{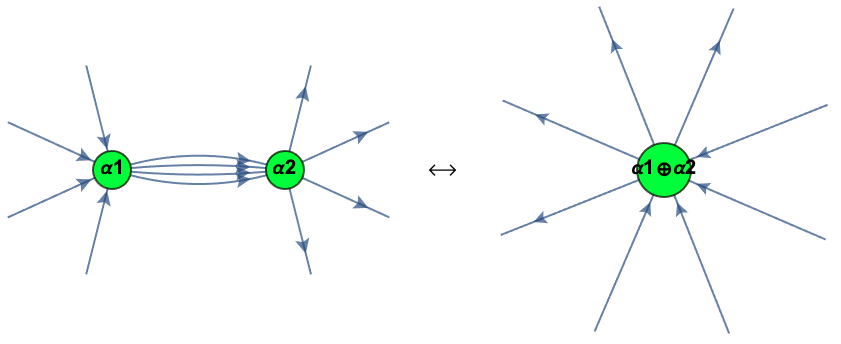}
\includegraphics[width=0.495\textwidth]{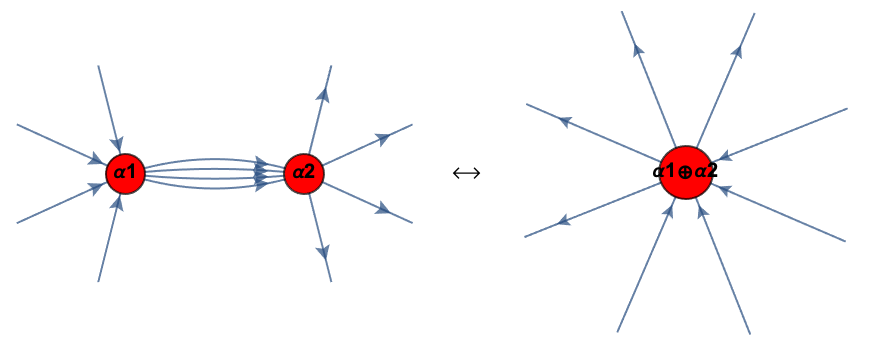}
\includegraphics[width=0.495\textwidth]{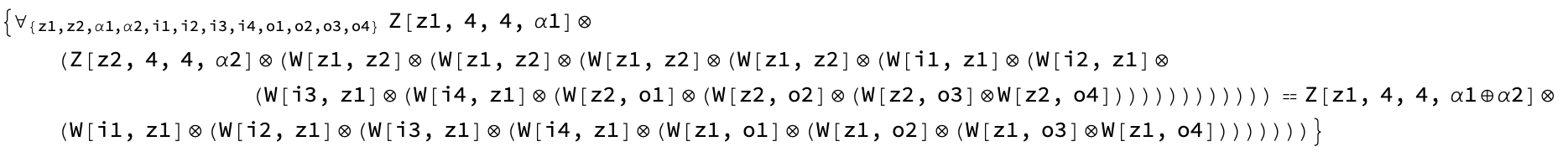}
\includegraphics[width=0.495\textwidth]{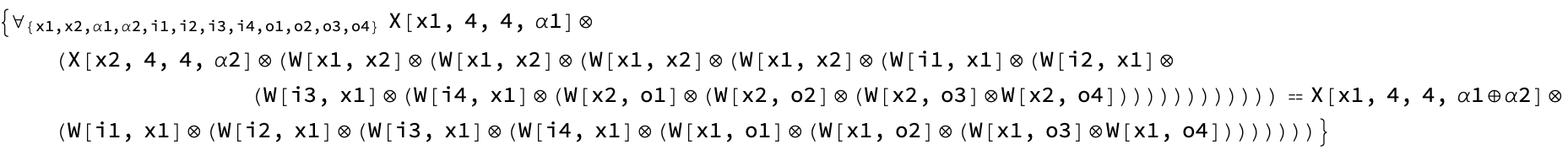}
\caption{The S1 rules (i.e. the Z- and X-spider fusion rules) for the case in which the input and output arities of the two spiders are both equal to 4, with the two spiders connected by exactly 4 wires, along with their associated operator-theoretic axiom forms.}
\label{fig:Figure18}
\end{figure}

\begin{figure}[ht]
\centering
\includegraphics[width=0.495\textwidth]{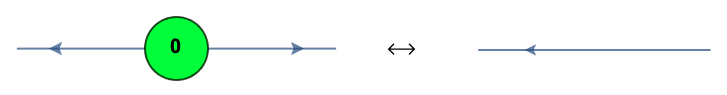}
\includegraphics[width=0.495\textwidth]{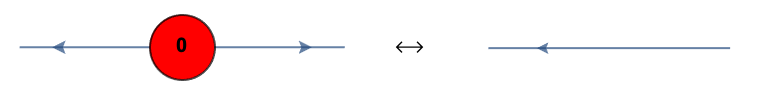}
\includegraphics[width=0.495\textwidth]{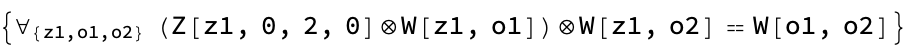}
\includegraphics[width=0.495\textwidth]{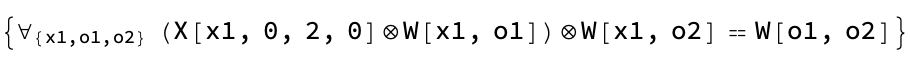}
\includegraphics[width=0.495\textwidth]{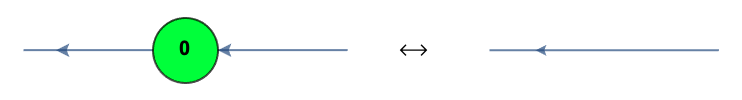}
\includegraphics[width=0.495\textwidth]{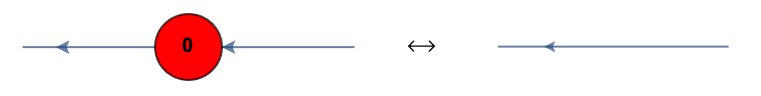}
\includegraphics[width=0.495\textwidth]{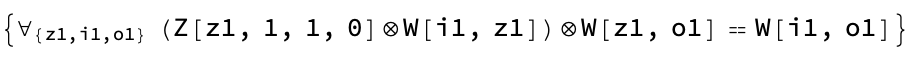}
\includegraphics[width=0.495\textwidth]{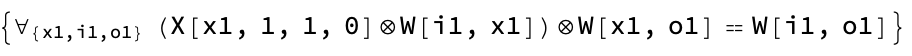}
\includegraphics[width=0.495\textwidth]{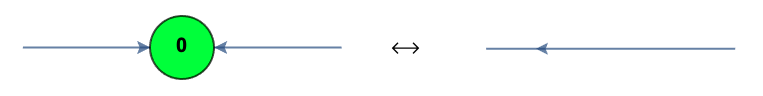}
\includegraphics[width=0.495\textwidth]{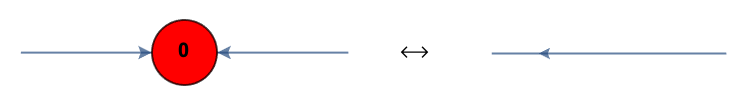}
\includegraphics[width=0.495\textwidth]{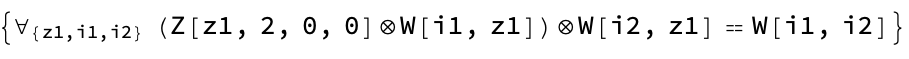}
\includegraphics[width=0.495\textwidth]{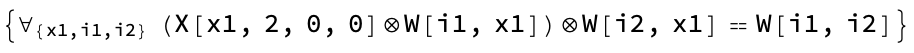}
\caption{The S2 rules (i.e. the Z- and X-spider identity rules) in the most general case, along with their associated operator-theoretic axiom forms.}
\label{fig:Figure19}
\end{figure}

From here, we are now able to employ the standard translation of the ZX-calculus of Coecke and Duncan\cite{coecke4} into the multiway operator system formalism\cite{gorard}, in order to perform fast diagrammatic reasoning over quantum circuits expressed as ZX-diagrams. We begin by translating the S1 rules (i.e. the Z- and X-spider fusion rules), which, for the case in which the input and output arities of the two spiders are both equal to 4, with the two spiders connected by exactly 4 wires, have the form shown in Figure \ref{fig:Figure18}, with the operator-theoretic axiom forms being:

\begin{multline}
\forall z_1, z_2, \alpha_1, \alpha_2, i_1, i_2, i_3, i_4, o_1, o_2, o_3, o_4 : Z \left[ z_1, 4, 4, \alpha_1 \right] \otimes \left( Z \left[ z_2, 4, 4, \alpha_2 \right] \otimes \left( W \left[ z_1, z_2 \right] \otimes \left( W \left[ z_1, z_2 \right] \otimes \right. \right. \right.\\
\left( W \left[ z_1, z_2 \right] \otimes \left( W \left[ z_1, z_2 \right] \otimes \left( W \left[ i_1, z_2 \right] \otimes \left( W \left[ i_2, z_1 \right] \otimes \left( W \left[ i_3, z_1 \right] \otimes \left( W \left[ i_4, z_1 \right] \otimes \left( W \left[ z_2, o_1 \right] \otimes \right. \right. \right. \right. \right. \right. \right.\\
\left. \left. \left. \left. \left. \left. \left. \left. \left( W \left[ z_2, o_2 \right] \otimes \left( W \left[ z_2, o_3 \right] \otimes \left( W \left[ z_2, o_4 \right] \right) \right) \right) \right) \right) \right) \right) \right) \right) \right) \right) = Z \left[ z_1, 4, 4, \alpha_1 \oplus \alpha_2 \right] \otimes \left( W \left[ i_1, z_1 \right] \otimes \left( W \left[ i_2, z_1 \right] \otimes \right. \right.\\
\left. \left( W \left[ i_3, z_1 \right] \otimes \left( W \left[ i_4, z_1 \right] \otimes \left( W \left[ z_1, o_1 \right] \otimes \left( W \left[ z_1, o_2 \right] \otimes \left( W \left[ z_1, o_3 \right] \otimes W \left[ z_1, o_4 \right] \right) \right) \right) \right) \right) \right),
\end{multline}
and:

\begin{multline}\\
\forall x_1, x_2, \alpha_1, \alpha_2, i_1, i_2, i_3, i_4, o_1, o_2, o_3, o_4 : X \left[ x_1, 4, 4, \alpha_1 \right] \otimes \left( X \left[ x_2, 4, 4, \alpha_2 \right] \otimes \left( W \left[ x_1, x_2 \right] \otimes \left( W \left[ x_1, x_2 \right] \otimes \right. \right. \right.\\
\left( W \left[ x_1, x_2 \right] \otimes \left( W \left[ x_1, x_2 \right] \otimes \left( W \left[ i_1, x_1 \right] \otimes \left( W \left[ i_2, x_1 \right] \otimes \left( W \left[ i_3, x_1 \right] \otimes \left( W \left[ i_4, x_1 \right] \otimes \left( W \left[ x_2, o_1 \right] \otimes \right. \right. \right. \right. \right. \right. \right.\\
\left. \left. \left. \left. \left. \left. \left. \left. \left( W \left[ x_2, o_2 \right] \otimes \left( W \left[ x_2, o_3 \right] \otimes \left( W \left[ x_2, o_4 \right] \right) \right) \right) \right) \right) \right) \right) \right) \right) \right) \right) = X \left[ x_1, 4, 4, \alpha_1 \oplus \alpha_2 \right] \otimes \left( W \left[ i_1, x_1 \right] \otimes \left( W \left[ i_2, x_1 \right] \otimes \right. \right.\\
\left. \left( W \left[ i_3 , x_1 \right] \otimes \left( W \left[ i_4, x_1 \right] \otimes \left( W \left[ x_1, o_1 \right] \otimes \left( W \left[ x_1, o_2 \right] \otimes \left( W \left[ x_1, o_3 \right] \otimes W \left[ x_1, o_4 \right] \right) \right) \right) \right) \right) \right),
\end{multline}
respectively. These rules are derived from the fact that the Z- and X-spiders represent orthonormal bases (specifically the computational basis and the Hadamard-transformed basis, respectively), and therefore whenever two spiders of the same type are adjacent, they can be merged, with their respective phases combining additively, due to matrix multiplication. We also translate the S2 rules (i.e. the Z and X-spider identity rules), which, in the most general case, have the form shown in Figure \ref{fig:Figure19}, with the operator-theoretic axiom forms being:

\begin{equation}
\forall z_1, o_1, o_2 : \left( Z \left[ z_1, 0, 2, 0 \right] \otimes W \left[ z_1, o_1 \right] \right) \otimes W \left[ z_1, o_2 \right] = W \left[ o_1, o_2 \right],
\end{equation}
\begin{equation}
\forall z_1, i_1, o_1 : \left( Z \left[ z_1, 1, 1, 0 \right] \otimes W \left[ i_1, z_1 \right] \right) \otimes W \left[ z_1, o_1 \right] = W \left[ i_1, o_1 \right],
\end{equation}
\begin{equation}
\forall z_1, i_1, i_2 : \left( Z \left[ z_1, 2, 0, 0 \right] \otimes W \left[ i_1, z_1 \right] \right) \otimes W \left[ i_2, z_1 \right] = W \left[ i_1, i_2 \right],
\end{equation}

and:

\begin{equation}
\forall x_1, o_1, o_2 : \left( X \left[ x_1, 0, 2, 0 \right] \otimes W \left[ x_1, o_1 \right] \right) \otimes W \left[ x_1, o_2 \right] = W \left[ o_1, o_2 \right],
\end{equation}
\begin{equation}
\forall x_1, i_1, o_1 : \left( X \left[ x_1, 1, 1, 0 \right] \otimes W \left[ i_1, x_1 \right] \right) \otimes W \left[ x_1, o_1 \right] = W \left[ i_1, o_1 \right],
\end{equation}
\begin{equation}
\forall x_1, i_1, i_2 : \left( X \left[ x_1, 2, 0, 0 \right] \otimes W \left[ i_1, x_1 \right] \right) \otimes W \left[ i_2, x_1 \right] = W \left[ i_1, i_2 \right],
\end{equation}
respectively. These rules are derived from the fact that any phaseless Z- or X-spider is equivalent to the identity map, since the Bell state is identical regardless of whether it is expressed with respect to the computational basis or the Hadamard-transformed basis; in category-theoretic terms, the Z- and X-spiders therefore induce the same compact structure.

\begin{figure}[ht]
\centering
\includegraphics[width=0.495\textwidth]{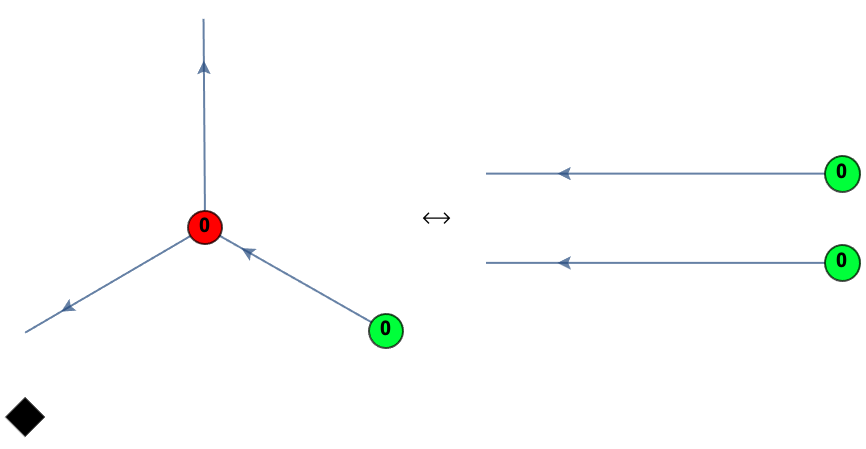}
\includegraphics[width=0.495\textwidth]{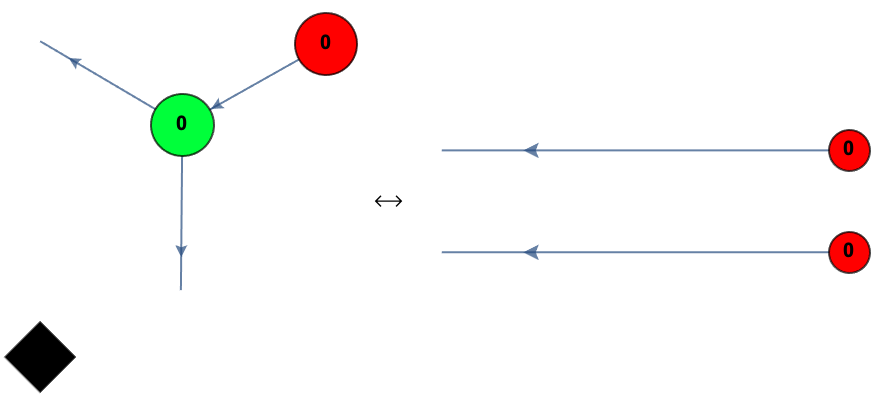}
\includegraphics[width=0.495\textwidth]{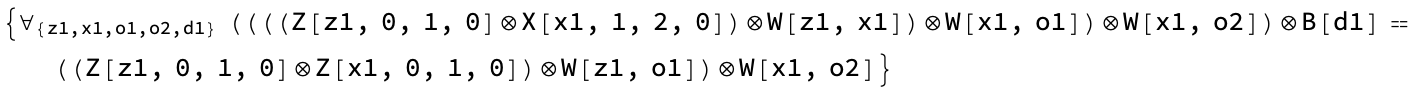}
\includegraphics[width=0.495\textwidth]{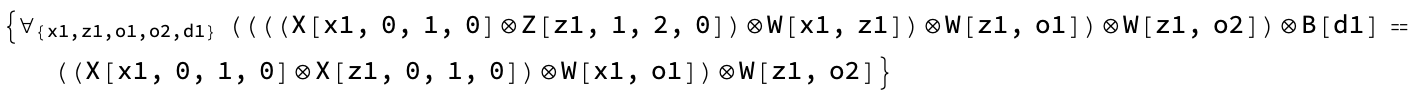}
\caption{The B1 rules (i.e. the Z- and X-spider copy rules) in the most general case, along with their respective operator-theoretic axiom forms.}
\label{fig:Figure20}
\end{figure}

\begin{figure}[ht]
\centering
\includegraphics[width=0.995\textwidth]{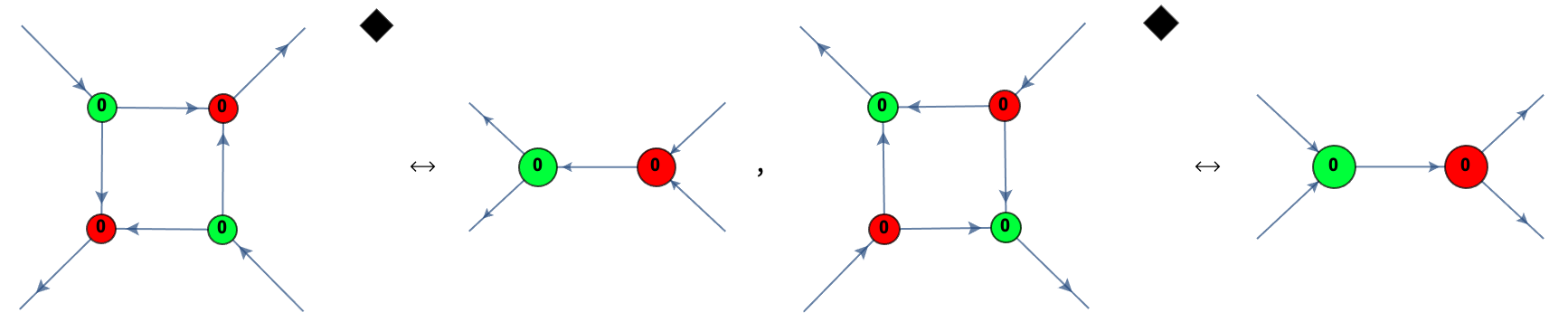}
\includegraphics[width=0.995\textwidth]{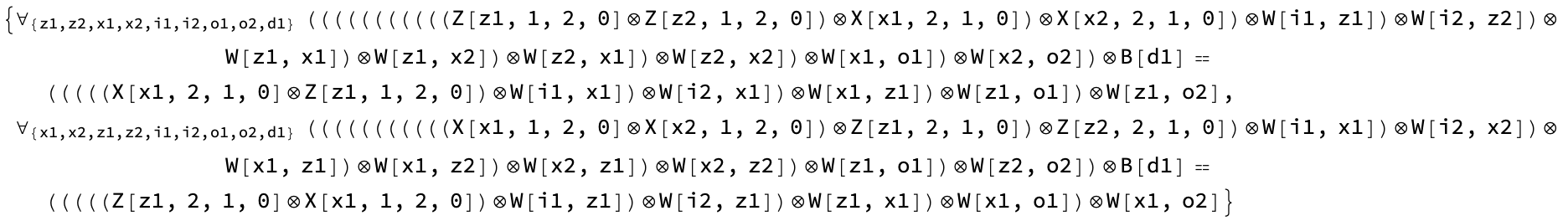}
\caption{The B2 rules (i.e. the bialgebra simplification rules) in the most general case, along with their respective operator-theoretic axiom forms.}
\label{fig:Figure21}
\end{figure}

Next, we translate the B1 rules (i.e. the Z- and X-spider copy rules), which, in the most general case, have the form shown in Figure \ref{fig:Figure20}, with the operator-theoretic axiom forms being:

\begin{multline}
\forall z_1, x_1, o_1, o_2, d_1 : \left( \left( \left( \left( Z \left[ z_1, 0, 1, 0 \right] \otimes X \left[ x_1, 1, 2, 0 \right] \right) \otimes W \left[ z_1, x_1 \right] \right) \otimes W \left[ x_1, o_1 \right] \right) \otimes W \left[ x_1, o_2 \right] \right) \otimes B \left[ d_1 \right]\\
= \left( \left( Z \left[ z_1, 0, 1, 0 \right] \otimes Z \left[ x_1, 0, 1, 0 \right] \right) \otimes W \left[ z_1, o_1 \right] \right) \otimes W \left[ x_1, o_2 \right],
\end{multline}
and:

\begin{multline}
\forall x_1, z_1, o_1, o_2, d_1 : \left( \left( \left( \left( X \left[ x_1, 0, 1, 0 \right] \otimes Z \left[ z_1, 1, 2, 0 \right] \right) \otimes W \left[ x_1, z_1 \right] \right) \otimes W \left[ z_1, o_1 \right] \right) \otimes W \left[ z_1, o_2 \right] \right) \otimes B \left[ d_1 \right] =\\
\left( \left( X \left[ x_1, 0, 1, 0 \right] \otimes X \left[ z_1, 0, 1, 0 \right] \right) \otimes W \left[ x_1, o_1 \right] \right) \otimes W \left[ z_1, o_2 \right],
\end{multline}
respectively. These rules are derived from the fact that any Z-spider with arity 1 can be ``copied through'' an X-spider, and vice versa, as a consequence of the fact that any Z-spider of arity 1 is proportional to a Hadamard-transformed basis state (namely the state ${\ket{+}}$), and any X-spider of arity 1 is proportional to a computational basis state (namely the state ${\ket{0}}$), up to a multiplicative scalar constant represented by the black diamond. We also translate the B2 rules (i.e. the bialgebra simplification rules), which, in the most general case, have the form shown in Figure \ref{fig:Figure21}, with the operator-theoretic axiom forms being:

\begin{multline}
\forall z_1, z_2, x_1, x_2, i_1, i_2, o_1, o_2, d_1 : \left( \left( \left( \left( \left( \left( \left( \left( \left( \left( \left( Z \left[ z_1, 1, 2, 0 \right] \otimes Z \left[ z_2, 1, 2, 0 \right] \right) \otimes X \left[ x_1, 2, 1, 0 \right] \right) \otimes \right. \right. \right. \right. \right. \right. \right. \right. \right.\\
\left. \left. \left. \left. \left. \left. \left. X \left[ x_2, 2, 1, 0 \right] \right) \otimes W \left[ i_1, z_1 \right] \right) \otimes W \left[ i_2, z_2 \right] \right) \otimes W \left[ z_1, x_1 \right] \right) \otimes W \left[ z_1, x_2 \right] \right) \otimes W \left[ z_2, x_1 \right] \right) \otimes W \left[ z_2, x_2 \right] \right) \otimes\\
\left. \left. W \left[ x_1, o_1 \right] \right) \otimes W \left[ x_2, o_2 \right] \right) \otimes B \left[ d_1 \right] = \left( \left( \left( \left( \left( X \left[ x_1, 2, 1, 0 \right] \otimes Z \left[ z_1, 1, 2, 0 \right] \right) \otimes W \left[ i_1, x_1 \right] \right) \otimes W \left[ i_2, x_1 \right] \right) \otimes \right. \right.\\
\left. \left. W \left[ x_1, z_1 \right] \right) \otimes W \left[ z_1, o_1 \right] \right) \otimes W \left[ z_1, o_2 \right],
\end{multline}
and:

\begin{multline}
\forall x_1, x_2, z_1, z_2, i_1, i_2, o_1, o_2, d_1 : \left( \left( \left( \left( \left( \left( \left( \left( \left( \left( \left( X \left[ x_1, 1, 2, 0 \right] \otimes X \left[ x_2, 1, 2, 0 \right] \right) \otimes Z \left[ z_1, 2, 1, 0 \right] \right) \otimes \right. \right. \right. \right. \right. \right. \right. \right. \right.\\
\left. \left. \left. \left. \left. \left. \left. Z \left[ z_2, 2, 1, 0 \right] \right) \otimes W \left[ i_1, x_1 \right] \right) \otimes W \left[ i_2, x_2 \right] \right) \otimes W \left[ x_1, z_1 \right] \right) \otimes W \left[ x_1, z_2 \right] \right) W \left[ x_2, z_1 \right] \right) \otimes W \left[ x_2, z_2 \right] \right) \otimes\\
\left. \left. W \left[ z_1, o_1 \right] \right) \otimes W \left[ z_2, o_2 \right] \right) \otimes B \left[ d_1 \right] = \left( \left( \left( \left( \left( Z \left[ z_1, 2, 1, 0 \right] \otimes X \left[ x_1, 1, 2, 0 \right] \right) \otimes W \left[ i_1, z_1 \right] \right) \otimes W \left[ i_2, z_1 \right] \right) \otimes \right. \right.\\
\left. \left. W \left[ z_1, x_1 \right] \right) \otimes W \left[ x_1, o_1 \right] \right) \otimes W \left[ x_1, o_2 \right],
\end{multline}
respectively. These rules are derived from the fact that any 2-cycle of Z-spiders and X-spiders must simplify, as a consequence of the computational and Hadamard-transformed bases being \textit{strongly complementary}.

\begin{figure}[ht]
\centering
\includegraphics[width=0.595\textwidth]{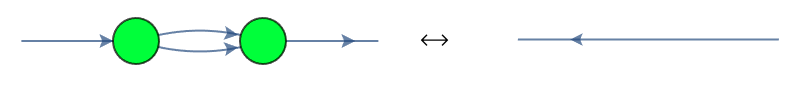}
\includegraphics[width=0.695\textwidth]{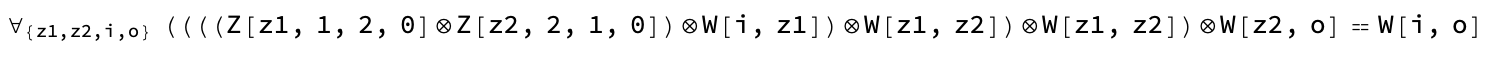}
\caption{The defining diagrammatic equality for dagger special (commutative) Frobenius algebras, along with its associated operator-theoretic axiom form.}
\label{fig:Figure22}
\end{figure}

\begin{figure}[ht]
\centering
\includegraphics[width=0.595\textwidth]{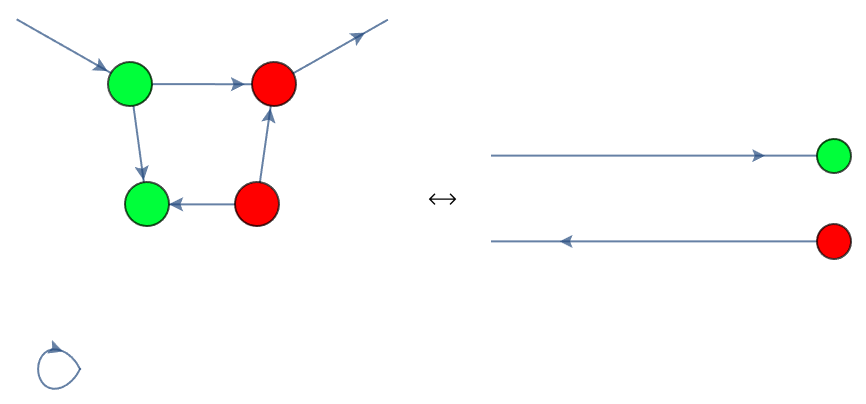}
\includegraphics[width=0.695\textwidth]{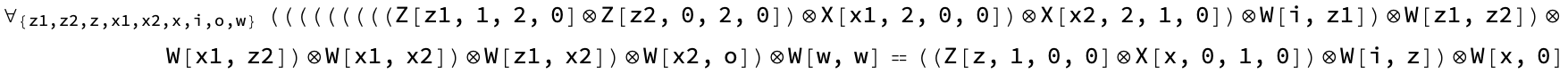}
\caption{The defining diagrammatic equality for complementarity of observable structures, along with its associated operator-theoretic axiom form.}
\label{fig:Figure23}
\end{figure}

\begin{figure}[ht]
\centering
\includegraphics[width=0.495\textwidth]{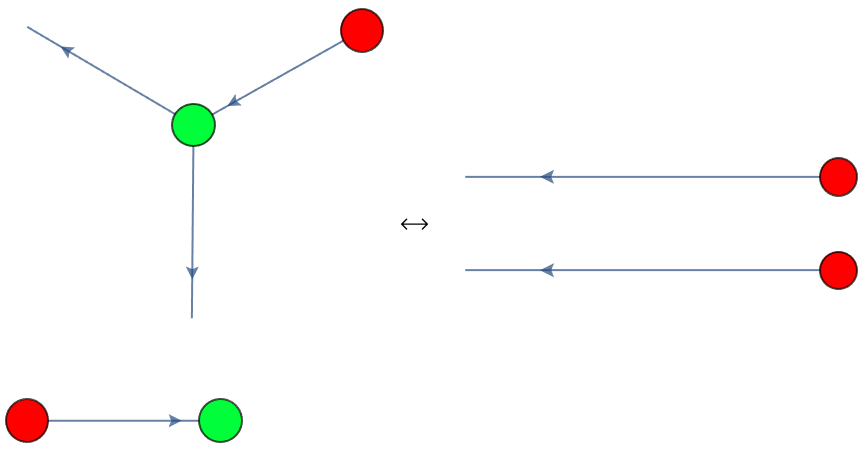}
\includegraphics[width=0.495\textwidth]{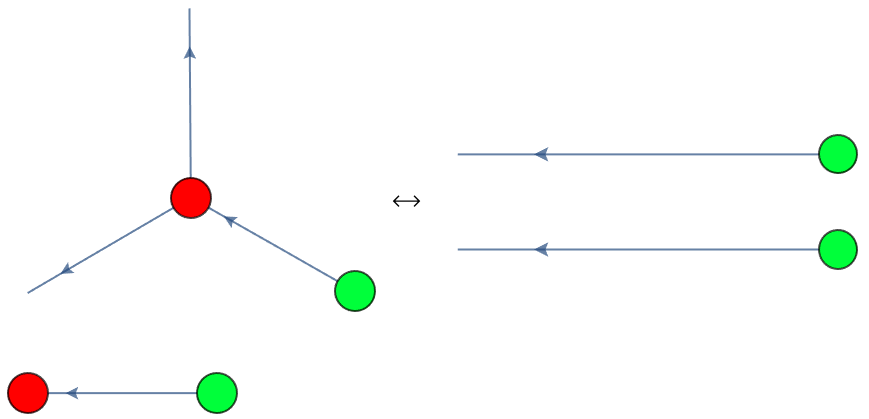}
\includegraphics[width=0.495\textwidth]{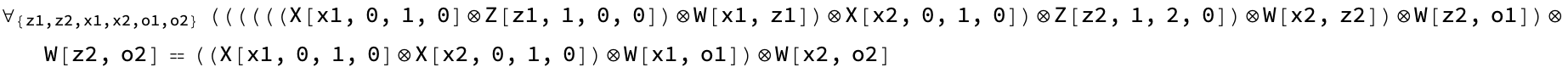}
\includegraphics[width=0.495\textwidth]{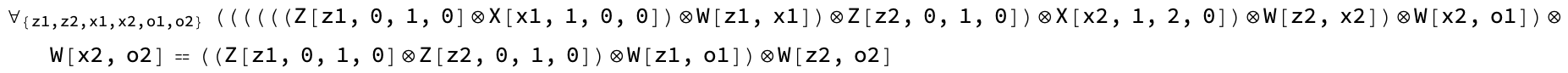}
\caption{The first and second defining diagrammatic equalities for the coherence of observable structures, along with their respective operator-theoretic forms.}
\label{fig:Figure24}
\end{figure}

\begin{figure}[ht]
\centering
\includegraphics[width=0.595\textwidth]{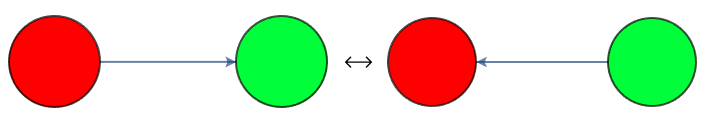}
\includegraphics[width=0.695\textwidth]{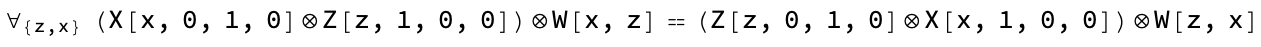}
\caption{The third defining diagrammatic equality for the coherence of observable structures, along with its associated operator-theoretic axiom form.}
\label{fig:Figure25}
\end{figure}

\begin{figure}[ht]
\centering
\includegraphics[width=0.595\textwidth]{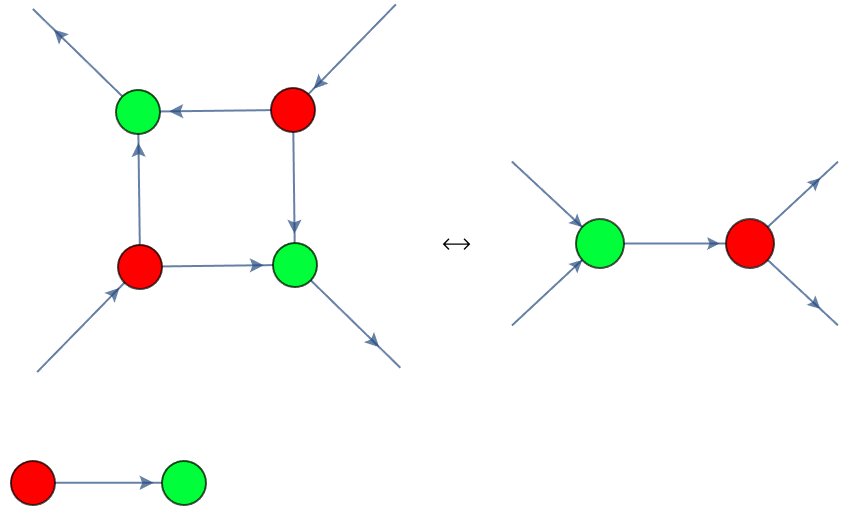}
\includegraphics[width=0.695\textwidth]{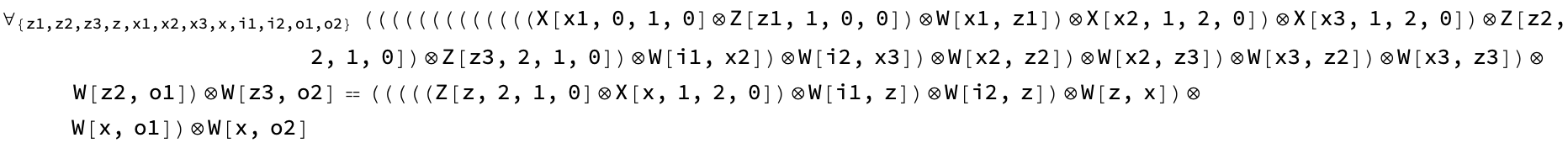}
\caption{The defining diagrammatic equality for the strong complementarity of observable structures, along with its associated operator-theoretic axiom form.}
\label{fig:Figure26}
\end{figure}

This notion of strong complementarity is a property of \textit{dagger special (commutative) Frobenius algebras}, first introduced by Coecke, Duncan, Kissinger and Wang\cite{coecke5}. In this context, a \textit{Frobenius algebra}, denoted ${\left( A, \mu, \eta, \delta, \epsilon \right)}$, associated with a monoidal category ${\left( \mathbf{C}, \otimes, I \right)}$, is simply an object $A$ in ${\mathrm{ob} \left( \mathbf{C} \right)}$ equipped with two pairs of morphisms, namely ${\left( \mu, \eta \right)}$ and ${\left( \delta, \epsilon \right)}$, of the form:

\begin{equation}
\mu : A \otimes A \to A, \qquad \text{ and } \qquad \eta : I \to A,
\end{equation}
and:

\begin{equation}
\delta : A \to A \otimes A, \qquad \text{ and } \qquad \epsilon : A \to I,
\end{equation}
in ${\mathrm{hom} \left( \mathbf{C} \right)}$, respectively, such that ${\left( A, \mu, \eta \right)}$ forms a monoid object in the monoidal category ${\mathbf{C}}$ (with ${\mu}$ being the \textit{multiplication} morphism and ${\eta}$ being the \textit{unit} morphism), i.e. such that the following diagrams:

\begin{equation}
\begin{tikzcd}
\left( A \otimes A \right) \otimes A \arrow[r, "\alpha"] \arrow[d, "\mu \otimes id"] & A \otimes \left( A \otimes A \right) \arrow[r, "id \otimes \mu"] & A \otimes A \arrow[d, "\mu"]\\
A \otimes A \arrow[rr, "\mu"] & & A
\end{tikzcd}, \qquad \text{ and } \qquad
\begin{tikzcd}
I \otimes A \arrow[r, "\eta \otimes id"] \arrow[dr, "\lambda"] & A \otimes A \arrow[d, "\mu"] & A \otimes I \arrow[l, "id \otimes \eta"] \arrow[dl, "\rho"]\\
& A &
\end{tikzcd},
\end{equation}
both commute, assuming associator isomorphism ${\alpha}$ and left and right unitor isomorphisms ${\lambda}$ and ${\rho}$ in the underlying monoidal category ${\mathbf{C}}$, as well as the monoidal identity object $I$, and also such that ${\left( A, \delta, \epsilon \right)}$ dually forms a comonoid object in the monoidal category ${\mathbf{C}}$, i.e. such that ${\left( A, \delta, \epsilon \right)}$ forms a monoid object in the opposite/dual category ${\mathbf{C}^{op}}$. Moreover, the monoid and comonoid objects must be specified in such a way that the following diagrams:

\begin{equation}
\begin{tikzcd}
A \otimes A \arrow[r, "\delta \otimes A"] \arrow[d, "\mu"] & A \otimes A \otimes A \arrow[d, "A \otimes \mu"]\\
A \arrow[r, "\delta"] & A \otimes A
\end{tikzcd}, \qquad \text{ and } \qquad
\begin{tikzcd}
A \otimes A \arrow[r, "A \otimes \delta"] \arrow[d, "\mu"] & A \otimes A \otimes A \arrow[d, "\mu \otimes A"]\\
A \arrow[r, "\delta"] & A \otimes A
\end{tikzcd},
\end{equation}
both commute. In this way, Frobenius algebras naturally generalize finite-dimensional unital associative algebras equipped with bilinear forms, provided that the underlying monoidal category ${\left( \mathbf{C}, \otimes, I \right)}$ is strict. The Frobenius algebra is \textit{commutative} if the associated monoid object ${\left( A, \mu, \eta \right)}$ (and, dually, the comonoid object ${\left( A, \delta, \epsilon \right)}$) is commutative, i.e. if the underlying monoidal category ${\left( \mathbf{C}, \otimes, I \right)}$ is symmetric, with symmetry isomorphism ${\sigma}$ in ${\mathrm{hom} \left( \mathbf{C} \right)}$, such that:

\begin{equation}
\mu \otimes \sigma = \mu.
\end{equation}
Furthermore, a \textit{dagger special (commutative) Frobenius algebra}\cite{coecke6}, denoted ${\mathcal{O}_{\color{green}{\circ}}}$, is simply a (commutative) Frobenius algebra $A$ associated with a dagger symmetric monoidal category ${\left( \mathbf{C}, \otimes, I, \dag \right)}$, whose pairs of morphisms ${\left( \mu_{\color{green}{\circ}}, \eta_{\color{green}{\circ}} \right)}$ and ${\left( \delta_{\color{green}{\circ}}, \epsilon_{\color{green}{\circ}} \right)}$ of the form:

\begin{equation}
\mu_{\color{green}{\circ}} : A \otimes A \to A, \qquad \text{ and } \qquad \eta_{\color{green}{\circ}} : I \to A,
\end{equation}
and:

\begin{equation}
\delta_{\color{green}{\circ}} : A \to A \otimes A, \qquad \text{ and } \qquad \epsilon_{\color{green}{\circ}} : A \to I,
\end{equation}
in ${\mathrm{hom} \left( \mathbf{C} \right)}$, respectively, are compatible with the underlying dagger structure ${\dag}$, such that the following equations hold:

\begin{equation}
\delta_{\color{green}{\circ}} = \left( \mu_{\color{green}{\circ}} \right)^{\dag}, \qquad \text{ and } \qquad \epsilon_{\color{green}{\circ}} = \left( \eta_{\color{green}{\circ}} \right)^{\dag},
\end{equation}
and also such that the diagrammatic equality shown in Figure \ref{fig:Figure22} holds, with the associated operator-theoretic axiom form:

\begin{multline}
\forall z_1, z_2, i, o : \left( \left( \left( \left( Z \left[ z_1, 1, 2, 0 \right] \otimes Z \left[ z_2, 2, 1, 0 \right] \right) \otimes W \left[ i, z_1 \right] \right) \otimes W \left[ z_1, z_2 \right] \right) \otimes W \left[ z_1, z_2 \right] \right) \otimes\\
W \left[ z_2, o \right] = W \left[ i, o \right].
\end{multline}
Since every non-degenerate quantum observable forms an orthonormal basis of eigenstates, and since, in the dagger symmetric monoidal category ${\mathbf{FdHilb}}$ of finite-dimensional Hilbert spaces, there exists a bijective correspondence between these orthonormal bases and dagger special commutative Frobenius algebras\cite{coecke7}\cite{coecke8}, we shall adopt the convention of referring to such dagger special commutative Frobenius algebras as \textit{observable structures} (since they naturally generalize the concept of quantum observables). \textit{Complementarity} of quantum observables can then be generalized to a notion of complementarity for a pair of observable structures ${\left( \mathcal{O}_{\color{green}{\circ}}, \mathcal{O}_{\color{red}{\circ}} \right)}$, acting on a common object $A$, defined by the diagrammatic equality shown in Figure \ref{fig:Figure23}, with the associated operator-theoretic axiom form:

\begin{multline}
\forall z_1, z_2, z, x_1, x_2, x, i, o, w : \left( \left( \left( \left( \left( \left( \left( \left( \left( Z \left[ z_1, 1, 2, 0 \right] \otimes Z \left[ z_2, 0, 2, 0 \right] \right) \otimes X \left[ x_1, 2, 0, 0 \right] \right) \otimes X \left[ x_2, 2, 1, 0 \right] \right) \otimes \right. \right. \right. \right. \right. \right.\\
\left. \left. \left. \left. \left. \left. W \left[ i, z_1 \right] \right) \otimes W \left[ z_1, z_2 \right] \right) \otimes W \left[ x_1, z_2 \right] \right) \otimes W \left[ x_1, x_2 \right] \right) \otimes W \left[ z_1, x_2 \right] \right) \otimes W \left[ x_2, o \right] \right) \otimes W \left[ w, w \right]\\
= \left( \left( Z \left[ z, 1, 0, 0 \right] \otimes X \left[ x, 0, 1, 0 \right] \right) \otimes W \left[ i, z \right] \right) \otimes W \left[ x, 0 \right].
\end{multline}
Finally a pair of observable structures ${\left( \mathcal{O}_{\color{green}{\circ}}, \mathcal{O}_{\color{red}{\circ}} \right)}$, acting on a common object $A$, are \textit{strongly complementary} if and only if they are \text{coherent}, meaning that the diagrammatic equalities shown in Figures \ref{fig:Figure24} and \ref{fig:Figure25} hold, with associated operator-theoretic axiom forms:

\begin{multline}
\forall z_1, z_2, x_1, x_2, o_1, o_2 : \left( \left( \left( \left( \left( \left( X \left[ x_1, 0, 1, 0 \right] \otimes Z \left[ z_1, 1, 0, 0 \right] \right) \otimes W \left[ x_1, z_1 \right] \right) \otimes X \left[ x_2, 0, 1, 0 \right] \right) \otimes \right. \right. \right.\\
\left. \left. \left. Z \left[ z_2, 1, 2, 0 \right] \right) \otimes W \left[ x_2, z_2 \right] \right) \otimes W \left[ z_2, o_1 \right] \right) \otimes W \left[ z_2, o_2 \right] = \left( \left( X \left[ x_1, 0, 1, 0 \right] \otimes X \left[ x_2, 0, 1, 0 \right] \right) \right.\\
\left. W \left[ x_1, o_1 \right] \right) \otimes W \left[ x_2, o_2 \right],
\end{multline}
\begin{multline}
\forall z_1, z_2, x_1, x_2, o_1, o_2 : \left( \left( \left( \left( \left( \left( Z \left[ z_1, 0, 1, 0 \right] \otimes X \left[ x_1, 1, 0, 0 \right] \right) \otimes W \left[ z_1, x_1 \right] \right) \otimes Z \left[ z_2, 0, 1, 0 \right] \right) \otimes \right. \right. \right.\\
\left. \left. \left. X \left[ x_2, 1, 2, 0 \right] \right) \otimes W \left[ z_2, x_2 \right] \right) \otimes W \left[ x_2, o_1 \right] \right) \otimes W \left[ x_2, o_2 \right] = \left( \left( Z \left[ z_1, 0, 1, 0 \right] \otimes Z \left[ z_2, 0, 1, 0 \right] \right) \right.\\
\left. W \left[ z_1, o_1 \right] \right) \otimes W \left[ z_2, o_2 \right],
\end{multline}
and:

\begin{equation}
\forall z, x : \left( X \left[ x, 0, 1, 0 \right] \otimes Z \left[ z, 1, 0, 0 \right] \right) \otimes W \left[ x, z \right] = \left( Z \left[ z, 0, 1, 0 \right] \otimes X \left[ x, 1, 0, 0 \right] \right) \otimes W \left[ z, x \right],
\end{equation}
respectively, and moreover the defining diagrammatic equality for strong complementarity shown in Figure \ref{fig:Figure26} holds, with the associated operator-theoretic axiom form:

\begin{multline}
\forall z_1, z_2, z_3, z, x_1, x_2, x_3, x, i_1, i_2, o_1, o_2 : \left( \left( \left( \left( \left( \left( \left( \left( \left( \left( \left( \left( \left( X \left[ x_1, 0, 1, 0 \right] \otimes Z \left[ z_1, 1, 0, 0 \right] \right) \otimes W \left[ x_1, z_1 \right] \right) \otimes \right. \right. \right. \right. \right. \right. \right. \right. \right. \right. \right.\\
\left. \left. \left. \left. \left. \left. X \left[ x_2, 1, 2, 0 \right] \right) \otimes X \left[ x_3, 1, 2, 0 \right] \right) \otimes Z \left[ z_2, 2, 1, 0 \right] \right) \otimes Z \left[ z_3, 2, 1, 0 \right] \right) \otimes W \left[ i_1, x_2 \right] \right) \otimes W \left[ i_2, x_3 \right] \right) \otimes\\
\left. \left. \left. \left. \left. W \left[ x_2, z_2 \right] \right) \otimes W \left[ x_2, z_3 \right] \right) \otimes W \left[ x_3, z_2 \right] \right) \otimes W \left[ x_3, z_3 \right] \right) \otimes W \left[ z_2, o_1 \right] \right) \otimes W \left[ z_3, o_2 \right]\\
= \left( \left( \left( \left( \left( Z \left[ z, 2, 1, 0 \right] \otimes X \left[ x, 1, 2, 0 \right] \right) \otimes W \left[ i_1, z \right] \right) \otimes W \left[ i_2, z \right] \right) \otimes W \left[ z, x \right] \right) \otimes W \left[ x, o_1 \right] \right) \otimes W \left[ x, o_2 \right].
\end{multline}
Thus, we confirm immediately that the computational and Hadamard-transformed orthonormal bases, when interpreted as observable structures, are indeed strongly complementary, as required.

\begin{figure}[ht]
\centering
\includegraphics[width=0.995\textwidth]{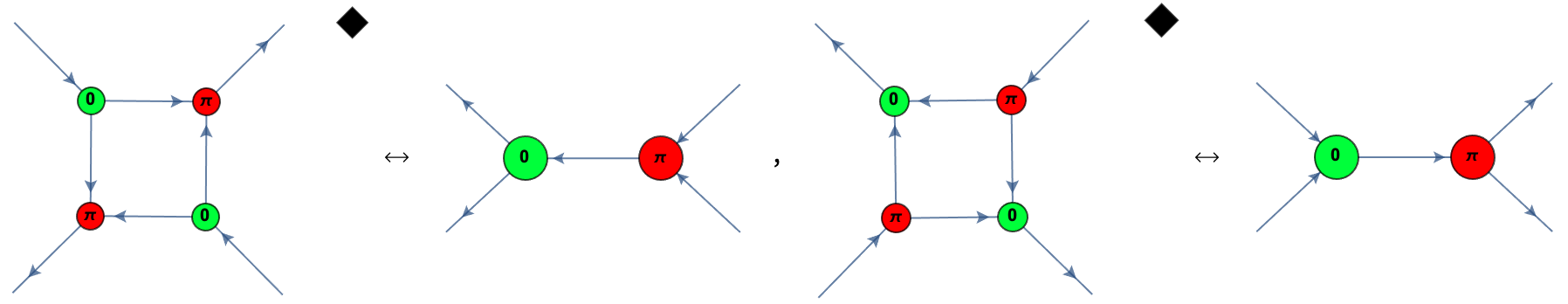}
\includegraphics[width=0.995\textwidth]{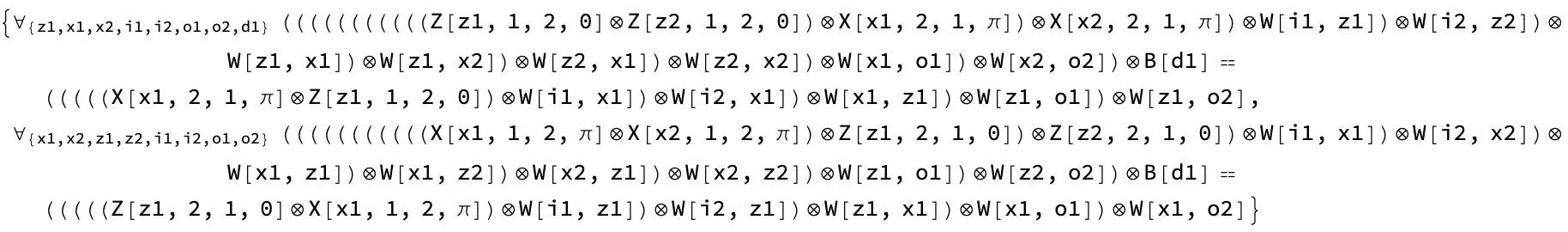}
\caption{The first pair of supplementary B2 rules (i.e. supplementary bialgebra simplification rules) in the most general case, in which the X-spider phases are equal to ${\pi}$, with the Z-spider phases equal to zero, along with their respective operator-theoretic axiom forms.}
\label{fig:Figure27}
\end{figure}

\begin{figure}[ht]
\centering
\includegraphics[width=0.995\textwidth]{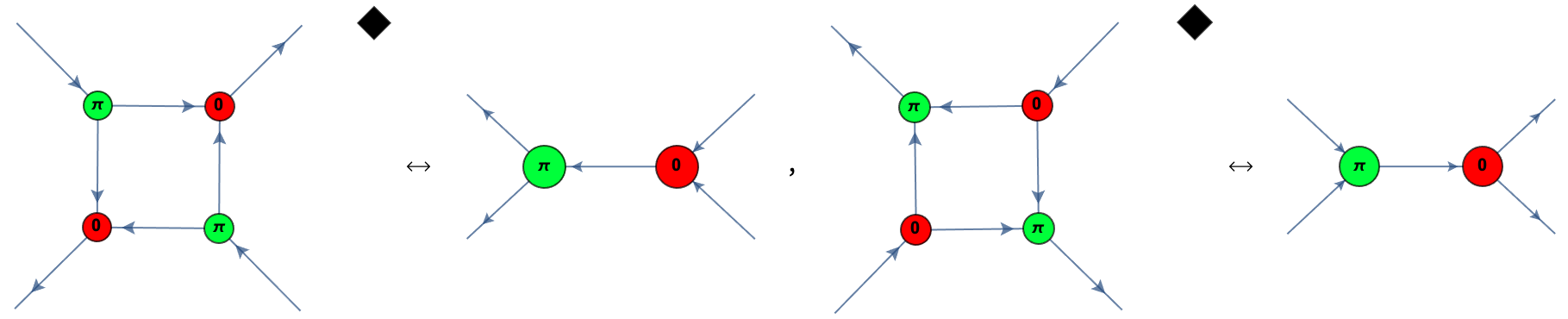}
\includegraphics[width=0.995\textwidth]{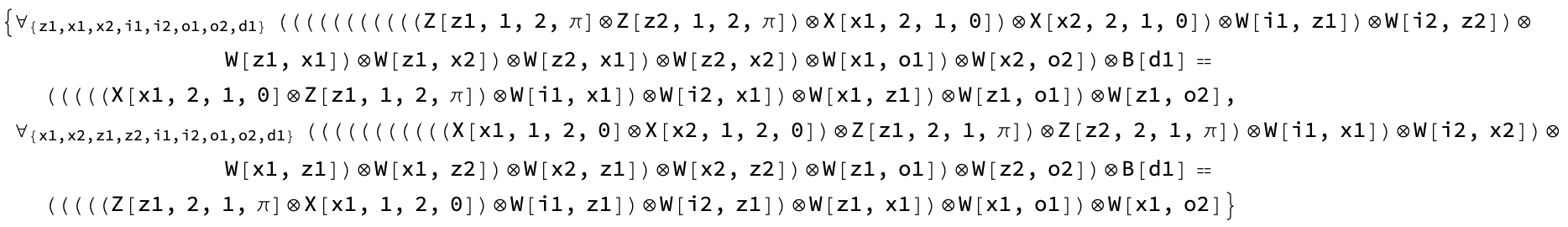}
\caption{The second pair of supplementary B2 rules (i.e. supplementary bialgebra simplification rules) in the most general case, in which the Z-spider phases are equal to ${\pi}$, with the X-spider phases equal to zero, along with their respective operator-theoretic axiom forms.}
\label{fig:Figure28}
\end{figure}

For the sake of axiomatic completeness, note that we must also add an additional set of B2 rules (beyond the purely phaseless rules presented above), to deal with the case in which the phase of one of the Z- or X-spiders is equal to ${\pi}$, with the other being zero. We can see easily that this additional set of equalities must hold, by noting that:

\begin{equation}
\sqrt{2} \left( \mathbf{Z} \left( \alpha \right) \otimes \mathbf{Z} \left( \alpha \right) \right) \left( \mathbf{I}_2 \otimes \mathbf{SWAP} \otimes \mathbf{I}_2 \right) \left( \mathbf{X} \left( \beta \right) \otimes \mathbf{X} \left( \beta \right) \right) = \mathbf{X} \left( \beta \right) \mathbf{Z} \left( \alpha \right),
\end{equation}
whenever ${\alpha = \pi, \beta = 0}$ or ${\alpha = 0, \beta = \pi}$, where ${\mathbf{I}_2}$ denotes the ${2 \times 2}$ identity matrix, ${\mathbf{H}}$, ${\mathbf{Z} \left( \alpha \right)}$ and ${\mathbf{X} \left( \beta \right)}$ correspond to explicit algebraic representations of the Hadamard, Z- and X-spiders (expressed in terms of the computational basis), respectively:

\begin{equation}
\mathbf{H} = \frac{1}{\sqrt{2}} \begin{bmatrix}
1 & 1\\
1 & -1
\end{bmatrix}, \qquad \mathbf{Z} \left( \alpha \right) = \begin{bmatrix}
1 & 0\\
0 & 0\\
0 & 0\\
0 & e^{i \alpha}
\end{bmatrix}, \qquad \mathbf{X} \left( \beta \right) = \mathbf{H} \begin{bmatrix}
1 & 0 & 0 & 0\\
0 & 0 & 0 & e^{i \beta}
\end{bmatrix} \left( \mathbf{H} \otimes \mathbf{H} \right),
\end{equation}
and where the wire swap operation ${\mathbf{SWAP}}$ is also represented explicitly as:

\begin{equation}
\mathbf{SWAP} = \begin{bmatrix}
1 & 0 & 0 & 0\\
0 & 0 & 1 & 0\\
0 & 1 & 0 & 0\\
0 & 0 & 0 & 1
\end{bmatrix}.
\end{equation}
The supplementary B2 rules (i.e. the supplementary bialgebra simplification rules), corresponding to the case in which the X-spider phases are equal to ${\pi}$ (with Z-spider phases equal to zero), and the case in which the Z-spider phases are equal to ${\pi}$ (with X-spider phases equal to zero), have the forms shown in Figures \ref{fig:Figure27} and \ref{fig:Figure28}, respectively, with the operator-theoretic axiom forms being:

\begin{multline}
\forall z_1, z_2, x_1, x_2, i_1, i_2, o_1, o_2, d_1 : \left( \left( \left( \left( \left( \left( \left( \left( \left( \left( \left( Z \left[ z_1, 1, 2, 0 \right] \otimes Z \left[ z_2, 1, 2, 0 \right] \right) \otimes X \left[ x_1, 2, 1, \pi \right] \right) \otimes \right. \right. \right. \right. \right. \right. \right. \right. \right.\\
\left. \left. \left. \left. \left. \left. \left. X \left[ x_2, 2, 1, \pi \right] \right) \otimes W \left[ i_1, z_1 \right] \right) \otimes W \left[ i_2, z_2 \right] \right) \otimes W \left[ z_1, x_1 \right] \right) \otimes W \left[ z_1, x_2 \right] \right) \otimes W \left[ z_2, x_1 \right] \right) \otimes W \left[ z_2, x_2 \right] \right) \otimes\\
\left. \left. W \left[ x_1, o_1 \right] \right) \otimes W \left[ x_2, o_2 \right] \right) \otimes B \left[ d_1 \right] = \left( \left( \left( \left( \left( X \left[ x_1, 2, 1, \pi \right] \otimes Z \left[ z_1, 1, 2, 0 \right] \right) \otimes W \left[ i_1, z_2 \right] \right) \otimes W \left[ i_2, x_1 \right] \right) \otimes \right. \right.\\
\left. \left. W \left[ x_1, z_1 \right] \right) \otimes W \left[ z_1, o_1 \right] \right) \otimes W \left[ z_1, o_2 \right],
\end{multline}
\begin{multline}
\forall x_1, x_2, z_1, z_2, i_1, i_2, o_1, o_2, d_1 : \left( \left( \left( \left( \left( \left( \left( \left( \left( \left( \left( X \left[ x_1, 1, 2, \pi \right] \otimes X \left[ x_2, 1, 2, \pi \right] \right) \otimes Z \left[ z_1, 2, 1, 0 \right] \right) \otimes \right. \right. \right. \right. \right. \right. \right. \right. \right.\\
\left. \left. \left. \left. \left. \left. \left. Z \left[ z_2, 2, 1, 0 \right] \right) \otimes W \left[ i_1, x_1 \right] \right) \otimes W \left[ i_2, x_2 \right] \right) \otimes W \left[ x_1, z_1 \right] \right) \otimes W \left[ x_1, z_2 \right] \right) \otimes W \left[ x_2, z_1 \right] \right) \otimes W \left[ x_2, z_2 \right] \right) \otimes\\
\left. \left. W \left[ z_1, o_1 \right] \right) \otimes W \left[ z_2, o_2 \right] \right) \otimes B \left[ d_1 \right] = \left( \left( \left( \left( \left( Z \left[ z_1, 2, 1, 0 \right] \otimes X \left[ x_1, 1, 2, \pi \right] \right) \otimes W \left[ i_1, z_1 \right] \right) \otimes W \left[ i_2, z_1 \right] \right) \otimes \right. \right.\\
\left. \left. W \left[ z_1, x_1 \right] \right) \otimes W \left[ x_1, o_1 \right] \right) \otimes \left[ x_1, o_2 \right],
\end{multline}
and:

\begin{multline}
\forall z_1, z_2, x_1, x_2, i_1, i_2, o_1, o_2, d_1 : \left( \left( \left( \left( \left( \left( \left( \left( \left( \left( \left( Z \left[ z_1, 1, 2, \pi \right] \otimes Z \left[ z_2, 1, 2, \pi \right] \right) \otimes X \left[ x_1, 2, 1, 0 \right] \right) \otimes \right. \right. \right. \right. \right. \right. \right. \right. \right.\\
\left. \left. \left. \left. \left. \left. \left. X \left[ x_2, 2, 1, 0 \right] \right) \otimes W \left[ i_1, z_1 \right] \right) \otimes W \left[ i_2, z_2 \right] \right) \otimes W \left[ z_1, x_1 \right] \right) \otimes W \left[ z_1, x_2 \right] \right) \otimes W \left[ z_2, x_1 \right] \right) \otimes W \left[ z_2, x_2 \right] \right) \otimes\\
\left. \left. W \left[ x_1, o_1 \right] \right) \otimes W \left[ x_2, o_2 \right] \right) \otimes B \left[ d_1 \right] = \left( \left( \left( \left( \left( X \left[ x_1, 2, 1, 0 \right] \otimes Z \left[ x_1, 1, 2, \pi \right] \right) \otimes W \left[ i_1, x_1 \right] \right) \otimes W \left[ i_2, x_1 \right] \right) \otimes \right. \right.\\
\left. \left. W \left[ x_1, z_1 \right] \right) \otimes W \left[ z_1, o_1 \right] \right) \otimes W \left[ z_1, o_2 \right],
\end{multline}
\begin{multline}
\forall x_1, x_2, z_1, z_2, i_1, i_2, o_1, o_2, d_1 : \left( \left( \left( \left( \left( \left( \left( \left( \left( \left( \left( X \left[ x_1, 1, 2, 0 \right] \otimes X \left[ x_2, 1, 2, 0 \right] \right) \otimes Z \left[ z_1, 2, 1, \pi \right] \right) \otimes \right. \right. \right. \right. \right. \right. \right. \right. \right.\\
\left. \left. \left. \left. \left. \left. \left. Z \left[ z_2, 2, 1, \pi \right] \right) \otimes W \left[ i_1, x_1 \right] \right) \otimes W \left[ i_2, x_2 \right] \right) \otimes W \left[ x_1, z_1 \right] \right) \otimes W \left[ x_1, z_2 \right] \right) \otimes W \left[ x_2, z_1 \right] \right) \otimes W \left[ x_2, z_2 \right] \right) \otimes\\
\left. \left. W \left[ z_1, o_1 \right] \right) \otimes W \left[ z_2, o_2 \right] \right) \otimes B \left[ d_1 \right] = \left( \left( \left( \left( \left( Z \left[ z_1, 2, 1, \pi \right] \otimes X \left[ x_1, 1, 2, 0 \right] \right) \otimes W \left[ i_1, z_1 \right] \right) \otimes W \left[ i_2, z_1 \right] \right) \otimes \right. \right.\\
\left. \left. W \left[ z_1, x_1 \right] \right) \otimes W \left[ x_1, o_1 \right] \right) \otimes W \left[ x_1, o_2 \right],
\end{multline}
respectively.

\begin{figure}[ht]
\centering
\includegraphics[width=0.495\textwidth]{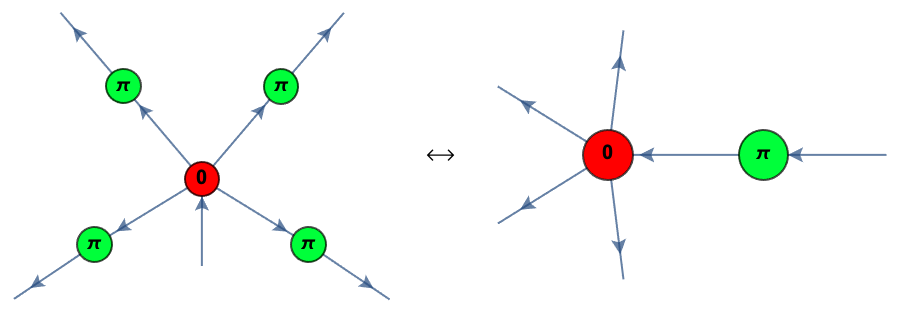}
\includegraphics[width=0.495\textwidth]{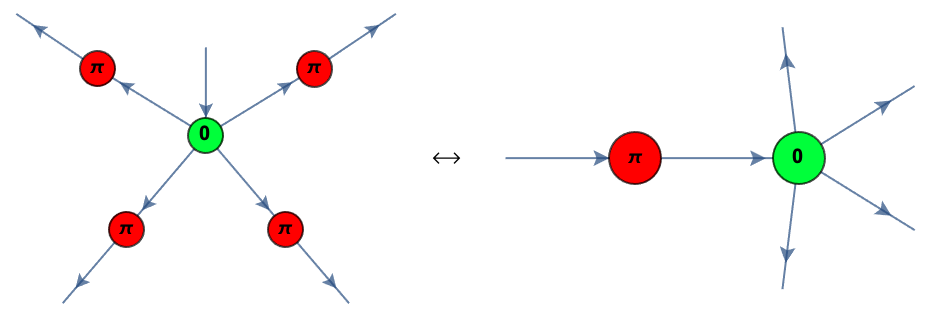}
\includegraphics[width=0.495\textwidth]{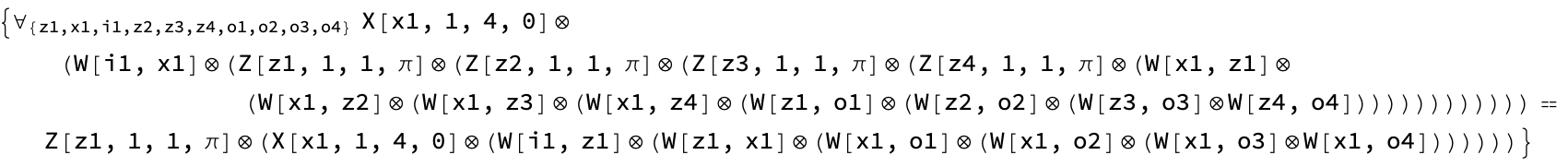}
\includegraphics[width=0.495\textwidth]{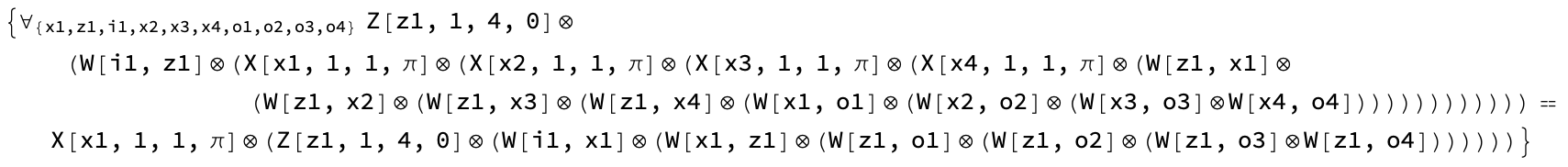}
\caption{The K1 rules (i.e. the Z- and X-spider ${\pi}$-copy rules) for the case in which the output arities of the Z- and X-spiders are respectively equal to 4, along with their associated operator-theoretic axiom forms.}
\label{fig:Figure29}
\end{figure}

\begin{figure}[ht]
\centering
\includegraphics[width=0.495\textwidth]{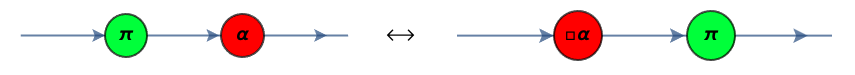}
\includegraphics[width=0.495\textwidth]{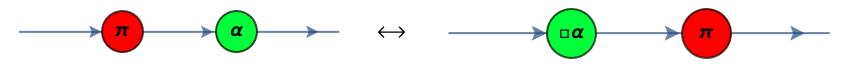}
\includegraphics[width=0.495\textwidth]{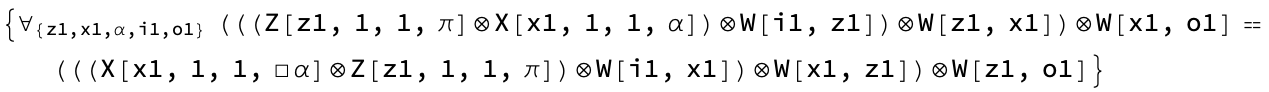}
\includegraphics[width=0.495\textwidth]{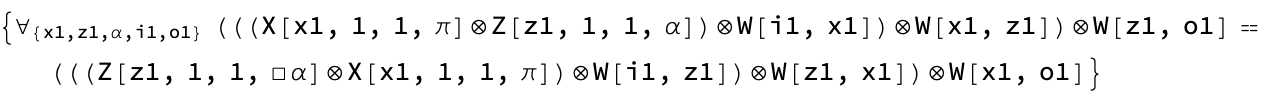}
\caption{The K2 rules (i.e. the Z- and X-spider phase flip rules) in the most general case, along with their associated operator-theoretic axiom forms.}
\label{fig:Figure30}
\end{figure}

Next, we translate the K1 rules (i.e. the Z- and X-spider ${\pi}$-copy rules), which, for the case in which the output arities of the Z- and X-spiders are respectively equal to 4, have the form shown in Figure \ref{fig:Figure29}, with the operator-theoretic axiom forms being:

\begin{multline}
\forall z_1, x_1, i_1, z_2, z_3, z_4, o_1, o_2, o_3, o_4 : X \left[ x_1, 1, 4, 0 \right] \otimes \left( W \left[ i_1, x_1 \right] \otimes \left( Z \left[ z_1, 1, 1, \pi \right] \otimes \left( Z \left[ z_2, 1, 1, \pi \right] \otimes \right. \right. \right.\\
\left( Z \left[ z_3, 1, 1, \pi \right] \otimes \left( Z \left[ z_4, 1, 1, \pi \right] \otimes \left( W \left[ x_1, z_1 \right] \otimes \left( W \left[ x_1, z_2 \right] \otimes \left( W \left[ x_1, z_3 \right] \otimes \left( W \left[ x_1, z_4 \right] \otimes \left( W \left[ z_1, o_1 \right] \otimes \right. \right. \right. \right. \right. \right. \right.\\
\left. \left. \left. \left. \left. \left. \left. \left. \left. \left( W \left[ z_2, o_2 \right] \otimes \left( W \left[ z_3, o_3 \right] \otimes W \left[ z_4, o_4 \right] \right) \right) \right) \right) \right) \right) \right) \right) \right) \right) \right) = Z \left[ z_1, 1, 1, \pi \right] \otimes \left( X \left[ x_1, 1, 4, 0 \right] \otimes \left( W \left[ i_1, z_1 \right] \otimes \right. \right.\\
\left. \left. \left( W \left[ z_1, x_1 \right] \otimes \left( W \left[ x_1, o_1 \right] \otimes \left( W \left[ x_1, o_2 \right] \otimes \left( W \left[ x_1, o_3 \right] \otimes W \left[ x_1, o_4 \right] \right) \right) \right) \right) \right) \right),
\end{multline}
and:

\begin{multline}
\forall x_1, z_1, i_1, x_2, x_3, x_4, o_1, o_2, o_3, o_4 : Z \left[ z_1, 1, 4, 0 \right] \otimes \left( W \left[ i_1, z_1 \right] \otimes \left( X \left[ x_1, 1, 1, \pi \right] \otimes \left( X \left[ x_2, 1, 1, \pi \right] \otimes \right. \right. \right.\\
\left( X \left[ x_3, 1, 1, \pi \right] \otimes \left( X \left[ x_4, 1, 1, \pi \right] \otimes \left( W \left[ z_1, x_1 \right] \otimes \left( W \left[ z_1, x_2 \right] \otimes \left( W \left[ z_1, x_3 \right] \otimes \left( W \left[ z_1, x_4 \right] \otimes \left( W \left[ x_1, o_1 \right] \otimes \right. \right. \right. \right. \right. \right. \right.\\
\left. \left. \left. \left. \left. \left. \left. \left. \left. \left( W \left[ x_2, o_2 \right] \otimes \left( W \left[ x_3, o_3 \right] \otimes W \left[ x_4, o_4 \right] \right) \right) \right) \right) \right) \right) \right) \right) \right) \right) \right) = X \left[ x_1, 1, 1, \pi \right] \otimes \left( Z \left[ z_1, 1, 4, 0 \right] \otimes \left( W \left[ i_1, x_1 \right] \otimes \right. \right.\\
\left. \left. \left( W \left[ x_1, z_1 \right] \otimes \left( W \left[ z_1, o_1 \right] \otimes \left( W \left[ z_1, o_2 \right] \otimes \left( W \left[ z_1, o_3 \right] \otimes W \left[ z_1, o_4 \right] \right) \right) \right) \right) \right) \right),
\end{multline}
respectively. These rules are derived from the fact that a Hadamard NOT gate, as represented by an X-spider with a phase of ${\pi}$ and with input and output arities both equal to 1, will always copy through a Z-spider (since the Hadamard NOT gate is a function map of the Hadamard-transformed basis, always mapping Hadamard-transformed basis states to Hadamard-transformed basis states), while a computational NOT gate, as represented by a Z-spider with a phase of ${\pi}$ and with input and output arities both equal to 1, will always copy through an X-spider (since the computational NOT gate is a function map of the computational basis, always mapping computational basis states to computational basis states). We also translate the K2 rules (i.e. the Z- and X-spider phase flip rules), which, in the most general case, have the form shown in Figure \ref{fig:Figure30}, with the operator-theoretic axiom forms being:

\begin{multline}
\forall z_1, x_1, \alpha, i_1, o_1 : \left( \left( \left( Z \left[ z_1, 1, 1, \pi \right] \otimes X \left[ x_1, 1, 1, \alpha \right] \right) \otimes W \left[ i_1, z_1 \right] \right) \otimes W \left[ z_1, x_1 \right] \right) \otimes W \left[ x_1, o_1 \right]\\
= \left( \left( \left( X \left[ x_1, 1, 1, \Box \alpha \right] \otimes Z \left[ z_1, 1, 1, \pi \right] \right) \otimes W \left[ i_1, x_1 \right] \right) \otimes W \left[ x_1, z_1 \right] \right) \otimes W \left[ z_1, o_1 \right],
\end{multline}
and:

\begin{multline}
\forall x_1, z_1, \alpha, i_1, o_1 : \left( \left( \left( X \left[ x_1, 1, 1, \pi \right] \otimes Z \left[ z_1, 1, 1, \alpha \right] \right) \otimes W \left[ i_1, x_1 \right] \right) \otimes W \left[ x_1, z_1 \right] \right) \otimes W \left[ z_1, o_1 \right]\\
= \left( \left( \left( Z \left[ z_1, 1, 1, \Box \alpha \right] \otimes X \left[ x_1, 1, 1, \pi \right] \right) \otimes W \left[ i_1, z_1 \right] \right) \otimes W \left[ z_1, x_1 \right] \right) \otimes W \left[ x_1, o_1 \right],
\end{multline}
respectively. These rules are derived from the fact that a Hadamard NOT gate, as represented by an X-spider with a phase of ${\pi}$ and with input and output arities both equal to 1, will induce a negated phase whenever it is commuted through a Z-rotation gate (since the rotation of the latter gate flips), while a computational NOT gate, as represented by a Z-spider with a phase of ${\pi}$ and with input and output arities both equal to 1, will induce a negated phase whenever it is commuted through an X-rotation gate (since the rotation of the latter gate flips).

\begin{figure}[ht]
\centering
\includegraphics[width=0.495\textwidth]{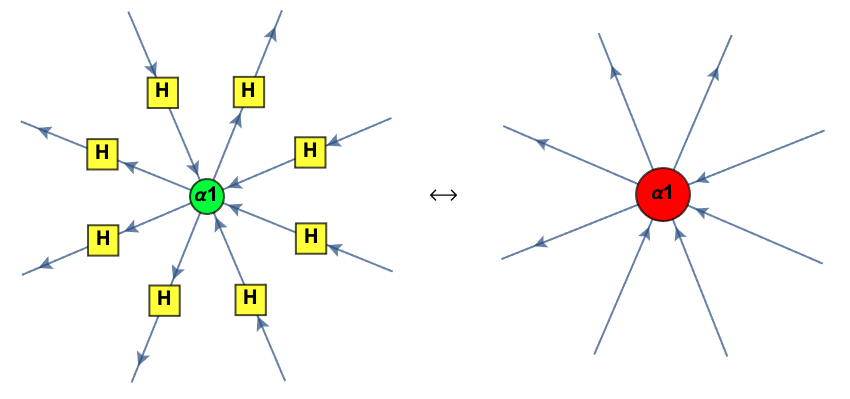}
\includegraphics[width=0.495\textwidth]{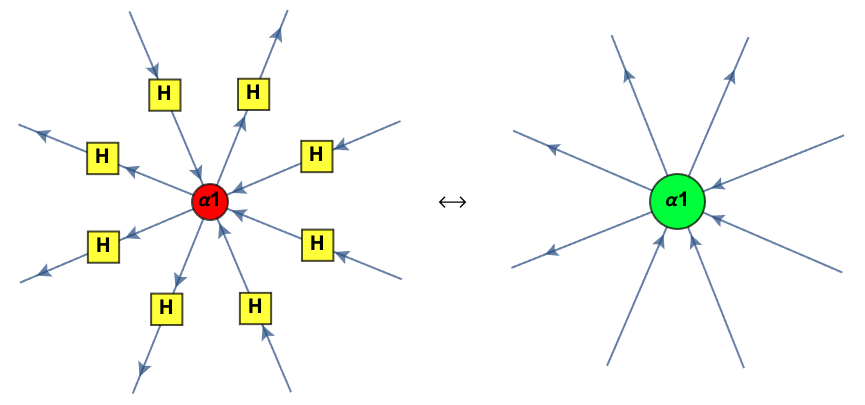}
\includegraphics[width=0.495\textwidth]{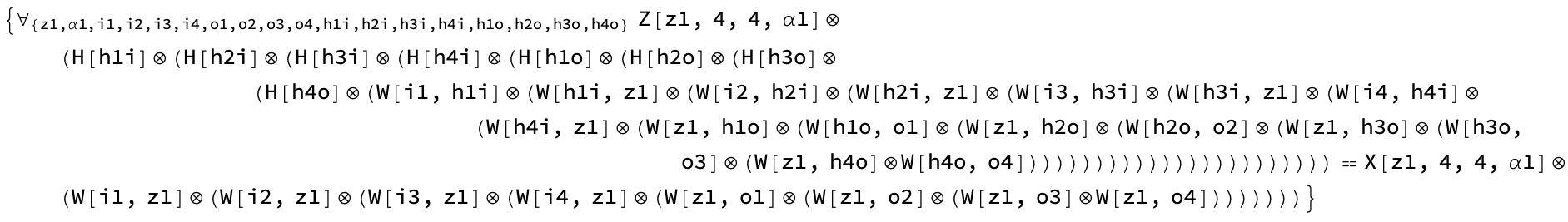}
\includegraphics[width=0.495\textwidth]{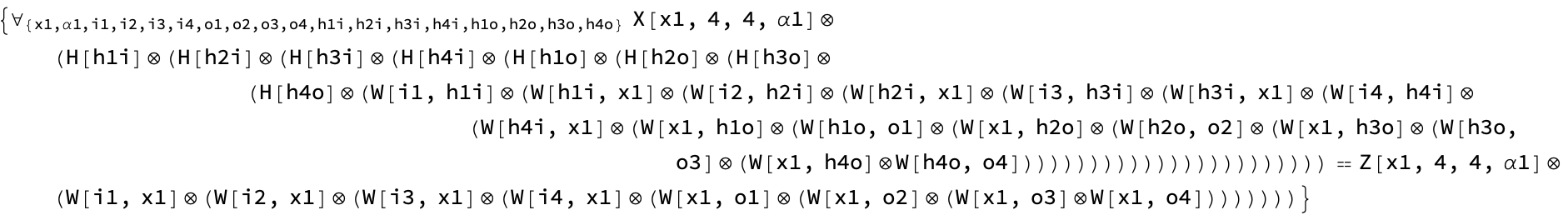}
\caption{The C rules (i.e. the Z- and X-spider color change rules) for the case in which the input and output arities of both spiders are equal to 4, along with their respective operator-theoretic axiom forms.}
\label{fig:Figure31}
\end{figure}

\begin{figure}[ht]
\centering
\includegraphics[width=0.595\textwidth]{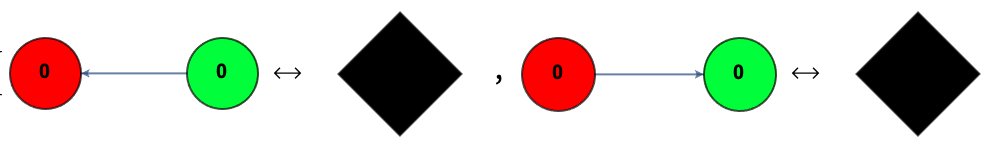}
\includegraphics[width=0.695\textwidth]{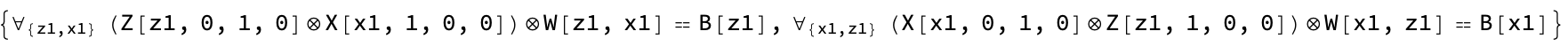}
\caption{The D1 rules (or spider cancellation rules) in the most general case, along with their associated operator-theoretic axiom forms.}
\label{fig:Figure32}
\end{figure}

Penultimately, we translate the C rules (i.e. the Z- and X-spider color change rules), which, for the case in which the input and output arities of both spiders are equal to 4, have the form shown in Figure \ref{fig:Figure31}, with the operator-theoretic axiom forms being:

\begin{multline}
\forall z_1, \alpha_1, i_1, i_2, i_3, i_4, o_1, o_2, o_3, o_4, h_{1}^{i}, h_{2}^{i}, h_{3}^{i}, h_{4}^{i}, h_{1}^{o}, h_{2}^{o}, h_{3}^{o}, h_{4}^{o} : Z \left[ z_1, 4, 4, \alpha_1 \right] \otimes \left( H \left[ h_{1}^{i} \right] \otimes \left( H \left[ h_{2}^{i} \right] \otimes \right. \right.\\
\left( H \left[ h_{3}^{i} \right] \otimes \left( H \left[ h_{4}^{i} \right] \otimes \left( H \left[ h_{1}^{o} \right] \otimes \left( H \left[ h_{2}^{o} \right] \otimes \left( H \left[ h_{3}^{o} \right] \otimes \left( H \left[ h_{4}^{o} \right] \otimes \left( W \left[ i_1, h_{1}^{i} \right] \otimes \left( W \left[ h_{1}^{i}, z_1 \right] \otimes \right. \right. \right. \right. \right. \right. \right. \right.\\
\left( W \left[ i_2, h_{2}^{i} \right] \otimes \left( W \left[ h_{2}^{i}, z_1 \right] \otimes \left( W \left[ i_3, h_{3}^{i} \right] \otimes \left( W \left[ h_{3}^{i}, z_1 \right] \otimes \left( W \left[ i_4, h_{4}^{i} \right] \otimes \left( W \left[ h_{4}^{i}, z_1 \right] \otimes \right. \right. \right. \right. \right. \right.\\
\left( W \left[ z_1, h_{1}^{o} \right] \otimes \left( W \left[ h_{1}^{o}, o_1 \right] \otimes \left( W \left[ z_1, h_{2}^{o} \right] \otimes \left( W \left[ h_{2}^{o}, o_2 \right] \otimes \left( W \left[ z_1, h_{3}^{o} \right] \otimes \left( W \left[ h_{3}^{o}, o_3 \right] \otimes \right. \right. \right. \right. \right. \right.\\
\left. \left. \left. \left. \left. \left. \left. \left. \left. \left. \left. \left. \left. \left. \left. \left. \left. \left. \left. \left. \left( W \left[ z_1, h_{4}^{o} \right] \otimes W \left[ h_{4}^{o}, o_4 \right] \right) \right) \right) \right) \right) \right) \right) \right) \right) \right) \right) \right) \right) \right) \right) \right) \right) \right) \right) \right) \right) = X \left[ z_1, 4, 4, \alpha_1 \right] \otimes \left( W \left[ i_1, z_1 \right] \otimes \left( W \left[ i_2, z_1 \right] \otimes \right. \right.\\
\left. \left. \left( W \left[ i_3, z_1 \right] \otimes \left( W \left[ i_4, z_1 \right] \otimes \left( W \left[ z_1, o_1 \right] \otimes \left( W \left[ z_1, o_2 \right] \otimes \left( W \left[ z_1, o_3 \right] \otimes W \left[ z_1, o_4 \right] \right) \right) \right) \right) \right) \right) \right),
\end{multline}
and:

\begin{multline}
\forall x_1, \alpha_1, i_1, i_2, i_3, i_4, o_1, o_2, o_3, o_4, h_{1}^{i}, h_{2}^{i}, h_{3}^{i}, h_{4}^{i}, h_{1}^{o}, h_{2}^{o}, h_{3}^{o}, h_{4}^{o} : X \left[ x_1, 4, 4, \alpha_1 \right] \otimes \left( H \left[ h_{1}^{i} \right] \otimes \left( H \left[ h_{2}^{i} \right] \otimes \right. \right.\\
\left( H \left[ h_{3}^{i} \right] \otimes \left( H \left[ h_{4}^{i} \right] \otimes \left( H \left[ h_{1}^{o} \right] \otimes \left( H \left[ h_{2}^{o} \right] \otimes \left( H \left[ h_{3}^{o} \right] \otimes \left( H \left[ h_{4}^{o} \right] \otimes \left( W \left[ i_1, h_{1}^{i} \right] \otimes \left( W \left[ h_{1}^{i}, x_1 \right] \otimes \right. \right. \right. \right. \right. \right. \right. \right.\\
\left( W \left[ i_2, h_{2}^{i} \right] \otimes \left( W \left[ h_{2}^{i}, x_1 \right] \otimes \left( W \left[ i_3, h_{3}^{i} \right] \otimes \left( W \left[ h_{3}^{i}, x_1 \right] \otimes \left( W \left[ i_4, h_{4}^{i} \right] \otimes \left( W \left[ h_{4}^{i}, x_1 \right] \otimes \right. \right. \right. \right. \right. \right.\\
\left( W \left[ x_1, h_{1}^{o} \right] \otimes \left( W \left[ h_{1}^{o}, o_1 \right] \otimes \left( W \left[ x_1, h_{2}^{o} \right] \otimes \left( W \left[ h_{2}^{o}, o_2 \right] \otimes \left( W \left[ x_1, h_{3}^{o} \right] \otimes \left( W \left[ h_{3}^{o}, o_3 \right] \otimes \right. \right. \right. \right. \right. \right.\\
\left. \left. \left. \left. \left. \left. \left. \left. \left. \left. \left. \left. \left. \left. \left. \left. \left. \left. \left. \left. \left( W \left[ x_1, h_{4}^{o} \right] \otimes W \left[ h_{4}^{o}, o_4 \right] \right) \right) \right) \right) \right) \right) \right) \right) \right) \right) \right) \right) \right) \right) \right) \right) \right) \right) \right) \right) \right) = Z \left[ x_1, 4, 4, \alpha_1 \right] \otimes \left( W \left[ i_1, x_1 \right] \otimes \left( W \left[ i_2, x_1 \right] \otimes \right. \right.\\
\left. \left. \left( W \left[ i_3, z_1 \right] \otimes \left( W \left[ i_4, x_1 \right] \otimes \left( W \left[ x_1, o_1 \right] \otimes \left( W \left[ x_1, o_2 \right] \otimes \left( W \left[ x_1, o_3 \right] \otimes W \left[ x_1, o_4 \right] \right) \right) \right) \right) \right) \right) \right),
\end{multline}
respectively. These rules are derived from the fact that a Hadamard gate inverts the color of an arbitrary Z- or X-spider (since, by definition, the Hadamard transform maps from the computational basis to the Hadamard-transformed basis, and back again). Finally, we translate the D1 and D2 rules (i.e. the spider cancellation and scalar multiplication rules), which, in the most general case, have the form shown in Figures \ref{fig:Figure31} and \ref{fig:Figure32}, respectively, with the operator-theoretic axiom forms being:

\begin{equation}
\forall z_1, x_1 : \left( Z \left[ z_1, 0, 1, 0 \right] \otimes X \left[ x_1, 1, 0, 0 \right] \right) \otimes W \left[ z_1, x_1 \right] = B \left[ z_1 \right],
\end{equation}
\begin{equation}
\forall x_1, z_1 : \left( X \left[ x_1, 0, 1, 0 \right] \otimes Z \left[ z_1, 1, 0, 0 \right] \right) \otimes W \left[ x_1, z_1 \right] = B \left[ x_1 \right],
\end{equation}
and:

\begin{equation}
\forall d_1, d_2 : B \left[ d_1 \right] \otimes B \left[ d_2 \right] = W \left[ d_1, d_2 \right],
\end{equation}
respectively. The D1 rules are derived from the fact that Z- and X-spiders cancel to yield a single black diamond (due to the reduction of a pair of basis states down  to a single scalar factor), while the D2 rules are derived from the fact that two black diamonds cancel to yield a single loop of wire (due to the reduction of two ${\sqrt{D}}$ multiplicative scalar factors to a single $D$ multiplicative scalar factor).

\begin{figure}[ht]
\centering
\includegraphics[width=0.395\textwidth]{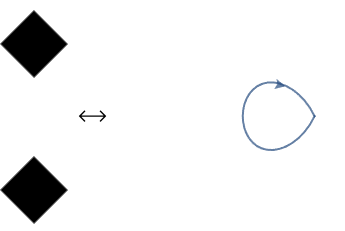}\\
\includegraphics[width=0.395\textwidth]{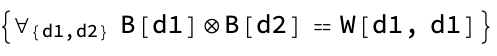}
\caption{The D2 rule (or scalar multiplication rule) in the most general case, along with its associated operator-theoretic axiom form.}
\label{fig:Figure33}
\end{figure}

\begin{figure}[ht]
\centering
\includegraphics[width=0.695\textwidth]{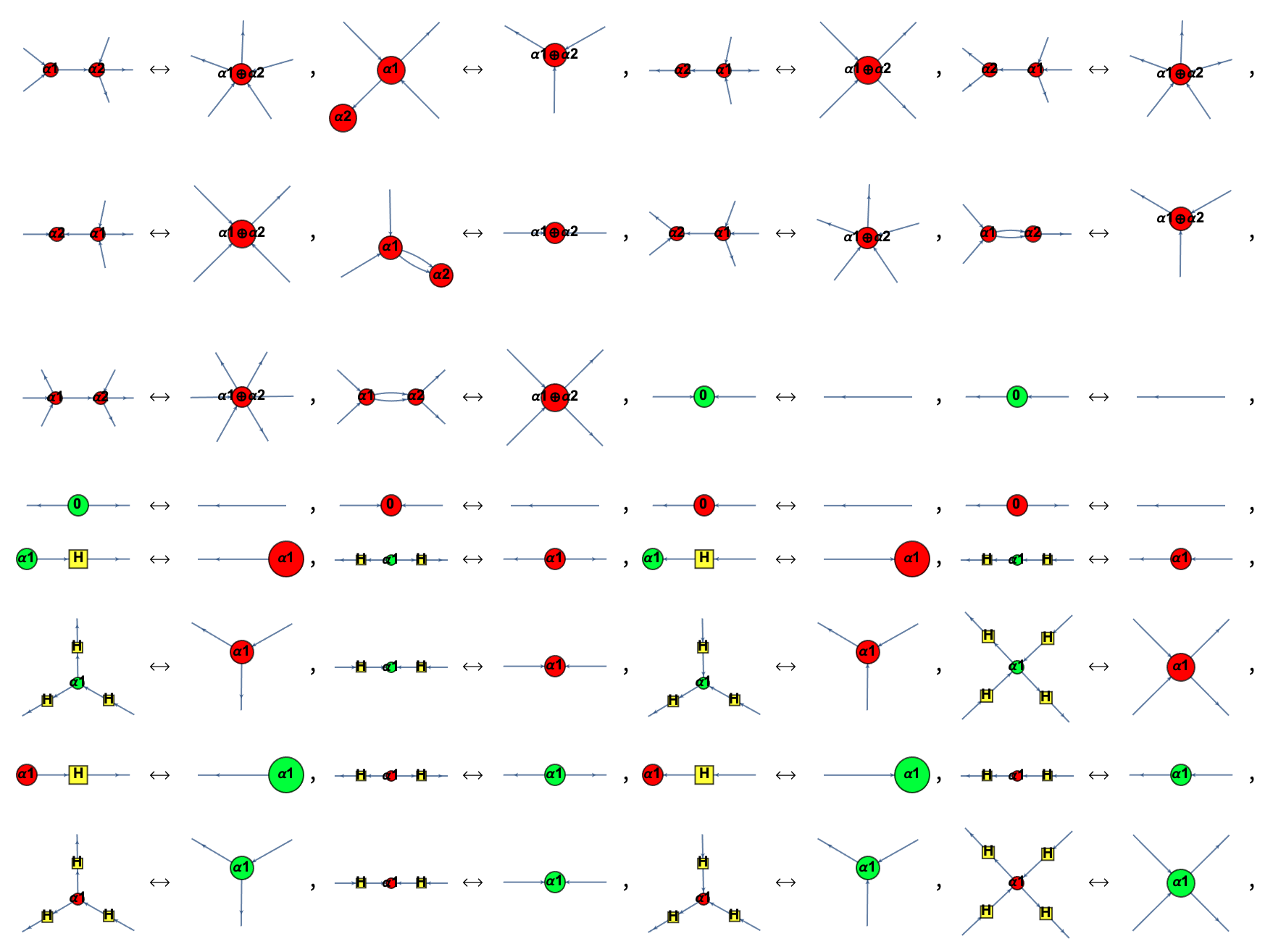}
\caption{A sample of the complete rule enumeration for the ZX-calculus, assuming input and output arities up to 2 across all generators.}
\label{fig:Figure34}
\end{figure}

\begin{figure}[ht]
\centering
\includegraphics[width=0.475\textwidth]{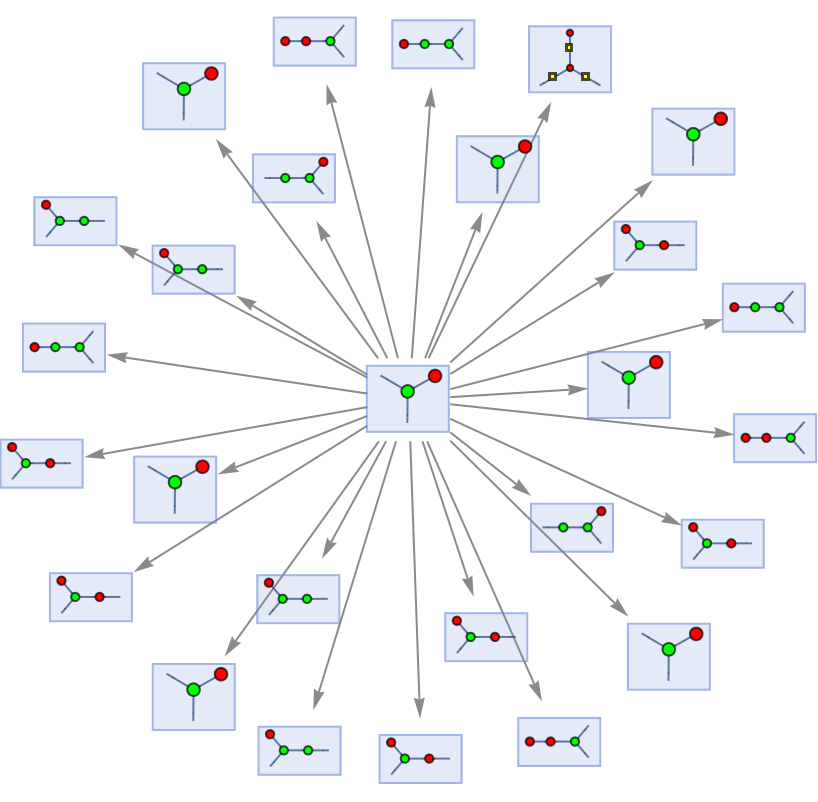}
\includegraphics[width=0.515\textwidth]{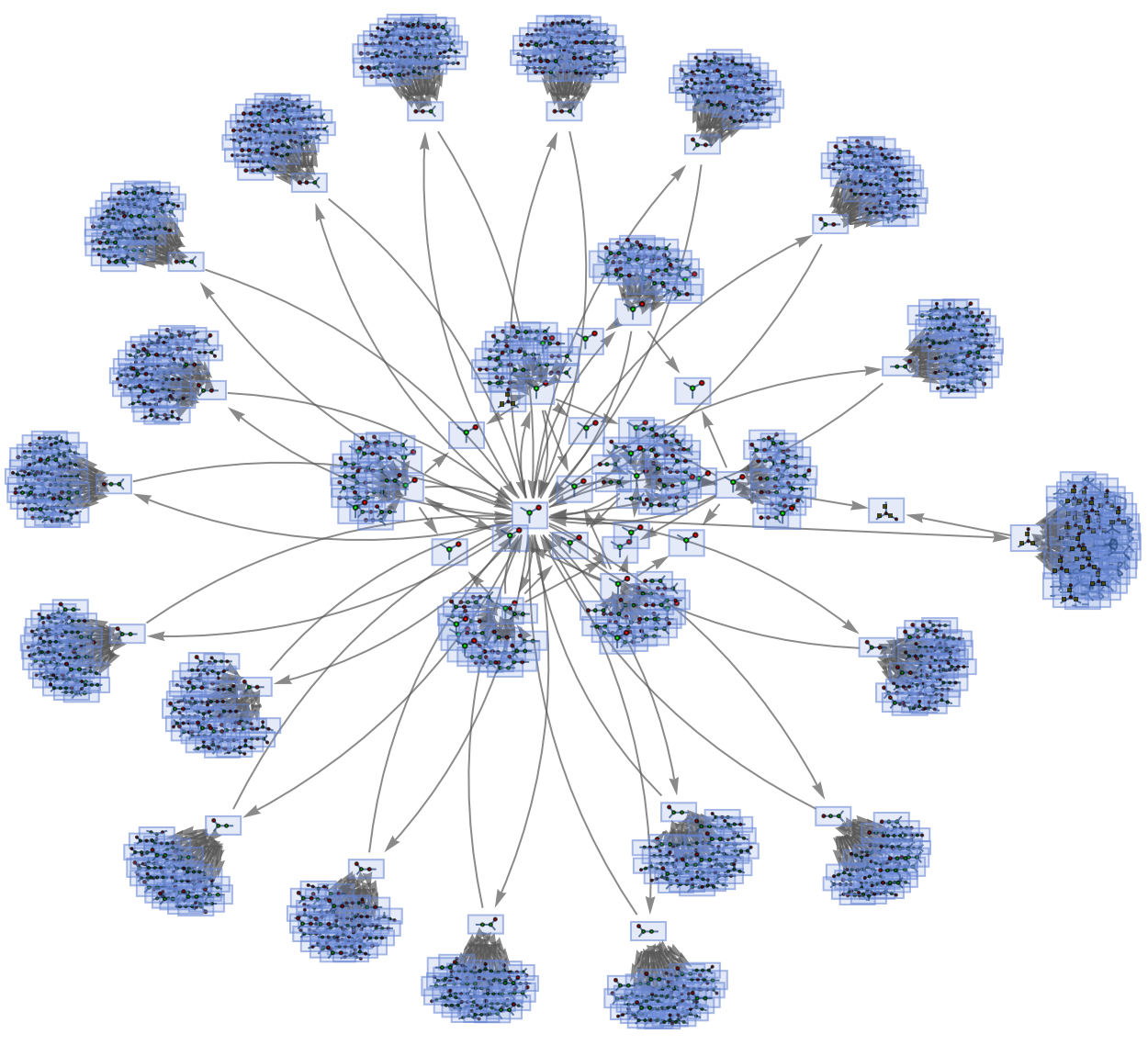}
\caption{The multiway evolution/states graph corresponding to the first 1 and 2 steps in the non-deterministic evolution of the ZX-calculus multiway operator system, respectively, assuming an initial condition consisting of a single Z/X-spider pair.}
\label{fig:Figure35}
\end{figure}

All that remains is to incorporate additional (non-diagrammatic) axioms representing the commutativity and associativity of the spider composition operator ${\otimes}$, namely:

\begin{equation}
\forall x, y : x \otimes y = y \otimes x,
\end{equation}
and:

\begin{equation}
\forall x, y, z : x \otimes \left( y \otimes z \right) = \left( x \otimes y \right) \otimes z,
\end{equation}
respectively, and one now has a complete specification of the multiway operator rules and higher-order rule schemas for the ZX-calculus; a sample of the enumeration of all possible rules up to a fixed arity across all generators is shown in Figure \ref{fig:Figure34}. This specification thus allows us to evolve the multiway operator system of the ZX-calculus for a simple initial diagram, such as the two-spider diagram:

\begin{equation}
X \left[ x_1, 0, 1, 0 \right] \otimes \left( Z \left[ z_1, 1, 2, 0 \right] \otimes \left( W \left[ x_1, z_1 \right] \otimes \left( W \left[ z_1, o_1 \right] \otimes W \left[ z_1, o_2 \right] \right) \right) \right),
\end{equation}
yielding a multiway evolution/state graph representing the embedding space of all possible proofs of equivalence between such diagrams, as shown in Figure \ref{fig:Figure35}.

\clearpage

\section{Causal Optimization, with an Application to Quantum Teleportation Protocols}
\label{sec:Section5}

\begin{figure}[ht]
\centering
\includegraphics[width=0.525\textwidth]{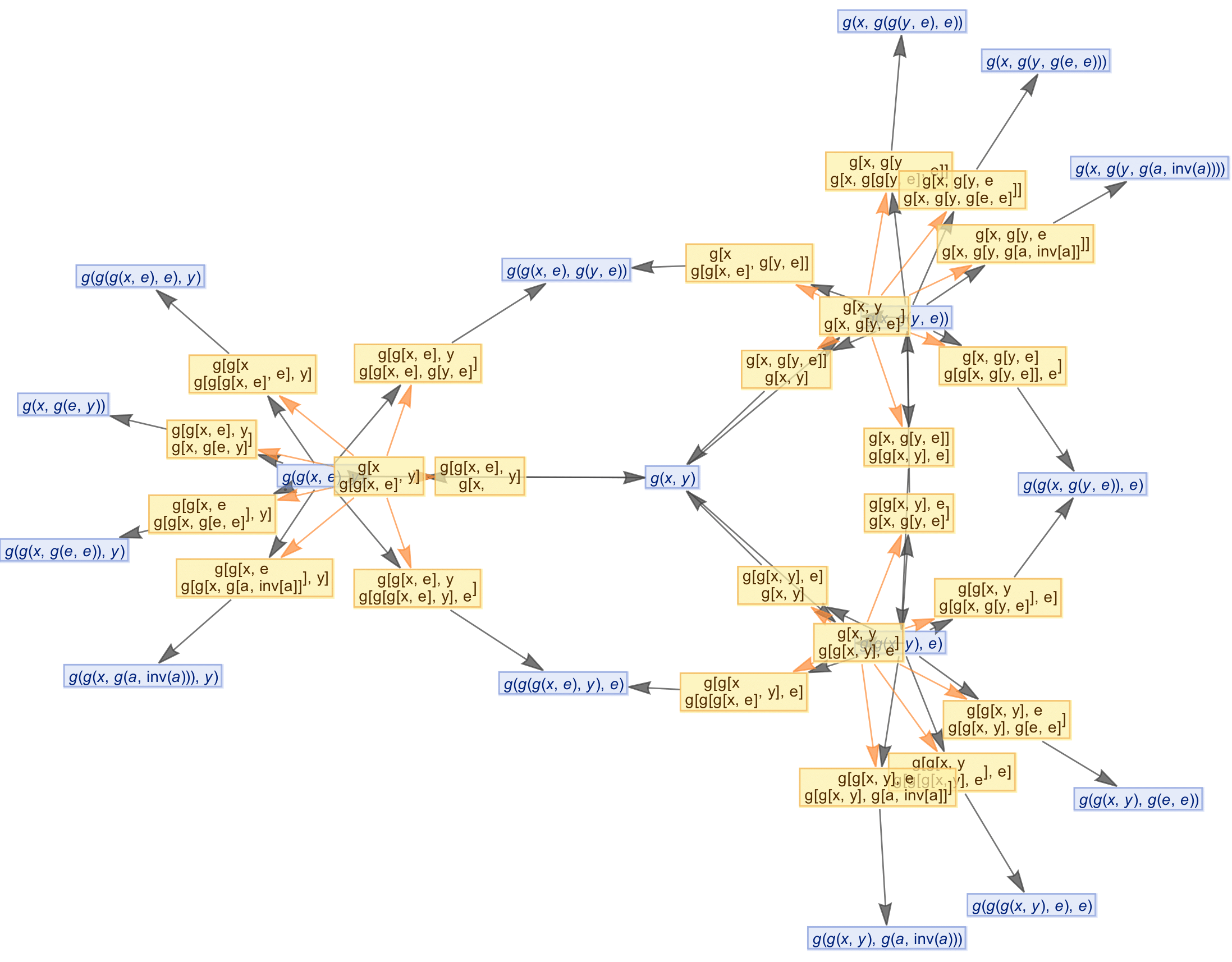}
\includegraphics[width=0.465\textwidth]{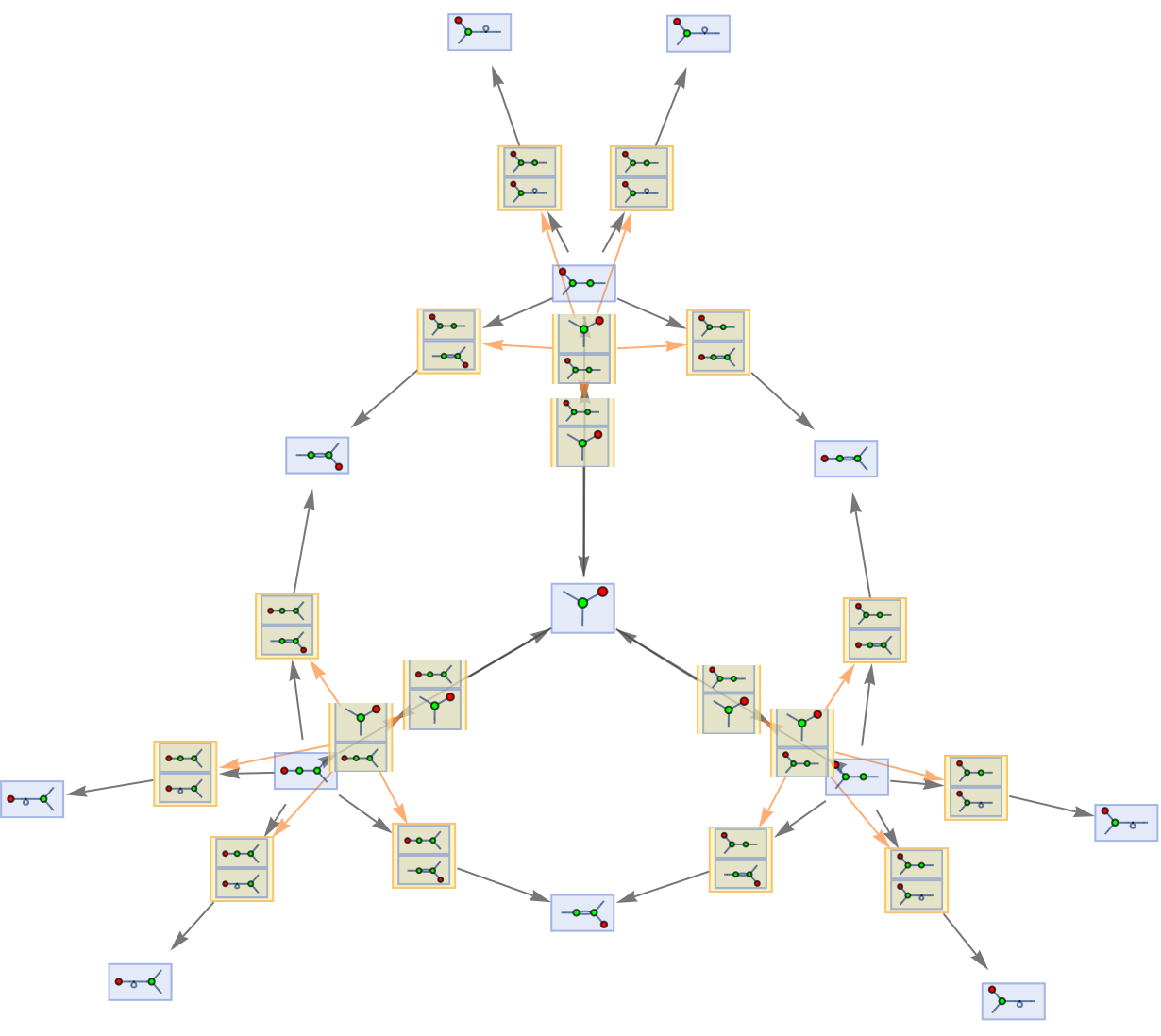}
\caption{Multiway evolution causal graphs corresponding to the first 2 steps in the non-deterministic evolution history for multiway operator systems corresponding to the standard axioms of group theory (left) and the restricted axioms of the ZX-calculus (right), with state vertices shown in blue, updating event vertices shown in yellow, evolution edges shown in gray and causal edges shown in orange.}
\label{fig:Figure36}
\end{figure}

As a general optimization method for our automated theorem-proving algorithm, we choose the selection function $S$ that appears within the definition of our deductive inference rules of selective resolution:

\begin{equation}
\infer{\Lambda \sigma \implies \Pi \sigma}{\Lambda \cup \left\lbrace u \approx c \right\rbrace \implies \Pi},
\end{equation}
selective superposition:

\begin{equation}
\infer{\left\lbrace u \left[ t \right] \sigma \approx v \sigma \right\rbrace \cup \Gamma \sigma \cup \Lambda \sigma \implies \Delta \sigma \cup \Pi \sigma}{\Gamma \implies \Delta \cup \left\lbrace s \approx t \right\rbrace \qquad \left\lbrace u \left[ s^{\prime} \right] \approx v \right\rbrace \cup \Lambda \implies \Pi},
\end{equation}
and ordered resolution:

\begin{equation}
\infer{\Gamma \sigma \cup \Lambda \sigma \implies \Delta \sigma \cup \Pi \sigma}{\Gamma \implies \Delta \cup \left\lbrace P \left( s_1, \dots, s_n \right) \approx tt \right\rbrace \qquad \left\lbrace P \left( t_1, \dots, t_n \right) \approx tt \right\rbrace \cup \Lambda \implies \Pi},
\end{equation}
such that all terms are ordered on the basis of their causal edge density. In other words, lemmas which correspond to the paths through a multiway evolution causal graph (as illustrated in Figure \ref{fig:Figure36} for the case of multiway operator systems corresponding to the standard axioms of group theory and the ZX-calculus) with higher-than-average causal connectivity are preferentially selected over lemmas associated with lower-than-average causal connectivity. This guarantees that those lemmas that are likely to exert the greatest causal influence in shortening the proofs of subsequent propositions are selected first, which, as we shall see, has the welcome effect of significantly reducing both the time complexity of the theorem-proving algorithm, and the proof complexity of the generated proofs (at least in the particular case with which we are concerned in the present paper, namely diagrammatic simplification of quantum circuits via the ZX-calculus, although there is strong reason to believe that the causal optimization method is actually much more general than this, and that it could in principle be applied within a wide range of contexts in both automated theorem-proving and diagrammatic rewriting).

\begin{figure}[ht]
\centering
\includegraphics[width=0.495\textwidth]{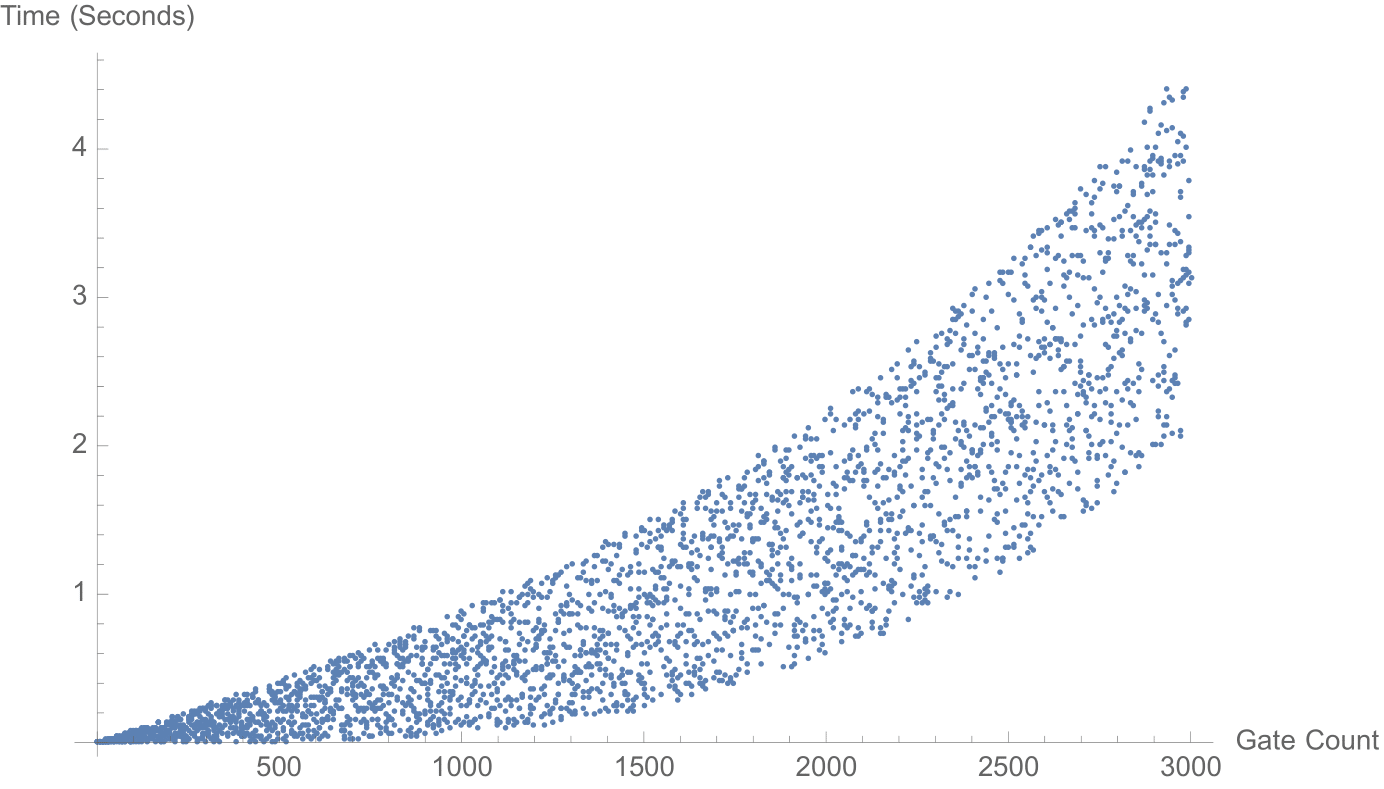}
\includegraphics[width=0.495\textwidth]{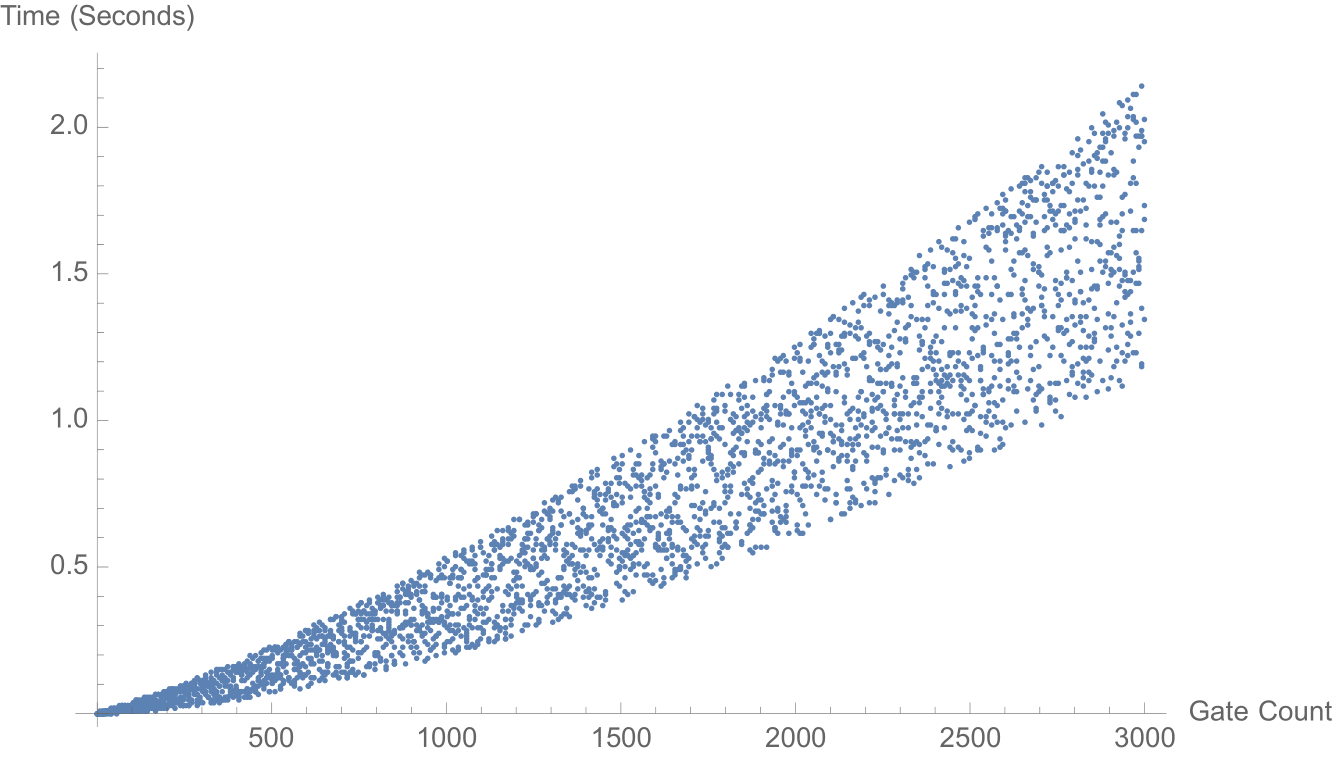}
\caption{Plots showing the time complexity (in seconds) of the automated theorem-proving algorithm when reducing randomly-generated Clifford circuits with sizes up to 3000 gates down to pseudo-normal form, both with (right) and without (left) causal optimization, showing approximately quadratic speedup in the optimized case.}
\label{fig:Figure37}
\end{figure}

\begin{figure}[ht]
\centering
\includegraphics[width=0.495\textwidth]{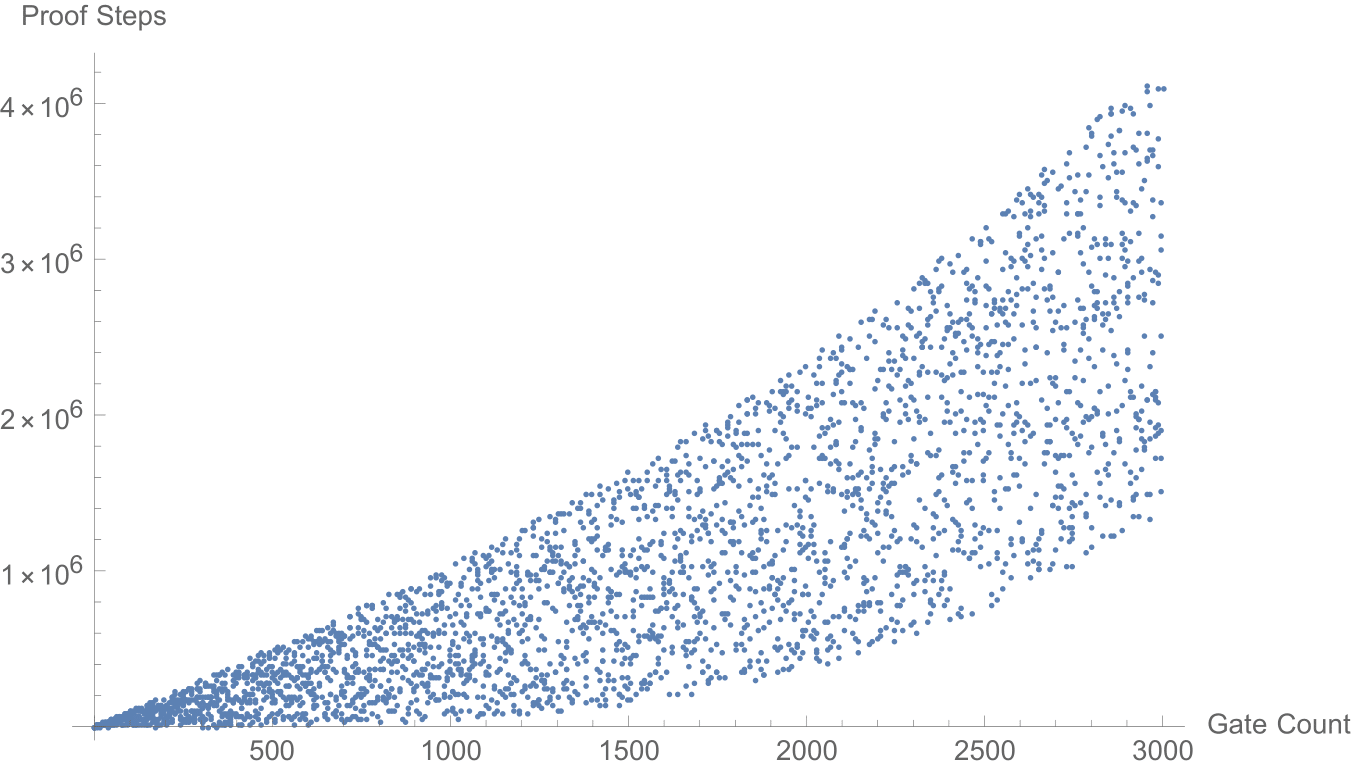}
\includegraphics[width=0.495\textwidth]{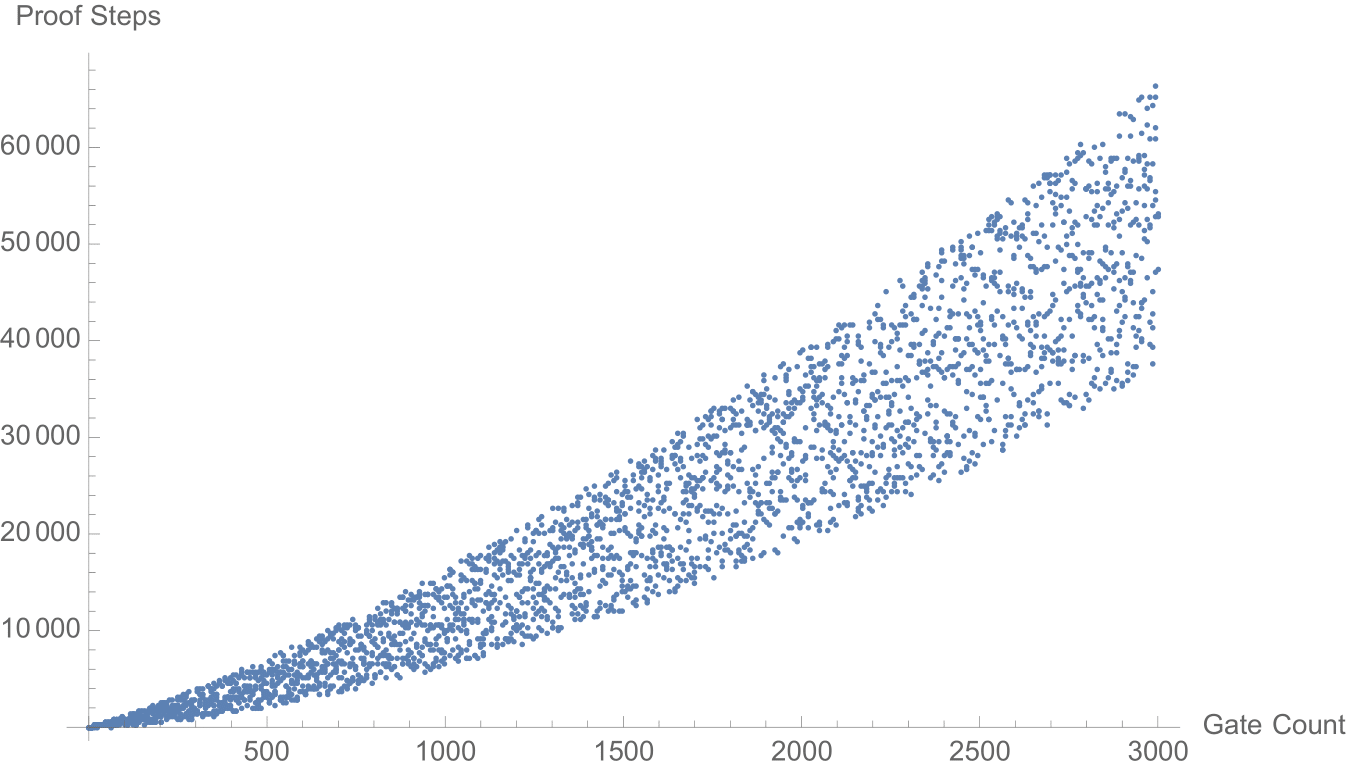}
\caption{Plots showing the proof complexity (in proof steps) of the proofs generated by the automated theorem-proving algorithm when reducing randomly-generated Clifford circuits with sizes up to 3000 gates down to pseudo-normal form, both with (right) and without (left) causal optimization, showing approximately quadratic reduction in proof complexity in the optimized case.}
\label{fig:Figure38}
\end{figure}

\begin{figure}[ht]
\centering
\includegraphics[width=0.495\textwidth]{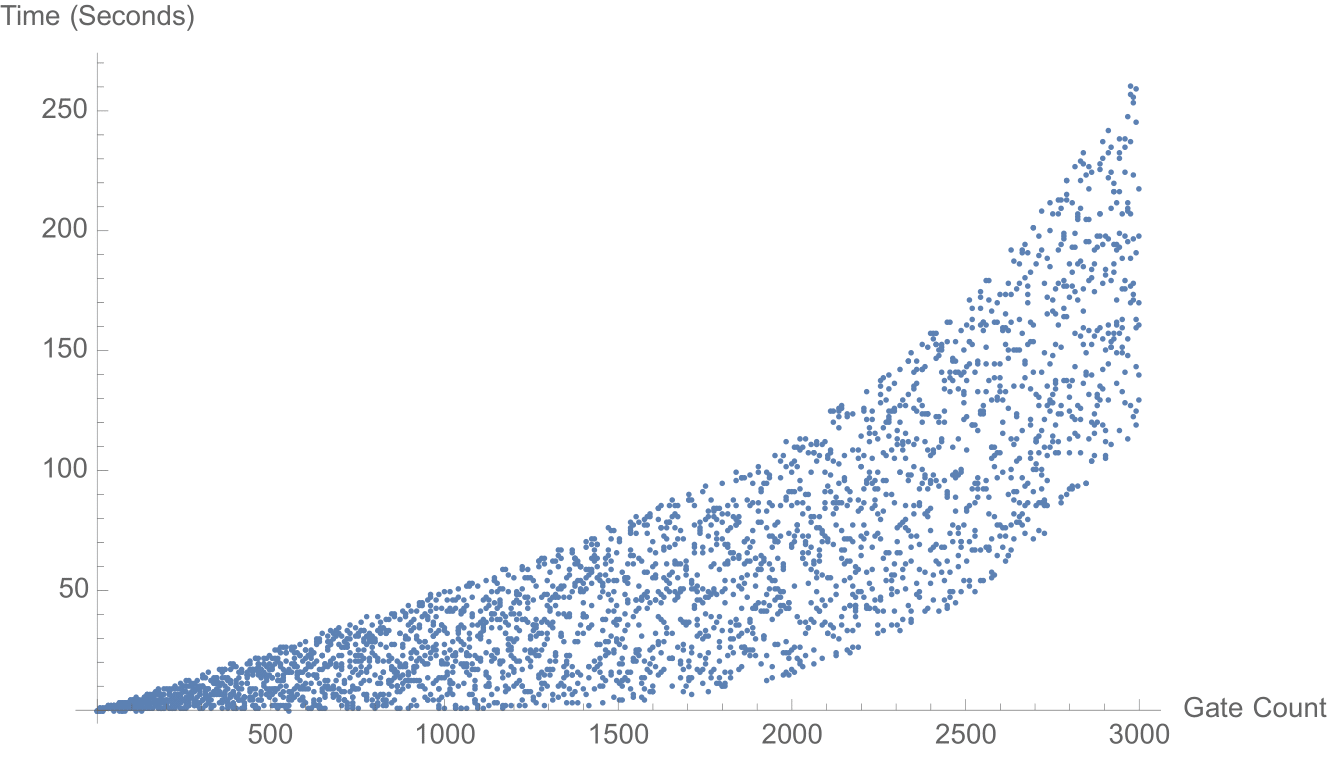}
\includegraphics[width=0.495\textwidth]{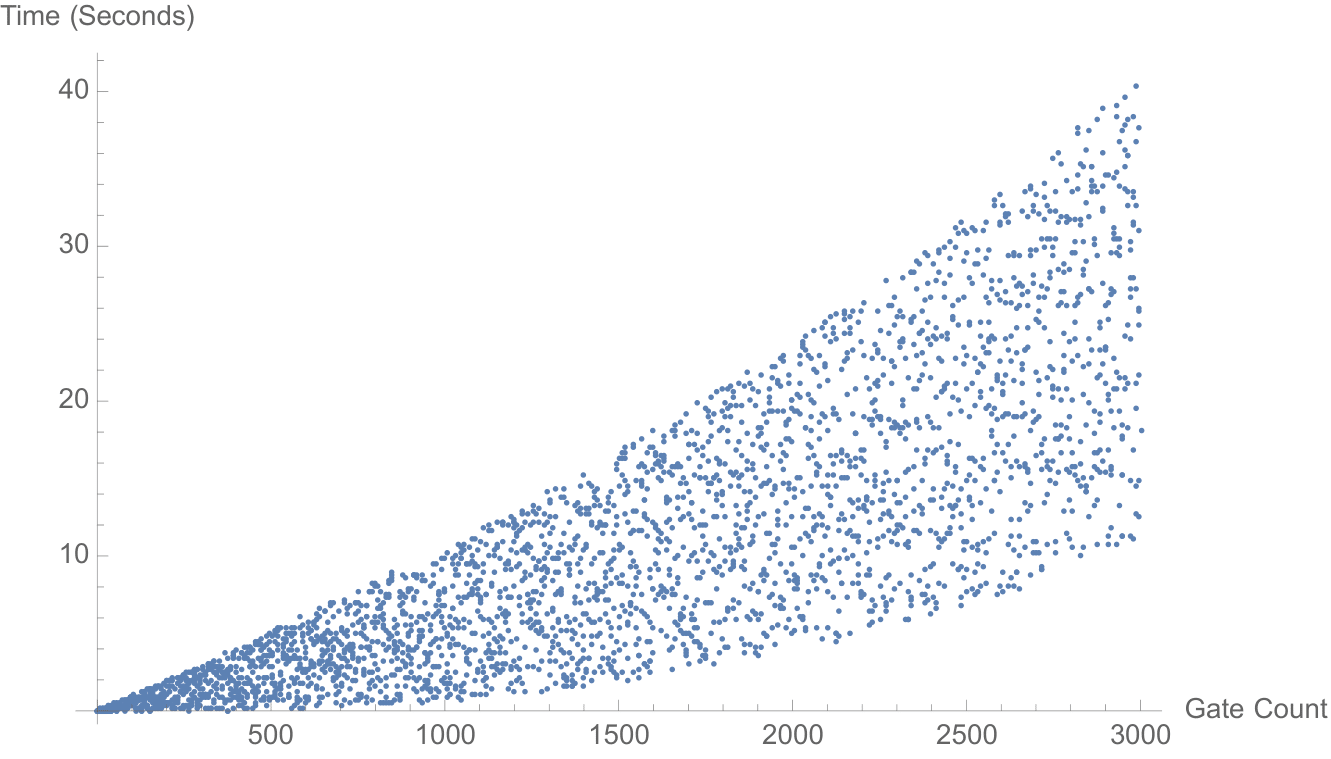}
\caption{Plots showing the time complexity (in seconds) of the automated theorem-proving algorithm when minimizing the number of T-gates in randomly-generated non-Clifford quantum circuits with sizes up to 3000 gates, both with (right) and without (left) causal optimization, showing approximately quadratic speedup in the optimized case.}
\label{fig:Figure39}
\end{figure}

\begin{figure}[ht]
\centering
\includegraphics[width=0.495\textwidth]{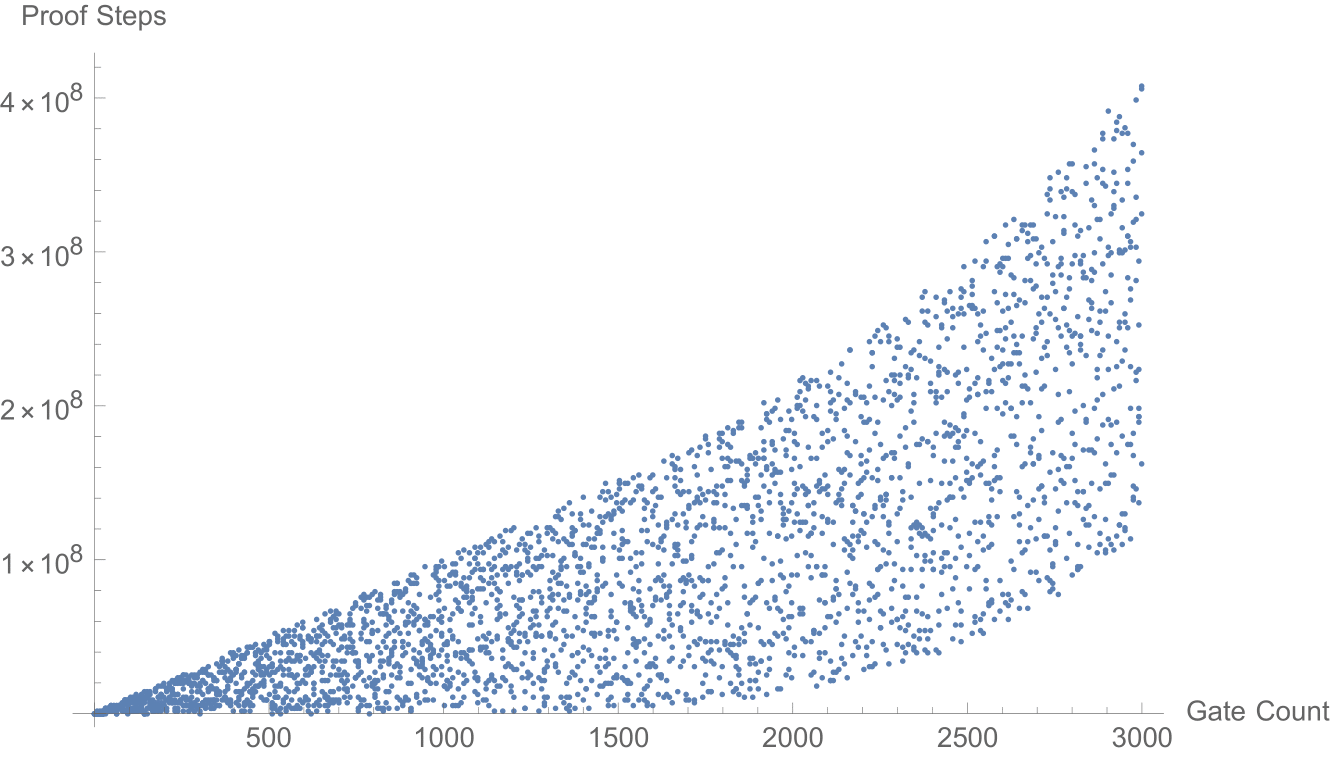}
\includegraphics[width=0.495\textwidth]{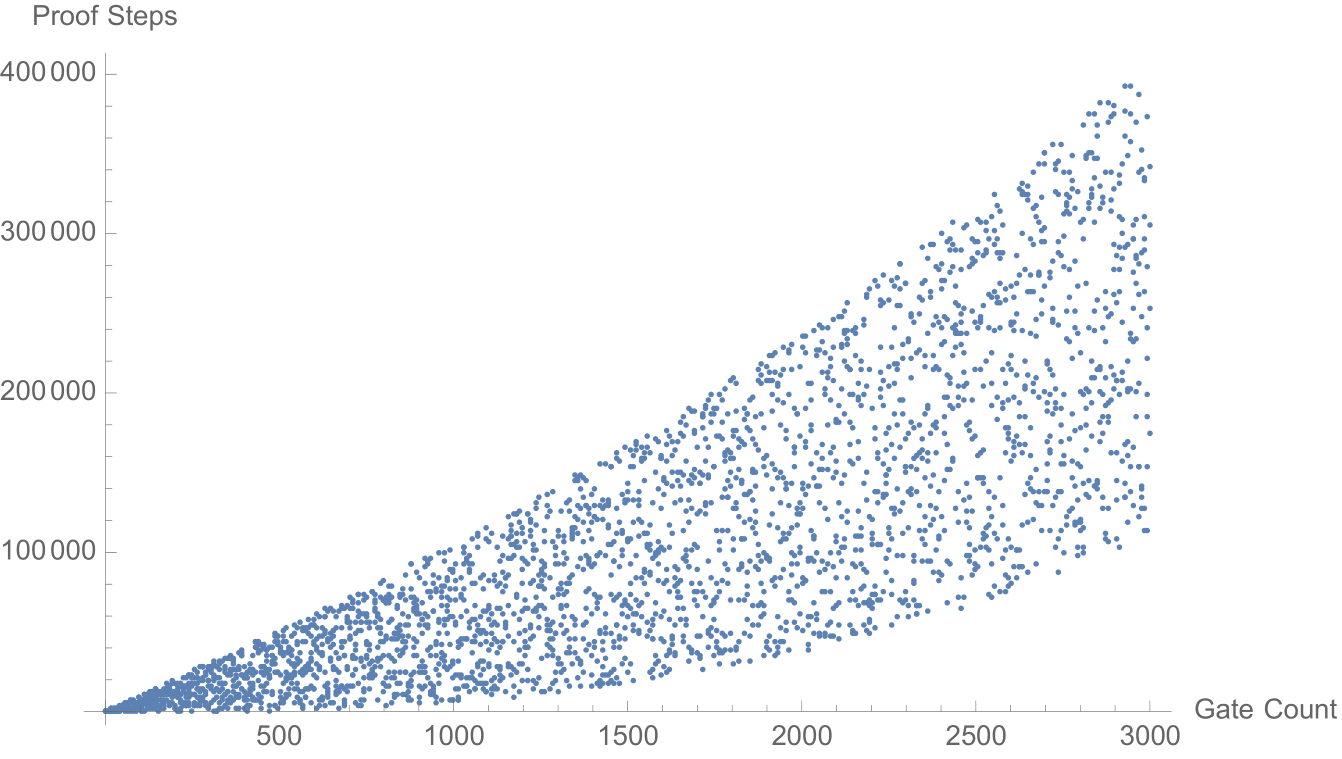}
\caption{Plots showing the proof complexity (in proof steps) of the proofs generated by the automated theorem-proving algorithm when minimizing the number of T-gates in randomly-generated non-Clifford quantum circuits with sizes up to 3000 gates, both with (right) and without (left) causal optimization, showing approximately quadratic reduction in proof complexity in the optimized case.}
\label{fig:Figure40}
\end{figure}

As a concrete illustration of these efficiency benefits, we begin by considering the case of reducing \textit{Clifford circuits}\cite{duncan}\cite{fagan}, i.e. stabilizer quantum circuits, obtained via arbitrary composition of CNOT gates ${\mathbf{CNOT}}$, Hadamard gates ${\mathbf{H}}$ and S/phase gates ${\mathbf{S}}$:

\begin{equation}
\mathbf{CNOT} = \begin{bmatrix}
1 & 0 & 0 & 0\\
0 & 1 & 0 & 0\\
0 & 0 & 0 & 1\\
0 & 0 & 1 & 0
\end{bmatrix}, \qquad \mathbf{H} = \frac{1}{\sqrt{2}} \begin{bmatrix}
1 & 1\\
1 & -1
\end{bmatrix}, \qquad \mathbf{S} = \begin{bmatrix}
1 & 0\\
0 & i
\end{bmatrix},
\end{equation}
which, as a consequence of the Gottesman-Knill theorem\cite{gottesman}, can be simulated in polynomial time using a probabilistic classical computer. If $T$-gates of the form ${\mathbf{T}}$ are also permitted within the composition:

\begin{equation}
\mathbf{T} = \begin{bmatrix}
1 & 0\\
0 & \exp \left( \frac{i \pi}{4} \right)
\end{bmatrix},
\end{equation}
then the resulting \textit{Clifford+T circuits} can be used to simulate an arbitrary unitary transformation on $n$ qubits to an arbitrary degree of precision; if general ${\mathbf{Z}_{\alpha}}$ gates are also included:

\begin{equation}
\mathbf{Z}_{\alpha} = \begin{bmatrix}
\exp \left( -i \frac{\alpha}{2} \right) & 0\\
0 & \exp \left( i \frac{\alpha}{2} \right)
\end{bmatrix},
\end{equation}
for all possible phase values ${\alpha}$, then such arbitrary unitary transformations may be constructed exactly\cite{nielsen}. The \textit{pseudo-normal form} to which we intend to reduce such Clifford circuits is a standard \textit{graph-state with local Cliffords} (or GS-LC) diagram; here, a \textit{graph state}\cite{hein} is any diagram in which every spider is a Z-spider with zero phase (modulo \textit{Hadamard edges}), every Z-spider is connected directly to an output (i.e. there are no interior spiders), every input and output in the diagram is connected directly to a Z-spider such that every Z-spider is connected to no more than one input or output overall, all Z-spiders are connected via Hadamard edges (i.e. edges consisting of a single Hadamard gate), and there are no self-loops or parallel Hadamard edges. \textit{Local Cliffords} are unitary transformations constructed from Clifford circuits acting on single qubits, i.e. they are circuits consisting of arbitrary compositions of Hadamard gates ${\mathbf{H}}$ and S/phase gates ${\mathbf{S}}$ only. Using a standard consumer laptop, we thus constructed several thousand randomly-generated Clifford circuits with sizes up to 3000 gates and performed automated diagrammatic simplifications of them down to pseudo-normal form using our automated theorem-proving algorithm, investigating both the time complexity of the algorithm itself (measured in seconds) and the proof complexity of the generated proofs (measured in standardized proof steps), comparing the algorithm both with and without causal optimizations, as shown in Figures \ref{fig:Figure37} and \ref{fig:Figure38}. We also performed an analogous test on several thousand randomly-generated \textit{non-Clifford} circuits with sizes up to 3000 gates, performing automated diagrammatic simplifications of them with the goal of reducing the total number of T-gates\cite{kissinger6}\cite{amy}\cite{amy2}, as shown in Figures \ref{fig:Figure39} and \ref{fig:Figure40}. In both cases, we find that the causal optimization method offers approximately quadratic speedup of the algorithm with respect to time complexity, as well as approximately quadratic reduction in the generated proof complexity. At least for these two particular classes of circuit simplification tasks, our algorithm performs favorably in comparison with \textit{Quantomatic}\cite{kissinger}, \textit{PyZX}\cite{kissinger2} and other existing theorem-proving and circuit simplification algorithms.

\begin{figure}[ht]
\centering
\includegraphics[width=0.395\textwidth]{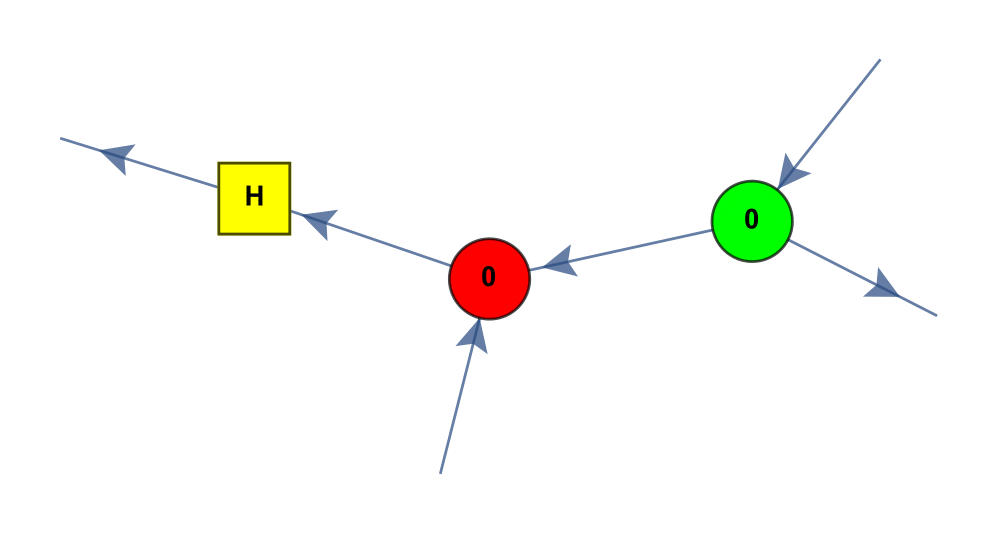}\hspace{0.1\textwidth}
\includegraphics[width=0.395\textwidth]{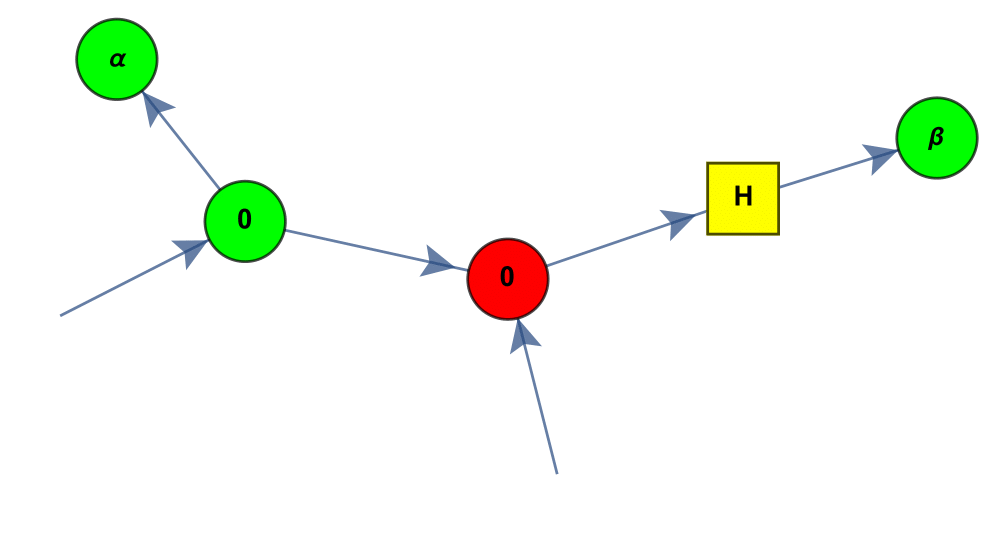}
\caption{On the left, the ZX-diagram for a circuit to rotate the Bell basis onto the X/computational basis. On the right, this diagram is connected to a pair of X-spiders with phases ${\alpha, \beta \in \left\lbrace 0, \pi \right\rbrace}$ to represent the 4 possible outcomes of the measurement in the Bell basis, namely ${\bra{\Psi_{+}}}$, ${\bra{\Psi_{-}}}$, ${\bra{\Phi_{+}}}$, and ${\bra{\Phi_{-}}}$.}
\label{fig:Figure41}
\end{figure}

\begin{figure}[ht]
\centering
\includegraphics[width=0.695\textwidth]{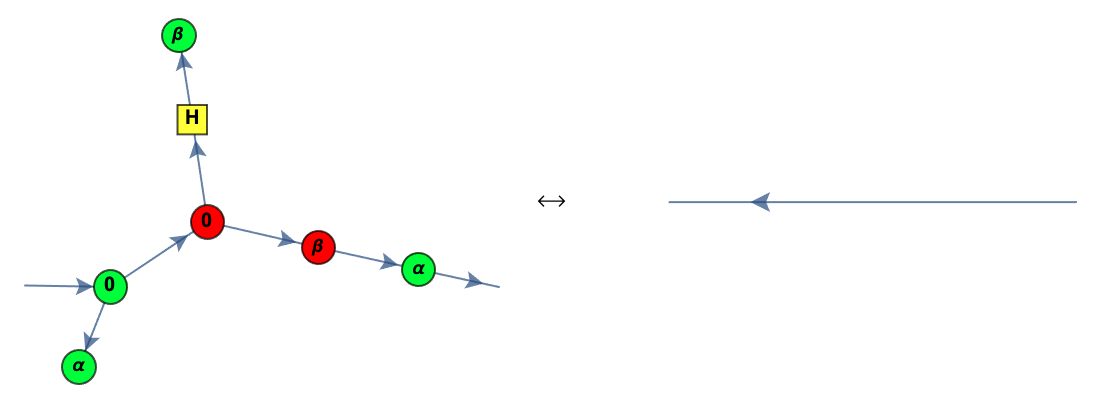}
\caption{The statement of correctness for the complete quantum teleportation protocol, represented as a theorem in the ZX-calculus.}
\label{fig:Figure42}
\end{figure}

\begin{figure}[ht]
\centering
\includegraphics[width=0.995\textwidth]{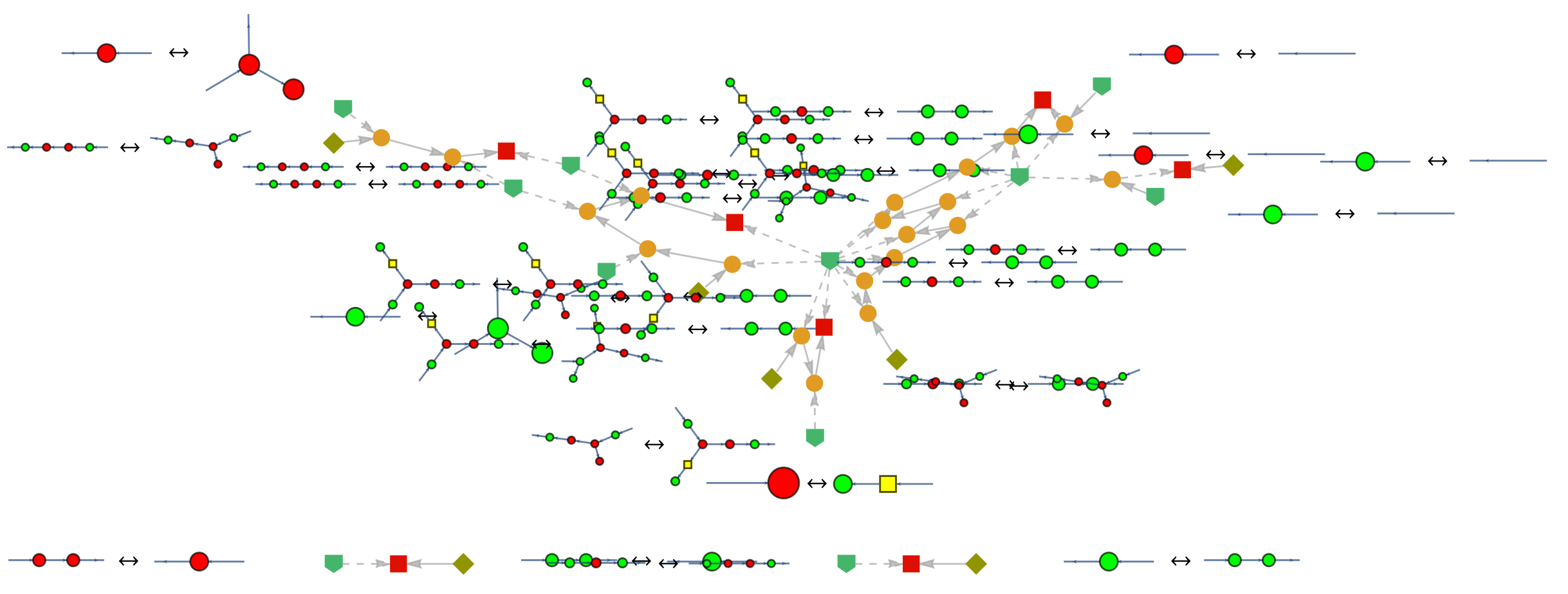}
\caption{The proof graph corresponding to the proof of correctness for the complete quantum teleportation protocol, subject to the rules of the ZX-calculus. Here, pointed light green boxes represent axioms, dark orange triangles represent critical pair lemmas (i.e. instances of completions/superpositions/paramodulations), light orange circles represent substitution lemmas (i.e. instances of resolutions/factorings), and dark green diamonds represent hypotheses. Solid lines represent substitutions, and dashed lines represent derived inference rules.}
\label{fig:Figure43}
\end{figure}

As a potentially illuminating example of how this circuit optimization algorithm can be employed in practice, we construct an automated proof of correctness for a simple quantum teleportation protocol, in which Alice and Bob communicate by preparing a single Bell state ${\ket{0 0} + \ket{1 1}}$ (represented by a ``cup'' wire), and then perform a measurement in the Bell basis using the projection ${\bra{0 0} + \bra{1 1}}$ (represented by a ``cap'' wire). We can represent the 4 possible measurement outcomes in the Bell basis, namely ${\bra{\Psi_{+}}}$, ${\bra{\Psi_{-}}}$, ${\bra{\Phi_{+}}}$ and ${\bra{\Phi_{-}}}$, by first constructing a circuit to rotate the Bell basis onto the X/computational basis, as given by the following two-spider, single-Hadamard diagram:

\begin{multline}
Z \left[ z_1, 1, 2, 0 \right] \otimes \left( X \left[ x_1, 1, 2, 0 \right] \otimes \left( W \left[ i_1, z_1 \right] \otimes \left( W \left[ z_1, o_1 \right] \otimes \left( W \left[ z_1, x_1 \right] \otimes \left( W \left[ i_2, x_1 \right] \otimes \left( W \left[ x_1, h_1 \right] \otimes \right. \right. \right. \right. \right. \right.\\
\left. \left. \left. \left. \left. \left. \left( H \left[ h_1 \right] \otimes W \left[ h_1, o_1 \right] \right) \right) \right) \right) \right) \right) \right),
\end{multline}
and then connecting this diagram to a pair of X-spiders with phases ${\alpha, \beta \in \left\lbrace 0, \pi \right\rbrace}$:

\begin{multline}
Z \left[ z_1, 1, 2, 0 \right] \otimes \left( X \left[ x_1, 2, 1, 0 \right] \otimes \left( W \left[ i_1, z_1 \right] \otimes \left( W \left[ z_1, z_2 \right] \otimes \left( W \left[ z_1, x_1 \right] \otimes \left( W \left[ i_2, x_1 \right] \otimes \left( W \left[ x_1, h_1 \right] \otimes \right. \right. \right. \right. \right. \right.\\
\left. \left. \left. \left. \left. \left. \left( H \left[ h_1 \right] \otimes \left( W \left[ h_1, z_3 \right] \otimes \left( Z \left[ z_2, 1, 0, \alpha \right] \otimes Z \left[ z_3, 1, 0, \beta \right] \right) \right) \right) \right) \right) \right) \right) \right) \right),
\end{multline}
as shown in Figure \ref{fig:Figure41}. This description makes manifest the role of classical communication within the protocol, as well as incorporating explicitly the Pauli errors generated by Alice in measuring any of the other possible outcomes. Thus, if we also include an additional pair of Z/X-spiders with phases ${\alpha, \beta \in \left\lbrace 0, \pi \right\rbrace}$ to represent Bob's corrections (which are correlated classically with Alice's measurement outcomes), then we obtain the following formal statement of correctness for the complete quantum teleportation protocol:

\begin{multline}
X \left[ x_1, 1, 2, 0 \right] \otimes \left( H \left[ h_1 \right] \otimes \left( Z \left[ z_3, 1, 0, \beta \right] \otimes \left( W \left[ x_1, h_1 \right] \otimes \left( W \left[ h_1, z_3 \right] \otimes \left( X \left[ x_2, 1, 1, \beta \right] \otimes \right. \right. \right. \right. \right.\\
\left( Z \left[ z_4, 1, 1, \alpha \right] \otimes \left( W \left[ x_1, x_2 \right] \otimes \left( W \left[ x_2, z_4 \right] \otimes \left( W \left[ z_4, o_1 \right] \otimes \left( Z \left[ z_1, 1, 2, 0 \right] \otimes \left( Z \left[ z_2, 1, 0, \alpha \right] \otimes \right. \right. \right. \right. \right. \right.\\
\left. \left. \left. \left. \left. \left. \left. \left. \left. \left. \left. \left( W \left[ z_1, z_2 \right] \otimes \left( W \left[ i_1, z_1 \right] \otimes W \left[ z_1, x_1 \right] \right) \right) \right) \right) \right) \right) \right) \right) \right) \right) \right) \right) \right) = W \left[ i_1, o_1 \right],
\end{multline}
as shown in Figure \ref{fig:Figure42}, with the associated proof graph shown in Figure \ref{fig:Figure43}.

\begin{figure}[ht]
\centering
\includegraphics[width=0.695\textwidth]{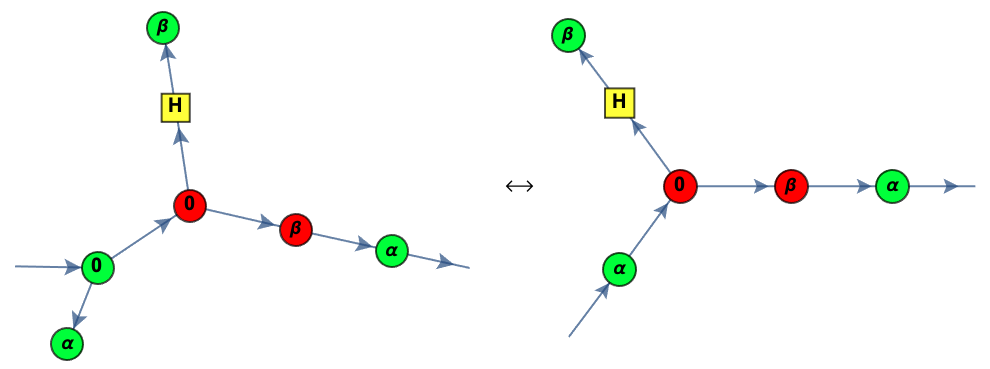}
\caption{The lemma obtained by applying the Z-spider fusion rule (S1) in order to fuse the phaseless Z-spider in the rotation circuit with the phase-${\alpha}$ Z-spider in Alice's Bell measurement.}
\label{fig:Figure44}
\end{figure}

\begin{figure}[ht]
\centering
\includegraphics[width=0.695\textwidth]{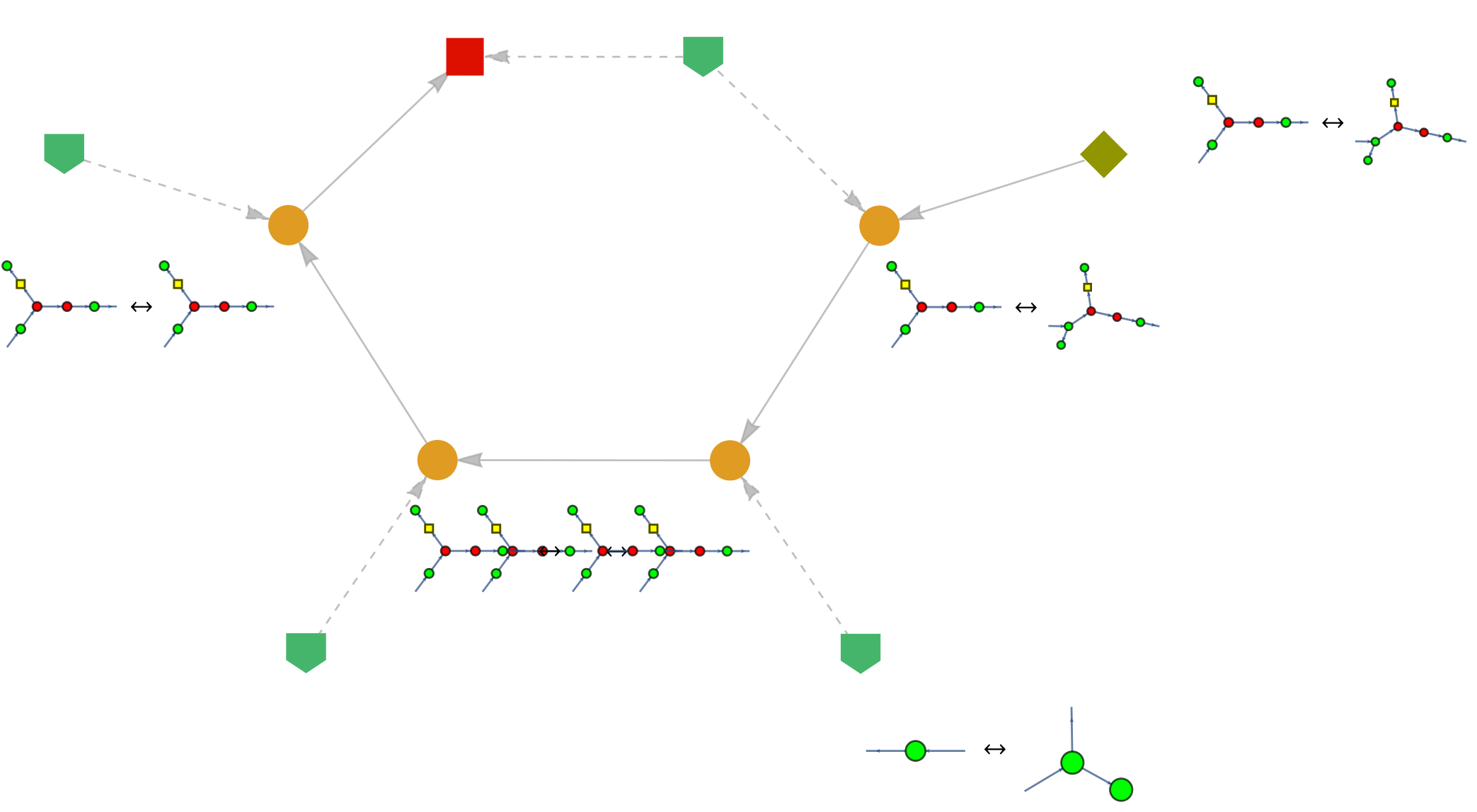}
\caption{The proof graph corresponding to the proof of the lemma obtained by applying the Z-spider fusion rule (S1) in order to fuse the phaseless Z-spider in the rotation circuit with the phase-${\alpha}$ Z-spider in Alice's Bell measurement. Here, pointed light green boxes represent axioms, light orange circles represent substitution lemmas (i.e. instances of resolutions/factorings), and dark green diamonds represent hypotheses. Solid lines represent substitutions, and dashed lines represent derived inference rules.}
\label{fig:Figure45}
\end{figure}

\begin{figure}[ht]
\centering
\includegraphics[width=0.695\textwidth]{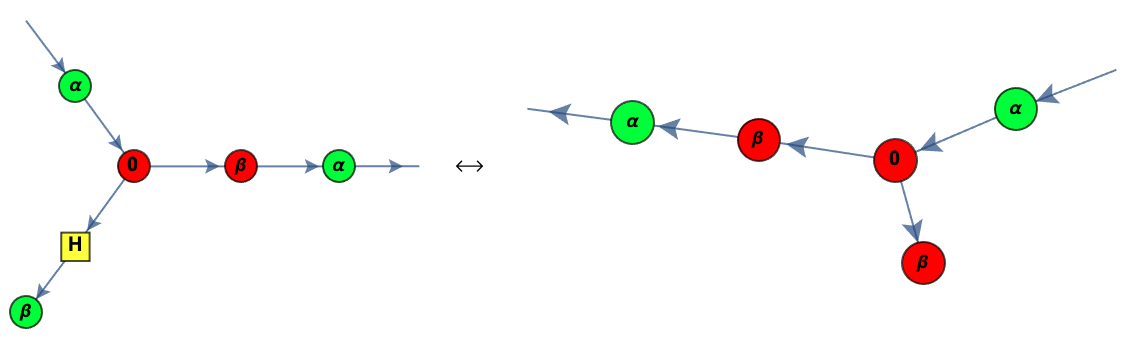}
\caption{The lemma obtained by applying the Z-spider color change rule (C) in order to swap the phase-${\beta}$ Z-spider in Alice's Bell measurement with a phase-${\beta}$ X-spider, using the Hadamard gate in the rotation circuit.}
\label{fig:Figure46}
\end{figure}

\begin{figure}[ht]
\centering
\includegraphics[width=0.695\textwidth]{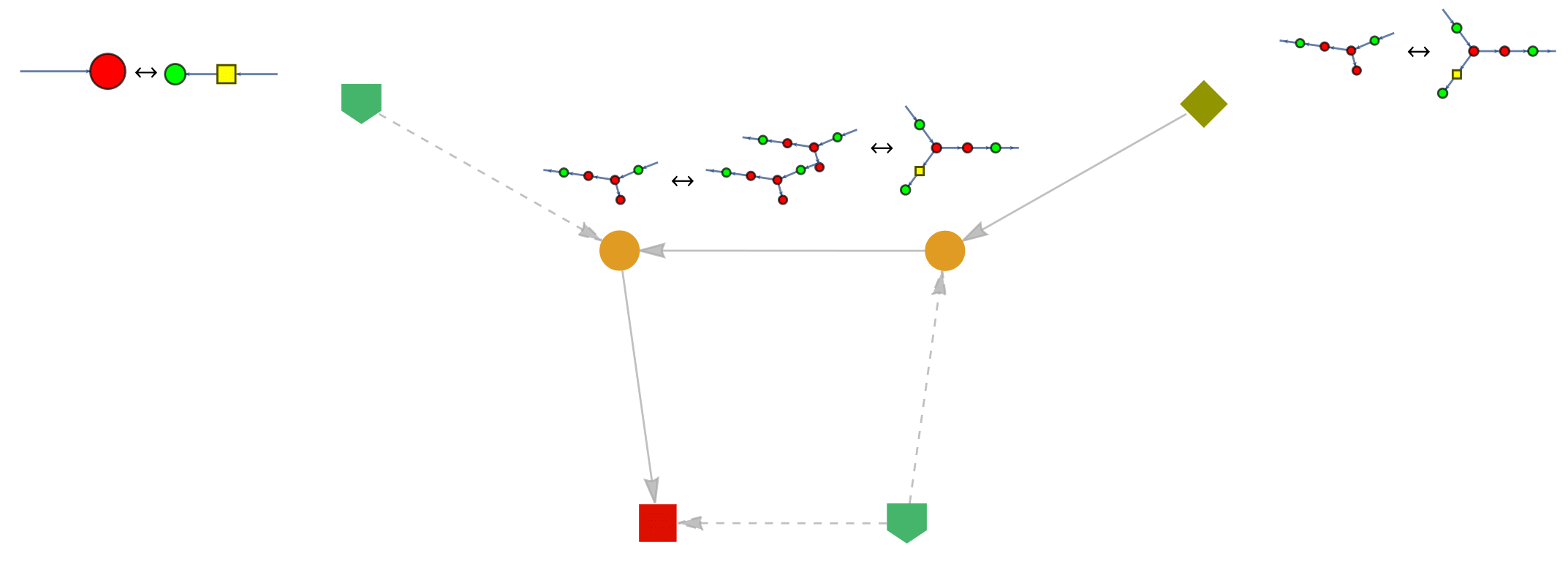}
\caption{The proof graph corresponding to the proof of the lemma obtained by applying the Z-spider color change rule (C) in order to swap the phase-${\beta}$ Z-spider in Alice's Bell measurement with a phase-${\beta}$ X-spider, using the Hadamard gate in the rotation circuit. Here, pointed light green boxes represent axioms, light orange circles represent substitution lemmas (i.e. instances of resolutions/factorings), and dark green diamonds represent hypotheses. Solid lines represent substitutions, and dashed lines represent derived inference rules.}
\label{fig:Figure47}
\end{figure}

\begin{figure}[ht]
\centering
\includegraphics[width=0.695\textwidth]{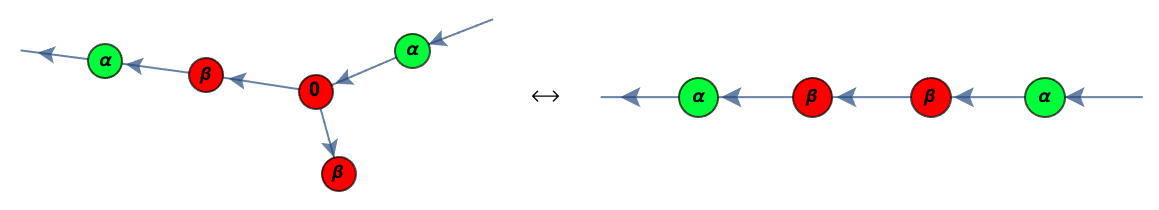}
\caption{The lemma obtained by applying the X-spider fusion rule (S1) in order to fuse the phaseless X-spider from the rotation circuit with the phase-${\beta}$ X-spider from Alice's Bell measurement.}
\label{fig:Figure48}
\end{figure}

\begin{figure}[ht]
\centering
\includegraphics[width=0.695\textwidth]{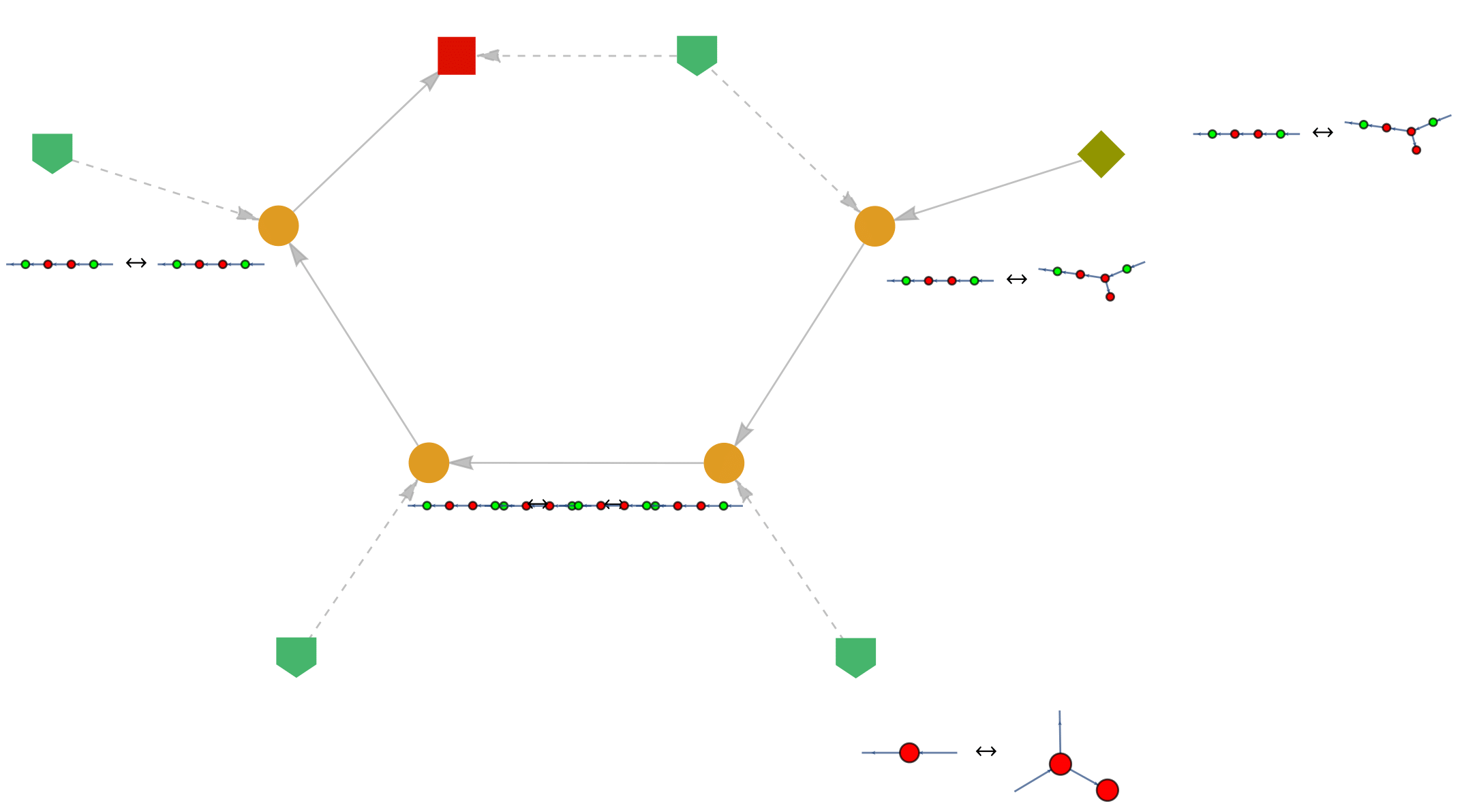}
\caption{The proof graph corresponding to the proof of the lemma obtained by applying the X-spider fusion rule (S1) in order to fuse the phaseless X-spider from the rotation circuit with the phase-${\beta}$ X-spider from Alice's Bell measurement. Here, pointed light green boxes represent axioms, light orange circles represent substitution lemmas (i.e. instances of resolutions/factorings), and dark green diamonds represent hypotheses. Solid lines represent substitutions and dashed lines represent derived inference rules.}
\label{fig:Figure49}
\end{figure}

To illustrate in detail how this proof is constructed, we begin by applying the Z-spider fusion rule (S1) to the initial diagram in order to fuse the phaseless Z-spider in the rotation circuit from the Bell basis to the computational basis with the phase-${\alpha}$ Z-spider in Alice's measurement in the Bell basis, yielding the lemma:

\begin{multline}
X \left[ x_1, 1, 2, 0 \right] \otimes \left( H \left[ h_1 \right] \otimes \left( Z \left[ z_3, 1, 0, \beta \right] \otimes \left( W \left[ x_1, h_1 \right] \otimes \left( W \left[ h_1, z_3 \right] \otimes \left( X \left[ x_2, 1, 1, \beta \right] \otimes \right. \right. \right. \right. \right.\\
\left( W \left[ z_4, 1, 1, \alpha \right] \otimes \left( W \left[ x_1, x_2 \right] \otimes \left( W \left[ x_2, z_4 \right] \otimes \left( W \left[ z_4, o_1 \right] \otimes \left( Z \left[ z_1, 1, 2, 0 \right] \otimes \left( Z \left[ z_2, 1, 0, \alpha \right] \otimes \right. \right. \right. \right. \right. \right.\\
\left. \left. \left. \left. \left. \left. \left. \left. \left. \left. \left. \left( Z \left[ z_1, z_2 \right] \otimes \left( W \left[ i_1, z_1 \right] \otimes W \left[ z_1, x_1 \right] \right) \right) \right) \right) \right) \right) \right) \right) \right) \right) \right) \right) \right) = X \left[ x_1, 1, 2, 0 \right] \otimes \left( H \left[ h_1 \right] \otimes \right.\\
\left( Z \left[ z_3, 1, 0, \beta \right] \otimes \left( W \left[ x_1, h_1 \right] \otimes \left( W \left[ h_1, z_3 \right] \otimes \left( X \left[ x_2, 1, 1, \beta \right] \otimes \left( Z \left[ z_4, 1, 1, \alpha \right] \otimes \left( W \left[ x_1, x_2 \right] \otimes \right. \right. \right. \right. \right. \right.\\
\left. \left. \left. \left. \left. \left. \left. \left( W \left[ x_2, z_4 \right] \otimes \left( W \left[ z_4, o_1 \right] \otimes \left( Z \left[ z_1, 1, 1, \alpha \right] \otimes \left( W \left[ i_1, z_1 \right] \otimes W \left[ z_1, x_1 \right] \right) \right) \right) \right) \right) \right) \right) \right) \right) \right) \right),
\end{multline}
as shown in Figure \ref{fig:Figure44}, with the associated proof graph shown in Figure \ref{fig:Figure45}. Next, we apply the Z-spider color change rule (C) in order to swap the phase-${\beta}$ Z-spider in Alice's measurement in the Bell basis with a phase-${\beta}$ X-spider, using the Hadamard gate in the rotation circuit from the Bell basis to the computational basis, yielding the lemma:

\begin{multline}
Z \left[ z_1, 1, 1, \alpha \right] \otimes \left( W \left[ i_1, z_1 \right] \otimes \left( W \left[ z_1, x_1 \right] \otimes \left( X \left[ x_1, 1, 2, 0 \right] \otimes \left( X \left[ x_2, 1, 1, \beta \right] \otimes \left( Z \left[ z_4, 1, 1, \alpha \right] \otimes \right. \right. \right. \right. \right.\\
\left( W \left[ x_1, x_2 \right] \otimes \left( W \left[ x_2, z_4 \right ] \otimes \left( W \left[ z_4, o_1 \right] \otimes \left( Z \left[ z_3, 1, 0, \beta \right] \otimes \left( H \left[ h_1 \right] \otimes \left( W \left[ x_1, h_1 \right] \otimes \right. \right. \right. \right. \right. \right.\\
\left. W \left[ h_1, z_3 \right] \right) = Z \left[ z_1, 1, 1, \alpha \right] \otimes \left( W \left[ i_1, z_1 \right] \otimes \left( W \left[ z_1, x_1 \right] \otimes \left( X \left[ x_1, 1, 2, 0 \right] \otimes \left( X \left[ x_2, 1, 1, \beta \right) \otimes \right. \right. \right. \right.\\
\left. \left. \left. \left. \left( Z \left[ z_4, 1, 1, \alpha \right] \otimes \left( W \left[ x_1, x_2 \right] \otimes \left( W \left[ x_2, z_4 \right] \otimes \left( W \left[ z_4, o_1 \right] \otimes \left( X \left[ z_3, 1, 0, \beta \right] \otimes W \left[ x_1, z_3 \right] \right) \right) \right) \right) \right) \right) \right) \right) \right),
\end{multline}
as shown in Figure \ref{fig:Figure46}, with the associated proof graph shown in Figure \ref{fig:Figure47}. Then, we apply the X-spider fusion rule (S1) in order to fuse the phaseless X-spider from the rotation circuit from the Bell basis to the computational basis with the phase-${\beta}$ X-spider from Alice's measurement in the Bell basis, yielding the lemma:

\begin{multline}
Z \left[ z_1, 1, 1, \alpha \right] \otimes \left( W \left[ i_1, z_1 \right] \otimes \left( X \left[ x_2, 1, 1, \beta \right] \otimes \left( Z \left[ z_4, 1, 1, \alpha \right] \otimes \left( W \left[ x_2, z_4 \right] \otimes \left( W \left[ z_4, o_1 \right] \otimes \right. \right. \right. \right. \right.\\
\left. \left. \left. \left. \left. \left( X \left[ x_1, 1, 2, 0 \right] \otimes \left( X \left[ z_3, 1, 0, \beta \right] \otimes \left( W \left[ x_1, z_3 \right] \otimes \left( W \left[ z_1, x_1 \right] \otimes W \left[ x_1, x_2 \right] \right) \right) \right) \right) \right) \right) \right) \right) \right) = Z \left[ z_1, 1, 1, \alpha \right] \otimes\\
\left( W \left[ i_1, z_1 \right] \otimes \left( X \left[ x_2, 1, 1, \beta \right] \otimes \left( Z \left[ z_4, 1, 1, \alpha \right] \otimes \left( W \left[ x_2, z_4 \right] \otimes \left( W \left[ z_4, o_1 \right] \otimes \left( X \left[ x_1, 1, 1, \beta \right] \otimes \right. \right. \right. \right. \right. \right.\\
\left. \left. \left. \left. \left. \left. \left( W \left[ z_1, x_1 \right] \otimes W \left[ x_1, x_2 \right] \right) \right) \right) \right) \right) \right) \right),
\end{multline}
as shown in Figure \ref{fig:Figure48}, with the associated proof graph shown in Figure \ref{fig:Figure49}.

\begin{figure}[ht]
\centering
\includegraphics[width=0.695\textwidth]{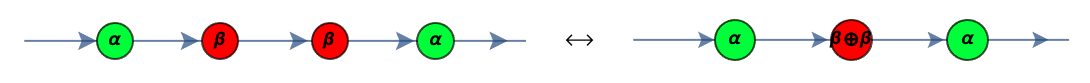}
\caption{The lemma obtained by applying the X-spider fusion rule (S1) in order to fuse the phase-${\beta}$ X-spider from Alice's Bell measurement with the phase-${\beta}$ X-spider from Bob's classically-correlated corrections.}
\label{fig:Figure50}
\end{figure}

\begin{figure}[ht]
\centering
\includegraphics[width=0.695\textwidth]{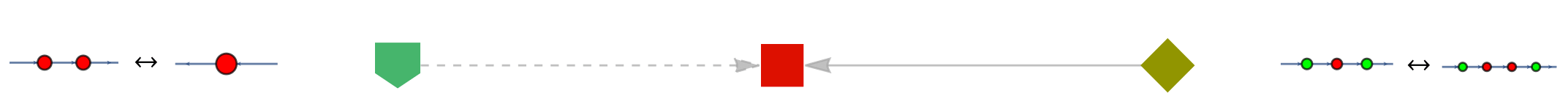}
\caption{The proof graph corresponding to the proof of the lemma obtained by applying the X-spider fusion rule (S1) in order to fuse the phase-${\beta}$ X-spider from Alice's Bell measurement with the phase-${\beta}$ X-spider from Bob's classically-correlated corrections. Here, pointed light green boxes represent axioms, light orange circles represent substitution lemmas (i.e. instances of resolutions/factorings), and dark green diamonds represent hypotheses. Solid lines represent substitutions and dashed lines represent derived inference rules.}
\label{fig:Figure51}
\end{figure}

\begin{figure}[ht]
\centering
\includegraphics[width=0.695\textwidth]{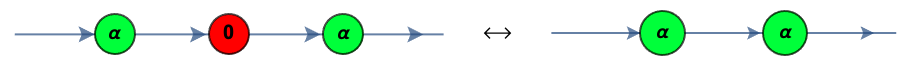}
\caption{The lemma obtained by applying the X-spider identity rule (S2) in order to eliminate the phaseless X-spider from the interaction between Alice's Bell measurement and Bob's classically-correlated corrections.}
\label{fig:Figure52}
\end{figure}

\begin{figure}[ht]
\centering
\includegraphics[width=0.695\textwidth]{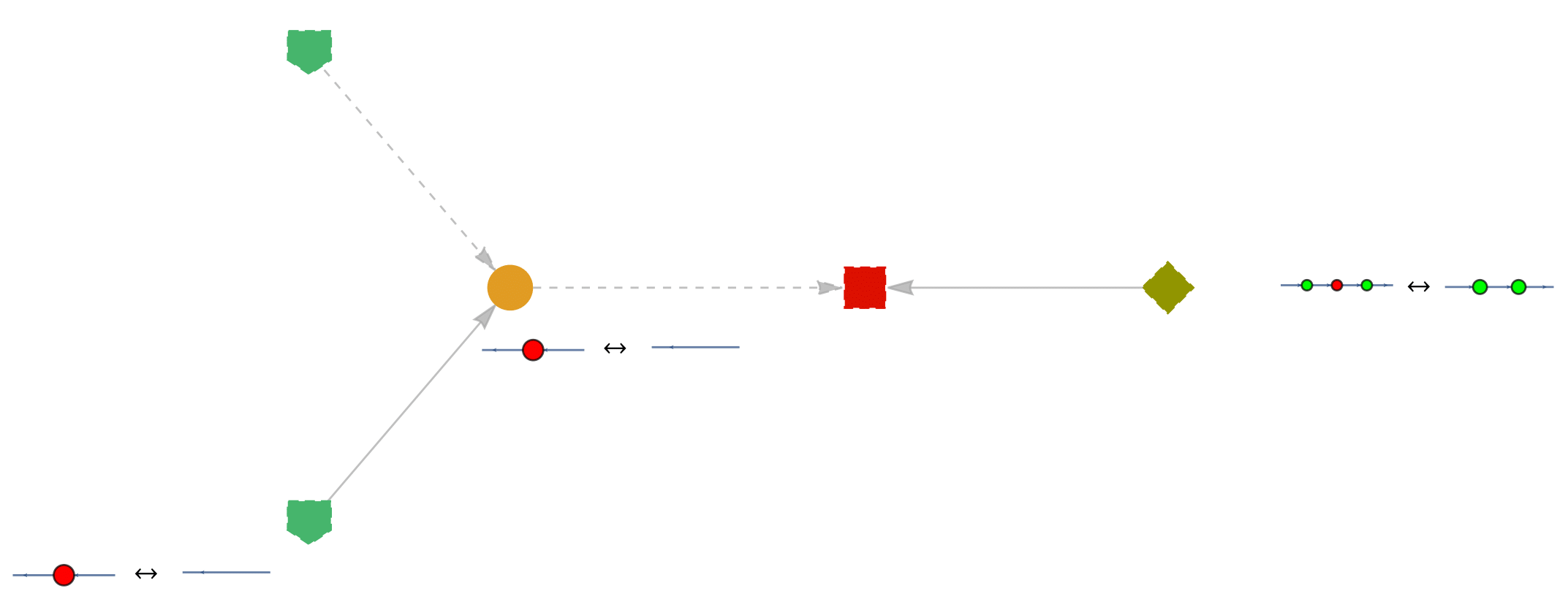}
\caption{The proof graph corresponding to the proof of the lemma obtained by applying the X-spider identity rule (S2) in order to eliminate the phaseless X-spider from the interaction between Alice's Bell measurement and Bob's classically-correlated corrections. Here, pointed light green boxes represent axioms, light orange circles represent substitution lemmas (i.e. instances of resolutions/factorings), and dark green diamonds represent hypotheses. Solid lines represent substitutions and dashed lines represent derived inference rules.}
\label{fig:Figure53}
\end{figure}

\begin{figure}[ht]
\centering
\includegraphics[width=0.695\textwidth]{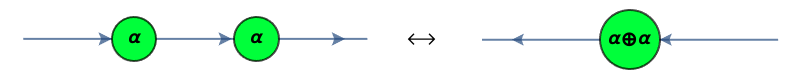}
\caption{The lemma obtained by applying the Z-spider fusion rule (S1) in order to fuse the phase-${\alpha}$ Z-spider from Alice's Bell measurement with the phase-${\alpha}$ Z-spider from Bob's classically-correlated corrections.}
\label{fig:Figure54}
\end{figure}

\begin{figure}[ht]
\centering
\includegraphics[width=0.695\textwidth]{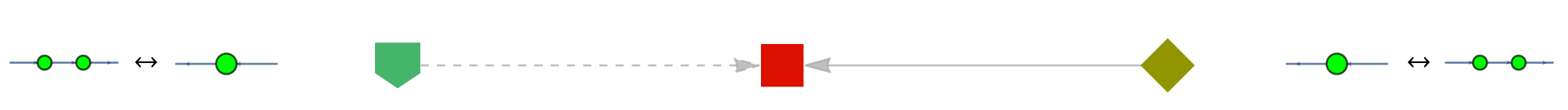}
\caption{The proof graph corresponding to the proof of the lemma obtained by applying the Z-spider fusion rule (S1) in order to fuse the phase-${\alpha}$ Z-spider from Alice's Bell measurement with the phase-${\alpha}$ Z-spider from Bob's classically-correlated corrections. Here, pointed light green boxes represent axioms, light orange circles represent substitution lemmas (i.e. instances of resolutions/factorings), and dark green diamonds represent hypotheses. Solid lines represent substitutions and dashed lines represent derived inference rules.}
\label{fig:Figure55}
\end{figure}

\begin{figure}[ht]
\centering
\includegraphics[width=0.695\textwidth]{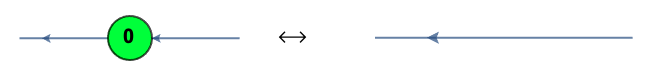}
\caption{The theorem obtained by applying the Z-spider identity rule (S2) in order to eliminate the phaseless Z-spider from the interaction between Alice's Bell measurement and Bob's classically-correlated corrections.}
\label{fig:Figure56}
\end{figure}

\begin{figure}[ht]
\centering
\includegraphics[width=0.695\textwidth]{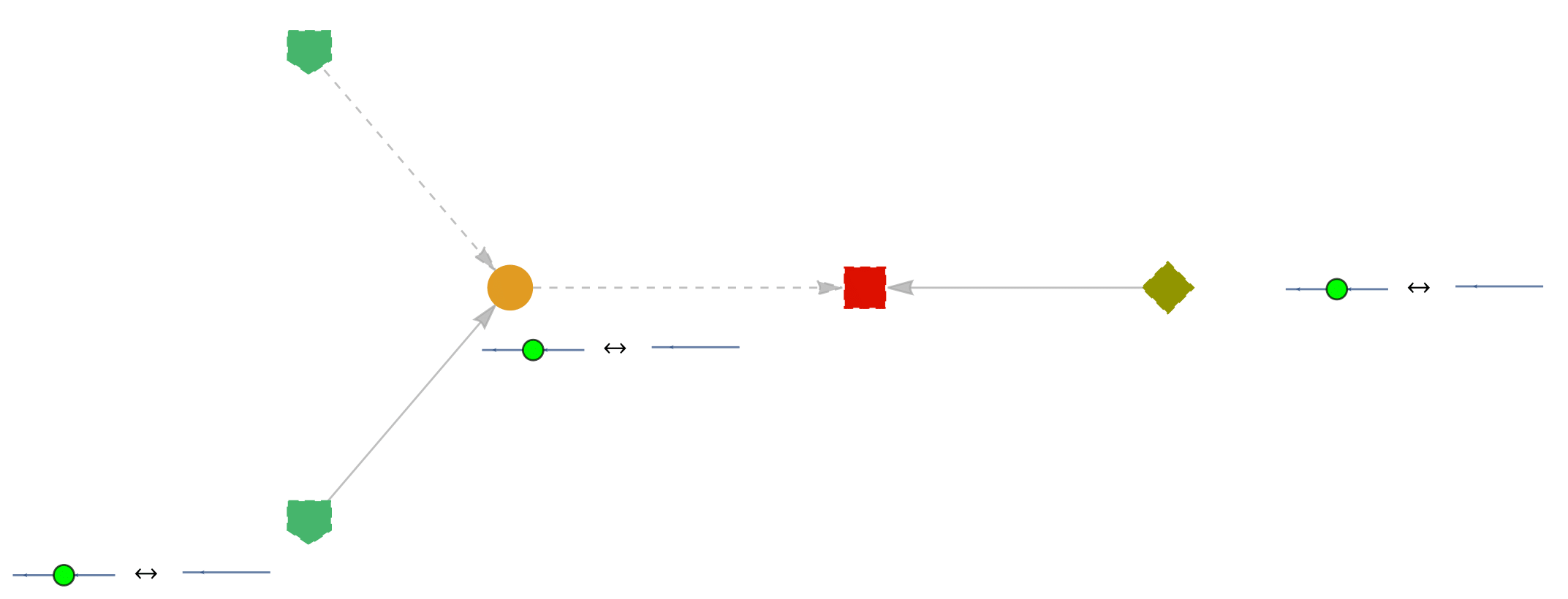}
\caption{The proof graph corresponding to the proof of the theorem obtained by applying the Z-spider identity rule (S2) in order to eliminate the phasless Z-spider from the interaction between Alice's Bell measurement and Bob's classically-correlated corrections. Here, pointed light green boxes represent axioms, light orange circles represent substitution lemmas (i.e. instances of resolutions/factorings), and dark green diamonds represent hypotheses. Solid lines represent substitutions and dashed lines represent derived inference rules.}
\label{fig:Figure57}
\end{figure}

From here, we proceed to apply the X-spider fusion rule (S1) in order to fuse the phase-${\beta}$ X-spider from Alice's measurement in the Bell basis with the phase-${\beta}$ X-spider from Bob's corrections which are classically-correlated with Alice's measurement outcomes, yielding the lemma:

\begin{multline}
Z \left[ z_1, 1, 1, \alpha \right] \otimes \left( W \left[ i_1, z_1 \right] \otimes \left( Z \left[ z_4, 1, 1, \alpha \right] \otimes \left( W \left[ z_4, o_1 \right] \otimes \left( X \left[ x_1, 1, 1, \beta \right] \otimes \left( X \left[ x_2, 1, 1, \beta \right] \otimes \right. \right. \right. \right. \right.\\
\left. \left. \left. \left. \left. \left( W \left[ x_1, x_2 \right] \otimes \left( W \left[ z_1, x_1 \right] \otimes W \left[x_2, z_4 \right] \right) \right) \right) \right) \right) \right) \right) = Z \left[ z_1, 1, 1, \alpha \right] \otimes \left( W \left[ i_1, z_1 \right] \otimes \left( Z \left[ z_4, 1, 1, \alpha \right] \otimes \right. \right.\\
\left. \left. \left( W \left[ z_4, o_1 \right] \otimes \left( X \left[ x_1, 1, 1, \beta \oplus \beta \right] \otimes \left( W \left[ z_1, x_1 \right] \otimes W \left[ x_1, z_4 \right] \right) \right) \right) \right) \right),
\end{multline}
as shown in Figure \ref{fig:Figure50}, with the associated proof graph shown in Figure \ref{fig:Figure51}. Now, we apply the X-spider identity rule (S2) in order to eliminate the phaseless X-spider obtained from the interaction between Alice's measurement in the Bell basis and Bob's corrections which are classically-correlated with Alice's measurement outcomes, yielding the lemma:

\begin{multline}
Z \left[ z_1, 1, 1, \alpha \right] \otimes \left( W \left[ i_1, z_1 \right] \otimes \left( Z \left[ z_4, 1, 1, \alpha \right] \otimes \left( W \left[ z_4, o_1 \right] \otimes \left( X \left[ x_1, 1, 1, 0 \right] \otimes \left( W \left[ z_1, x_1 \right] \otimes \right. \right. \right. \right. \right.\\
\left. \left. \left. \left. \left. W \left[ x_1, z_4 \right] \right) \right) \right) \right) \right) = Z \left[ z_1, 1, 1, \alpha \right] \otimes \left( W \left[ i_1, z_1 \right] \otimes \left( Z \left[ z_4, 1, 1, \alpha \right] \otimes \left( W \left[ z_4, o_1 \right] \otimes W \left[ z_1, z_4 \right] \right) \right) \right),
\end{multline}
as shown in Figure \ref{fig:Figure52}, with the associated proof graph shown in Figure \ref{fig:Figure53}. Penultimately, we apply the Z-spider fusion rule (S1) in order to fuse the phase-${\alpha}$ Z-spider from Alice's measurement in the Bell basis with the phase-${\alpha}$ Z-spider from Bob's corrections which are classically-correlated with Alice's measurement outcomes, yielding the lemma:

\begin{multline}
Z \left[ z_1, 1, 1, \alpha \right] \otimes \left( Z \left[ z_4, 1, 1, \alpha \right] \otimes \left( W \left[ z_1, z_4 \right] \otimes \left( W \left[ i_1, z_1 \right] \otimes W \left[ z_4, o_1 \right] \right) \right) \right) = Z \left[ z_1, 1, 1, \alpha \oplus \alpha \right] \otimes\\
 \left( W \left[ i_1, z_1 \right] \otimes W \left[ z_4, o_1 \right] \right),
\end{multline}
as shown in Figure \ref{fig:Figure54}, with the associated proof graph shown in Figure \ref{fig:Figure55}. Finally, we apply the Z-spider identity rule (S2) in order to eliminate the phaseless Z-spider obtained from the interaction between Alice's measurement in the Bell basis and Bob's corrections which are classically-correlated with Alice's measurement outcomes, yielding the correctness theorem:

\begin{equation}
Z \left[ z_1, 1, 1, 0 \right] \otimes \left( W \left[ i_1, z_1 \right] \otimes W \left[ z_1, o_1 \right] \right) = W \left[ i_1, o_1 \right],
\end{equation}
as shown in Figure \ref{fig:Figure56}, with the associated proof graph shown in Figure \ref{fig:Figure57}.

\clearpage

\section{Concluding Remarks and Future Work}

The article has succeeded in its stated aim of developing a novel automated theorem-proving algorithm for diagrammatic rewriting systems based upon a higher-order formulation of extended Wolfram model multiway operator systems, equipped with causal optimization, and has applied this algorithm to the problem of simplifying a variety of quantum circuits specified in the ZX-calculus formalism, demonstrating in most cases comparable or superior performance in comparison to existing software frameworks. However, there still exist many avenues for future development, extension and improvement of these general techniques. Perhaps the most obvious is the extension of the same underlying rewriting framework to accommodate similar diagrammatic calculi, such as the ZW-calculus\cite{coecke9} for describing W-state quantum computing, which would therefore allow for a more direct description of both quantum entanglement processes\cite{hadzihasanovic2} and Fermionic quantum computing\cite{defelice}, or alternatively the ZH-calculus\cite{backens3} for describing classically non-linear computation, which would therefore allow for the simulation of Toffoli-Hadamard circuits, among many other things. Another possibility would be to extend the same algorithm to enable automated diagrammatic reasoning over possible sequences of lattice surgery operations that could theoretically be performed on a surface code, due to the ZX-calculus' known applicability to reasoning over such quantum error correcting codes\cite{beaudrap}. It would be extremely interesting to compare how effective (if at all) the causal optimization approach is at reducing both time complexity and proof complexity across more general classes of diagrammatic calculi.

Another, arguably much more ambitious, possible direction for future investigation would be a synthesis of the techniques developed within this article with the recent relativistic methods developed in \cite{gorard6}, as well as the general approach to reasoning about quantum computations in discrete spacetimes presented in \cite{shah}, in order to obtain a fully ``spatially-extended'' analog of the ZX-calculus using Wolfram model evolution. This could, in principle, be achieved by effectively ``blending'' the labeled open graph rewriting rules of the ZX-calculus with the unlabeled spatial hypergraph rewriting rules of (relativistic) Wolfram model evolution, thus deriving a general diagrammatic formalism for reasoning about quantum processes in curved spacetimes (within which it might, for instance, be possible to compute from first principles the expected loss of fidelity of quantum teleportation due to Hawking radiation in a Schwarzschild background). Another important point of consideration is that all of the methods thus far developed within this article have applied only at the first level of the \textit{rulial hierarchy} of multiway systems, since they have made use of ordinary inductive types only. If these same methods could be extended to apply also to higher inductive types, and hence if they could be lifted to higher levels of the \textit{rulial hierarchy}, then they would allow one to construct a complete and sound diagrammatic reasoning language for homotopy type theory\cite{univalent} (in which the objects being reasoned over would no longer be ZX-diagrams or Wolfram model hypergraphs, but rather the proof graphs themselves, corresponding to homotopies between paths in some associated space representing a higher inductive type). This would potentially permit a systematic investigation of the formal relationship between the combinatorial structures of the Wolfram model and the spatial structures invoked within Shulman's cohesive homotopy type theory\cite{shulman}, with the rulial multiway system playing the role of an ${\infty}$-groupoid\cite{arsiwalla}.

\section*{Acknowledgments}

The authors would like to thank the Foundations, Structures, and Quantum group at the University of Oxford, as well as Cambridge Quantum Computing, for their hospitality in inviting us to present the early stages of this research and receive valuable feedback. JG would particularly like to thank Bob Coecke, Ross Duncan, Hector Miller-Bakewell and Konstantinos Meichanetzidis for useful early conversations and suggestions, as well as Aleks Kissinger for pointing out several relevant items of literature (especially in relation to performance comparisons). At all stages during the conducting of this research, the encouragement and advice of Stephen Wolfram proved invaluable.

\end{document}